%% file: dynreview.tex
\documentclass[10pt,a4paper]{article}
\usepackage[utf8]{inputenc}

\usepackage{graphicx,float}
\graphicspath{{figures/}}

\usepackage{amsmath,amssymb,amsfonts}
\usepackage{epsfig,color}
\usepackage{dsfont}
\usepackage[margin=3cm]{geometry}
\usepackage{multirow}

\usepackage{natbib}

\def\rf#1{(\ref{#1})}
\def\DerN#1{\frac{d #1}{d \eta}}
\def\G{{\cal G}}
\newcommand{\ts}{\mathsf{t}}
\newcommand{\xs}{\mathsf{x}}

\def \A {{\mathbb A}}

\usepackage{tikz}
\usetikzlibrary{arrows,shapes,positioning,shadows,trees}

\numberwithin{equation}{section}

\usepackage{amsthm}
\theoremstyle{plain}
\newtheorem{theorem}{Theorem}

\theoremstyle{definition}
\newtheorem{definition}{Definition}
\theoremstyle{remark}

\newcommand{\grad}{\operatorname{{\mathrm grad}}}

\usepackage[breaklinks=true]{hyperref}  

\begin{document}

\title{Dynamical systems applied to cosmology:\\ dark energy and modified gravity}
\author{
  Sebastian Bahamonde\footnote{sebastian.beltran.14@ucl.ac.uk}~${}^{1}$,
  Christian G.~B\"ohmer\footnote{c.boehmer@ucl.ac.uk}~${}^{1}$,
  Sante Carloni\footnote{sante.carloni@tecnico.ulisboa.pt}~${}^{2}$,\\
  Edmund J. Copeland\footnote{ed.copeland@nottingham.ac.uk}~${}^{3}$,
  Wei Fang\footnote{wfang@shnu.edu.cn}~${}^{4}$,
  Nicola Tamanini\footnote{nicola.tamanini@aei.mpg.de}~${}^{5,6,7}$\\[2ex]
  ${}^1$Department of Mathematics, University College London\\
  Gower Street, London, WC1E 6BT, United Kingdom\\[0.5ex]
  ${}^2$Centro Multidisciplinar de Astrofisica - CENTRA\\
  Instituto Superior Tecnico, Universidade de Lisboa\\
  Av. Rovisco Pais, 1049-001 Lisboa, Portugal\\[1ex]
  ${}^3$School of Physics and Astronomy, University of Nottingham\\
  Nottingham, NG7 2RD, United Kingdom\\[1ex]
  ${}^4$Department of Physics, Shanghai Normal University and\\
  The Shanghai Key Lab for Astrophysics,\\
  100 Guilin Rd., Shanghai, 200234, P.R.China\\[1ex]
  ${}^5$Institut de Physique Th\'eorique, CEA-Saclay, CNRS UMR 3681,\\
  Universit\'e Paris-Saclay, F-91191 Gif-sur-Yvette, France\\[1ex]
  ${}^6$Laboratoire Astroparticule et Cosmologie, CNRS UMR 7164,\\ Universit\'e Paris-Diderot, 10 rue Alice Domon et L\'eonie Duquet, 75013 Paris, France\\[1ex]
  ${}^7$Max-Planck-Institut f\"{u}r Gravitationsphysik, Albert-Einstein-Institut,\\ Am M\"{u}hlenberg 1, 14476 Potsdam-Golm, Germany
}
\date{29 January 2018}
\maketitle

\clearpage

\begin{abstract}
The Nobel Prize winning confirmation in 1998 of the accelerated expansion of our Universe put into sharp focus the need of a consistent theoretical model to explain the origin of this acceleration. As a result over the past two decades there has been a huge theoretical and observational effort into improving our understanding of the Universe. The cosmological equations describing the dynamics of a homogeneous and isotropic Universe are systems of ordinary differential equations, and one of the most elegant ways these can be investigated is by casting them into the form of dynamical systems.
This allows the use of powerful analytical and numerical methods to gain a quantitative understanding of the cosmological dynamics derived by the models under study.
In this review we apply these techniques to cosmology. We begin with a brief introduction to dynamical systems, fixed points, linear stability theory, Lyapunov stability, centre manifold theory and more advanced topics relating to the global structure of the solutions. Using this machinery we then analyse a large number of cosmological models and show how the stability conditions allow them to be tightly constrained and even ruled out on purely theoretical grounds. We are also able to identify those models which deserve further in depth investigation through comparison with observational data. This review is a comprehensive and detailed study of dynamical systems applications to cosmological models focusing on the late-time behaviour of our Universe, and in particular on its accelerated expansion.
In self contained sections we present a large number of models ranging from canonical and non-canonical scalar fields, interacting models and non-scalar field models through to modified gravity scenarios. Selected models are discussed in detail and interpreted in the context of late-time cosmology.
\end{abstract}

\clearpage
\tableofcontents

\clearpage

\input{chapters/01_intro/intro.tex}
\clearpage
\input{chapters/02_dyn/dyn.tex}
\clearpage
\input{chapters/03_cosmo/cosmo.tex}
\clearpage
\input{chapters/04_quint/quint.tex}
\clearpage
\input{chapters/05_noncanonical/noncanonical.tex}

\clearpage
\input{chapters/06_coupled/coupled.tex}

\clearpage
\input{chapters/07_higherfields/higherfields.tex}
\clearpage
\input{chapters/08_othermodels/othermodels.tex}
\clearpage
\input{chapters/09_conclusions/conclusions.tex}

\subsection*{Acknowledgements}

We would like to thank Matthew Wright for his contribution in the initial phases of this project, and Stephen Baigent for useful comments regarding the theory of dynamical systems.

We furthermore thank Mustafa Amin, John Barrow, Daniele Bertacca, Shantanu Desai, Lavinia Heisenberg, Purnendu Karmakar, David Langlois, Genly Leon, Lucas Lombriser, Prado Martin-Moruno, José P.~Mimoso, Behrouz Mirza, Davood Momeni, João Morais, G.G.L.~Nashed, Shin'ichi Nojiri, Evgeny Novikov, Fatemeh Oboudiat, Sergei Odintsov, Vasilis Oikonomou, George Pantazis, Anthony J.~Roberts, Mahmood Roshan, Emmanuel N.~Saridakis, Robert J.~Scherrer, Glenn D.~Starkman, Paul K.~Townsend, Maurice H.P.M.~van Putten, Alexander Vikman, Sergio Zerbini, Ying-li Zhang, Wen Zhao and Miguel Zumalacarregui for useful comments and suggestions for references.

N.T.~acknowledges partial support from the Labex P2IO and an Enhanced Eurotalents Fellowship. E.J.C.~acknowledges financial support from STFC consolidated Grant No.~ST/L000393/1 and ST/P000703/1. Part of his work was performed at the Aspen Center for Physics, which is supported by National Science Foundation grant PHY-1607761. S.B.~is supported by the Comisi{\'o}n Nacional de Investigaci{\'o}n Cient{\'{\i}}fica y Tecnol{\'o}gica (Becas Chile Grant No.~72150066). S.C.~is supported by the Funda\c{c}\~{a}o para a Ci\^{e}ncia e Tecnologia through project IF/00250/2013 and acknowledges financial support provided under the European Union's H2020 ERC Consolidator Grant ``Matter and strong-field gravity: New frontiers in Einstein's theory'' Grant Agreement No.~MaGRaTh646597. W.F. is supported by the Chinese National Nature Science Foundation under Grant No. 11333001 and No. 11433003.

This article is partly based upon work from COST Action CA15117 (Cosmology and Astrophysics Network for Theoretical Advances and Training Actions), supported by COST (European Cooperation in Science and Technology).

All authors contributed to the development and writing of this paper.

\bibliographystyle{apa-good-2}
\bibliography{bibfiles/masterrefs}

\end{document}

%% file: chapters/01_intro/intro.tex
\section{Introduction}

\subsection{Aim and style of this review}

The goal of this review is to provide an overview on the applications of the powerful approach of dynamical systems to the plethora of models that have been proposed to describe the observed cosmological evolution of our Universe. The mathematical theory of dynamical systems is extremely useful to understand the global dynamics of any cosmological model, especially its late time-asymptotic behaviour. Through the careful choice of the dynamical variables, a given cosmological model can be written as an autonomous system of differential equations. In this way the analysis of the features of the phase space of this system, i.e.~the analysis of the fixed points and the determination of the general behaviour of the orbits, provides insight on the global behaviour of the cosmological model. This kind of analysis is particularly useful when we are trying to establish whether a given model which presents complicated governing equations can reproduce the observed expansion of the Universe. The (semi-qualitative) understanding of the dynamics of such cosmologies is the main reason behind the rapid growth of dynamical systems techniques in the last few years.
In this review we will be investigating the applications of dynamical systems to such models, starting from the ones having general relativity as the common framework at the heart of them, and subsequently exploring models beyond general relativity.

\tikzset{
  basic/.style  = {draw, text width=2.1cm, font=\sffamily, rectangle},
  root/.style   = {basic, text width=5cm, rounded corners=3pt, thin, align=center, fill=lightgray!30},
  level 2/.style = {basic, rounded corners=6pt, align=center, fill=lightgray!60,text width=8em,sibling distance=40mm},
  level 3/.style = {basic, align=left, fill=gray!60, text width=7em}
}

\begin{figure}[!ht]
\centering
\resizebox{15cm}{!}{\begin{tikzpicture}[
      level 1/.style={sibling distance=80mm},
      edge from parent/.style={->,draw,black},
      >=latex
    ]
    \node[root] {Cosmological dynamical systems \\ Secs.~\ref{sec:dynamicalsystems} \& \ref{chap:Cosmology} }
    child {
      node[level 2] (a1) {General Relativity}
      child {node[level 2] (a2) {non-interacting}}
      child {node[level 2] (a3) {interacting}}
    }
    child {
      node[level 2] (b1) {Beyond GR}
      child {node[level 2] (b2) {modified gravity}}
      child {node[level 2] (b3) {other models}}
    };
%
\begin{scope}[every node/.style={level 3}]
  \node [below of = a2, xshift=15pt, yshift=0pt] (a20) {$\Lambda\text{CDM}$\\ Sec.~\ref{sec:LCDM_dynam}};
  \node [below of = a20, yshift=-10pt] (a21) {{\it canonical scalar fields}\\ Sec.~\ref{chap:scalarfields}};
  \node [below of = a21, yshift=-15pt] (a22) {{\it non-canonical scalar fields}\\ Sec.~\ref{chap:noncanonicalscalarfields}};
  \node [below of = a22, yshift=-15pt] (a23) {{\it non-scalar fields}\\ Sec.~\ref{sec:non-scalar_models}};
  \node [below of = a3, xshift=15pt] (a31) {{\it coupled fluids}\\ Sec.~\ref{sec:coupled_fluids}};
  \node [below of = a31, yshift=-10pt] (a32) {{\it coupled quintessence}\\ Sec.~\ref{sec:coupled_quintessence}};
  \node [below of = a32, yshift=-15pt] (a33) {{\it non-canonical coupled scalars}\\ Sec.~\ref{sec:coupled_noncanonical_scalars}};
  \node [below of = a33, yshift=-15pt] (a34) {{\it scalar-fluid models}\\ Sec.~\ref{sec:scalar_fluid_models}};
  \node [below of = b2, xshift=15pt] (b21) {{\it Brans-Dicke}\\ Sec.~\ref{sec:BD_theory}};
  \node [below of = b21, yshift=-10pt] (b22) {{\it scalar-tensor models}\\ Sec.~\ref{sec:ST_theories}};
  \node [below of = b22, yshift=-10pt] (b23) {{\it $f(R)$ theories}\\ Secs.~\ref{sec:f(R)} \& \ref{sec:Palatini_f(R)_gravity}};
  \node [below of = b23, yshift=-10pt] (b24) {{\it higher order theories}\\ Sec.~\ref{otherhigher}};
  \node [below of = b24, yshift=-15pt] (b25) {{\it teleparallel models}\\ Sec.~\ref{sec:f(T)_gravity}};
  \node [below of = b3, xshift=18pt, yshift=-5pt] (b31) {{\it string/ brane cosmology}\\ Sec.~\ref{sec:string_and_brane}};
  \node [below of = b31, yshift=-15pt] (b32) {{\it quant.~gravity phenomenology}\\ Sec.~\ref{sec:quantum_grav}};
  \node [below of = b32, yshift=-15pt] (b33) {{\it other modified gravity models}\\ Sec.~\ref{sec:other_modified_theories}};
\end{scope}
%
\foreach \value in {0,1,2,3}
\draw[->] (a2.195) |- (a2\value.west);
\foreach \value in {1,2,3,4}
\draw[->] (a3.195) |- (a3\value.west);
\foreach \value in {1,2,3,4,5}
\draw[->] (b2.195) |- (b2\value.west);
\foreach \value in {1,2,3}
\draw[->] (b3.195) |- (b3\value.west);
\end{tikzpicture}}
\caption{Classification of cosmological models analysed with dynamical systems techniques.}
\label{fig:models}
\end{figure}
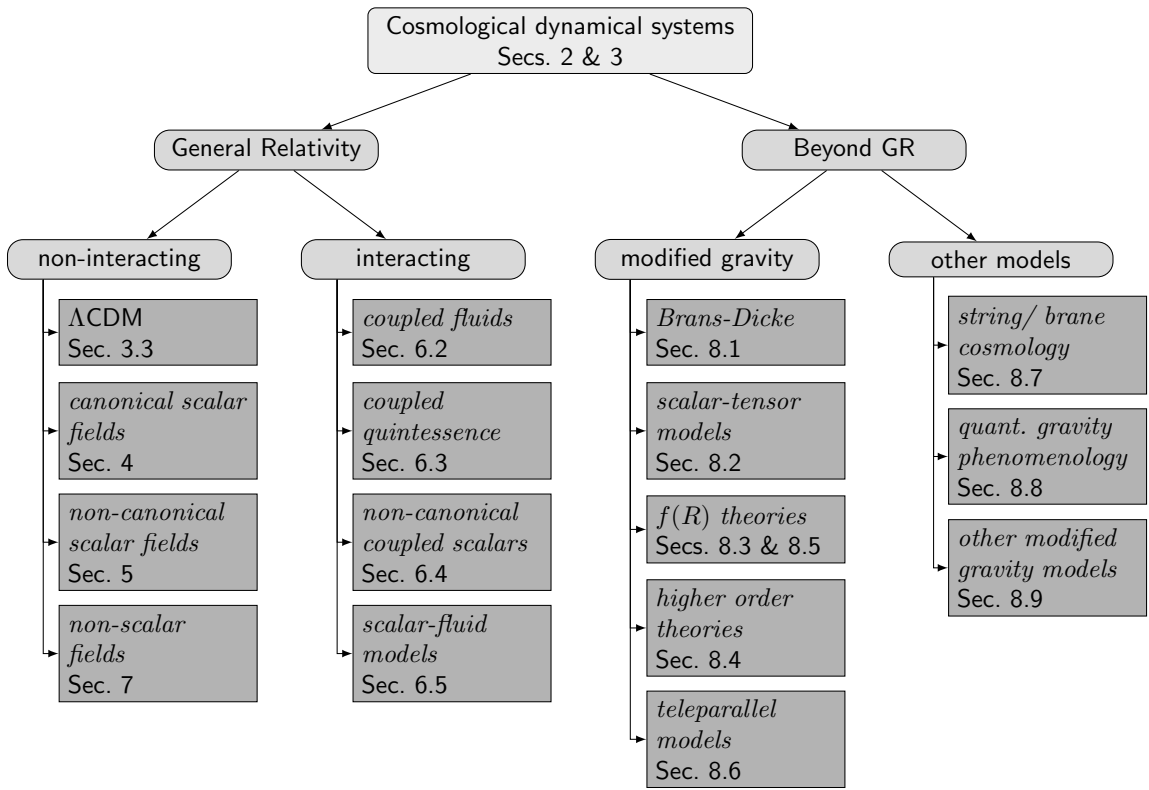

Hence we begin by considering the broad distinctions between the models based on their relation to general relativity, see Fig.~\ref{fig:models}. Some of them present modifications only in the matter sector (the r.h.s.~of the Einstein equations) but retain general relativity in the geometrical sector (the l.h.s.). Within that class we can distinguish between non-interacting and interacting models, according to the absence or presence of a non-gravitational interaction between dark matter and dark energy. Common non-interacting models are those based on a scalar field to characterise dark energy: within them we can distinguish between those with canonical and non-canonical kinetic terms. Other models rely instead on different kinds of fields to describe dark energy, or even consider some phenomenological assumptions to modify the matter sector.
On the other hand, models allowing for modifications of general relativity are grouped under a different class. Such modifications or extensions of general relativity will be distinguished between theories with non-minimal couplings and higher order theories, for example $f(R)$ gravity.
Within this class we can further consider theories based on the teleparallel equivalent of general relativity and its modifications. Moreover we will discuss models inspired by string theory, extra dimensions and quantum gravity phenomenology. Such a classification scheme, with references to the respective sections where the models are discussed, is represented in Fig.~\ref{fig:models}. This guide should allow readers to quickly find those sections which are most relevant to their work.

The review is thus aimed at presenting several dynamical systems applications to dark energy models introduced to characterise the late-time evolution of the universe, specifically the observed accelerated expansion. It mainly collects results presented in the literature of the last two decades, and some important parts (particularly Secs.~\ref{sec:dynamicalsystems}, \ref{chap:Cosmology} and \ref{chap:scalarfields}) are presented in a pedagogical manner to help the reader unfamiliar with the subject. More specifically Sec.~\ref{sec:dynamicalsystems} and \ref{chap:Cosmology} are, respectively, built as simple introductions to dynamical systems methods and standard cosmology. Sec.~\ref{chap:scalarfields} instead reviews the applications of dynamical systems to the simplest models of dark energy, namely canonical scalar field models. In this section many details of the calculations are stated explicitly to help the reader understand how a cosmological model can be analysed with dynamical system techniques. The remaining sections, on the other hand, are constructed to present the dynamical features of advanced cosmological models, and thus the emphasis is switched towards discussing the results in the literature rather than following through the details of each computation. Furthermore in these advanced sections some original results, appearing for the first time in this review, are also reported. These are the result of new investigations or alternative analyses on the dynamics of well known models.

Historically the first applications of dynamical systems techniques to cosmology date back to the 1970s \citep{doi:10.1093/mnras/153.4.419,collins1971,collins1972,Bogoyavlenskii:1973,shikin1975}. These studies were not necessarily concerned with homogeneous and isotropic models but generally studied anisotropic models; see for instance \citet{Barrow:1986} and references therein.

The idea for this project was born from the PhD thesis of one of us \citep{Tamanini:2014nvd}, where many of the topics presented here have been discussed with a similar approach and philosophy. In particular we have chosen to focus on the late-time cosmological dynamics, and to neglect other possible cosmological applications of dynamical systems, as for example the early universe, mainly for two reasons. On the one hand, dark energy is an important subject of modern cosmology, with forthcoming astronomical observations which will possibly provide an increasing amount of information about its fundamental nature. It is of great interest thus to write a review on the dynamical properties of several dark energy models. Although this has already been the subject of many different reviews, notably the work of \citet{Copeland:2006wr}, none of them considers dynamical systems applications to be the central topic of the discussion, as we do here. The original approach adopted in this review distinguishes it from previous literature and aims at rendering the presented material useful to both theoretical physicists and mathematicians interested in cosmology. On the other hand, dynamical systems applications to cosmology in a broad sense have already been collected in two well-known books written by \citet{WainwrightEllis} and \citet{Coley:2003mj}. The approach and the choices of arguments in those books are however more mathematical in nature than the ones presented in this review, where the main discussions focus on the phenomenological applications to dark energy models rather than on the formal issues of dynamical systems in cosmology.

Those books also assume general relativity as the only fundamental description of gravity and thus do not treat all possible alternative theories which nowadays are extensively considered to build dark energy models. In this review, Sec.~\ref{chap:modifiedgrav} is completely dedicated to alternative theories of gravity motivated by both phenomenological ideas as well as high energy and quantum physics. For these reasons the topics discussed here have never been properly reviewed anywhere else (see however \citet{Leon:2009} and \citet{Leon:2014rra} where some models beyond general relativity have been collectively studied).

Since the aim of this work is to provide a review on dynamical systems applications to dark energy models, only literature involving dynamical systems techniques has been considered. An effort has been made in order to include as much literature as possible in order to refer to all the works employing dynamical systems methods in late-time cosmology. In general the most important models are analysed in detail, while for more complex and technical models only brief discussions and references to the relevant literature are provided. Note that publications which have not used dynamical systems for the main, or at least relevant, part of the calculations have not been considered in what follows. In other words this review must not be confused for a general review on dark energy phenomenology, including for example detailed comparison with observational data. These considerations extend also to the issue of the physical validity of the cosmological models we have analysed. We make no attempt here to judge the merits or flaws of such models, unless this is strictly necessary to explain the use of particular dynamical systems techniques. Many of the cosmologies we will deal with, particularly in Sec.~\ref{sec:non-scalar_models} and Sec.~\ref{chap:modifiedgrav}, have not been developed enough to be compared with experimental data and therefore it is not possible to determine if they are compatible with them. Hence, the presence of a models in this review does not imply compatibility with current observations nor an endorsement by any of the authors.

\subsection{How to use this review}

We suggest reading through Secs.~\ref{sec:dynamicalsystems} and~\ref{chap:Cosmology} for a quick introduction to dynamical systems and cosmology. These introductory sections fix our notation, introduce some more advanced methods and discuss our approach in the framework of modern cosmology. These sections might also turn out particularly useful to the reader who is not familiar with dynamical systems methods and/or basic cosmology. The subsequent sections are self-contained and can be read largely independently from each other, see again Fig.~\ref{fig:models}. The arrangement of the material follows roughly the relative size of the specific topic. Topics mentioned towards the latter part of the review have received less attention than topics discussed earlier (this does not imply any judgement on the importance of the subject). Specifically Sec.~\ref{chap:scalarfields} presents many calculations in details in order to thoroughly show how dynamical systems methods apply to cosmological models.

We stress that the cited references to the state of the art research in dark energy are only the ones explicitly employing dynamical systems techniques. Furthermore some of the analyses in this review have never been presented anywhere else, and in this respect they constitute the result of new investigations. In particular the sections containing these original results are:
\begin{itemize}
  \item Sec.~\ref{sec:power_law_potential}: thorough dynamical investigation of quintessence with a power-law potential, including centre manifold analysis and Lyapunov functions;
  \item Sec.~\ref{sec:phantom}: phase space compactification and analysis at infinity for the phantom scalar field;
  \item Sec.~\ref{sec:coupled_fluids}: simple dynamical system investigation of an interacting dark energy fluid model;
  \item Sec.~\ref{sec:spinors}: dynamical system investigation of a new coupled ELKO spinor model;
  \item Sec.~\ref{sec:ST_theories}: introduction of new dimensionless variables for the dynamical study of general scalar-tensor theories;
  \item Sec.~\ref{sec:Palatini_f(R)_gravity}: dynamical systems analysis of generalised hybrid metric-Palatini gravity in the Jordan frame.
\end{itemize}

Finally in this review we present several figures depicting the phase space and its flow for some dynamical systems corresponding to different cosmological models. We moreover include some plots of the cosmic evolution of physically interesting quantities, such as the energy densities and equation of state of various cosmic components. All the figures that appear in this review have been created with the software \textit{Mathematica}. Given the partly pedagogical nature of this work, for the reader interested in understanding how these figures have been produced we made available on line three examples of \textit{Mathematica notebooks} that we used to create three of the figures in the review. These files are available at the following links:
\begin{itemize}
\item  \url{https://github.com/cgboehmer/dynamical-systems-review}
\item \url{https://www.christianboehmer.co.uk/review.html}
\end{itemize}

\noindent
In particular the three notebooks have been chosen in order to explain how 2D and 3D phase space pictures are made and how evolution plots have been computed and drawn. These notebooks have been integrated with comments to help the reader understand every passage of the code used to make the figures.

\subsection{Notation and conventions}

An attempt has been made to keep the basic notation clean and simple. The meaning of every symbol will always be defined at its first occurrence in the text and in all places where ambiguities might arise. The signature of the metric tensor is assumed to be $(-,+,+,+)$. The coupling constant appearing in the Einstein field equations is denoted by $\kappa^2=8\pi G/c^4$, where $c$ is the speed of light and $G$ the Newton's gravitational constant. Throughout this review, we use natural units with $c=1$. Greek indices are spacetime indices which take values $(0,1,2,3)$, Latin indices are used for other notation as necessary. In some parts of the review, the logic symbols are used, for example $\lor$ and $\land$ represent the logical conjunction (or) and the logical disjunction (and) respectively. Generally the variable $\rho$ denotes the energy density of a generic fluid sourcing the cosmological equations, while $\rho_{\rm m}, \rho_{\rm r}, \rho_{\Lambda}$ and $\rho_{\rm de }$ denote, respectively, the energy density for dust-like fluid (or matter), radiation-like fluid, a cosmological constant like fluid and a dark energy-like fluid, and are used to distinguish models with more than one fluid. The latter rules generally apply to all quantities (e.g.~$w$) except $\Omega_{\rm m}$, which will always refer to the relative energy density of any type of fluid irrespectively of its equation of state. This exception will help in distinguishing among different relative energy densities (e.g.~$\Omega_\phi$, $\Omega_{\rm de}$, ...).

The variables of a two-dimensional dynamical system will be denoted by $x$ and $y$, for three-dimensional dynamical systems we work with $x,y,z$. The definition of these variables will change when different models are considered, but it will always be stated explicitly when the concrete model is introduced.

A dot generally means differentiation with respect to a time parameter, which will coincide with the cosmological time from Sec.~\ref{chap:Cosmology} onward. A prime will refer to differentiation with respect to a newly defined time parameter, which will always be $\eta = \log a$ unless otherwise specified. We will sometimes encounter systems where the new time parameter is more complicated than $\eta$.



%% file: chapters/02_dyn/dyn.tex
\section{Dynamical systems}
\label{sec:dynamicalsystems}

The purpose of this section is to give a succinct introduction to those parts of the dynamical systems approach which are most relevant to applications in cosmology. The topics covered in this part reflect the majority of techniques used by researchers in the field, and does not cover other, equally interesting, mathematical techniques applied to dynamical systems elsewhere. The reader interested in more mathematical details and further applications is referred to well known textbooks on the subject (e.g.~\citet{arrowsmith,Wiggins,Perko}). For an introduction that focuses on cosmological applications see~\citet{Boehmer:2014vea}.

\subsection{An introduction to dynamical systems}

Let us start this section with the following question: What are dynamical systems? It is a framework which can be used to study something as simple as a single pendulum to something as complicated as the dynamics of the universe. In general, we can think of a dynamical system as any abstract system consisting of
\begin{enumerate}
  \item a space (\textit{state space} or \textit{phase space}), and
  \item a mathematical rule describing the evolution of points in that space.
\end{enumerate}
It is the second point which needs to be emphasised. Deriving from first principles, a mathematical rule describing a complex process in nature might simply be impossible. Therefore, we need a mathematical rule as an input. In order to characterise the system, we need a set of quantities describing the state of the system. We call the state space the set of all possible values of these quantities. For ecological models, for instance, these quantities tend to be population sizes, where predator-prey models are a typical example. When considering the universe, one is tempted to choose the energy densities $\rho_i$ of the constituents of the universe as our quantities to describe its state. However, it turns out that these densities are not a particularly good choice in this context and one can identify much better variables.

Let us begin by denoting $\mathbf{x} = (x_1,x_2,\ldots,x_n) \in X$ to be an element of the state space $X\subseteq\mathbb{R}^n$. A dynamical system is generally written in the form \citep{Wiggins}
\begin{align}
  \dot{\mathbf{x}} = \mathbf{f}(\mathbf{x})\,,
  \label{dy1}
\end{align}
where the function $\mathbf{f}:X\rightarrow X$ and where the dot denotes differentiation with respect to some suitable time parameter $t\in \mathbb{R}$, which in general does not have to be related to physical time. The function $\mathbf{f}$ is viewed as a vector field on $\mathbb{R}^n$ such that
\begin{align}
  \mathbf{f}(\mathbf{x}) = (f_1(\mathbf{x}),\cdots,f_n(\mathbf{x}))\,.
\end{align}
This means we have $n$ equations which describe the dynamical behaviour of the $n$ variables.
Let us choose a point $\mathbf{x} \in X$ at some particular time $t$, then $\mathbf{f}(\mathbf{x})$ defines a vector field in $\mathbb{R}^n$. When discussing a particular solution to~(\ref{dy1}) this will often be denoted by $\psi(t)$ to simplify the notation.
Any solution $\psi(t)$ of the system \eqref{dy1} is commonly called an \textit{orbit} or a \textit{trajectory} of the phase space. We restrict ourselves to systems which are finite dimensional and continuous. In fact, we will require the function $f$ to be at least differentiable in $X$. Generally, dynamical systems which appear in cosmology are finite dimensional and continuous. Indeed, the function $f$ tends to only contain elementary functions and in general is smooth almost everywhere. However we will see that in the most complex cases the function $f$ might present some singularities. In those cases we will use the standard tools presented in the following only in the parts of the phase space in which $f$ is continuous.

\begin{definition}[\textit{Critical point} or \textit{fixed point} or \textit{equilibrium point}]
  The autonomous equation $\dot{\mathbf{x}} = \mathbf{f}(\mathbf{x})$ is said to have a critical (or fixed or equilibrium) point at $\mathbf{x} = \mathbf{x}_0$ if and only if $\mathbf{f}(\mathbf{x}_0) = 0$.
  \label{def1}
\end{definition}

According to Definition~\ref{def1}, the critical points of system~(\ref{dy1}) correspond to those points $\mathbf{x}$ where the system is at rest. In principle, the system could remain in this (steady) state indefinitely. However, one needs to clarify whether or not the system can in fact attain such a state and whether or not this state is stable with respect to small perturbations.

This naturally leads to the question of stability of a critical point or fixed point. The following two definitions will clarify what is meant by stable and asymptotically stable. In simple words a fixed point $\mathbf{x}_0$ of the system~(\ref{dy1}) is called stable if all solutions $\mathbf{x}(t)$ starting near $\mathbf{x}_0$ stay close to it.

\begin{definition}[\textit{Stable fixed point}]
  Let $\mathbf{x}_0$ be a fixed point of system~(\ref{dy1}). It is called stable if for every $\varepsilon > 0$ we can find a $\delta$ such that if $\psi(t)$ is any solution of~(\ref{dy1}) satisfying $\|\psi(t_0)-\mathbf{x}_0\| < \delta$, then the solution $\psi(t)$ exists for all $t \geq t_0$ and it will satisfy $\|\psi(t)-\mathbf{x}_0\| < \varepsilon$ for all $t \geq t_0$.
  \label{def2}
\end{definition}

The point is called asymptotically stable if it is stable and the solutions approach the critical point for all nearby initial conditions.

\begin{definition}[\textit{Asymptotically stable fixed point}]
  Let $\mathbf{x}_0$ be a stable fixed point of system~(\ref{dy1}). It is called asymptotically stable if there exists a number $\delta$ such that if $\psi(t)$ is any solution of~(\ref{dy1}) satisfying $\|\psi(t_0)-\mathbf{x}_0\| < \delta$, then $\lim_{t \rightarrow \infty} \psi(t) = \mathbf{x}_0$.
  \label{def3}
\end{definition}

The subtle difference between these two definitions is that all trajectories near an asymptotically stable fixed point will eventually reach that point, while trajectories near a stable point could for instance circle around that point. Almost all fixed points in cosmology which are stable are also asymptotically stable. Therefore, when discussing cosmological models we will not distinguish between these two possibilities unless there is a need for further clarification. \textit{Unstable fixed points} are critical points which are not stable: the solutions starting near the fixed point move away from it. Understanding the critical or fixed points of a dynamical system allows us to (almost) completely understand the properties of the time evolution of the physical system being considered. This can be done without studying explicit solutions for given initial conditions, rather, we are able to make qualitative statements for all possible solutions.

Two concepts that will turn out to be useful in the following sections are the definitions of \textit{invariant set} and \textit{heteroclinic orbit}.
A subset $S\subset X$ is an invariant set of $X$ if for all $\mathbf{x}\in S$ and all $t\in\mathbb{R}$ then $\psi(t)\in S$ if $\psi(t_0)=\mathbf{x}$ at some time $t_0\in \mathbb{R}$.
In other words any solution within an invariant set will never leave the invariant set. An orbit connecting two distinct critical points is called a heteroclinic orbit. In more mathematical terms, a heteroclinic orbit is a solution $\psi(t)$ for which there exist two critical points $\mathbf{x}_i$ and $\mathbf{x}_f$ such that $\lim_{t\rightarrow+\infty}\psi(t) = \mathbf{x}_f$ and $\lim_{t\rightarrow-\infty}\psi(t) =\mathbf{x}_i$.

Having defined a concept of stability, we are now able to introduce methods which can be used to study the stability properties of critical points. The most common technique is the so-called linear stability theory which for most cosmological applications suffices to gain a good understanding of the physical properties of the cosmological model in question. We will also introduce Lyapunov stability and centre manifold theory. Note that other notions of stability do exist, for instance Kosambi-Cartan-Chern theory which has been applied to cosmology, see~\citet{Bohmer:2010re}.

\subsection{Linear stability theory}

The basic idea of linear stability theory is to linearise the system near a fixed point in order to understand the dynamics of the entire system near that point. Let $\dot{\mathbf{x}} = \mathbf{f}(\mathbf{x})$ be a given dynamical system with fixed point at $\mathbf{x}_0$. Since we assume $\mathbf{f}$ to be sufficiently regular, we can linearise the system around its critical point. We have $\mathbf{f}(\mathbf{x}) = (f_1(\mathbf{x}),\ldots,f_n(\mathbf{x}))$, so that each $f_i(x_1,x_2,\ldots,x_n)$ can be Taylor expanded near $\mathbf{x}_0$ which yields
\begin{align}
  f_i(\mathbf{x}) = f_i(\mathbf{x}_0) +
  \sum_{j=1}^{n} \frac{\partial f_i}{\partial x_j}(\mathbf{x}_0) y_j +
  \frac{1}{2!} \sum_{j,k=1}^{n} \frac{\partial^2 f_i}{\partial x_j \partial x_k}(\mathbf{x}_0) y_j y_k + \ldots\,,
\end{align}
where the vector $\mathbf{y}$ is defined by $\mathbf{y} = \mathbf{x} - \mathbf{x}_0$. In linear stability theory one only considers the first partial derivatives. Therefore, of particular importance is the object $\partial f_i/\partial x_j$ which corresponds to the \textit{Jacobian matrix} of vector calculus. One can define
\begin{align}
  J = \frac{\partial f_i}{\partial x_j} =
  \begin{pmatrix}
    \frac{\partial f_1}{\partial x_1} &
    \ldots &
    \frac{\partial f_1}{\partial x_n} \\
    \vdots & \ddots & \vdots \\
    \frac{\partial f_n}{\partial x_1} &
    \ldots &
    \frac{\partial f_n}{\partial x_n}
  \end{pmatrix} \,,
  \label{Jac}
\end{align}
which is also called the \textit{stability matrix}. It is the eigenvalues of the Jacobian matrix $J$, evaluated at the critical points $\mathbf{x}_0$, which contain the information about stability.

The Jacobian is an $n \times n$ matrix with $n$, possibly complex, eigenvalues (counting repeated eigenvalues accordingly). Linear stability theory faces some limitation which motivate the definition of hyperbolic points (see~\cite{Wiggins}).

\begin{definition}[\textit{Hyperbolic point}]
  Let $\mathbf{x} = \mathbf{x}_0\in X \subset \mathbb{R}^n$ be a fixed point (critical point) of the system $\dot{\mathbf{x}} = \mathbf{f}(\mathbf{x})$. Then $\mathbf{x}_0$ is said to be hyperbolic if none of the eigenvalues of the Jacobian matrix $J(\mathbf{x}_0)$ have zero real part. Otherwise the point is called \textit{non-hyperbolic}.
\end{definition}

Linear stability theory fails for non-hyperbolic points and other methods have to be employed to study the stability properties. In Secs.~\ref{sec:lyapunov} and \ref{SecCMT} we provide two alternative approaches to deal with such points.

For cosmological dynamical systems we need to distinguish three broad cases when classifying fixed points. Firstly, if all eigenvalues of the Jacobian matrix have positive real parts, trajectories are repelled from the fixed point and we speak in this case of an \textit{unstable point} (or a \textit{repeller} or a \textit{repelling node}). Secondly, if all eigenvalues have negative real parts, the point would attract all nearby trajectories and is regarded as \textit{stable} and sometimes called \textit{attractor} or \textit{attracting node}. Lastly, if at least two eigenvalues have real parts with opposite signs, then the corresponding fixed point is called a \textit{saddle} point, which attracts trajectories in some directions but repels them along others. In 2 and 3 dimensions one can classify all possible critical points, however, in more than 3 dimensions this becomes very difficult. For all practical purposes of cosmology the above broad classifications are sufficient for the majority of models.

\subsubsection{Example: 2D dynamical systems}

We will briefly consider a generic two dimensional dynamical system given by
\begin{align}
  \dot{x} = f(x,y)\,, \qquad
  \dot{y} = g(x,y)\,,
  \label{2dsys}
\end{align}
where $f$ and $g$ are (smooth) functions of $x$ and $y$. We assume $(x_0,y_0)$ to be a hyperbolic critical point so that $f(x_0,y_0)=0$ and $g(x_0,y_0)=0$. The Jacobian matrix of the system is simply given by
\begin{align}
  J =
  \begin{pmatrix}
    f_{,x} & f_{,y} \\
    g_{,x} & g_{,y}
  \end{pmatrix}\,,
  \label{Jacobian:2d}
\end{align}
where the the comma in $f_{,x}$ stands for the partial derivative with respect to the argument. Since this system is two dimensional, the Jacobian has two eigenvalues which we denote by $\lambda_{1,2}$. Explicitly, these are given by
\begin{align}
  \lambda_{1} &=
  \frac{1}{2}(f_{,x}+g_{,y})
  + \frac{1}{2}\sqrt{(f_{,x}-g_{,y})^2 + 4 f_{,y}g_{,x}}\,,\\
  \lambda_{2} &=
  \frac{1}{2}(f_{,x}+g_{,y})
  - \frac{1}{2}\sqrt{(f_{,x}-g_{,y})^2 + 4 f_{,y}g_{,x}}\,,
\end{align}
and are to be evaluated at the fixed point $(x_0,y_0)$. For any two dimensional system it is straightforward to compute the eigenvalues and determine the stability properties of the fixed points. We will see this approach in action for example in Sec.~\ref{sec:LCDM_dynam} and Sec.~\ref{sec:quintessence} where dark energy is modelled in the first case as a cosmological constant and in the second case as a canonical scalar field.

\subsection{Lyapunov's method}\label{sec:lyapunov}

There exist a large array of applied mathematics literature which explores dynamical systems beyond linear stability theory. Although the majority of this review is concerned with applying linear stability theory, for completeness we will mention a very powerful technique which has the potential of showing both local and global stability and is also applicable to non-hyperbolic points. The method is due to Lyapunov, it does not rely on linear stability and can be applied directly to the dynamical system. This makes the method very powerful, however, it has a drawback: one needs to find a so-called Lyapunov function and there is no systematic way of doing so, therefore one needs to guess a suitable function. Let us begin  with the definition of a Lyapunov function.

\begin{definition}[\textit{Lyapunov function}]
Let $\dot{\mathbf{x}} = \mathbf{f}(\mathbf{x})$ with $\mathbf{x} \in X \subset \mathbb{R}^n$ be a smooth dynamical system with fixed point $\mathbf{x}_0$. Let $V : \mathbb{R}^n \rightarrow \mathbb{R}$ be a continuous function in a neighbourhood $U$ of $\mathbf{x}_0$, then $V$ is called a Lyapunov function for the point $\mathbf{x}_0$ if
\begin{enumerate}
\item $V$ is differentiable in $U \setminus \{\mathbf{x}_0\}$.
\item $V(\mathbf{x}) > V(\mathbf{x}_0)$.
\item $\dot{V} \leq 0 \quad \forall\, x\in U \setminus \{\mathbf{x}_0\}$.
\end{enumerate}
\end{definition}

Note that the third requirement is the crucial one. It implies
\begin{align}
  \frac{d}{dt} V(x_1,x_2,\ldots,x_n) & = \frac{\partial V}{\partial x_1} \dot{x}_1 + \ldots
  + \frac{\partial V}{\partial x_n} \dot{x}_n \,,
  \nonumber \\
  &= \frac{\partial V}{\partial x_1} f_1 + \ldots
  + \frac{\partial V}{\partial x_n} f_n\,,
 \nonumber \\
  &= \grad V \cdot \mathbf{f}(\mathbf{x}) \leq 0\,,
 \label{gradV}
\end{align}
which required repeated use of the chain rule and substitution of the dynamical system equations to eliminate the terms $\dot{x}_i$ for $i=1,\ldots,n$.
In the following we state the main theorem which connects a Lyapunov function to the stability of a fixed point of a dynamical system. One can also find some instability results, see e.g.~\cite{Brauer:1989}, however, these have hardly been applied so far in the context of cosmology, see \cite{Charters:2001hi} for one such example.

\begin{theorem}[\textit{Lyapunov stability}]\label{lyapunovtheorem}
  Let $\mathbf{x}_0$ be a critical point of the system $\dot{\mathbf{x}} = \mathbf{f}(\mathbf{x})$, and let $U$ be a domain containing $\mathbf{x}_0$. If there exists a Lyapunov function $V(\mathbf{x})$ for which $\dot{V} \leq 0$, then $\mathbf{x}_0$ is a stable fixed point. If there exists a Lyapunov function $V(\mathbf{x})$ for which $\dot{V} < 0$, then $\mathbf{x}_0$ is an asymptotically stable fixed point.

Furthermore, if $\|\mathbf{x}\| \rightarrow \infty$ and $V(\mathbf{x})\rightarrow\infty$ for all $\mathbf{x}$, then $\mathbf{x}_0$ is said to be globally stable or globally asymptotically stable, respectively.
\end{theorem}

Therefore, if we were able to find a suitable Lyapunov function satisfying the criteria of the Lyapunov stability theorem, we could establish (asymptotic) stability without any reference to a solution of the dynamical system. However, the converse is not quite true. Failing to find a Lyapunov function at a particular point does not imply instability, we simply might not have been clever enough to find one.

\subsection{Centre manifold theory}\label{SecCMT}

Linear stability theory cannot determine the stability of critical points with Jacobian having eigenvalues with zero real parts. The method of centre manifold theory reduces the dimensionality of the dynamical system and the stability of this reduced system can then be investigated. The stability properties of the reduced system determine the stability of the critical points of the full system. Here the essential basics of centre manifold theory are discussed following~\cite{Wiggins} and~\cite{jackcarr}.

Let us begin by considering the dynamical system\footnote{To avoid confusion we denote here an element of $\mathbb{R}^n$ with $\mathbf{z}$ and the function defining the system with $\mathbf{F}$ since $\mathbf{x}$, $\mathbf{y}$ and $f$ will be used in what follows; see Eqs.~\eqref{cmexvec}.}
\begin{align}
  \dot{\mathbf{z}} = \mathbf{F}(\mathbf{z})
  \label{cdy1}
\end{align}
with $\mathbf{F}$ a (regular) function of $\mathbf{z} \in \mathbb{R}^n$. Let us assume that the system has a fixed point $\mathbf{z}_0$. Following the linear stability approach, near this point we can linearise the system using~(\ref{Jac}).

Denoting $\mathbf{z_*}=\mathbf{z}-\mathbf{z}_0$, at linear order we can rewrite~(\ref{cdy1}) as follows
\begin{align}
  \dot{\mathbf{z}}_* = J\, \mathbf{z}_* \,.
  \label{cdy2}
\end{align}
Since $J$ is an $n \times n$ matrix, it will have $n$ eigenvalues (counting multiple eigenvalues accordingly) which motivates the introduction of the following three spaces. The space $\mathbb{R}^n$ is the direct sum of three subspaces which will be denoted by $\mathbb{E}^s$, $\mathbb{E}^u$ and $\mathbb{E}^c$, where the superscripts stand for (s) stable, (u) unstable and (c) centre, respectively.

The `stable' space $\mathbb{E}^s$ is spanned by the eigenvectors of $J$ associated to eigenvalues with negative real part. The `unstable' space $\mathbb{E}^u$ is spanned by the eigenvectors of $J$ associated to eigenvalues with positive real part, and $\mathbb{E}^c$ is spanned by the eigenvectors of $J$ associated to eigenvalues with zero real part. The dynamics of the phase space trajectories in $\mathbb{E}^s$ and $\mathbb{E}^u$ can be understood using linear stability theory, centre manifold theory allows us to understand the dynamics of trajectories in $\mathbb{E}^c$.

If the unstable space is not empty, i.e.~$J$ has at least one eigenvalue with positive real part, then the corresponding critical point cannot be stable, irrespectively of it being hyperbolic or non-hyperbolic.
If instead the unstable space of a non-hyperbolic critical point is empty, i.e.~$J$ has no eigenvalues with positive real part, then stability can be decided applying centre manifold techniques. In this last case there always exists a coordinate transformation which allows us to rewrite the dynamical system~(\ref{cdy1}) in the form
\begin{subequations}
  \label{cmexvec}
  \begin{align}
    \dot{\mathbf{x}} &= A\mathbf{x} + f(\mathbf{x},\mathbf{y})\,,
    \label{cenxdot}\\
    \dot{\mathbf{y}} &= B\mathbf{y} + g(\mathbf{x},\mathbf{y})\,,
    \label{cenydot}
  \end{align}
\end{subequations}
where $(\mathbf{x},\mathbf{y}) \in\mathbb{R}^c\times\mathbb{R}^s$, with $c$ the dimension of $\mathbb{E}^c$ and $s$ the dimension of $\mathbb{E}^s$, and the two functions $f$ and $g$ satisfy
\begin{align}
  f(0,0) &= 0\,, \qquad \nabla f(0,0) = 0 \,,\\
  g(0,0) &= 0\,, \qquad \nabla g(0,0) = 0 \,.
\end{align}
In the system~(\ref{cmexvec}), $A$ is a $c \times c$ matrix having eigenvalues with zero real parts, while $B$ is an $s \times s$ matrix whose eigenvalues have negative real parts.

\begin{definition}[\textit{Centre Manifold}]\label{cmdef}
  A geometrical space is a centre manifold for~(\ref{cmexvec}) if it can be locally represented as
  \begin{align}
    W^{c}(0) =\{(\mathbf{x},\mathbf{y})\in \mathbb{R}^c\times\mathbb{R}^s\, |\, \mathbf{y} = h(\mathbf{x}), |\mathbf{x}|<\delta, h(0) = 0, \nabla h(0)= 0 \} \,,
  \end{align}
  for $\delta$ sufficiently small and $h(\mathbf{x})$ a (sufficiently regular) function on $\mathbb{R}^s$.
\end{definition}
The conditions $h(0) = 0$ and $\nabla h(0) = 0$ from the definition imply that the space $W^c(0)$ is tangent to the eigenspace $E^c$ at the critical point $(\mathbf{x},\mathbf{y}) = (0,0)$.

In the following we present three theorems that form the basis of centre manifold theory which will turn out useful in applications to cosmology. The first shows the existence of the centre manifold while the second one addresses the issue of stability. The final theorem shows how to locally construct the actual centre manifold and that this local construction is sufficient to investigate the stability. The interested reader is referred to~\cite{jackcarr} where proofs of the following theorems can be found.

\begin{theorem}[Existence]\label{cmexist}
  There exists a centre manifold for~(\ref{cmexvec}). The dynamics of the system~(\ref{cmexvec}) restricted to the centre manifold is given by
  \begin{align}
    \dot{\mathbf{u}} = A \mathbf{u} + f(\mathbf{u},h(\mathbf{u})) \,,
    \label{cmexisteqn}
  \end{align}
  for $\mathbf{u}\in\mathbb{R}^c$ sufficiently small.
\end{theorem}

\begin{theorem}[Stability]
  Suppose the zero solution of~(\ref{cmexisteqn}) is stable (asymptotically stable or unstable). Then the zero solution of~(\ref{cmexvec}) is also stable (asymptotically stable or unstable).
  Furthermore, if $(\mathbf{x}(t),\mathbf{y}(t))$ is also a solution of~(\ref{cmexvec}) with $(\mathbf{x}(0),\mathbf{y}(0))$ sufficiently small, there exists a solution $\mathbf{u}(t)$ of~(\ref{cmexisteqn}) such that
  \begin{align}
    \mathbf{x}(t) &= \mathbf{u}(t) +\mathcal{O}(e^{-\gamma t})\,,\\
    \mathbf{y}(t) &= h(\mathbf{u}(t)) + \mathcal{O}(e^{-\gamma t})\,,
  \end{align}
  as $t\rightarrow\infty$, where $\gamma>0$ is a constant.
\end{theorem}

Both these theorems rely on the knowledge of the function $h(\mathbf{x})$ which needs to be found. In the following we derive a differential equation for the function $h(\mathbf{x})$.
Following Definition~\ref{cmdef}, we have that $\mathbf{y} = h(\mathbf{x})$. We differentiate this with respect to time and apply the chain rule which yields
\begin{align}
  \dot{\mathbf{y}} = \nabla h(\mathbf{x}) \cdot \dot{\mathbf{x}}\,.
  \label{cmydot}
\end{align}
Since $W^c(0)$ is based on the dynamics generated by the system~(\ref{cmexvec}), we can substitute for $\dot{\mathbf{x}}$ the right-hand side of~(\ref{cenxdot}) and for $\dot{\mathbf{y}}$ the right-hand side of~(\ref{cenydot}). This yields
\begin{align}
  Bh(\mathbf{x}) + g(\mathbf{x},h(\mathbf{x})) = \nabla h(\mathbf{x}) \cdot \left[A\mathbf{x} + f(\mathbf{x},h(\mathbf{x}))\right]\,,
\end{align}
where we also used the fact that $\mathbf{y} = h(\mathbf{x})$. The latter equation can be re-arranged into the quasilinear partial different equation
\begin{align}
  \mathcal{N}(h(\mathbf{x})) :=
  \nabla h(\mathbf{x})\left[A\mathbf{x} + f(\mathbf{x},h(\mathbf{x}))\right] - Bh(\mathbf{x}) - g(\mathbf{x},h(\mathbf{x})) = 0\,,
  \label{cmn}
\end{align}
which must be satisfied by $h(\mathbf{x})$ for it to be the centre manifold. In general, we cannot find a solution to this equation, even for relatively simple systems this is often impossible. It is the third and final theorem which explain that we do not need to know the entire function.

\begin{theorem}[Approximation]\label{apptheorem}
  Let $\phi:\mathbb{R}^c\rightarrow\mathbb{R}^s$ be a mapping with $\phi(0) = \nabla \phi(0) = 0$ such that $\mathcal{N}(\phi(\mathbf{x})) = \mathcal{O}(|\mathbf{x}|^q)$ as $\mathbf{x}\rightarrow 0$ for some $q>1$. Then
  \begin{align}
    |h(\mathbf{x}) - \phi(\mathbf{x})| = \mathcal{O}(|\mathbf{x}|^q) \quad\text{as}\quad \mathbf{x}\rightarrow 0 \,.
  \end{align}
\end{theorem}
The main point here is that an approximate knowledge of the centre manifold returns the same information about stability as the exact solution of Eq.~(\ref{cmn}). Moreover, an approximation for the centre manifold can often be found straightforwardly by assuming a series expansion of $h$. The coefficients in this series are then determined by satisfying Eq.~(\ref{cmn}) for each order in the series.

We will discuss cosmological applications of this approach in Secs.~\ref{sec:power_law_potential} and \ref{chap:modifiedgrav}. Further examples, together with executive instructions on how to apply centre manifold techniques to cosmological models, can be found in the appendix of \citet{Dutta:2017wfd}.
Below we provide a simple mathematical example.

Interestingly, it is also possible to construct the centre manifold directly without first diagonalising the system, going back to the work of \citet{doi:10.1137/0143052}; see also \cite{roberts_1989,ROBERTS1997215,Roberts:2015}. While we are not exploring these techniques in our review, they may be of potential interest in future applications.

\subsection{Stability of non-hyperbolic critical points: an explicit example}

\subsubsection{Changing variables}
\label{sub:changing_vars_example}

Let us begin with the complicated looking dynamical system given by
\begin{align}
  \dot{u} &= \frac{1}{8} \Bigl[ \gamma -26 +36 v -(6\gamma+4) u -(8 \gamma +1) u^3+u^2 (12 \gamma -v+3)+u\left(v^2-2 v\right)+v^3-13 v^2 \Bigr] \,, \\
  \dot{v} &= \frac{1}{8}\Bigl[18+\gamma -20 v -(6\gamma-4) u -(8 \gamma +1) u^3+u^2 (12 \gamma -v+3)+3 v^2+v^3+u\left(v^2-2 v\right)\Bigr] \,.
\end{align}
Here $\gamma$ is an arbitrary parameter. We will use this example to show explicitly how to transform these equations into the form required for the use of centre manifold theory. One can verify that this system has a critical point at $(u,v)=(1/2,3/2)$ with eigenvalues $0$ and $-1$. First, we will shift this point to become the origin, this means we will introduce the new set of variables given by
\begin{align}
  U = u-\frac{1}{2}\,, \qquad V=v-\frac{3}{2} \,.
\end{align}
In these variables the system becomes
\begin{align}
  \dot{U} &= \frac{1}{8} \Bigl[-4U + 4V + (-8 \gamma -1) U^3-U^2 V+U V^2+V^3-8 V^2\Bigr]\,, \\
  \dot{V} &= \frac{1}{8} \Bigl[-4V + 4U + (-8 \gamma -1) U^3-U^2 V+U V^2+V^3+8 V^2 \Bigr]\,.
\end{align}
While the critical point is now the origin, these equations are not of the correct form stated in (\ref{cmexvec}). To see this, let us compute the Jacobian at the origin which gives
\begin{align}
  J\Bigr|_{U=0,V=0} = \frac{1}{2} \begin{pmatrix} -1 & 1 \\ 1 & -1\end{pmatrix} \,.
\end{align}
However, the equation in the first variable $U$ should not contain any terms linear in our variables. This means, we must diagonalise the system by diagonalising the Jacobian matrix. This is a standard method used in linear algebra, we can write
\begin{align}
  \frac{1}{2} \begin{pmatrix} -1 & 1 \\ 1 & -1\end{pmatrix} =
    \begin{pmatrix} 1 & -1 \\ 1 & 1\end{pmatrix}
      \begin{pmatrix} 0 & 0 \\ 0 & -1\end{pmatrix}
        \begin{pmatrix} 1/2 & 1/2 \\ -1/2 & 1/2\end{pmatrix} =
          S D S^{-1} \,.
\end{align}
Here $D$ is the diagonal matrix which contains the eigenvalues, $0$ and $-1$ in this case, the matrix $S$ is the matrix of eigenvectors and $S^{-1}$ denotes its inverse. It follows that the system can be brought into the correct form by introducing another set of new variables given by
\begin{align}
  \begin{pmatrix} x \\ y \end{pmatrix} =
  S^{-1} \begin{pmatrix} U \\ V \end{pmatrix} =
  \begin{pmatrix} 1/2 & 1/2 \\ -1/2 & 1/2\end{pmatrix}
    \begin{pmatrix} U \\ V \end{pmatrix} \,,
\end{align}
or, in simpler form, $U = x - y$ and $V = x + y$. Making this final substitution, our equations take the form
\begin{align}
  \dot{x} = x^2 y - \gamma (x-y)^3 \,, \qquad
  \dot{y} = -y + (x+y)^2 \,.
  \label{eqn:centre_ex_pre}
\end{align}
This is the desired form of the equations and will be the starting point for our Lyapunov stability and centre manifold theory example.

As we have seen, transforming equations into the correct form can be a somewhat tedious task which is why these steps are generally omitted in the literature. However, this might be a useful addition for readers less familiar with these techniques.

\subsubsection{Lyapunov stability}
\label{sec:Lyapunovexample}

Let us begin with the simple Lyapunov `candidate' function
\begin{align}
  V = \frac{1}{2}x^2 + \alpha y^4 \,,
  \label{Lyap:ex}
\end{align}
for the system~(\ref{eqn:centre_ex_pre}), where $\alpha$ is a positive constant. We are interested in the stability of the critical point $(x,y)=(0,0)$. A function of this type clearly satisfies the first two conditions of the Lyapunov function definition. It also satisfies $V \rightarrow \infty$ as $\|(x,y)\|\rightarrow \infty$. It remains to check whether or not $\dot{V} \leq 0$ near the critical point. Computing the time derivative of this function gives
\begin{align}
  \dot{V} = x \dot{x} + 4\alpha y^3 \dot{y} =
  x^3 y + 4 \alpha y^3 ((x + y)^2 -y) - \gamma x (x - y)^3 \,.
\end{align}
For points near the origin, the quartic terms dominate and so this function might indeed satisfy $\dot{V} < 0$ for some $\alpha$. To see this more explicitly, one could introduce polar coordinates $x=r\cos(\phi)$ and $y=r\sin(\phi)$ to find that
\begin{align}
  \dot{V} = \Bigl(
  -4 \alpha \sin^4\phi - \gamma \cos\phi(\cos\phi-\sin\phi)^3 + \sin\phi\cos^3\phi
  \Bigr) r^4 + \mathcal{O}(r^5)\,.
\end{align}
Our first observation is that for negative values of $\gamma < 0$ and $\gamma = 0$, the function (\ref{Lyap:ex}) will not be a suitable Lyapunov function as $\dot{V}$ will not be negative for all values of $\phi$. On the other hand, for $\gamma > 0$ it suffices to choose a sufficiently large $\alpha$ to ensure the $\dot{V}$ is indeed negative. For example, one easily verifies that for the specific choice $\gamma=1$ the parameter choice $\alpha=2$ makes $\dot{V}$ negative definite for all $\phi$.

This quick and direct calculation proves that the origin is a globally asymptotically stable critical point when $\gamma > 0$. We will see in the following that centre manifold theory applied to the same critical point is considerably more involved and challenging. However, one will always arrive at a definite answer in that case. The method of Lyapunov stability requires the ability to find a suitable function and failure to do so does implies neither stability or instability.

It is somewhat surprising though that this method is not used more frequently when studying cosmological dynamical systems.

\subsubsection{Centre manifold theory}
\label{sec:centre_manifold_example}

Centre manifold theory has been used infrequently in the context of cosmological dynamical systems. This motivates us to present an example of this technique. Let us consider the previous system
\begin{align}
  \dot{x} = x^2 y - \gamma (x-y)^3 \,, \qquad
  \dot{y} = -y + (x+y)^2 \,.
  \label{eqn:centre_ex}
\end{align}
Here $\gamma$ is an arbitrary parameter. One notes that the origins $(0,0)$ is a critical point of this system, moreover, the eigenvalues of the stability matrix are $\lambda_1 = 0$ and $\lambda_2 = -1$.

The system~(\ref{eqn:centre_ex}) is of the correct form which was assumed in~(\ref{cmexvec}).
Let us repeat here that it is always possible to write a dynamical system in this way by choosing new dynamical variables which diagonalise the stability matrix at the critical point in question (see Sec.~\ref{sub:changing_vars_example} above).
Comparison with~(\ref{cmexvec}) gives the following explicit expressions for the involved quantities: $A=0$, $B=-1$, $f=x^2 y - \gamma (x-y)^3$ and $g=(x+y)^2$. The next step is to state Eq.~(\ref{cmn}) which now becomes
\begin{align}
  h'(x)\left[x^2 h(x) - \gamma(x-h(x))^3\right] + h(x) - (x+h(x))^2 = 0 \,,
  \label{eqn:centre_ex2}
\end{align}
which is a non-linear first order ODE in the unknown function $h(x)$. One cannot find an explicit solution to this equation using standard methods. Therefore, we make a series expansion of $h(x)$ in $x$, that this is sufficient for the purpose of stability is implied by Theorem~\ref{apptheorem}. We write
\begin{align}
  h(x) = a_2 x^2 + a_3 x^3 + a_4 x^4 \,,
  \label{eqn:centre_ex3}
\end{align}
and substitute this into (\ref{eqn:centre_ex2}), keeping only terms up to fourth power in $x$. This yields
\begin{align}
  (a_2-1)x^2 + (a_3-2a_2) x^3 + (a_4 - 2a_3 - a_2^2 - 2\gamma a_2) x^4 = 0 \,.
  \label{eqn:centre_ex4}
\end{align}
Since this has to be true for all values of $x$, we can deduce the solution
\begin{align}
  a_2 =1\,, \qquad a_3 = 2\,, \qquad a_4 = 5 + 2\gamma \,.
  \label{eqn:centre_ex5}
\end{align}
Therefore, we find that the centre manifold is locally given by the equation $h(x)=x^2 + 2x^3 + (5+2\gamma)x^4$. This information can be used in Theorem~\ref{cmexist} to study the dynamics of the system reduced to the centre manifold, which is given by
\begin{align}
  \dot{u} = u^2 h(u) - \gamma (u-h(u))^3 = -\gamma u^3 + (1+3\gamma) u^4 + \mathcal{O}(u^5) \,.
  \label{eqn:centre_ex6}
\end{align}
Finally, we deduce the stability properties of the critical point $(0,0)$. For $\gamma > 0$, solutions of~(\ref{eqn:centre_ex6}) are stable while for $\gamma < 0$ are unstable. When $\gamma=0$, the cubic terms vanishes and one must consider the next term in the series, the quartic term in this case. As this term is of even power, we deduce instability.
More generally for any equation of the type $\dot{u} = \beta u^n$, where $\beta$ is a constant and $n$ is a positive integer number, stability is achieved only if $\beta < 0$ and $n$ is odd-parity, while any other case will be unstable (see e.g.~\citet{Perko}).

\subsection{Further advanced dynamical systems methods}
\label{sec:advanced_dynamical_systems_methods}

\subsubsection{Poincar\'e sphere and behaviour at infinity}
\label{sec:poincarecomp}

In this section we will study the properties of the flow at infinity, i.e.~when $||\mathbf{x}||=\sqrt{x^2+y^2}\rightarrow+\infty$, for 2D dynamical systems.
In $\mathbb{R}^2$ (i.e.~on a plane) it is indeed possible to make use of certain projection techniques needed to analyse the behaviour of the flow at infinity.
If this procedure is successfully performed, then one is able to draw the {\it global portrait} of the phase space, including the asymptotic behaviour as $||\mathbf{x}||\rightarrow+\infty$. What follows can be found in different books treating dynamical systems such as \citet{Lefschetz,Lynch,Perko}.

In order to compactify the phase space we will introduce the so-called {\it Poincar\'{e} sphere} which maps points at infinity onto its equator.
\begin{definition}[\textit{Poincar\'{e} sphere}]
The Poincar\'{e} sphere is defined to be the unit sphere
\begin{align}
  S^2 = \{(X,Y,Z)\in\mathbb{R}^3\,|\,X^2+Y^2+Z^2=1\} \,,
\end{align}
such that its north (or south) pole is tangent to the $(x,y)$-plane at the origin. Points on the $(x,y)$-plane can be mapped on the surface of the upper hemisphere by projecting lines passing by the centre of the sphere. This mapping is provided by the change of variables
\begin{align}
X = x\,Z \,,\qquad Y= y\,Z \,,\qquad Z=\frac{1}{\sqrt{1+x^2+y^2}} \,. \label{eq:Poincare_sphere}
\end{align}
\end{definition}

The following two theorems can be used to determine the flow at infinity. Only the statements of the theorem will be provided, but the reader interested in the details can find them in \citet{Lefschetz} or \citet{Perko}. Consider the following dynamical systems defined in $\mathbb{R}^2$
\begin{align}
\dot{x}&=P(x,y) \,,\label{018}\\
\dot{y}&=Q(x,y) \,,\label{019}
\end{align}
where $P$ and $Q$ are {\it polynomial} functions in $x$ and $y$. Let $m$ denote the maximum polynomial degree of the terms in $P$ and $Q$ and let $P_m$ and $Q_m$ be the higher terms of the corresponding polynomial functions $P$ and $Q$.

\begin{theorem}[\textit{Critical points at infinity}]
The critical points at infinity of the systems (\ref{018})--(\ref{019}) lie on the points~$(X,Y,0)$ of the equator of the Poincar\'{e} sphere where $X^2+Y^2=1$ and
\begin{align}
  X\,Q_m(X,Y) - Y\,P_m(X,Y) = 0 \,,
\end{align}
or equivalently at the polar angles $\theta_j$ and $\theta_j+\pi$ satisfying
\begin{align}
  G_{m+1}(\theta) = \cos\theta\,Q_m(\cos\theta,\sin\theta) -\sin\theta\,P_m(\cos\theta,\sin\theta) =0 \,,
  \label{020}
\end{align}
which, if not identically zero, has at most $m+1$ pairs $\theta_j$ and $\theta_j+\pi$. Moreover, if $G_{m+1}(\theta)$ is not identically zero, the flow on the equator of the Poincar\'{e} sphere is clockwise (counter-clockwise) at points corresponding to polar angles $\theta$ where $G_{m+1}(\theta)<0$ ($G_{m+1}(\theta)>0$).
\end{theorem}

Note that the points at infinity of $\mathbb{R}^2$ will always come in pairs since the projective lines intersect the equator of the Poincar\'{e} sphere twice as $||x||\rightarrow+\infty$.

The behaviour of the flow near critical points at infinity, i.e.~the {\it stability properties of critical points at infinity}, can then be described projecting the flow on the Poincar\'{e} sphere onto the two planes $(x,z)$ and $(y,z)$ tangent to the equator points~$Y=1$ and $X=1$ respectively. This is summarised in the following.

\begin{theorem}[\textit{Stability at infinity}]
The flow on the Poincar\'{e} sphere in the neighbourhood of any critical point on the equator, except the points~$(0,\pm 1,0)$, is topologically equivalent to the flow defined by the system
\begin{align}
  \pm \dot{y} &= y\,z^m\,P\Big(\frac{1}{z},\frac{y}{z}\Big)-z^m\,Q\Big(\frac{1}{z},\frac{y}{z}\Big) \,,\label{021}\\
  \pm \dot{z} &= z^{m+1}\,P\Big(\frac{1}{z},\frac{y}{z}\Big) \,, \label{022}
\end{align}
where the sign is determined by the flow on the equator of $S^2$ as provided by the sign of (\ref{020}).

Similarly, the flow on the Poincar\'{e} sphere in the neighbourhood of any critical point on the equator, except the points~$(\pm 1,0,0)$, is topologically equivalent to the flow defined by the system
\begin{align}
\pm \dot{x} &= x\,z^m\,Q\Big(\frac{x}{z},\frac{1}{z}\Big)-z^m\,P\Big(\frac{x}{z},\frac{1}{z}\Big) \,,\label{023}\\
\pm \dot{z} &= z^{m+1}\,Q\Big(\frac{x}{z},\frac{1}{z}\Big) \,, \label{024}
\end{align}
where the sign is determined by the flow on the equator of $S^2$ as provided by the sign of (\ref{020}).
\end{theorem}

This means that if $(0,\pm 1,0)$ is not a critical point at infinity we can use (\ref{021})--(\ref{022}) to find the stability of all critical points at infinity. Similarly if $(\pm 1,0,0)$ is not a critical point at infinity we can use (\ref{023})--(\ref{024}) to find the stability of all critical points at infinity. If both $(0,\pm 1,0)$ and $(\pm 1,0,0)$ are critical points at infinity, then we must analyse both (\ref{021})--(\ref{022}) and (\ref{023})--(\ref{024}). An application of this technique in the context of cosmology is provided in Sec.~\ref{sec:phantom}.

\subsubsection{The concept of invariant submanifold}

In the sections above we have introduced the so called invariant sets (or manifolds) i.e.~the parts of the phase space that are not connected to the rest of the phase space by any orbit. The simplest member (zero dimension) of this set are the fixed points, but there are other invariant manifolds of interest like e.g.~the periodic orbits we will examine briefly later.

In literature one also encounter sometimes the term {\it invariant submanifolds}. Those objects  are indeed an invariant set, but are dubbed ``submanifolds'' to highlight the fact that  their dimension is lower than the one of the phase space (but greater than zero) and span all the phase space. By definition an orbit belonging to an invariant submanifold in a certain instant will always belong into the invariant submanifold.

Invariant submanifolds are very important in terms of the characterisation of the phase space. The reason is that, as far as the dynamical system is of order $C(1)$, no orbit can cross such submanifolds. If this were the case the orbit would have to abruptly change direction at the intersection and this would mean a discontinuity in its first derivative. Hence, invariant submanifolds separate the phase space in independent sections which are not connected by orbits.

The presence of invariant submanifolds also implies that an attractor cannot, in general, be  global. Since the phase space is separated in independent sections, there will always be orbits that do not lead to it. In a phase space with invariant submanifolds  the only point which can be global attractors are the ones that lie in the intersection (if any) of all the invariant submanifolds.

There is a simple way to determine the presence of invariant submanifolds by looking at the structure of the dynamical system equations. More specifically if the r.h.s.~of a dynamical equation for a given variable $y$ can be factorised in such a way to present  the constant root $\alpha$ for $y$, then $y=\alpha$ will be an invariant submanifold. In fact for orbits with $y=\alpha$,  one has $y'=0$ and therefore the coordinate $y$ of the orbit will always be constant.

We can see this concretely in a simple example in two dimensions. Consider the system
\begin{align}
  \dot{x} &=6x^3+ y^2\,,\\
  \dot{y} &=3y +4 y^2+ 2yx \,,
\end{align}
The second dynamical equation can be written as
\begin{equation}
  \dot{y}=y(3 +4 y+ 2x) \,.
\end{equation}
If an orbit has initial condition $y_0=0$ it will preserve the value of this coordinate. Therefore the plane $(x,y)$ will be divided in two independent sections: $y>0$ and $y<0$. No orbit will be able to cross the $y=0$ line and an attractor will be truly global only if it has coordinates $(x=\bar{x},y=0)$ for some value $\bar{x}$.

\subsubsection{Global behaviour of the phase space and monotonic functions}\label{globalPS}

Some other techniques are also useful to study the global properties of the phase space.  Such properties include, for example, periodic orbits, homoclinic loops, separatrix attractors and strange attractors.  There are very few papers in which the global analysis of the phase space is pursued  and the tools to perform this analysis can be quite technical (e.g.~\citet{Perko,Wiggins}). For this reason, we will not delve too much into the theory behind this kind of analysis. In fact we will focus only on two simple methods to determine the absence of periodic orbits and fixed points, which are connected with the construction of a function of the phase space variables with certain properties. These methods will be useful in Sec.~\ref{chap:modifiedgrav}.

We start with the so called {\it Dulac's criterion}, which is used to determine the absence of periodic orbits and, indirectly, of fixed points in a phase space. Consider again a system of the form $\dot{\mathbf{x}} = \mathbf{f}(\mathbf{x})$. Dulac's criterium \citep{Perko} has similarities with the Lyapunov function approach as it relies on our ability to find a function with suitable properties.

\begin{theorem}[\textit{Dulac's criterion}]
  Let us assume there exists a scalar function $H(\mathbf{x})$ such that in a simply connected domain the quantity $H\,\mathbf{f}$ satisfies the condition
  \begin{align}
    \nabla\cdot (H\, \mathbf{f}) \geq0 \,.
    \end{align}
Then the phase space does not contain periodic orbits.
\end{theorem}

We can sketch a proof  in two dimensions, but the criteria can be easily demonstrated in general. Consider the two dimensional system
\begin{align}
  \dot{x} &=P(x,y) \,,\\
  \dot{y} &=Q(x,y) \,.
\end{align}
Suppose that there exist, in a simply connected region of the real plane, a closed orbit $\Gamma$. The dynamical system guarantees that the integral
\begin{equation}
  \int_\Gamma H \left(P d y-Q d x\right) = \int_\Gamma H \left( \dot{x} d y-\dot{y} d x \right)\,,
\end{equation}
is identically zero. However, using Stokes theorem, we can write
\begin{equation}
  \int_\Gamma H \left( P d y-Q d x \right)= \int_S \left( \frac{\partial (H P)}{\partial x}+ \frac{\partial (H Q)}{\partial y}\right) dx\, dy\,,
\end{equation}
where $S$ is the interior bounded by $\Gamma$ (which we also assumed oriented positively). The last integral, however, cannot be zero by definition. This leads to a contradiction and proves that there cannot be closed orbits. Note that the criterion is only a sufficient condition for the absence of isotropic orbits. There are other theorems that help the analysis of this kind of phase space structures e.g.~the Poincar\'e-Bendixon theorem, see~\citet{arrowsmith,Wiggins,Perko}.

The construction of specific functions of the phase space variables can also help to determine whether a set in the phase space is devoid of fixed points \citep{Wainwright:1989nm}. Suppose we have a positive semi-definite function $W$. One can use such a function to show that no fixed points can exist in the phase space. To see this, one notes that the derivative of $W$ ($\dot{W} = \grad W \cdot \mathbf{f}$, computed using the chain rule) is a linear combination of the derivatives of the variables, see Eq.~(\ref{gradV}). Therefore $\dot{W}$ can only vanish at fixed points of the system. Consequently, a positive definite $\dot{W}$ would imply the absence of fixed points.

As an elementary example, we consider the system
\begin{align}
  \dot{x} &=\alpha x+ \beta^2 y\,,\\
  \dot{y} &=-\alpha x +\gamma^2 y \,,
\end{align}
It is clear that the system only has a fixed point at the origin of the phase plane. Suppose, for the moment, that for some reason we were not able to show directly that no other fixed points exist. We can approach the problem using the function $W = x^2+y^2$ which gives
\begin{align}
  \dot{W}= 2\beta^2 x^2+ 2\gamma^2 y^2\,,
\end{align}
which is positive definite for $(x\neq 0, y\neq 0)$. Therefore no fixed points other than the origin exist.

%% file: chapters/03_cosmo/cosmo.tex
\section{Standard cosmology}
\label{chap:Cosmology}

In this section we introduce the basic principles of cosmology and analyse the dynamical behaviour of the standard cosmological model, namely $\Lambda$CDM. We will also discuss the theoretical problems associated with the cosmological constant and explore cosmological models beyond spatial flatness. This section should be helpful to readers unfamiliar with the subject of cosmology, while the reader not used to dynamical systems techniques will find here some simple applications in the cosmological setting. Only the topics directly relevant to this review will be discussed here, while more detailed presentations about cosmology can be found in well known textbooks, e.g.~\citet{DodelsonModCosmo} or \citet{WeinbergCosmology}.

\subsection{Elements of FLRW cosmology}
\label{sec:FRW}

Modern cosmology is based on the so-called {\it cosmological principle} which states that at sufficiently large scales ($\sim 10^8$ light years) the universe is assumed to be homogeneous and isotropic\footnote{As seen by comoving observers; more below.}. In other words the principle asserts that the Earth, or any other location, does not occupy a special position in the universe. A consequence of this principle is that the spacetime describing the universe must be highly symmetric in its spatial part. The constant time hyper-surfaces are spaces of constant curvature. It is not too complicated to prove that the most general four dimensional metric  which is also maximally spatially symmetric is the  {\it Friedmann-Lema\^{i}tre-Robertson-Walker (FLRW) metric} $g_{\mu\nu}$; see \citet{Wald} or \citet{Weinberg}. In pseudo-spherical\footnote{The coordinates $(r,\theta,\varphi)$ coincides with actual spherical coordinates only for the case $k=0$, but for $k\neq 0$ they represent a more general system of coordinates.} coordinates $x^\mu = (t,r,\theta,\varphi)$ centred at any point of the universe, the line element of the FLRW metric reads
\begin{align}
  ds^2 = g_{\mu\nu} dx^\mu dx^\nu = -dt^2 + a(t)^2 \left(\frac{dr^2}{1-k\,r^2} +r^2d\theta^2+r^2\sin^2\theta\, d\varphi^2\right) \,,
  \label{eq:FRWmetric}
\end{align}
where $k=-1,0,+1$ is the spatial curvature and $a(t)>0$ is a function of the time coordinate called the {\it scale factor}. For, $k=1$ we say that the universe is spatially closed, for $k=-1$ it is spatially open and if $k=0$ it is spatially flat. The coordinates $(r,\theta,\varphi)$ are referred to as comoving coordinates: an observer at rest in these coordinates remains at rest, i.e.~at constant $r$, $\theta$, and $\varphi$ for all time $t$.

The dynamics of the metric tensor $g_{\mu\nu}$, i.e.~of the gravitational field, is described by the {\it Einstein field equations}, which in the absence of a cosmological constant are given by
\begin{align}
  R_{\mu\nu} -\frac{1}{2}R\,g_{\mu\nu} =\kappa^2\, T_{\mu\nu} \,,
  \label{eq:Einsteinfieldeq}
\end{align}
where $R_{\mu\nu}$ is the {\it Ricci tensor}, $R=g^{\mu\nu}R_{\mu\nu}$ is the {\it Ricci scalar} or {\it curvature scalar}, $T_{\mu\nu}$ is the {\it energy-momentum tensor} of matter sources and $\kappa^2=8\pi G/c^4$. Matter inside a homogeneous and isotropic universe can be described,
at large scales and with high precision, as a {\it perfect fluid}.
In particular its energy-momentum tensor is solely determined by its energy density $\rho(t)$ and isotropic (no shear nor viscosity) pressure $p(t)$:
\begin{align}
  T_{\mu\nu} = p\,g_{\mu\nu} + (\rho+p)\,u_\mu u_\nu \,,
  \label{eq:fluidEMtensor}
\end{align}
where the vector $u^\mu$ denotes the four-velocity of an observer comoving with
the fluid and in comoving coordinates is given by $u^\mu=(-1,0,0,0)$.
The energy density and pressure of a perfect fluid are related by an {\it equation of state} $p=p(\rho)$ (referred to as EoS). For {\it barotropic} perfect fluids this is a linear relation
\begin{align}
p= w \,\rho \,,
\label{eq:ofstate}
\end{align}
where $w$ is called the {\it equation of state parameter}. For a non-relativistic (dust-like) perfect fluid $w=0$, while for a relativistic (radiation-like) fluid $w=1/3$. Values of $w$ outside the $[0,1]$ range are not permitted by known macroscopic physics, though, as we will see, some phenomenological models rely on non-physical values of $w$ in order to match the astronomical observations.

The {\it cosmological equations} arising from the Einstein field equations (\ref{eq:Einsteinfieldeq}) with the FLRW metric ansatz (\ref{eq:FRWmetric}) consist of two coupled differential equations for the scale factor $a(t)$ and the matter variables $\rho(t)$ and $p(t)$. The {\it Friedmann equation} (or {\it Friedmann constraint}) follows from the time-time component of the Einstein field equation and can be written as
\begin{align}
\frac{k}{a^2}+H^2 = \frac{\kappa^2}{3}\rho \,,
\label{eq:Friedmann}
\end{align}
where the {\it Hubble rate} (or {\it parameter}) is defined as
\begin{align}
H=\frac{\dot a}{a} \,,
\end{align}
with an over-dot denoting differentiation with respect to $t$.
On the other hand, from the spatial (diagonal) components of the Einstein field equations we obtain the {\it acceleration equation}
\begin{align}
\frac{k}{a^2}+2\dot{H} +3H^2 = -\kappa^2p \,.
\label{eq:accel}
\end{align}
The cosmological equations (\ref{eq:Friedmann}) and (\ref{eq:accel}) determine the evolution of the scale factor $a(t)$ once an equation of state relating $\rho$ and $p$ has been assumed.

Using the Friedmann equation (\ref{eq:Friedmann}), the acceleration equation (\ref{eq:accel}) can be rewritten as
\begin{align}
\frac{\ddot a}{a} = -\frac{\kappa^2}{6} \left(\rho+3p\right) \,,
\label{eq:Raychaudhuri}
\end{align}
which is sometimes called the {\it Raychaudhuri equation}. Note that from (\ref{eq:Raychaudhuri}) we can obtain a condition on the matter variables that discriminates between a universe with an expansion rate which is increasing ($\ddot a>0$) or decreasing ($\ddot a<0$). It is customary in the literature to refer to the above cases as a {\it accelerating} or {\it decelerating} universe (or expansion). If $\rho+3p>0$ the universe is decelerating, while if $\rho+3p<0$ the universe is accelerating. The first inequality can be shown to coincide with the strong energy condition. If the linear equation of state (\ref{eq:ofstate}) holds, the condition can be transferred to the equation of state parameter implying $w>-1/3$ for deceleration and $w<-1/3$ for acceleration. It follows that the type of physically meaningful matter that we experience with equations of state lying between 0 and 1/3 therefore always describe a decelerating universe when they are the dominant contributors to the energy density budget.

From the conservation of the energy-momentum tensor, $\nabla_\mu T^{\mu\nu}=0$, or equivalently from equations (\ref{eq:Friedmann}) and (\ref{eq:accel}), we can derive the {\it energy conservation equation} for the matter fluid
\begin{align}
\dot\rho + 3H\left(\rho+p\right) = 0 \,,
\label{eq:conserveq}
\end{align}
expressing the conservation of energy through the evolution of the universe. Substituting (\ref{eq:ofstate}) into (\ref{eq:conserveq}) leads to the solution for $\rho(a)$:
\begin{align}
\rho\propto a^{-3(w+1)} \,,
\label{036}
\end{align}
which is valid for all $w \neq -1$. It follows that for a dust-like (or {\it matter}) fluid with $p_{\rm m}=0$ (i.e.~$w = w_{\rm m} = 0$)
\begin{align}
\rho_{\rm m} \propto a^{-3}  \qquad\mbox{({\it matter})}\,,
\label{030}
\end{align}
while for a radiation-like fluid $p_{\rm r}=\rho_{\rm r}/3$ (i.e.~$w = w_{\rm r} = 1/3$)
\begin{align}
\rho_{\rm r} \propto a^{-4}  \qquad\mbox{({\it radiation})}\,.
\label{031}
\end{align}
A particularly simple set of solutions for $a(t)$ are found for the case of a flat ($k=0$) universe. Substituting (\ref{036}) into  (\ref{eq:Friedmann}) we obtain the general result
\begin{align}
  a(t)\propto t^{\frac{2}{3(w+1)}} \,,
  \label{eq:mattersolw}
\end{align}
showing that the scale factor evolves as a power-law function of time, in particular as $t^{2/3}$ for matter and $t^{1/2}$ for radiation domination.

Eq.~\eqref{036} holds only if there is one single perfect fluid, with constant EoS parameter $w$, appearing in the r.h.s.~of the Einstein field equations. The situation might be different if multiple fluids source the cosmological equations. If the total energy-momentum tensor $T_{\mu\nu}$ in the Einstein field equations (\ref{eq:Einsteinfieldeq}) is composed of more than one matter component, e.g.~$T_{\mu\nu}=T_{\mu\nu}^{(1)}+T_{\mu\nu}^{(2)}$, the conservation equation $\nabla^\mu T_{\mu\nu}=0$ will only imply the conservation of the total energy and momentum of the fluids. The energy and momentum of a single fluid component might not be conserved due to possible interactions with the other fluid components. For two fluids $T_{\mu\nu}^{(1)}$ and $T_{\mu\nu}^{(2)}$ sourcing the Einstein field equations we can generally write
\begin{align}
\nabla^\mu T_{\mu\nu}^{(1)}=Q_\nu \quad\mbox{and}\quad \nabla^\mu T_{\mu\nu}^{(2)}=-Q_\nu \,,
\label{032}
\end{align}
where $Q_\nu$ denotes the energy-momentum exchanged between the two fluids. If $Q_\nu=0$ there is no exchange between the two fluids and they evolve without interacting. Note that the total energy-momentum is always conserved
\begin{align}
\nabla^\mu T_{\mu\nu} = \nabla^\mu \left(T_{\mu\nu}^{(1)}+T_{\mu\nu}^{(2)}\right)=0 \,,
\end{align}
even if $Q_\nu \neq 0$. The specification of $Q_\nu$ is an assumption regarding the physical properties of the two fluids that must be taken into account in order to solve the field equations. Without this assumption the dynamics of the single components of the fluid cannot be found from the Einstein field equations alone. Of course, if there are more than two fluids sourcing the Einstein field equations we will need more than one exchange vector $Q_\nu$. In general if there are $n$ fluids we must specify $n-1$ exchange vectors in order to fully determine the dynamics of the system.

In what follows we will assume negligible interactions between the matter sources appearing in the cosmological equations. This assumption will hold true for all the cosmological models analysed in this review, unless otherwise explicitly specified. On the other hand, Sec.~\ref{chap:IDE} will be completely dedicated to the study of cosmological models with interactions in the matter sector.

\subsection{Dark energy and the cosmological constant}
\label{sec:cosmoconst}

What does our Universe look like on large scales? Perhaps our best guide to date comes from the latest release of parameter estimates arising from the Planck satellite observations \citep{Ade:2015xua} of the cosmic microwave background (CMB). These and other cosmological observations tell us that the universe is old, at around 13.8 Gyrs, it appears to be consistent with the cosmological principle, and it is currently undergoing a period of accelerated expansion. As shown in Sec.~\ref{sec:FRW}, standard matter components, such as dust or radiation, can only provide decelerated dynamics. In order to explain the observed accelerated expansion, the concept of \textit{dark energy} was thus introduced. It is defined as  matter component with EoS characterised by negative pressure such that $p_{\rm de} < -\rho_{\rm de}/3$ (or equivalently $w_{\rm de} < -1/3$).

The simplest model of dark energy is represented by the \textit{cosmological constant} $\Lambda$, which can be associated with a cosmological source with EoS $p_\Lambda = - \rho_\Lambda$ (i.e.~$w_\Lambda = -1$). The cosmological constant appears as a simple modification of the Einstein field equations and was first introduced by \citet{Einstein:1917ce} himself in order to construct a cosmological model which would lead to a static universe. Following the discovery that the universe is expanding, and the realisation that the static solution he obtained was unstable, the classical cosmological constant was dropped from the field equations as it was not required to deliver the type of dynamical expansion that at the time was consistent with observations.

Ignoring the cosmological constant was  considered acceptable until the late 1980s and early 1990s when observations of the angular correlations of galaxies in the APM data suggested that there were issues fitting those correlations with the standard paradigm of a completely matter dominated universe. In particular the data suggested that there was more cosmological structure than expected on very large scales, $l > 10 h^{-1}$ Mpc \citep{Maddox:1990hb}. It actually led \citet{Efstathiou:1990xe} to propose the presence of a cosmological constant contributing 80\% of the critical density as a solution to the discrepancy with the data. There were a few papers in the early 1990s that also suggested there was a problem with the standard matter dominated cosmology, but it is fair to say the dramatic shift in our interpretation of the role of $\Lambda$ in cosmology took place in 1998 when \citet{Perlmutter:1998} and \citet{Riess:1998} reported that the universe was accelerating and must be dominated by some form of energy density resembling a cosmological constant. Since then the issue of the cosmological constant problem has once again become of paramount importance as has the question of why the universe is accelerating, or equivalently what is the fundamental nature of dark energy? Is it caused by the presence of a $\Lambda$ term? Or perhaps some other form of dynamical dark energy (see \citet{Copeland:2006wr} for a review)? Or maybe because of a modification of General Relativity (for a review see \citet{Clifton:2011jh})?

The latest astronomical observations \citep{Ade:2015xua} suggest that roughly 70\% of the energy budget of the universe is composed by dark energy, with the cosmological constant model accurately fitting all data with an energy density parameter $\Omega_{\Lambda 0} = 0.692 \pm 0.012$ characterising its current energy fraction. In fact, when the Planck data are combined with other astrophysical data, including Type-Ia supernovae, the EoS parameter of dark energy is constrained to be $w = -1.006 \pm 0.045$, a result consistent with that of a cosmological constant. The remaining part of the cosmic energy budget is dominated by a non-relativistic (dust) matter component, divided into standard baryonic matter and another invisible entity called \textit{dark matter}, which is needed to explain discrepancies in the observed rotation curves of galaxies and in the dynamics of galaxy clusters \citep{Bertone:2004pz}. The present value of the matter density parameter is $\Omega_{\textrm{m}0} = 0.308 \pm 0.012$. The baryonic content of the universe inferred by Planck is perfectly consistent with nucleosynthesis observations $\Omega_{\textrm{b}0} h^2 = 0.02234 \pm 0.00023$, and the cold dark matter (CDM) component, where cold means non-relativistic ($w_{\rm dm} = 0$), is $\Omega_{\textrm{C}0} h^2 = 0.1189 \pm 0.0022$, where $h = 0.7324 \pm 0.0174$ is the conventional way we present our uncertainty in the Hubble parameter. Cosmological structure has arisen from a power-law spectrum of adiabatic scalar perturbations which has a tilted scalar spectral index with $n_s = 0.968 \pm 0.006$. There is as yet no evidence of any tensor contributions to the primordial density fluctuations, the limit of the tensor to scalar ratio of amplitudes at a scale $k_* = 0.002$ is $r<0.11$ \citep{Ade:2015tva}. There is no indication of the presence of isocurvature fluctuations or of cosmic defects. Our universe appears to be consistent with being spatially flat, $|\Omega_{k0}| < 0.005$, implying $k=0$ in the FLRW metric~\eqref{eq:FRWmetric}. Moreover it is expanding with a Hubble rate $H_0 = 67.8 \pm 0.9 {\rm km s^{-1} Mpc^{-1}}$ (although we mention the possible tension with the recent local value of the Hubble parameter, $H_0 =  73.24 \pm 1.74 {\rm km s^{-1} Mpc^{-1}}$ determined by observations of Type-Ia supernovae \citep{Riess:2016jrr}).

Let us initially consider the effect a cosmological constant has on cosmology. It is introduced as a new term in the Einstein field Eqs.~(\ref{eq:Einsteinfieldeq}), which now read
\begin{align}
R_{\mu\nu} -\frac{1}{2}R\,g_{\mu\nu} + \Lambda\,g_{\mu\nu} =\kappa^2\, T_{\mu\nu} \,,
\label{eq:EinsteinfieldeqLambda}
\end{align}
where $\Lambda$ is the cosmological constant. The value of a positive $\Lambda$ needed to match the cosmological observations is of the order
\begin{align}
\Lambda\simeq 10^{-52}\,{\rm m}^{-2} \,,
\label{num:Lambda}
\end{align}
which is sufficiently small to not produce detectable effects at solar system distances \citep{Clifton:2011jh}. Considering again the FLRW metric (\ref{eq:FRWmetric}) with vanishing spatial curvature ($k=0$), from the new field equations (\ref{eq:EinsteinfieldeqLambda}) we now obtain
\begin{align}
3H^2 &= \kappa^2\rho + \Lambda \,,
\label{eq:FriedmannLambda} \\
2\dot{H} +3H^2 &= -\kappa^2p + \Lambda \,,
\label{eq:accelLambda}
\end{align}
which generalise the Friedmann equation (\ref{eq:Friedmann}) and the acceleration equation (\ref{eq:accel}). Note that from these equations we can equally see the contribution of the cosmological constant as a constant energy fluid with $\rho_\Lambda=\Lambda/\kappa^2$ and $p_\Lambda=-\rho_\Lambda$. The EoS parameter of the cosmological constant thus has the constant value $w_\Lambda=-1$, which as we have just seen is consistent with current observations. The physical motivations and implications of this new matter component will be briefly discussed in Sec.~\ref{sec:cosmoconstpbm}. For the moment we will focus on the dynamics arising from a universe with a non-vanishing $\Lambda$.

If the cosmological constant completely dominates the evolution equations (\ref{eq:FriedmannLambda}) and (\ref{eq:accelLambda}), meaning that the other matter contributions can be neglected ($\rho=0$ and $p=0$), then one immediately obtains the solution
\begin{align}
a(t)\propto e^{Ht} \quad\mbox{with}\quad H=\sqrt{\frac{\Lambda}{3}} \,,
\label{eq:cosmoconstsol}
\end{align}
which is known as the {\it de Sitter solution}. In such a universe we have that the scale factor expands exponentially, meaning that the condition $\ddot a>0$ is always satisfied and there is a never-ending accelerating phase. Of course this solution cannot be used as a realistic model for our universe, since we know that at early times a radiation and then a matter dominated phase must have occurred in order for cosmic structure to have formed. However it can be used as an asymptotic solution at early and late times (with vastly different values of $\Lambda$ in each case). A universe evolving according to matter domination for a sufficiently long time and then switching to a de Sitter expansion could be an accurate description for the observed dynamics on cosmological scales.

For this reason we will now solve the cosmological equations (\ref{eq:FriedmannLambda}) and \eqref{eq:accelLambda} for a non-vanishing matter contribution with a linear equation of state ($\rho> 0$ and $p=w\rho$). There are two ways to solve these equations. The first strategy is to rely on the conservation equation (\ref{eq:conserveq}) which again follows from combining (\ref{eq:FriedmannLambda}) and (\ref{eq:accelLambda}). The solution of $\rho$ in terms of $a$ is thus given again by (\ref{036}), i.e.~
\begin{align}
\rho\propto a^{-3(1+w)} \,,
\end{align}
unless $w=-1$ for which solution \eqref{eq:cosmoconstsol} applies. Plugging this back into (\ref{eq:FriedmannLambda}) will provide a differential equation for $a$ which must be solved in order to find the solution. The second way consists in eliminating $\rho$ from Eqs.~(\ref{eq:FriedmannLambda})--(\ref{eq:accelLambda}) and then solve the resulting differential equation for $H$. Once the solution of $H$ in terms of $t$ has been found, one can obtain the evolution of $a$ solving $H=\dot a/a$.

No matter which way one follows, the physical solution (no negative energies and $a(0)=0$) for the scale factor will eventually be
\begin{align}
a(t)\propto \left[\sinh\left(Ct\right)\right]^{\frac{2}{3(w+1)}} \,,
\label{037}
\end{align}
where $C$ is a constant. Note that at early and late times this solution has the correct asymptotic behaviour expected from matter and cosmological constant domination, respectively. In more mathematical terms we have that
\begin{alignat}{2}
  a(t)&\propto t^{\frac{2}{3(w+1)}} \quad & \mbox{as} 
  \quad &t\rightarrow 0 \,,\\
  a(t)&\propto \exp\left[\frac{2Ct}{3(w+1)}\right]\quad & \mbox{as} 
  \quad &t\rightarrow+\infty \,,
\end{alignat}
which correspond, respectively, to the perfect fluid solution (\ref{eq:mattersolw}) and to the cosmological constant solution (\ref{eq:cosmoconstsol}), where $H=2C/(3(w+1))$.
Solution (\ref{037}) can thus well describe the observed universe which must decelerate at early times and accelerate at late times. This means that a universe filled with a cosmological constant and some matter fluid provides the general features of the observed Universe.

Note that in (\ref{037}) the matter EoS parameter has been left arbitrary. We can choose to have an early universe dominated by matter ($w=0$), radiation ($w=1/3$) or any other kind of fluid. Usually $w$ is set to zero in order to have a matter dominated phase followed by a dark energy dominated epoch, which aims at characterising the late time behaviour of our universe. It is not possible to describe a two fluid universe with the solution (\ref{037}) and in fact the radiation to matter transition happening at early times is overlooked in this model. However using dynamical system techniques we will be able to analyse the evolution of a cosmological constant universe filled with both matter and radiation (see Sec.~\ref{sec:LCDM_dynam}).

\subsection{Dynamics of $\Lambda$CDM}
\label{sec:LCDM_dynam}

In this section we present the complete cosmological dynamics derived from the $\Lambda$CDM model, where dark energy is modelled by the cosmological constant $\Lambda$ and dark matter is assumed to be a non-relativistic matter component ($w_{\rm dm} = 0$), namely cold dark matter (CDM).
We will show that an early-time epoch dominated by radiation can be followed by a matter dominated epoch and eventually by a dark energy dominated late-time epoch.
In agreement with the $\Lambda$CDM paradigm in what follows we will assume that dark energy has a constant EoS parameter equal to $-1$ (namely the cosmological constant).
Similar dynamical analyses have been performed by \citet{Garcia-Salcedo:2015ora}, while further investigations with more general dark energy EoSs have been performed by \cite{Fay:2013ola}.

As discussed in Section~\ref{sec:dynamicalsystems}, critical points of dynamical systems are key to understanding the time evolution of the system, in this case a cosmological model. Should we find that there are no stable (or saddle) dark energy dominated point, for example, we could immediately rule out such a model as it would not be able to correctly describe the late time acceleration of the Universe. This means cosmological dynamical systems are required to display certain mathematical features through their critical points. Unstable points are relevant as early time attractors while stable points act as late time attractors. Saddle points, on the other hand, attract some trajectories while repelling others which makes them interesting for instance for matter and radiation dominated epochs as these are neither the initial nor the final epochs of the Universe's evolution.

We start from the cosmological equations (\ref{eq:FriedmannLambda}) and (\ref{eq:accelLambda}) with both matter ($p_{\rm m}=0$) and radiation ($p_{\rm r}=\rho_{\rm r}/3$):{}
\begin{align}
3H^2 &= \kappa^2\rho_{\rm m} +\kappa^2\rho_{\rm r} + \Lambda \,,\label{038}\\
2\dot{H} +3H^2 &= -\frac{\kappa^2}{3}\rho_{\rm r} + \Lambda \,,\label{040}
\end{align}
The system above suggests the introduction of the dimensionless variables
\begin{align}
x=\Omega_{\rm m}= \frac{\kappa^2\rho_{\rm m}}{3H^2},\quad y=\Omega_{\rm r}=\frac{\kappa^2\rho_{\rm r}}{3H^2} \quad\mbox{and}\quad \Omega_{\rm \Lambda}=\frac{\kappa^2\rho_{\rm \Lambda}}{3H^2} \,.
\label{eq:defxyOmega}
\end{align}
which represent, respectively, the relative energy densities of matter, radiation and cosmological constant with respect to the total energy density $\rho_{\rm tot} = \rho_{\rm m} + \rho_{\rm r} + \rho_\Lambda = 3H^2/\kappa^2$. The variables \eqref{eq:defxyOmega} are usually referred to as the {\it expansion normalised variables} \citep{WainwrightEllis}, and in what follows we will denote them as the {\it EN variables}. 

\begin{figure}
\begin{center}
\includegraphics[width=0.6\textwidth]{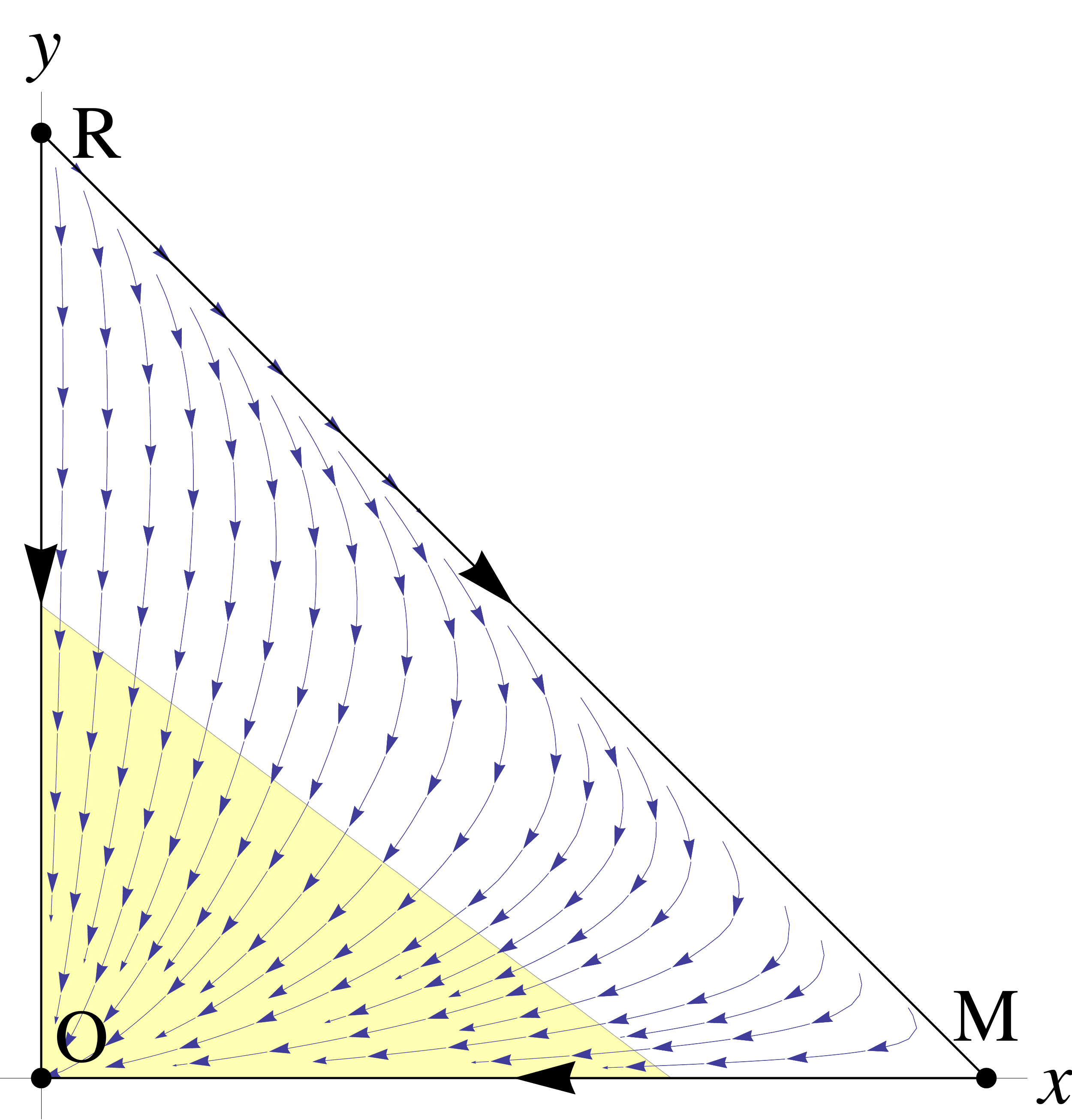}
\end{center}
\caption{Phase space portrait of the dynamical system (\ref{043})--(\ref{044}). The yellow/shaded area denotes the region of the phase space where the universe is accelerating.}
\label{fig:twofluidLambda}
\end{figure}

Thanks to the assumptions $\rho_{\rm m}>0$ and $\rho_{\rm r}>0$ (positive energy), we obtain the constraints $x>0$ and $y>0$ which restrict the physical phase space in the $(x,y)$-plane. Moreover we can now rewrite the Friedmann equation (\ref{038}) as
\begin{align}
1=x+y+\Omega_\Lambda \,.
\label{039}
\end{align}
Note that because of \eqref{039}, $\Omega_\Lambda$ can always be substituted with $x$ and $y$ in the following equations, meaning that only a 2D dynamical system is required to characterise the evolution of the universe\footnote{The choice to eliminate $\Omega_\Lambda$ is of course purely arbitrary. There is no loss of generality in eliminating any other variable. }. Furthermore given that we are assuming a positive $\Lambda$, Eq.~\eqref{039} implies that the constraint $x+y\leq 1$ must hold. Adding the fact that $x\geq 0$ and $y\geq 0$, the physically meaningful dynamics in the $(x,y)$-plane happens inside the triangle with vertices at the origin $(0,0)$, Point~$(1,0)$ and Point~$(0,1)$; see Fig.~\ref{fig:twofluidLambda}. This triangle constitutes an invariant set of the whole phase space which will be called the {\it physical invariant set} or simply the {\it physical phase space}.

\begin{table}
\begin{center}
\begin{tabular}{|c|c|c|c|c|c|}
\hline
Point & $x$ & $y$ & $w_{\rm eff}$ & Eigenvalues & Stability \\
\hline
$O$ & 0 & 0 & -1 & $\{-4,-3\}$ & Stable point \\
$R$ & 0 & 1 & 1/3 & $\{1,4\}$ & Unstable point \\
$M$ & 1 & 0 & 0 & $\{-1,3\}$ & Saddle point\\
\hline
\end{tabular}
\end{center}
\caption{Critical points of the dynamical system (\ref{043})--(\ref{044}) and their properties.}
\label{tab:twofluidLambda}
\end{table}

To obtain the dynamical system, we differentiate the variable $x$ and $y$ with respect to $\eta=\log a$ ($d\eta=Hdt$), which represents our dimensionless time variable.
We obtain
\begin{align}
  x'&=\frac{dx}{d\eta}=\frac{1}{H} \frac{dx}{dt} = \frac{\kappa^2\dot\rho_{\rm m}}{3H^3}-\frac{2 \kappa^2 \rho_{\rm m}}{3H^2}\frac{\dot H}{H} \,,\label{118}\\
  y'&=\frac{dy}{d\eta}=\frac{1}{H} \frac{dy}{dt} = \frac{\kappa^2\dot\rho_{\rm r}}{3H^3}-\frac{2 \kappa^2 \rho_{\rm r}}{3H^2}\frac{\dot H}{H} \,.\label{119}
\end{align}
Assuming that (dark) matter and radiation do not interact ($Q_\nu=0$), from the conservation equation (\ref{eq:conserveq}), for the two fluids we obtain
\begin{alignat}{3}
\dot\rho_{\rm m} +3H\rho_{\rm m} &= 0 \qquad &\rightarrow &\qquad \dot\rho_{\rm m} &=-3H\rho_{\rm m} \,,\\
\dot\rho_{\rm r} +4H\rho_{\rm r} &= 0 \qquad &\rightarrow &\qquad \dot\rho_{\rm r} &=-4H\rho_{\rm r} \,,
\end{alignat}
where the assumptions $p_{\rm m}=0$ and $p_{\rm r}=1/3 \rho_{\rm r}$ have been considered.
The acceleration equation (\ref{040}) yields
\begin{align}
  \frac{\dot H}{H^2}= -\frac{3}{2} -\frac{\kappa^2\rho_{\rm r}}{6H^2} + \frac{\Lambda}{2H^2} = -\frac{1}{2} \left(3-y-3\Omega_\Lambda\right) \,.
\end{align}
Substituting these results into (\ref{118})--(\ref{119}) and using the Friedmann constraint (\ref{039}) produces the following 2D dynamical system
\begin{align}
x'&= x \left(3 x + 4 y-3\right) \,,\label{043}\\
y'&= y \left(3 x + 4 y-4\right) \,.\label{044}
\end{align}
We note that the variables $x$ and $y$ together with the new time variable $\eta$ require $H \neq 0$. The condition $H=0$ is usually associated with an Einstein static universe and hence is not captured by this choice of variables. 

The system (\ref{043})--(\ref{044}) presents the  invariant submanifolds $x=0$ and $y=0$ which make the physical phase space compact. There are three critical points: $O=(0,0)$, $R=(0,1)$ and $M=(1,0)$. Performing linear stability analysis near the critical points we find that the eigenvalues of the Jacobian are $\{-4,-3\}$ at $O$, $\{1,4\}$ at $R$ and $\{-1,3\}$ at $M$. This implies that the origin is a stable point (simple attracting node), $R$ is an unstable point (simple repelling node) and $M$ is a saddle point.
The effective EoS parameter is defined as
\begin{align}
  w_{\rm eff} \equiv \frac{p_{\rm tot}}{\rho_{\rm tot}} \,.
\end{align}
For our $\Lambda$CDM model it reads
\begin{align}
  w_{\rm eff} = \frac{\Omega_{\rm r}}{3}-\Omega_\Lambda=-1+x+ \frac{4}{3}y \,,
\end{align}
and takes the values $-1$ in $O$, $1/3$ in $R$ and $0$ in $M$, meaning that these points correspond to a cosmological constant (dark energy) dominated universe, a radiation dominated universe and a matter dominated universe, respectively. For the sake of simplicity all the properties of the critical points have been summarised in Tab.~\ref{tab:twofluidLambda}.

The physical phase space for the system (\ref{043})--(\ref{044}) has been plotted in Fig.~\ref{fig:twofluidLambda}. Point~$R$ is clearly the past attractor while the origin represents the future attractor. Every solution is thus a heteroclinic orbit starting from $R$ as $\eta\rightarrow-\infty$ and ending in $O$ as $\eta\rightarrow+\infty$. The only exceptions are the heteroclinic orbits on the $x$-axis and the $y=1-x$ line, which connect $M$ to $O$ and $R$ to $M$, respectively. However these trajectories correspond either to a vanishing cosmological constant or to a universe without any radiation. From Fig.~\ref{fig:twofluidLambda} it is clear that for every initial condition in the physical phase space, the universe was radiation dominated as $a\rightarrow 0$ and dark energy dominated as $a\rightarrow+\infty$. The yellow/shaded region in Fig.~\ref{fig:twofluidLambda} denotes the area of the phase space where $w_{\rm eff}<-1/3$, i.e.~where the universe undergoes an accelerated expansion\footnote{The effective EoS parameter is connected to the acceleration equation (\ref{eq:Raychaudhuri}) as $\ddot{a}/a = -\kappa^2\rho(1+3w_{\rm eff})/6$ where $\rho=\rho_{\rm m}+\rho_{\rm r}+\Lambda/\kappa^2$ is the total energy density. It is then clear that whenever $w_{\rm eff}<-1/3$ the universe undergoes accelerated expansion ($\ddot a>0$).}. All the trajectories will eventually enter this region so that the radiation (or matter) to dark energy transition always happens at some moment in the history of the universe.

Note that there is also an heteroclinic sequence connecting $R\rightarrow M\rightarrow O$. This heteroclinic sequence is of fundamental importance since it is the path our Universe follows. We can understand it with the following reasoning. In Fig.~\ref{fig:twofluidLambda} the line corresponding to a vanishing cosmological constant is the $y=1-x$ line connecting $R$ with $M$. Since the measured value of the cosmological constant is actually positive but extremely small (see (\ref{num:Lambda})), we expect that the evolution of our universe corresponds to a trajectory passing exceptionally close to the $y=1-x$ line. Such a solution will shadow the heteroclinic orbit $R\rightarrow M\rightarrow O$, implying a universe which undergoes first a radiation dominated phase, followed by a matter dominated phase and finally reaching a dark energy dominated phase. This is exactly the expected behaviour of the observed universe which is well modelled by a universe filled with radiation, dark matter and a small positive cosmological constant. Such a theoretical description of the universe in known as the {\it $\Lambda$CDM model} after the cosmological constant $\Lambda$ and the cold (non-relativistic) dark matter fluid.

\begin{figure}
\begin{center}
\includegraphics[width=0.8\textwidth]{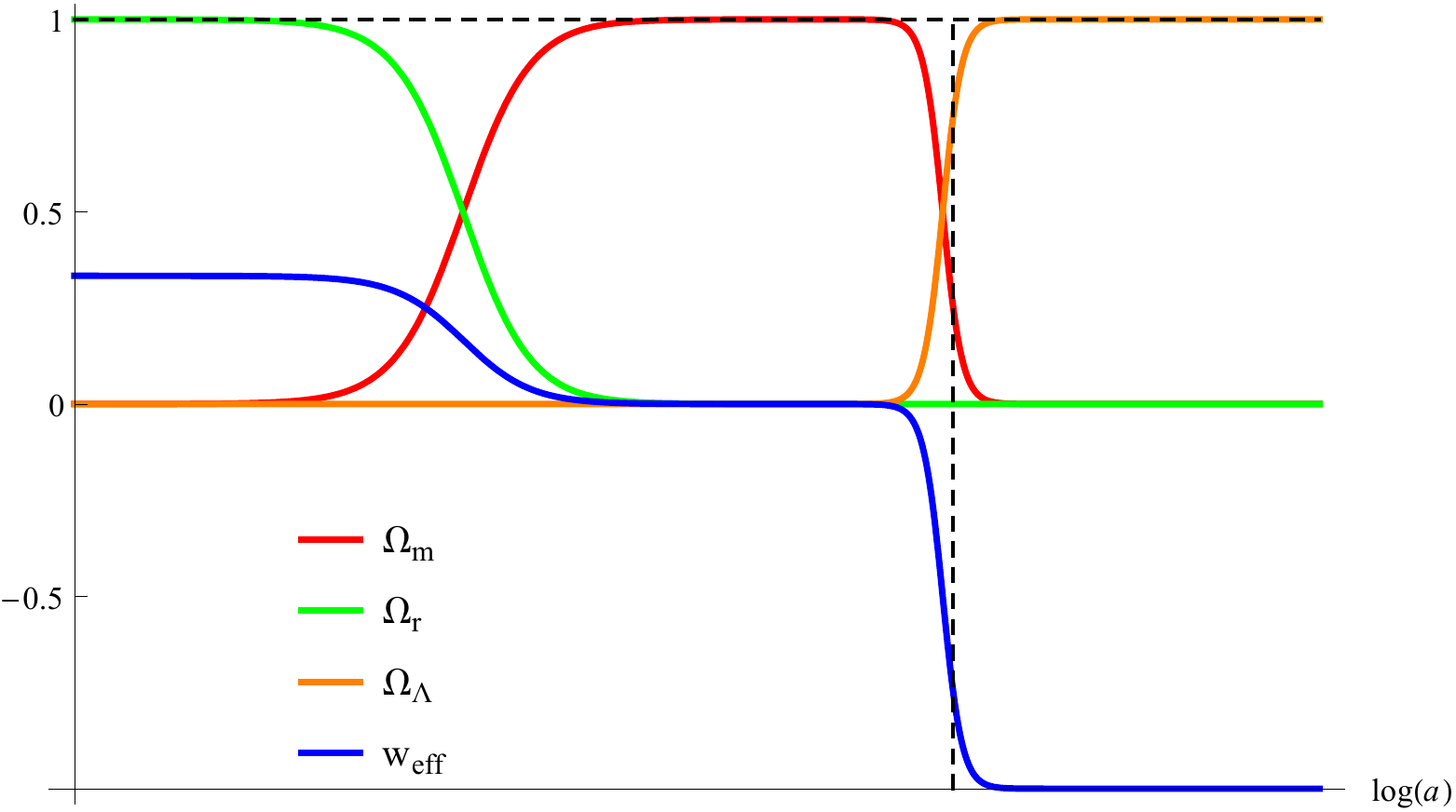}
\end{center}
\caption{Evolution of the relative energy density of dark matter ($\Omega_{\rm m}$), radiation ($\Omega_{\rm r}$) and dark energy ($\Omega_\Lambda$), together with the effective EoS parameter ($w_{\rm eff}$) in the $\Lambda$CDM model. The vertical dashed line indicates the present cosmological time.}
\label{fig:twofluidLambdaOmega}
\end{figure}

The relative energy densities of dark matter ($\Omega_{\rm m}$), radiation ($\Omega_{\rm r}$) and dark energy ($\Omega_\Lambda$) have been plotted in Fig~\ref{fig:twofluidLambdaOmega}, together with the effective EoS of the universe ($w_{\rm eff}$), for a solution shadowing the $R\rightarrow M\rightarrow O$ heteroclinic orbit. As we can see from the picture, at early times radiation dominates, then there is a transient period of dark matter domination and eventually the universe becomes dominated by the cosmological constant. Note the vertical dashed line denoting the present cosmological time. As suggested by the observations, today we are in the transition period between dark matter and dark energy domination. The relative energy density of dark energy is indeed around $0.7$, while the remaining $0.3$ is composed by dark (and baryonic) matter. The effective EoS parameter starts from the radiation value of $1/3$, drops to $0$ during the matter dominated era and eventually reaches $-1$ as the effects of dark energy becomes important.

To conclude we have seen in this section that adding a simple cosmological constant term to the Einstein field equations leads to the desired cosmological acceleration at late times. However it is questionable why such a constant must possess its extremely small measured value. From Fig.~\ref{fig:twofluidLambda} one can immediately notice that a slightly greater value for $\Lambda$, corresponding to trajectories more distant from the $y=1-x$ line, would immediately lead to a fast transition from radiation to dark energy domination without allowing the intermediate matter epoch to happen. This would result in a completely different universe where all the cosmological structure, and thus also life as we know it, would be absent. The problem with the observed value of the cosmological constant, as we are going to see in Sec.~\ref{sec:cosmoconstpbm}, is an issue which does not yet have a satisfactory solution from a theoretical point of view.

\subsection{Standard cosmology beyond spatial flatness}
\label{sec:standard_cosmology_beyond_spatial_flatness}

In the analysis of the $\Lambda$CDM model given above we have neglected the presence of spatial curvature. This choice is usually justified in literature by the fact that observations point towards a very small amount of spatial curvature in the observed universe (see e.g.~\citet{Adam:2015rua}). Here we will show how dynamical systems can be used to consider the effect of the presence of  spatial curvature in the $\Lambda$CDM model. Spatial curvature has also an important role in modified theories of gravity which we will treat in Sec.~\ref{chap:modifiedgrav}, where however we will limit ourselves to consider only flat cosmologies, providing also the references in which the non-spatially flat generalisation can be found. The very first treatment of the $\Lambda$CDM model with expansion normalised variables (cf.~Sec.~\ref{sec:choosing-variables}) was given in \citet{Goliath:1998na}. Since the treatment has interesting mathematical and physical aspects, we will follow it here briefly.

Starting from the cosmological equations (\ref{eq:Friedmann}) and (\ref{eq:accel}), adding the cosmological constant and considering only a single additional fluid with energy density $\rho$ and barotropic index $w$ suggests the definition of the EN variables 
\begin{align}
  x=\Omega= \frac{\kappa^2\rho}{3H^2} \qquad\mbox{and}\qquad 
  y= \Omega_k= \frac{k}{a^2H^2} \,.
  \label{eq:defxykOmega}
\end{align}
The Friedmann constraint now reads
\begin{align}
1=x-y+ \Omega_\Lambda \,,
\label{039k}
\end{align}
and as before can be used to replace $\Omega_\Lambda$ in terms of $x$ and $y$ everywhere in the equations that follow.  Let us consider this case first.  The dynamical equations  read
\begin{align}
x'&= x [3 (w+1) (x-1)-2 y]\,,\label{043k}\\
y'&= y [3 (w+1) x-2 (y+1)] \,,\label{044k}
\end{align}
where again a prime denotes differentiation with respect to $\eta = \log a$. As for its spatially flat counterpart the system presents two invariant submanifolds. Therefore if we assume $y<0$ (i.e.~$k<0$), the physical phase space is {\em compact}, i.e.~all the variables are defined on a compact interval ($0\leq x\leq1$, $-1\leq y\leq 0$, $0\leq \Omega_\Lambda\leq 1$). It is not difficult to show that  this system presents three fixed points as given in Tab. \ref{tab:LCDM_k1}. They respectively correspond to the domination over the others of each of the dynamical variables. Their nature is clear. The first represents a Friedmann evolution (matter domination with EoS $p = w \rho$), the second a Milne universe (curvature domination) and the third a de Sitter solution. The stability of the fixed points gives information on the global behaviour of these types of cosmologies. As expected, only one attractor appears, Point~$dS$, which is global.

What about the case in which $k=1$? In this case the phase space is not compact and a simple analysis of the stability of the finite fixed point is not sufficient to characterise the global flow of the orbits. In fact there could be fixed points in the asymptotic part of the phase space which might be global attractors. There are a number of approaches that can be used to tackle this problem, for example the Poincar\'{e} projection methods described in Sec.~\ref{sec:poincarecomp}. The analysis at infinity will be a major issue in Sec.~\ref{chap:modifiedgrav}. In this case the problem can be solved easily defining the new set of variables:
\begin{align}
\bar{x}= \frac{H}{\bar\eta},\quad\quad\bar{y}=\frac{\rho}{3 \bar{\eta}^2}, \quad\quad \bar{z}=\frac{\Lambda}{3 \bar{\eta}^2}\, \quad\mbox{where}\quad \bar{\eta}=\sqrt{H^2+\frac{k}{a^2}}\quad \mbox{and}\quad k=1 \,.
\label{eq:defxykOmega2}
\end{align}
Setting a new time variable  so that $x'=\bar\eta^{-1} \frac{d x}{d t}$ the cosmological equations are equivalent to the dynamical equations
\begin{align}
\bar{x}'&= (1-\bar{x}^2) \left[1- \frac{3}{2}(w-1)(1-\bar{z})\right]\,,\label{043k2}\\
\bar{z}'&= 3\bar{z}\bar{x}  (w+1) (1-\bar{z}) \,,\label{044k2}\\
\bar{y}&=1-\bar{z} \,.
\end{align}
The system admits five fixed points, four of which correspond to the Friedmann and Milne solutions, however, since $H$ is now a dynamical variable, they appear in both the expanding and contracting versions. In addition to these points, an additional point $E$ appears which satisfies $H=0$, i.e.~the \textit{static Einstein universe}. Tab.~\ref{tab:LCDM_k2} summarises these results.

\begin{figure}
\centering
\includegraphics[width=0.8\columnwidth]{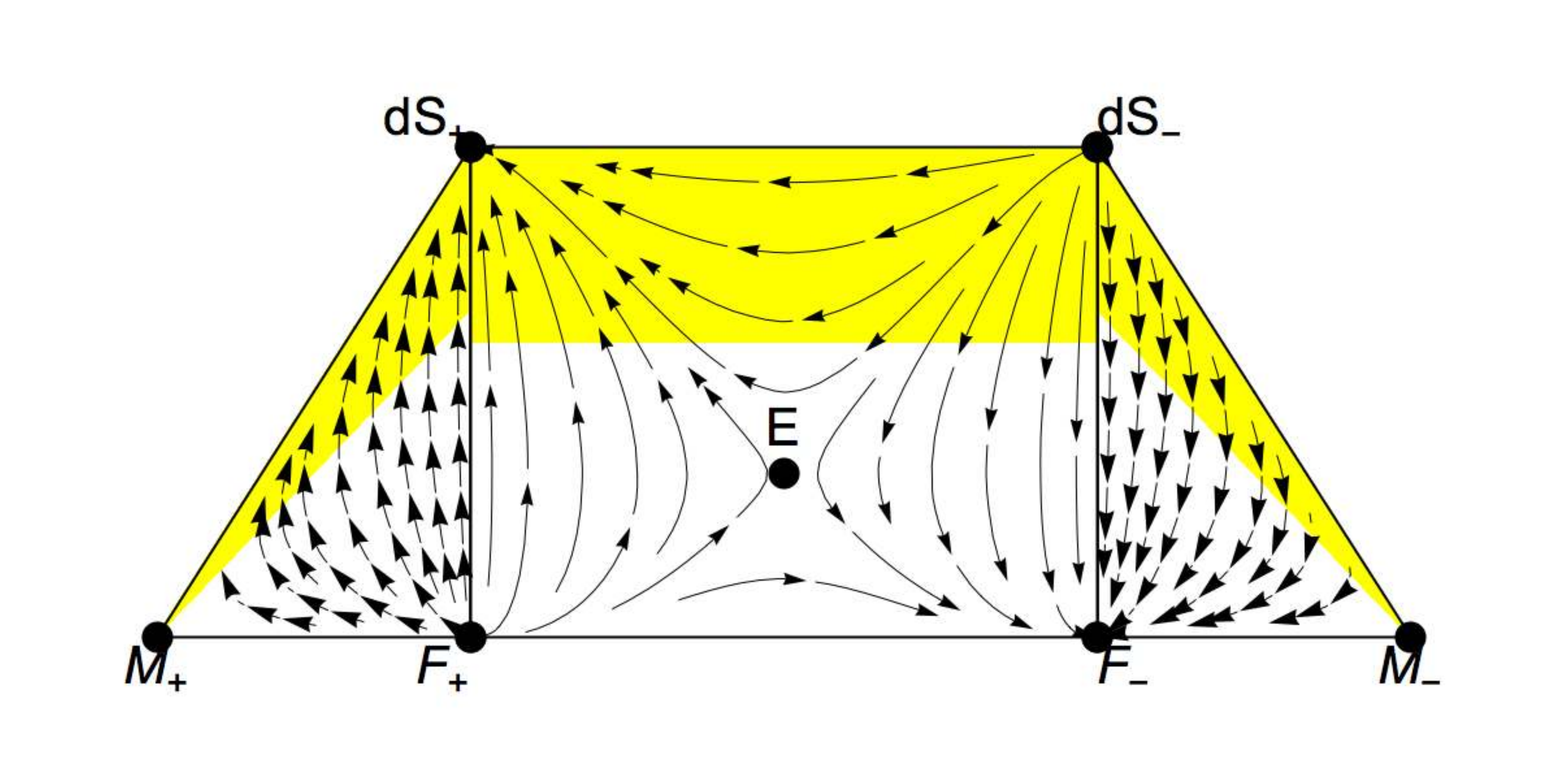}
\caption{Global phase space of the spatially curved $\Lambda$CDM model in the single fluid case \citep{Goliath:1998na}. The global space has been obtained by patching the $k<0$ (rectangular part) and the $k>0$ phase spaces (triangular parts) along the $k=0$ invariant submanifold that connects the points $F$ and $dS$. Note that the different sections of this plot represent different phase spaces defined by different variables. For this reason the yellow area representing the accelerating regime is not delimited by a  smooth boundary.}
\label{fig:PlotLCDM-k}
\end{figure}

Since $k=0$ is an invariant submanifold we can  glue together these phase spaces along these manifolds. The results are given in Fig.~\ref{fig:PlotLCDM-k}.  It is clear that the spatial curvature influences significantly the evolution of the cosmology giving rise to new types of cosmic histories. For example, for $k=1$ one obtains ``loitering'' solutions, i.e.~cosmologies in which the expansion goes through an almost static phase and successively starts to expand exponentially.

There are other ways to explore the role of the spatial curvature. One interesting approach is based on the realisation that the dynamical equations (\ref{043k})--(\ref{044k}) represent a Lokta-Volterra system. \citet{Perez:2013zya} considered the case above in great generality including a number of different perfect fluids interacting with each other. Their result shows that non-trivial phenomena may arise like limit cycles and in some cases even chaos.

\begin{table}
\begin{center}
\begin{tabular}{|c|c|c|c|c|c|c|}
 \hline
 Point & $\{x,y, \Omega_\Lambda\}$  & Solution &Stability \\
 \hline  &&& \\
 $F$ & $\left\{1,0,0\right\}$ & Friedmann  &  Repeller\\ & & & \\
 $M$ & $\left\{0,-1,0\right\}$ & Milne  &  Saddle\\ & & & \\
 $dS$ & $\left\{0,0,1\right\}$ &  deSitter &  Attractor\\ & & & \\
  \hline
\end{tabular}
\end{center}
\caption{Critical points of the negatively spatially curved $\Lambda$CDM model.}
\label{tab:LCDM_k1}
\end{table}

\begin{table}
\begin{center}
\begin{tabular}{|c|c|c|c|c|c|c|}
 \hline
 Point & $\{\bar{x},\bar{y},\bar{z}\}$  & Solution & Stability \\
 \hline  &&& \\
 $F_{+}$ & $\left\{1,1,0\right\}$ & Expanding Friedmann  &  Repeller\\ & & & \\
 $F_{-}$ & $\left\{-1,1,0\right\}$ &  Contracting Friedmann  &  Attractor\\ & & & \\
 $dS_{+}$ & $\left\{1,0,1\right\}$ & Expanding  deSitter &  Attractor\\ & & & \\
 $dS_{-}$ & $\left\{-1,0,1\right\}$ & Contracting deSitter &  Repeller\\ & & & \\
  $E$ & $\left\{0,\frac{2}{3 (w+1)},\frac{3 w+1}{3 (w+1)}\right\}$ &  Einstein Static &  Attractor\\ & & & \\
  \hline
\end{tabular}
\end{center}
\caption{Critical points of the positively spatially curved $\Lambda$CDM model.}
\label{tab:LCDM_k2}
\end{table}

\subsection{Problems with the cosmological constant}
\label{sec:cosmoconstpbm}

The extremely small value (\ref{num:Lambda}) of the observed effective cosmological constant is at odds with theoretical predictions. It is well known that a non-vanishing cosmological constant brings with it theoretical and philosophical problems at both the classical and quantum level. They arise because we are allowed to identify a cosmological constant term in the Einstein field equations (\ref{eq:EinsteinfieldeqLambda}) with the vacuum energy of (quantum) fields. In what follows we will briefly review the major problems that follows from a positive cosmological constant. For more details we refer the reader to the classic reviews of \citet{Weinberg:1988cp,Weinberg:2000yb}, and more recent reviews by \citet{Carroll:2000fy}, \citet{Martin:2012bt} and \citet{Padilla:2015aaa}.

In Sec.~\ref{sec:cosmoconst} we saw that the cosmological constant term in the Einstein equations (\ref{eq:EinsteinfieldeqLambda}) can be seen as a matter fluid contribution with constant energy density $\rho_\Lambda$ and negative pressure $p_\Lambda=-\rho_\Lambda$. The same type of contribution to the right hand side of the Einstein field equations arises from the {\it vacuum energy} of matter fields. The energy-momentum tensor of a field in its vacuum state $\left|0\right>$ is given by
\begin{align}
\left<0|T_{\mu\nu}|0\right> = -\rho_{\rm vac}\,g_{\mu\nu} \,,
\label{045}
\end{align}
where $\rho_{\rm vac}$ is the {\it constant} energy density of the vacuum.
This can be derived from both classical and quantum mechanical considerations. From the classical point of view the term (\ref{045}) can be identified with the value of matter fields when they are at rest in their minimal energy state, i.e.~the vacuum state.
However at the quantum level the Heisenberg uncertainty principle prevents the kinetic and potential energies from vanishing at the same time. In fact taking into account the quantum mechanical fluctuations from the zero point energy of the quantum fields, they provide another source of energy which contributes to the Einstein equations with a term of the form (\ref{045}). There are thus two different contributions of the form (\ref{045}) coming from considerations on the vacuum state of matter fields: one is classical and the other quantum mechanical.

Let us first consider the classical contribution. The so-called {\it classical cosmological constant problem} can be understood through a simple scalar field example. The energy-momentum tensor of a scalar field $\phi$ is given by
\begin{align}
T_{\mu\nu}^{(\phi)} = \partial_\mu\phi\partial_\nu\phi -g_{\mu\nu}\left[g^{\alpha\beta}\partial_\alpha\phi\partial_\beta\phi +V(\phi)\right] \,,
\label{046}
\end{align}
where $V(\phi)$ is the self-interacting potential of the scalar field. At the classical level the vacuum state corresponds to the state of minimum energy where the kinetic energy of the field vanishes and the potential takes its minimum value $V_{\rm min}$. This means that in the vacuum state the energy-momentum tensor (\ref{046}) takes the form
\begin{align}
\left<0\right|T_{\mu\nu}^{(\phi)}\left|0\right> = -V_{\min}\,g_{\mu\nu} \,,
\label{047}
\end{align}
which indeed matches (\ref{045}) since $V_{\rm min}$ is constant. Every matter field whose vacuum energy does not vanish, will source the Einstein field equations with a term of the form (\ref{047}). In the Standard Model of particle physics a non-vanishing value of the vacuum energy is present after (or before) a {\it (symmetry breaking) phase transition}. Without going into the details we mention that in the Standard Model we expect two such possible phase transitions: the Electro-Weak phase transition and the QCD phase transition. The first one leads to a value of the (Higgs field's) vacuum energy density of
\begin{align}
\rho_{\rm vac}^{\rm EW} \simeq 10^8\, {\rm GeV}^4 \,,
\label{048}
\end{align}
while the vacuum energy coming from QCD is
\begin{align}
\rho_{\rm vac}^{\rm QCD} \simeq 10^{-2}\, {\rm GeV}^4 \,.
\label{049}
\end{align}
These two values should be added and compared with the measured value of $\rho_\Lambda$, which, in GeV units, is
\begin{align}
\rho_\Lambda \simeq 10^{-47}\, {\rm GeV}^4 \,.
\label{050}
\end{align}
Comparing (\ref{048}) and (\ref{049}) with the measured value (\ref{050}) immediately gives the severity of the problem we are facing with. The observed value of $\rho_\Lambda$ is 55 and 45 orders of magnitude away from the numbers predicted by the Electro-Weak and QCD phase transitions, respectively. Theoretically this is a catastrophe since the predicted vacuum energy is so high that it should have been observed a long time ago and would have led to a completely different cosmology from the one we experience.

The problem is not ameliorated if quantum considerations are included, as we will now show. As we have argued quantum fluctuations in the vacuum state of matter fields contribute a sourcing term of the form (\ref{045})  in Einstein's field equations. This effect also needs to be added when evaluating the net value of the cosmological constant. The contribution arising from these quantum fluctuations leads to what is called the {\it quantum cosmological constant problem}. Again we will explain the problem using a scalar field as example.
Consider a massive ($V=m^2\phi^2/2$) scalar field in Minkowski spacetime. From quantum-mechanical considerations, the energy density of the field in its vacuum state is given by
\begin{align}
  \rho_{\rm vac}^{\rm QM} = \left<0\right|\rho_\phi\left|0\right> = 
  \frac{1}{2(2\pi)^3} \int \sqrt{k^2+m^2}\, d^3k \,,
\label{051}
\end{align}
with the integral performed over all 3-dimensional momentum space. Clearly the integral diverges and the energy result is formally infinite. This is however the kind of divergence that in quantum field theory can be handled with the concept of {\it renormalisation}. There are various techniques that can be employed to regularise the integral (\ref{051}), but the one working properly in our case is {\it dimensional regularisation}\footnote{Introducing a cut-off at some higher energy breaks Lorentz invariance and thus leads to a wrong result \citep{Martin:2012bt}.}. Without going into the details, this procedure gives the result
\begin{align}
\rho_{\rm vac}^{\rm QM}= \frac{m^4}{64\pi^2}\log\left(\frac{m^2}{\mu^2}\right) \,,
\label{052}
\end{align}
where $\mu$ is a constant scale introduced to fix the dimensionality of the equation. All massive matter fields in the universe contribute with a term similar to (\ref{052}) in the vacuum energy. Summing the contribution from all the particles of the Standard Model, and choosing a suitable value for $\mu$ \citep{Martin:2012bt}, gives the number
\begin{align}
\rho_{\rm vac}^{\rm QM} \simeq - 10^{8}\, {\rm GeV}^4 \,,
\end{align}
which, regardless of the sign, is still 55 order of magnitude away from the measured value (\ref{050}). So from the quantum side of matter fields we predict another contribution which completely disagrees with observations.

Of course, since physically and mathematically nothing prevents it, one can also suppose that a {\it bare} cosmological constant $\Lambda_{\rm B}$ is present in the Einstein field equations and it adds its energy contribution $\rho_{\rm B}$ to the vacuum energies we have just computed. This implies that in general the total vacuum energy will be given by
\begin{align}
\rho_\Lambda = \rho_{\rm B} + \rho_{\rm vac}^{\rm QM} +\rho_{\rm vac}^{\rm EW} +\rho_{\rm vac}^{\rm QCD} + \dots \,,
\label{053}
\end{align}
where for completeness every contribution from either unknown phase transitions or quantum fluctuations of particles beyond the Standard Model should be added. We know the measured value of $\rho_\Lambda$ and have estimated the values of all the other vacuum energy appearing on the right hand side of (\ref{053}) except for $\rho_{\rm B}$. According to quantum field theory, the value of $\rho_{\rm B}$ cannot be determined from theoretical arguments and it is a number that must be chosen in order to renormalise $\rho_\Lambda$ to let it agree with experiments. In our case this means that $\rho_{\rm B}$ must be chosen such that it cancels all the other vacuum energy contributions on the right hand side of (\ref{053}) leaving exactly the small measured value on the left hand side. This is clearly absurd since one should adjust $\rho_{\rm B}$ up to fifty orders of magnitude or more. In these terms the cosmological constant problem is nothing but a problem of {\it fine tuning}. It makes no sense to assume that the bare cosmological constant is exactly the one needed to cancel all the vacuum energy sources giving only the small amount we need to match the observations.

The issues related to the vacuum energy of matter fields are not the only problems plaguing the cosmological constant. As we have seen in Sec.~\ref{sec:cosmoconst}, even if we put aside all the theoretical explanations for $\Lambda$, an extremely small value of the cosmological constant is needed in order to have a sufficiently long period of matter domination during the history of the universe. A slightly bigger value of $\Lambda$ would lead to a direct transition from radiation to dark energy domination, preventing in this way the formation of galaxies, stars and all other cosmic structures. Moreover the observed value of the cosmological constant is the right one needed for the transition from dark matter to dark energy to happen exactly today, i.e.~during the relatively small time when humanity has evolved. This is known as the {\it cosmic coincidence problem}. It can equivalently be formulated as follows: how is it possible that we are observing the universe exactly when the relative energy densities of dark energy and dark matter are comparable in magnitude? If we take a look back to Fig.~\ref{fig:twofluidLambdaOmega} it is easy to realise that the present cosmological time, denoted by the vertical dashed line, could be placed anywhere\footnote{Of course life as we know needs an Earth-like planet to prosper and thus the human race could have appeared only after (dark) matter dominated for a sufficiently long time.} in the history of the universe and there is no apparent reason for it to be exactly where the transition from matter to dark energy domination happens. If the cosmological constant dominated phase is the final state of our universe, it was much more likely that we would have lived during a period of dark energy domination rather than matter domination, or even less probably, exactly during the transition phase. This rather philosophical problem is not specific to the cosmological constant, though, as we will see in the following sections, in other models of dark energy it could be relaxed.

%% file: chapters/04_quint/quint.tex
\section{Quintessence: dark energy from a canonical scalar field}
\label{chap:scalarfields}

The cosmology of minimally coupled, canonical scalar fields is the main subject of this section. We will discuss the main dynamical features of the background cosmology of a canonical scalar field with a self-interacting potential and illustrate a choice of suitable dimensionless variables for the dynamical systems analysis. In contrast to the other sections in this review, we will provide here a complete dynamical systems analysis  of the specific cases of quintessence with an exponential and power-law potential, outlining the interesting phenomenological properties of these models at both early and late times. This is motivated by the fact that  on one hand these two examples represent the most relevant canonical scalar field models of dark energy; and on the other hand that they offer an easy framework to illustrate in a pedagogical way some of the more advanced techniques outlined in Sec.~\ref{sec:dynamicalsystems}.

\subsection{Dark energy as a canonical scalar field}
\label{sec:quintessence}

Let us begin by considering a scalar field minimally coupled to gravity. The action which will then represent our physical system is
\begin{align}
  S = \int d^4x \sqrt{-g}\left(\frac{R}{2\kappa^2}+\mathcal{L}_{\rm m}+\mathcal{L}_\phi\right) \,,
  \label{eq:action_R_Lm_Lphi}
\end{align}
where $\mathcal{L}_\phi$ is the {\it canonical Lagrangian of a scalar field} $\phi$ given by
\begin{align}
  \mathcal{L}_\phi = -\frac{1}{2}g^{\mu\nu}\partial_\mu\phi\partial_\nu\phi -V(\phi) \,,
  \label{def:can_scalar_field_lagrangian}
\end{align}
with $V(\phi)$ being a general self-coupling potential for $\phi$ which must be positive for physically acceptable fields. The variation with respect to $g_{\mu\nu}$ yields the gravitational field equations
\begin{align}
  R_{\mu\nu}-\frac{1}{2}g_{\mu\nu}R = \kappa^2 \left(T_{\mu\nu}+T_{\mu\nu}^{(\phi)}\right) \,,
  \label{eq:EinsteineqsPhi}
\end{align}
where
\begin{align}
  T_{\mu\nu}^{(\phi)}=\partial_\mu\phi\partial_\nu\phi -\frac{1}{2}g_{\mu\nu}\partial\phi^2 -g_{\mu\nu}V(\phi) \,,
  \label{def:EM_tensor_scalar_field}
\end{align}
is the energy-momentum tensor of the scalar field and we introduce the notation
\begin{align}
  \partial\phi^2 \equiv g_{\alpha\beta} \partial^\alpha\phi\partial^\beta\phi \,,
\label{eq:definition-delsq}
\end{align}
which will be used throughout the review.
The variation with respect to $\phi$ gives the {\it Klein-Gordon equation}
\begin{align}
  \Box\phi-V_{,\phi}=0\,,
  \label{eq:KGeq}
\end{align}
with $\Box\phi=\nabla_\mu\nabla^\mu\phi$ and $V_{,\phi}=\partial V/\partial \phi$.

For cosmological applications we consider the homogeneous, isotropic, spatially flat ($k=0$) FLRW metric (\ref{eq:FRWmetric}), which in Cartesian coordinates reads
\begin{align}
  ds^2 = -dt^2 + a^2(t) \left(dx^2+dy^2+dz^2\right) \,.
\end{align}
Furthermore, the linear equation of state (EoS) $p=w\rho$ is assumed for the matter field.
With these assumptions the Einstein field equations (\ref{eq:EinsteineqsPhi}) reduce to the following Friedmann and acceleration equations
\begin{align}
  3H^2 &= \kappa^2 \left(\rho+\frac{1}{2}\dot\phi^2+V\right) \,,\label{eq:FriedmannPhi}\\
  2\dot H+3H^2 &= -\kappa^2 \left(w\rho+\frac{1}{2}\dot\phi^2-V\right) \,,\label{eq:accPhi}
\end{align}
while the Klein-Gordon equation (\ref{eq:KGeq}) takes the simple form
\begin{align}
  \ddot\phi+3H\dot\phi +V_{,\phi} = 0 \,.
  \label{eq:KGFRW}
\end{align}
We can define the energy density and pressure of the scalar field as follows
\begin{align}
  \rho_{\phi} &= \frac{1}{2} \dot{\phi}^2+V \,,
  \label{def:energy_scalar_field} \\
  p_{\phi} &= \frac{1}{2}\dot{\phi}^2-V \,,
  \label{def:pressure_scalar_field}
\end{align}
so that its equation of state becomes
\begin{align}
  w_\phi =\frac{p_\phi}{\rho_\phi} = \frac{\frac{1}{2}\dot\phi^2-V}{\frac{1}{2}\dot\phi^2+V} \,.
  \label{eq:EoSPhi}
\end{align}
Note that $w_\phi$ is a dynamically evolving parameter which can take values in the range $[-1,1]$. Whenever the potential energy $V$ dominates over the kinetic energy $\dot\phi^2/2$ the EoS (\ref{eq:EoSPhi}) becomes $w_\phi=-1$, recovering in this way a cosmological constant EoS capable of accelerating the universe. This is the feature that renders a canonical scalar field the simplest dynamical framework for describing dark energy.
On the other hand, if $V \ll \dot\phi^2/2$, then $w_\phi \simeq 1$.
Models based on a canonical scalar field for explaining the late time cosmic acceleration, are collectively denoted with the name {\it quintessence} and different models are distinguished by the form of their potential $V$.

\subsection{Dynamical systems approach: choosing variables}
\label{sec:choosing-variables}

Since we are concerned with possible dynamical systems applications to such models, we need first to rewrite the cosmological equations (\ref{eq:FriedmannPhi}), (\ref{eq:accPhi}) and (\ref{eq:KGFRW}) into an autonomous system of equations. In general there are many possible ways to achieve this task, but the most common one is to introduce the EN  variables\footnote{Note that the definitions of $x$ and $y$ are not the same as in (\ref{eq:defxyOmega}). For applications of dynamical systems to scalar field cosmology using different variables see e.g.~\citet{Halliwell:1986ja,Faraoni:2012bf}. The EN variables (\ref{def:ENvars}) are of physical interest since the energy densities of matter and dark energy can be easily visualised in terms of them.}
\begin{align}
  x=\frac{\kappa\dot\phi}{\sqrt{6}H} \quad\mbox{and}\quad y=\frac{\kappa\sqrt{V}}{\sqrt{3}H} \,.
	\label{def:ENvars}
\end{align}
For a scalar field in the presence of barotropic matter, they were first introduced in a seminal paper by \citet{Copeland:1997et}. Note that the definition above assumes that we are dealing with a positive defined scalar field potential. If this is not the case one can define
\begin{align}
  y=\frac{\kappa\sqrt{|V|}}{\sqrt{3}H}\,.
  \label{sec4:eqydef}
\end{align}
Hence, $y$ can be considered of fixed sign on a given orbit\footnote{The case of potentials which switch signs requires, of course, a different choice of variable. On the other hand, physically motivated potentials normally have a constant sign. For this reason we will not consider this case here. }. Using the EN variables, the Friedmann equation (\ref{eq:FriedmannPhi}) can be rewritten as
\begin{align}
  1=\Omega_{\rm m}+x^2+y^2 \,,
  \label{eq:Friedmann_constr_xy}
\end{align}
where $\Omega_{\rm m}=\kappa^2\rho/(3H^2)$ is the {\it relative energy density of matter}. From~(\ref{eq:Friedmann_constr_xy}) the meaning of the EN variables (\ref{def:ENvars}) is clear: $x^2$ stands for the relative kinetic energy density of $\phi$ while $y^2$ stands for its relative amount of potential energy density. The total relative energy density of the scalar field is given by
\begin{align}
  \Omega_\phi = x^2+y^2 \,,
  \label{eq:rel_energy_phi}
\end{align}
while its EoS \eqref{eq:EoSPhi} becomes
\begin{align}
  w_\phi = \frac{x^2-y^2}{x^2+y^2} \,.
   \label{eq:EoS_phi_xy}
\end{align}
It is now easy to see that in the limit $x\ll y$ one obtains $w_\phi\simeq -1$ while for $x \gg y$ one obtains $w_\phi\simeq 1$. From~(\ref{eq:rel_energy_phi}) it is also clear that the further away we are from the origin on the $(x,y)$-plane, the higher is the energy of the scalar field, with the origin corresponding to a completely matter dominated universe ($\Omega_{\rm m}=1$). The quantities $\Omega_\phi$ and $w_\phi$ are sometimes used as dynamical variables to replace the EN ones (see e.g.~\citet{Scherrer:2007pu,Gong:2014dia,Fang:2014bta,Qi:2015pqa}). The transformation $(x,y)\mapsto(\Omega_\phi,w_\phi)$ however is not convenient for mathematical and computational reasons and it is usually employed only to better parametrise the dynamical properties of dark energy when comparison with observational data is performed.

We define the effective EoS parameter of the universe as
\begin{align}
  w_{\rm eff} \equiv \frac{p_{\rm tot}}{\rho_{\rm tot}} = \frac{p+p_\phi}{\rho+\rho_\phi} = w\Omega_{\rm m}+w_\phi\Omega_\phi \,,
\end{align}
which in the EN variables reads
\begin{align}
	w_{\rm eff} = x^2-y^2 + w \left(1-x^2-y^2\right) \,.
   \label{eq:eff_EoS}
\end{align}
The effective EoS parameter $w_{\rm eff}$ is of fundamental importance because it tells us whether the universe undergoes an accelerating ($w_{\rm eff}<-1/3$) or decelerating ($w_{\rm eff}>-1/3$) expansion. For example, if $x=y=0$ then $w_{\rm eff}=w$ and the universe is matter dominated, if $x=1$ and $y=0$ then $w_{\rm eff}=1$ and the universe is dominated by the kinetic energy of the scalar field which behaves as a {\it stiff matter fluid} ($w=1$), finally if $x=0$ and $y=1$ then $w_{\rm eff}=-1$ and the universe is dominated by the potential energy of the scalar field which behaves as an effective cosmological constant driving an accelerated expansion.

Since the energy density of matter fields $\rho$ is always positive, we also have that $\Omega_{\rm m}>0$. This implies that the EN variables must satisfy the constraint
\begin{align}
	0\leq x^2+y^2 = 1-\Omega_{\rm m} \leq 1 \,,
	\label{eq:xy_canonical_constraint}
\end{align}
for physically viable solutions. In the $(x,y)$-plane, the constraint (\ref{eq:xy_canonical_constraint}) reduces the phase space of physically sensible trajectories to the unit disc centred at the origin\footnote{See \citet{Roy:2013wqa} for the analysis in polar coordinates.}. If we add also the fact that $y>0$ then the {\it physical phase space}\footnote{The physical phase space is the invariant set composed by physically meaningful orbits of the phase space; see Sec.~\ref{sec:cosmoconst}.} reduced to $(x,y)$-planes is represented by the positive $y$ half-unit disk centred at the origin. Note that points on the unit circle correspond to scalar field dominated universes ($\Omega_\phi=1$). The Friedmann constraint (\ref{eq:Friedmann_constr_xy}) can also be used to replace $\Omega_{\rm m}$ in favour of $x$ and $y$ in the following equations, reducing in this way the dimensionality of the phase space. Furthermore from the acceleration equation (\ref{eq:accPhi}) we obtain
\begin{align}
\frac{\dot H}{H^2}=\frac{3}{2} \left[(w-1) x^2+(w+1) \left(y^2-1\right)\right]
\label{055} \,,
\end{align}
which at any fixed point~$(x_*,y_*)$ of the phase space can be integrated to give
\begin{align}
a\propto (t-t_0)^{\frac{2}{3[(w+1)(1-x_*^2-y_*^2)+2x_*^2]}}\,,
\label{056}
\end{align}
where $t_0$ is a constant of integration.
This corresponds to a power-law solution, i.e.~a solution for which the scale factor $a$ evolves as a power of the cosmological time $t$. Again if $x_*=0$ and $y_*=0$ the universe is matter dominated and its evolution coincides with the standard $w$-dependent scaling solution (\ref{eq:mattersolw}). If $x_*=0$ and $y_*=1$ the denominator of (\ref{056}) vanishes and the universe undergoes a de Sitter expansion as can be seen from (\ref{055}) which forces $H$ to be constant. Solution (\ref{056}) shows us that at any critical point of the phase space, the universe expands according to a power-law evolution. This means that even if we are not able to derive its complete evolution analytically, its asymptotic behaviour, provided it is given by critical points, will always be well characterised.

Employing the EN variables (\ref{def:ENvars}), from the acceleration equation (\ref{eq:accPhi}) and the scalar field equation (\ref{eq:KGFRW}) we can derive the following dynamical system\footnote{The derivation of these equations can proceed as follows. First take the derivative of $x$ and $y$ with respect to $d\eta=Hdt$, and then replace $\dot H$ using Eq.~(\ref{eq:accPhi}) and $\ddot\phi$ using Eq.~(\ref{eq:KGFRW}). Finally rewrite everything in terms of the variables $x$, $y$ and $\lambda$.}
\begin{align}
	x'&=-\frac{3}{2} \left[2x+(w-1) x^3+ x (w+1) \left( y^2-1\right)-\frac{\sqrt{2}}{\sqrt{3}} \lambda  y^2\right]\,, \label{eq:x_can_scalar_field_gen_V}\\
	y'&=-\frac{3}{2} y \left[(w-1) x^2+(w+1) \left(y^2-1\right)+\frac{\sqrt{2}}{\sqrt{3}} \lambda  x\right]\,,\label{eq:y_can_scalar_field_gen_V}
\end{align}
where a prime denotes differentiation with respect to $\eta=\log a$ and we have defined
\begin{align}
	\lambda = -\frac{V_{,\phi}}{\kappa V} \,.
	\label{def:lambda_dynamical}
\end{align}
Note that the EN variables fail to close the system of equations to an autonomous system since $\lambda$ still depends upon the scalar field $\phi$. In fact the EN variables were first introduced to study a scalar field with an exponential potential \citep{Copeland:1997et} for which $\lambda$ is indeed just a parameter and the system (\ref{eq:x_can_scalar_field_gen_V})--(\ref{eq:y_can_scalar_field_gen_V}) becomes autonomous; see Sec.~\ref{sec:exp_potential}. In order to close the system for a general potential we can regard $\lambda$ as another dynamical variable and look for an evolution equation governing its dynamics. This approach was first considered by \citet{Steinhardt:1999nw} and \citet{delaMacorra:1999ff}, and it has been pursued with the use of dynamical system techniques since the work of \citet{Ng:2001hs}.
The equation for the variable $\lambda$ follows from its definition and is given by
\begin{align}
	\lambda' = -\sqrt{6}\left(\Gamma-1\right)\lambda^2x \,,
	\label{eq:lambda}
\end{align}
where
\begin{align}
	\Gamma= \frac{V\,V_{,\phi\phi}}{V_{,\phi}^2} \,.
	\label{def:Gamma}
\end{align}
At first it seems that we gain nothing from this new equation since we still have a quantity ($\Gamma$) which explicitly depends on the scalar field $\phi$. However since both $\lambda$ and $\Gamma$ are functions of $\phi$, it is in principle possible to relate one to the other \citep{Zhou:2007xp,Fang:2008fw}. In other words, provided that the function $\lambda(\phi)$ is invertible so that we can obtain $\phi(\lambda)$, we can write $\Gamma$ as a function of $\lambda$, i.e.~$\Gamma(\phi(\lambda))$. The simplest case is the exponential potential where $\Gamma=1$ and $\lambda$ is a constant. However, also the power-law potential is easily treatable since it leads to a dynamical $\lambda$ but to a constant $\Gamma$. The exponential and power-law potentials will be studied in Secs.~\ref{sec:exp_potential} and \ref{sec:power_law_potential} respectively, while Sec.~\ref{sec:other_potentials} will be devoted to more complicated potentials. Of course, if the function $\lambda(\phi)$ is not invertible, this approach fails to close the equations to an autonomous system. Different choices of variables, which in practice will lead to a phase space with higher dimensions, could better represent the system in such cases.

Note that all the phenomenological properties of the universe, such as the relative energy density of the scalar field (\ref{eq:rel_energy_phi}), the EoS of the scalar field (\ref{eq:EoSPhi}) and the effective EoS (\ref{eq:eff_EoS}), are independent of $\lambda$. This means that different models of quintessence, i.e.~different choices of the potential $V$, do not directly change the physical features of the universe. It is only through the dynamical evolution of the $x$ and $y$ variables that different quintessence models are distinguished from each other. If two potentials lead to the same qualitative evolution of the EN variables, then the universes described by those two models will be physically indistinguishable\footnote{This is only valid at the background level. It might be that the two models give a different dynamics at the level of cosmological perturbations.}.

The dynamical system (\ref{eq:x_can_scalar_field_gen_V})--(\ref{eq:y_can_scalar_field_gen_V}) plus (\ref{eq:lambda}) has the invariant submanifold $y=0$ and is invariant under the transformation
\begin{align}
 	 y\mapsto -y \,,
 	 \label{057}
\end{align}
so even if we drop the $V>0$ assumption, the dynamics for negative values of $y$ would be a copy of the one for positive values. Note also that we are assuming $H>0$ in order to describe an expanding universe. However the dynamics of a contracting universe ($H<0$) would have the same features of our analysis in the negative $y$ region switching the direction of time because of the symmetry (\ref{057}).
On the other hand, provided that $\Gamma$ can be written as a function of $\lambda$ and that $\Gamma(\lambda)=\Gamma(-\lambda)$, the dynamical system (\ref{eq:x_can_scalar_field_gen_V})--(\ref{eq:y_can_scalar_field_gen_V}) plus (\ref{eq:lambda}) is also invariant under the simultaneous transformation
\begin{align}
	\lambda\mapsto-\lambda \quad\mbox{plus}\quad x\mapsto-x \,,
	\label{054}
\end{align}
which shows that the system is parity-odd invariant if restricted to planes of constant $y$. In other words the dynamics for opposite values of $\lambda$ are invariant after a reflection over the $(y,\lambda)$-plane. The symmetry (\ref{054}) implies that we can fully analyse the system by simply taking into account positive values of $\lambda$. Negative values would give the same dynamical properties reflected over the $(y,\lambda)$-plane. If $\Gamma(\lambda)$ is not an even function of $\lambda$, the symmetry (\ref{054}) is broken and one must study both positive and negative values of $\lambda$ separately. This depends on the model at hand, i.e.~on the form of the potential $V(\phi)$, but, as we will see, for the simplest examples it is always satisfied.

\subsection{Exponential potential}
\label{sec:exp_potential}

In this section the self-interacting potential of the scalar field is assumed to be of the exponential kind, namely
\begin{align}
	V(\phi) = V_0\,e^{-\lambda \kappa \phi} \,,
	\label{def:exp_potential}
\end{align}
where $V_0>0$ is a constant and $\lambda$ is now a constant parameter which agrees with definition (\ref{def:lambda_dynamical}). The exponential case (\ref{def:exp_potential}) is the simplest example of quintessence, and can be easily justified from high-energy phenomenology, such as string theory (see e.g.~\citet{Baumann:2014nda}).

Dynamical systems for cosmological scalar fields with an exponential potential have been studied long before the discovery of cosmic acceleration, mainly in relation with early universe inflation and high-energy physics phenomenology (see e.g.~\citet{Halliwell:1986ja,Burd:1988ss,WCL1993, Coley:1997nk,Coley:2003mj}). The reference work for such a system is the well known paper by \citet{Copeland:1997et} where a thorough dynamical analysis is performed (see also \citet{UrenaLopez:2011ur,Tamanini:2014mpa}). The arguments of this section are based on the results of that work.
Similar investigations extended to cosmologies with non vanishing spatial curvature have been delivered by \citet{vandenHoogen:1999qq} and \citet{Gosenca:2015qha}.

Equations (\ref{eq:x_can_scalar_field_gen_V}) and (\ref{eq:y_can_scalar_field_gen_V}) now represents a 2D autonomous dynamical system
\begin{align}
	x'&=-\frac{3}{2} \left[2x+(w-1) x^3+ x (w+1) \left( y^2-1\right)-\frac{\sqrt{2}}{\sqrt{3}} \lambda  y^2\right]\,, \label{eq:x_exp}\\
	y'&=-\frac{3}{2} y \left[(w-1) x^2+(w+1) \left(y^2-1\right)+\frac{\sqrt{2}}{\sqrt{3}} \lambda  x\right]\,,\label{eq:y_exp}
\end{align}
where we recall that, thanks to (\ref{def:exp_potential}), $\lambda$ is a parameter and the physical phase space, from now on referred to as simply the phase space, is the (closed) upper half unit disk in the $(x,y)$-plane.
Thanks to the symmetry (\ref{054}), which now holds trivially since $\Gamma=1$, we only need to analyse positive values of $\lambda$ since negative values would yield the same dynamics reflected along the $y$-axis\footnote{In the exponential case (\ref{def:exp_potential}), this symmetry is related to the invariance of the action under a sign redefinition of the scalar field: $\phi\mapsto-\phi$.}.

\begin{table}[t]
\begin{center}
\begin{tabular}{|c|c|c|c|c|c|c|c|}
\hline
Point & $x$ & $y$ & Existence & $w_{\rm eff}$ & Accel. & $\Omega_\phi$ & $w_\phi$  \\
\hline
\multirow{2}*{$O$} & \multirow{2}*{0} & \multirow{2}*{0} & \multirow{2}*{$\forall\;\lambda,w$} & \multirow{2}*{$w$} & \multirow{2}*{No} & \multirow{2}*{0} & \multirow{2}*{--} \\ & & & & & & & \\
\multirow{2}*{$A_\pm$} & \multirow{2}*{$\pm 1$} & \multirow{2}*{0} & \multirow{2}*{$\forall\;\lambda,w$} & \multirow{2}*{1} & \multirow{2}*{No} & \multirow{2}*{1} & \multirow{2}*{1} \\ & & & & & & & \\
\multirow{2}*{$B$} & \multirow{2}*{$\frac{\sqrt{3}}{\sqrt{2}}\frac{1+w}{\lambda}$} & \multirow{2}*{$\sqrt{\frac{3(1-w^2)}{2\lambda^2}}$} & \multirow{2}*{$\lambda^2\geq 3(1+w)$} & \multirow{2}*{$w$} & \multirow{2}*{No} & \multirow{2}*{$\frac{3(1+w)}{\lambda^2}$} & \multirow{2}*{$w$} \\ & & & & & & & \\
\multirow{2}*{$C$} & \multirow{2}*{$\frac{\lambda}{\sqrt{6}}$} & \multirow{2}*{$\sqrt{1-\frac{\lambda^2}{6}}$} & \multirow{2}*{$\lambda^2<6$}  & \multirow{2}*{$\frac{\lambda^2}{3}-1$} & \multirow{2}*{$\lambda^2<2$} & \multirow{2}*{1} & \multirow{2}*{$\frac{\lambda^2}{3}-1$} \\ & & & & & & & \\
\hline
\end{tabular}
\caption{Critical points of the system (\ref{eq:x_exp})--(\ref{eq:y_exp}) with existence and physical properties.}
\label{tab:exp_CP_physics}
\end{center}
\end{table}

\subsubsection{Critical points and phenomenology: scaling solutions}

We are now ready to find the critical points of the dynamical system (\ref{eq:x_exp})--(\ref{eq:y_exp}) and to perform the stability analysis. The results are summarised in Tab.~\ref{tab:exp_CP_physics}, where the existence and physical properties are outlined, and Tab.~\ref{tab:exp_CP_stability}, where details of the stability analysis are presented. There can be up to five critical points in the phase space depending on the numerical value of $\lambda$. In what follows we go through each critical point discussing its mathematical and physical features.

\begin{table}
\begin{center}
\begin{tabular}{|c|c|c|c|}
\hline
P & Eigenvalues & Eigenvectors & Stability \\
\hline
\multirow{3}*{$O$} & \multirow{3}*{$\{\frac{3}{2} (w\pm 1)\}$} & \multirow{3}*{$\{\left(
\begin{array}{c}
 1 \\ 0
\end{array}
\right),\left(
\begin{array}{c}
 0 \\ 1
\end{array}
\right)
\}$} & \multirow{3}*{Saddle} \\ & & & \\ & & & \\
\multirow{3}*{$A_-$} & \multirow{3}*{$\{3 - 3 w\,,\, 3 + \frac{\sqrt{3}}{\sqrt{2}} \lambda\}$} & \multirow{3}*{$\{\left(
\begin{array}{c}
 1 \\ 0
\end{array}
\right),\left(
\begin{array}{c}
 0 \\ 1
\end{array}
\right)
\}$}
&Unstable node if $\lambda\geq-\sqrt{6}$ \\ & & & Saddle if $\lambda<-\sqrt{6}$\\  & & &  \\
\multirow{3}*{$A_+$} & \multirow{3}*{$\{3 -3 w\,,\, 3 - \frac{\sqrt{3}}{\sqrt{2}} \lambda\}$} & \multirow{3}*{$\{\left(
\begin{array}{c}
 1 \\ 0
\end{array}
\right),\left(
\begin{array}{c}
 0 \\ 1
\end{array}
\right)\}$} &  Unstable node if $\lambda\leq \sqrt{6}$  \\ & & &  Saddle if $\lambda>\sqrt{6}$ \\  &&& \\
& & & Stable node if \\ \multirow{2}*{$B$} & \multirow{2}*{$\{\frac{3}{4\lambda}\left[(w-1)\lambda\pm\Delta\right]\}$} & \multirow{2}*{$\{\left(
\begin{array}{c}
 \frac{\lambda}{2}\frac{\sqrt{1-w}}{\sqrt{1+w}} \frac{[6(w+1)^2-\lambda\pm\Delta]} {[2(1-w^2)-\lambda^2]} \\ 1
\end{array}
\right)\}$} & $3(w+1)<\lambda^2<\frac{24(w+1)^2}{9w+7}$ \\ & & & Stable spiral if \\ & & & $\lambda^2\geq\frac{24(w+1)^2}{9w+7}$ \\
\multirow{4}*{$C$} & \multirow{2}*{$\{\frac{\lambda^2}{2}-3\,,$} & \multirow{4}*{$\{\left(
\begin{array}{c}
 \frac{\sqrt{6-\lambda^2}}{-\lambda} \\ 1
\end{array}
\right),\left(
\begin{array}{c}
 \frac{(w-1)\lambda}{w\sqrt{6-\lambda^2}} \\ 1
\end{array}
\right)\}$} & \\ & & & Stable if $\lambda^2<3(1+w)$ \\ & \multirow{2}*{$\lambda ^2-3w-3\}$} & & Saddle if $3(1+w)\leq\lambda^2<6$ \\ & & & \\
\hline
\end{tabular}
\caption{Stability properties for the critical points of the system (\ref{eq:x_exp})--(\ref{eq:y_exp}). Here $\Delta=\sqrt{(w-1)[(7+9w)\lambda^2-24(w+1)^2]}$.}
\label{tab:exp_CP_stability}
\end{center}
\end{table}

\begin{itemize}

\item {\it Point~$O$}. The origin of the phase space, corresponding to a matter dominated universe ($\Omega_{\rm m}=1$), is a critical point which exists for all values of $\lambda$. This point is always a saddle point attracting trajectories along the $x$-axis and repelling them towards the $y$-axis\footnote{For $-1<w<1$ the two eigenvalues of the Jacobian matrix have always opposite sign when evaluated at Point~$O$ and the two eigenvectors coincide with the $x$ and $y$ axis, with the latter one always corresponding to the positive eigenvalues, i.e.~to the unstable direction (cf.~Tab.~\ref{tab:exp_CP_stability}).}. Point~$O$ stands for the matter solution where the universe evolves according to~(\ref{eq:mattersolw}). Of course the effective EoS matches the matter EoS, $w_{\rm eff}=w$, and thus for physically admissible values of $w$ there is no acceleration. The dark EoS is undetermined in $O$ since both its kinetic and potential energies vanish. This is in any case physically unimportant since the total energy of the scalar field, kinetic plus potential, is zero.

\item {\it Points~$A_\pm$}. In the points~$(\pm 1,0)$ the universe is dominated by the scalar field kinetic energy ($x^2=\Omega_\phi=1$) and thus the effective EoS reduces to a stiff fluid with $w_{\rm eff}=w_\phi=1$ and no acceleration. Their existence is always guaranteed and they never represent stable points. They are unstable or saddle points depending on the value of $\lambda$ being greater or smaller than $\sqrt{6}$. Strictly speaking, a stiff-fluid EoS cannot be viable at the classical macroscopic level, however these solutions are expected to be relevant only at early times and thus are commonly ignored in dark energy applications. According to (\ref{056}), in Points~$A_\pm$ the universe expands as $a\propto t^{1/3}$.

\item {\it Point~$B$}. This point (see Tab.~\ref{tab:exp_CP_physics} for the coordinates) represents a so-called {\it scaling solution} where the effective EoS matches the matter EoS (for this reason sometimes called more accurately a matter scaling solution). These solutions were originally obtained by \citet{Wetterich:1987fm} and \citet{Ferreira:1997hj}. They derive their name from the fact that the scalar field energy density scales proportionally to the matter energy density: $\Omega/\Omega_\phi=\lambda^2/(3w+3)-1$. In other words we always have both $0<\Omega_\phi=3(1+w)/\lambda^2<1$ and $0<\Omega_{\rm m}=1-\Omega_\phi<1$, obtaining also $w_\phi=w$. This means that the universe evolves under both the matter and scalar field influence, but it expands as if it was completely matter dominated, i.e.~according to~(\ref{eq:mattersolw}). This solution is of great physical interest for the coincidence problem since according to it, a scalar field can or could be present in the universe hiding its effects on cosmological scales. However, since for Point~$B$ we have $w_{\rm eff}=w$ there cannot be accelerated expansion. When this point exists, i.e.~for $\lambda^2\geq 3(1+w)$, it always represents a stable point attracting the trajectories in the physical phase space.

\item {\it Point~$C$}. The last point stands for the cosmological solution where the universe is completely scalar field dominated (see Tab.~\ref{tab:exp_CP_physics} for the coordinates). This implies $\Omega_{\rm m}=0$ and $\Omega_\phi=x^2+y^2=1$, meaning that Point~$C$ will always lie on the unit circle. It exists for $\lambda^2<6$ and it is a stable attractor for $\lambda^2<3(1+w)$ (i.e.~when Point~$B$ does not appear) and a saddle point for $3(1+w)\leq\lambda^2<6$ (i.e.~when Point~$B$ is present). The effective EoS parameter assumes the value $w_{\rm eff}=w_\phi=\lambda^2/3-1$ which implies an accelerating universe for $\lambda^2<2$. This point represents the well-known power-law accelerated expansion driven by a sufficiently flat scalar field potential. In the limit $\lambda\rightarrow 0$ this solution reduces to a de Sitter expansion dominated by a cosmological constant.
\end{itemize}

On physical ground, if $\lambda^2<2$ a dark matter to dark energy transition can be achieved by the heteroclinic orbit connecting Point~$O$ with Point~$C$. However the origin is always a saddle point meaning that it cannot be the past attractor, and the early time behaviour of the universe is given by Points~$A_\pm$. The future attractor can either be Point~$C$ or Point~$B$, with the latter one never giving acceleration. The so-called scaling solution of Point~$B$ can be used to hide the presence of a scalar field in the cosmic evolution, at least at the background level. This behaviour can be used at early times in order to obtain negligible effects of dark energy when matter dominates, though CMB experiments impose strong constraints on dark energy at early times ($\Omega_\phi<0.0036$ from Planck \citep{Ade:2015rim}). The problem is that at late times dark energy should start to dominate, but the scaling solution of Point~$B$ never gives acceleration and, being a future attractor, once the universe reaches the scaling solution it never leaves it. For this reason, even if this behaviour is of great interest at early times, it cannot represent a viable model for the late time universe since it does not lead to dark energy domination. One needs a mechanism which allows the Universe to exit the scaling solution and to join the dark energy accelerating solution, but this cannot be achieved with a single canonical scalar field and more complex dynamics is required.

On a more mathematical note we mention that for Points~$B$ and $C$, Lyapunov functions can be easily constructed as shown by \citet{Bohmer:2010re,Boehmer:2014vea}. However the Lyapunov analysis does not lead to a conclusive result and the linear stability analysis seems to be more suited to determine the stability in this case.

\subsubsection{Phase space portraits}

\begin{figure}
\centering
\includegraphics[width=\columnwidth]{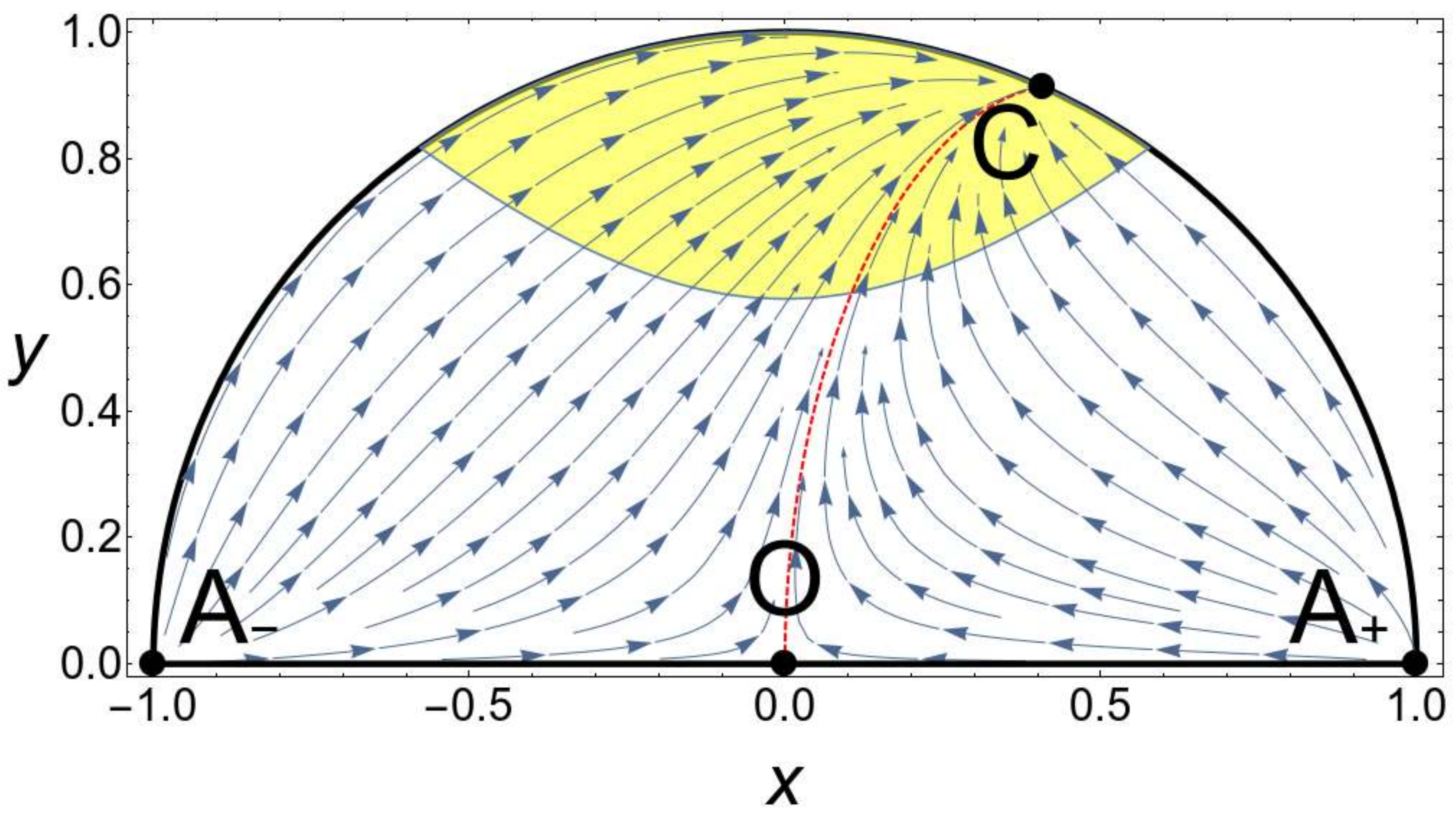}
\caption{Phase space with $\lambda=1$ and $w=0$. The only attractor is Point~$C$ which represents an accelerating solution. For values $\lambda^2>2$ Point~$C$ would lie outside the acceleration region (yellow/shaded) and would not be an inflationary solution. The red/dashed line highlights the heteroclinic orbit connecting Point~$O$ to Point~$C$. The yellow/shaded region denotes the part of the phase space where the universe is accelerating ($w_{\rm eff}<-1/3$).}
\label{fig:01}
\end{figure}

\begin{figure}
\centering
\includegraphics[width=\columnwidth]{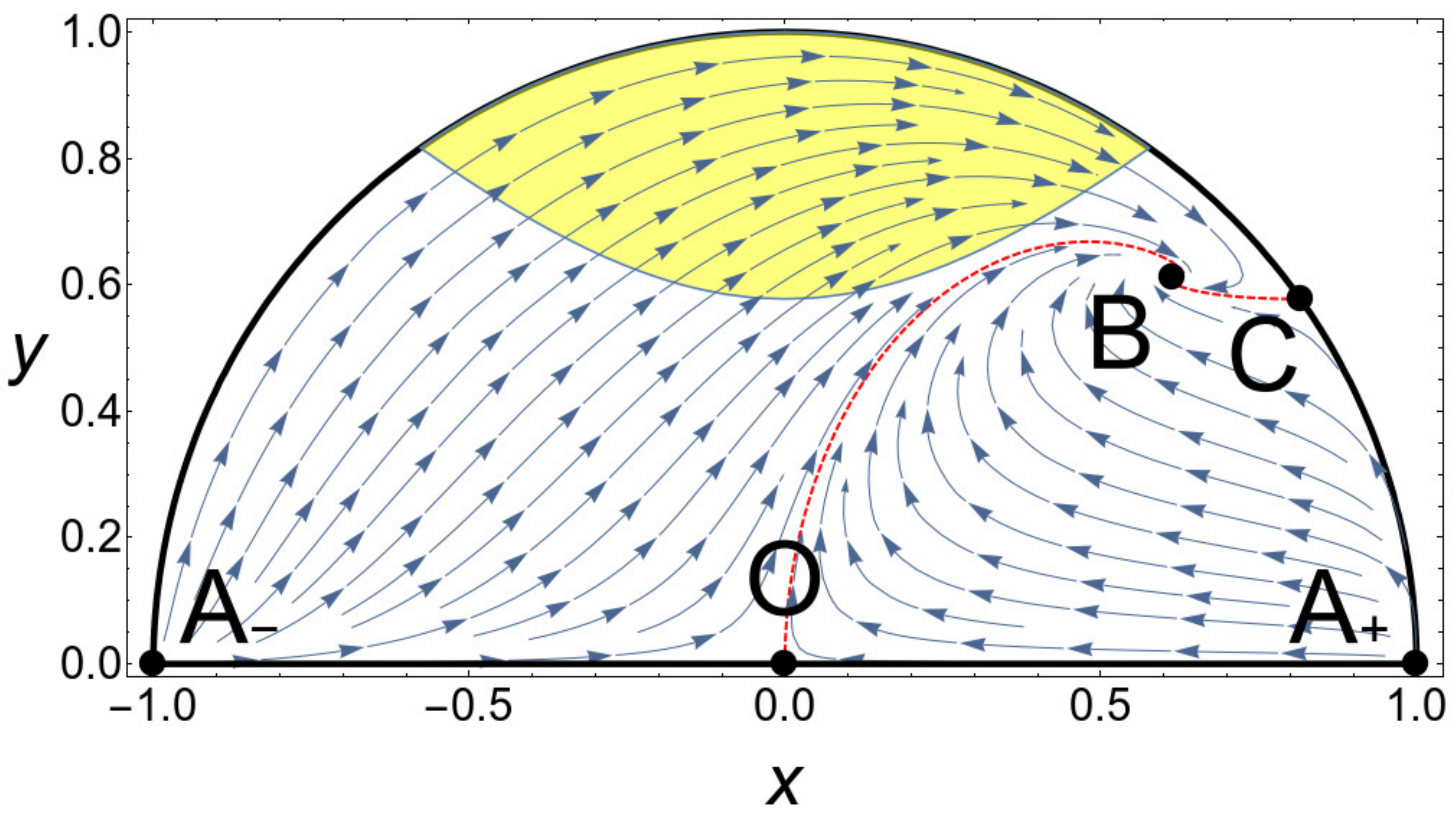}
\caption{Phase space with $\lambda=2$ and $w=0$. The only attractor is Point~$B$ where the universe expands as it was completely matter dominated (scaling solution), while Point~$C$ is a saddle point. The red/dashed lines highlight the heteroclinic orbits connecting Point~$O$ and Point~$C$ to Point~$B$. The yellow/shaded region denotes the part of the phase space where the universe is accelerating.
}
\label{fig:02}
\end{figure}

\begin{figure}
\centering
\includegraphics[width=\columnwidth]{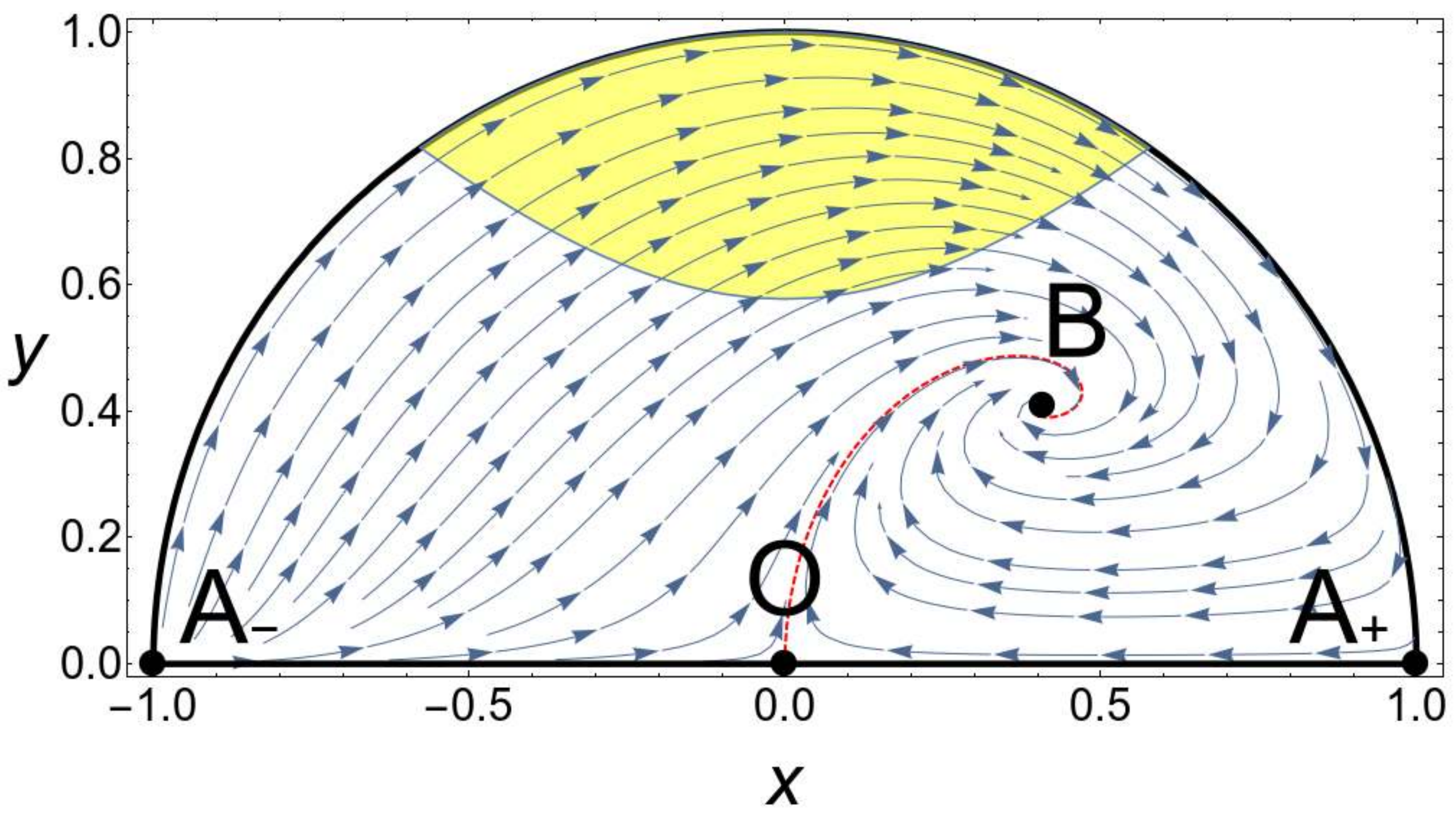}
\caption{Phase space with $\lambda=3$ and $w=0$. Point~$B$ is the only attractor describing a scaling solution with $w_{\rm eff}=w$. The red/dashed line highlights the heteroclinic orbit connecting Point~$O$ to Point~$B$. The yellow/shaded region denotes the part of the phase space where the universe is accelerating.}
\label{fig:03}
\end{figure}

We can now look at the phase space portrait for different values of the parameters. Looking at Tab.~\ref{tab:exp_CP_stability}, the qualitative behaviour of the phase space can be divided into three regions according to the value of $\lambda^2$: $0$ to $3(1+w)$, $3(1+w)$ to $6$ and $6$ to infinity. In what follows we will only consider positive values for $\lambda$ since, as we pointed out above, the dynamics for negative values coincides with the positive one after a reflection around the $y$ axis due to (\ref{054}). Since we are concerned only with dark energy applications and thus late time cosmology, we will restrict the following phase space plots (Figs.~\ref{fig:01}, \ref{fig:02} and \ref{fig:03}) to the case $w=0$ in order to better visualise possible dark matter to dark energy transitions\footnote{Different values of $w$ within the physically meaningful range $[0,1/3]$ lead to the same qualitative phase space.}. The yellow/shaded region in Figs.~\ref{fig:01}, \ref{fig:02} and \ref{fig:03} highlights the zone of the phase space where the universe undergoes an accelerated expansion, i.e.~where $w_{\rm eff}<-1/3$.

{\it Range 1}: If $\lambda^2<3(1+w)$ there are four critical points in the phase space. Points~$A_\pm$ are both unstable nodes, while Point~$O$ is a saddle point. For orbits with $y>0$ the only attractor is Point~$C$ which represents an inflationary cosmological solution if $\lambda^2<2$. The portrait of the phase space is depicted in Fig.~\ref{fig:01} where the value $\lambda=1$ has been chosen. Point~$C$ always lies on the unit circle and it happens to be outside the yellow/shaded acceleration region if $\lambda^2>2$. All the trajectories in the phase space are heteroclinic orbits starting from Points~$A_\pm$ and ending in Point~$C$. The only exceptions are the orbits on the $x$-axis which connects Points~$A_\pm$ with Point~$O$ and the orbit connecting Point~$O$ with Point~$C$. This last trajectory divides the phase space into two invariant sets: solutions on its right have Point~$A_+$ as the past attractor, while the past attractor of solutions on its left is Point~$A_-$.
There are two possible heteroclinic sequences: $A_\pm\rightarrow O\rightarrow C$. They can be used as physical models for dark matter to dark energy transition well characterising the late time evolution of the universe with a final effective EoS given by $w_{\rm eff}=-1+\lambda^2/3$. However at early times we always obtain a stiff-fluid domination represented by Points~$A_\pm$ which is phenomenologically disfavoured.

{\it Range 2}: In the range $3(1+w)\leq\lambda^2<6$ there are five critical points in the phase space. Points~$A_\pm$ and $O$ still behave as unstable nodes and a saddle point respectively. The future attractor is now Point~$B$ and Point~$C$ becomes a saddle point. The phase space portrait for $\lambda=2$ is drawn in Fig.~\ref{fig:02}. Point~$B$ always lies outside the acceleration region (yellow/shaded) and thus never describes an inflationary solution. The effective EoS parameter at this point coincides with the matter EoS parameter and thus the universe experiences a matter-like expansion even if it is not completely matter dominated (scaling solution). All the solutions are again heteroclinic orbits connecting Points~$A_\pm$ to Point~$B$. The exceptions are the orbits on the boundary of the phase space, which connect Points~$A_\pm$ to either Point~$O$ or Point~$C$, and the two heteroclinic orbits connecting Point~$O$ and Point~$C$ to Point~$B$. These last two orbits divide the phase space into two invariant sets: one with Point~$A_-$ and the other with Point~$A_+$ as past attractors.
The phase space depicted in Fig.~\ref{fig:02} can be used for applications to transient periods of dark energy. For many trajectories (the ones passing through the yellow/shaded region) a finite period of acceleration can be achieved, and for $\lambda$ sufficiently close to $\sqrt{3}$ the physically relevant heteroclinic sequence connecting the matter domination to the scaling solution experiences a transient accelerating phase.

{\it Range 3}: Finally if $\lambda^2\geq 6$ there are again only four critical points. Point~$A_-$ is the only unstable node, while Points~$A_+$ and $O$ behave as saddle points. Point~$C$ does not appear anymore and the future attractor is still Point~$B$, which again represents a scaling solution with $w_{\rm eff}=w$. The phase space dynamics for $\lambda=3$ is depicted in Fig.~\ref{fig:03}. Now all the orbits start from Point~$A_-$, the past  attractor, and end in Point~$B$, which is a simple attracting node for $3(w+1)<\lambda^2<24(w+1)^2/(9w+7)$ and an attracting spiral for $\lambda^2\geq 24(w+1)^2/(9w+7)$ as pointed out in Tab.~\ref{tab:exp_CP_stability}. There are a few special heteroclinic orbits connecting Point~$A_-$ to Point~$A_+$, Points~$A_\pm$ to Point~$O$ and Point~$O$ to Point~$B$. Exactly as before, no solution of Fig.~\ref{fig:03} can be used to model a dark energy dominated universe since the heteroclinic orbit connecting the origin to Point~$B$ never enters the yellow/shaded region. For increasing values of $\lambda$ the qualitative dynamics of the phase space does not change while Point~$B$ lies closer to the origin. In the limit $\lambda\rightarrow+\infty$, implying $V(\phi)\rightarrow 0$, Point~$B$ coincides with Point~$O$.

We can now draw our conclusions on the canonical scalar field with an exponential potential. From the mathematical perspective this model is of great interest because of its simplicity. The cosmological equations can be reduced to a 2D dynamical system with a compact phase space which is relatively easy to analyse. There are no periodic orbits and all the asymptotic behaviours are represented by critical points. All this allowed us to capture the whole dynamics of the system in the three plots of Figs.~\ref{fig:01}, \ref{fig:02} and \ref{fig:03}.

From the physics point of view instead, the cosmological dynamics of the exponential potential is interesting because of the appearance of late time accelerated solutions which can be employed to model dark energy. For these solutions to be cosmologically viable a sufficiently flat potential ($\lambda^2<2$) is expected and a strong fine tuning of initial conditions is required in order for matter domination to last enough time (the solution must shadow the sequence $A_\pm\rightarrow O\rightarrow C$ in Fig.~\ref{fig:01}). Moreover at early times the only possible solutions are the non-physical stiff-fluid universes which cannot represent a viable description of the universe.

\begin{figure}
\centering
\includegraphics[width=\columnwidth]{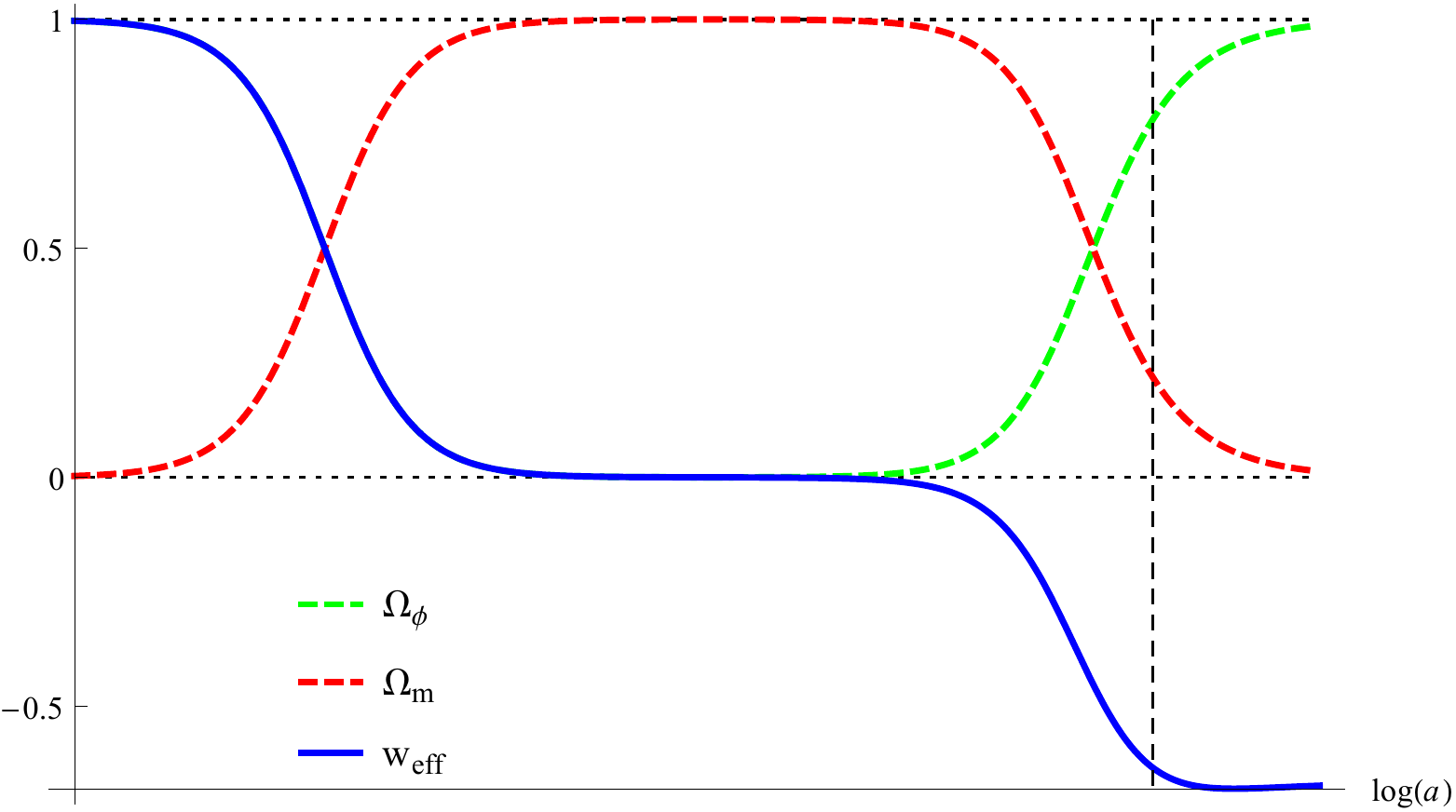}
\caption{Evolution of the effective EoS parameter ($w_{\rm eff}$), the matter ($\Omega_{\rm m}$) and dark energy ($\Omega_\phi$) relative energy densities for the quintessence model with an exponential potential. The vertical dashed line denotes the present cosmological time.}
\label{fig:quintessence_exp_w_plot}
\end{figure}

The effective EoS, together with the matter and scalar field relative energy densities, for a solution shadowing the heteroclinic sequence $A_-\rightarrow O\rightarrow C$ with $\lambda=1$ has been plotted in Fig.~\ref{fig:quintessence_exp_w_plot}. It is clear that in this model a sufficiently long period of matter domination followed by a never ending phase of dark energy domination can be achieved. Interestingly in this situation the final value of $w_{\rm eff}$ lies between $-1/3$ and $-1$ according to the value of $\lambda$. Only in the limit $\lambda\rightarrow 0$, for which the exponential potential becomes a cosmological constant, the value $w_{\rm eff}=-1$ represent the final state of the universe. From Fig.~\ref{fig:quintessence_exp_w_plot} it is also evident that before the matter domination era a period of scalar field kinetic domination must have occurred. This period however happens at very early times when the effective description provided by the quintessence model is expected to fail since new physics, such as inflation, should come into play. For this reason the early time stiff fluid solutions are usually ignored in this model and only the late time matter to dark energy transition is considered phenomenologically interesting.
Note also that the quintessence model with an exponential potential does not solve the cosmic coincidence problem since, as shown by the vertical dotted line in Fig.~\ref{fig:quintessence_exp_w_plot}, the present cosmological time still lies with no explanation exactly when the dark matter to dark energy transition happens.
Moreover it does not solve the problem of fine tuning of initial conditions since only the trajectories shadowing the heteroclinic orbit $O\rightarrow C$ can be used to describe the matter to dark energy transition, while for all other trajectories a sufficiently long matter domination era cannot be achieved.

In our analysis we have assumed a positive potential $V>0$ for physical reasons. However negative exponential potentials have been analysed using dynamical systems techniques by \citet{Heard:2002dr}. In that case a sufficiently flat potential ($\lambda^2<6$) always lead to a re-collapse, while with a steeper potential ($\lambda^2>6$) it is possible to achieve scaling solutions, but a strong dependence on initial conditions is present. Negative potentials have also been analysed by \citet{Copeland:2009be} who generalised the work of \citet{Heard:2002dr} to cosmologies with positive and negative spatial curvatures.

The scaling solutions we found in Point~$B$ are also phenomenologically important since in principle they allow the scalar field to hide its presence during the cosmological evolution. This situation can be used to postulate a scalar field which gives no contribution at early times but becomes relevant at late times. There are strong observational constraints for this situation \citep{Ade:2013zuv} and more complicated dynamics than the exponential potential is required in order for the scalar field to exit the scaling solution and eventually drive the cosmic acceleration.

It is impossible to achieve both the scaling and accelerating regimes with a canonical scalar field and an exponential potential, but with more complicated potentials, such as a double exponential potential \citep{Barreiro:1999zs}, a transition from the scaling to the dark energy solution can be achieved. Furthermore we mention that scaling solutions are in general unstable in anisotropic spacetimes, though they still represent critical points which can hide the scalar field for a sufficiently long time (see \citet{Coley:2003mj} and references therein). Scaling solutions for scalar fields with an exponential potential have also been studied in higher dimensional spacetimes \citep{Chang:2004pbb}.

Finally we note that the analysis of quintessence with an exponential potential can be generalised to include both radiation and dark matter with the introduction of two barotropic fluids \citep{Azreg-Ainou:2013jxa}. This approach breaks the degeneracy in the matter sector and the radiation to matter transition can be explicitly represented.

\subsection{Power-law potential}
\label{sec:power_law_potential}

In the following we will consider quintessence with an inverse power-law type potential, known as the {\it Ratra-Peebles potential} \citep{Ratra:1987rm,Peebles:1987ek}. It can be motivated from supersymmetry phenomenology, and as an explicit example we will consider
\begin{align}
  V(\phi) = \frac{M^{\alpha+4}}{\phi^\alpha} \,,
  \label{def:inv_power_law_pot}
\end{align}
where $\alpha$ is a dimensionless parameter and $M$ a positive constant with units of mass. In general one considers a positive potential $V>0$. For dynamical systems applications with negative power-law potentials see e.g.~\citet{Felder:2002jk}. Inverse power-law potentials are popular in quintessence models because of their behaviour at late time which allows for a solution, or at least an alleviation, of the fine tuning of initial conditions \citep{Zlatev:1998tr,Liddle:1998xm,Steinhardt:1999nw,delaMacorra:2001ay}. As we will see, models with $\alpha>0$ are physically more interesting while scalar field potentials of the kind (\ref{def:inv_power_law_pot}) with $\alpha<0$ are less attractive for dark energy phenomenology, though they are largely used in early universe inflation (see \citet{UrenaLopez:2007vz} and \citet{Alho:2014fha} for applications to inflationary quadratic potentials).

We will consider both positive and negative values of $\alpha$, though the main discussion will focus on the positive $\alpha>0$ case.

\subsubsection{Phase space compactification}

The dynamical system controlling the evolution of a universe filled by quintessence with a power-law potential is given by Eqs.~(\ref{eq:x_can_scalar_field_gen_V})--(\ref{eq:y_can_scalar_field_gen_V}) and Eq.~\eqref{eq:lambda}.
From Eq.~(\ref{def:Gamma}) we obtain
\begin{align}
  \Gamma= \frac{V\,V_{,\phi\phi}}{V_{,\phi}^2} = \frac{\alpha+1}{\alpha} \,,
  \label{eq:defgamma}
\end{align}
or equivalently $\Gamma-1=1/\alpha$. In this case $\Gamma$ is a constant depending on the parameter $\alpha$ and Eqs.~(\ref{eq:x_can_scalar_field_gen_V})--(\ref{eq:lambda}) become an autonomous 3D dynamical system.
Since $\Gamma=1$ corresponds to the exponential potential, we will assume $\Gamma \neq 1$ henceforth, which anyway according to Eq.~\eqref{eq:defgamma} corresponds to the limit $\alpha\rightarrow \infty$.
Note that the most general form of the scalar field potential obtained by requiring $\Gamma$ be a constant is actually $V(\phi) = V_0\left(\phi+\phi_0\right)^\beta$ with $V_0$, $\phi_0$ and $\beta$ being all constants. The dynamics with this latter potential is however equivalent to the one resulting from the potential \eqref{def:inv_power_law_pot} since the dynamical equations do not change \citep{Roy:2014yta}.

The relevant phenomenological properties of a power-law model have been extensively analysed using dynamical systems methods \citep{Ng:2001hs,UrenaLopez:2011ur,Gong:2014dia,Roy:2014yta}. However, some of the results that follow are the product of new and original analysis, for instance the use of centre manifold theory, the compactification of the phase space and the use of Lyapunov functions (see however \citet{Alho:2015ila} for similar approaches).

The (physical) phase space of the power-law system is represented by the positive-$y$ half-cylinder stretching from $\lambda=0$ to $\lambda=+\infty$ (negative values of $\lambda$ are included thanks to the symmetry \eqref{054}). This phase space is non-compact being infinite in the positive $\lambda$ direction. However, following \citet{Ng:2001hs}, we can compactify it defining a new variable $z$ as
\begin{align}
  z = \frac{\lambda}{\lambda+1} \,.{}
  \label{def:z}
\end{align}
When $\lambda=0$ we get $z=0$ and in the limit $\lambda\rightarrow+\infty$ we have $z=1$, meaning that the new variable $z$ is bounded as $0\leq z\leq 1$. We can invert (\ref{def:z}) in order to obtain $\lambda=z/(z-1)$. Definition (\ref{def:z}) would become problematic for negative $\lambda$ due to the point $\lambda=-1$.

In the current case we are only dealing with positive values of $\lambda$ due to the symmetry (\ref{054}). However, in more general cases where negative values of $\lambda$ need to be considered, i.e.~when $\Gamma(\lambda)\neq \Gamma(-\lambda)$ (see Sec.~\ref{sec:other_potentials}), definition (\ref{def:z}) must be changed accordingly.

In the new variable (\ref{def:z}) the complete dynamical system reads
\begin{align}
  x' &= \frac{1}{2} \left\{3 (1-w) x^3-3 x \left[w \left(y^2-1\right)+y^2+1\right]+\frac{\sqrt{6} y^2 z}{1-z}\right\} \,,
  \label{058}\\
  y' &= -\frac{1}{2} y \left[3 (w-1) x^2+3 (w+1) \left(y^2-1\right)+\frac{\sqrt{6} x z}{1-z}\right] \,,\\
  z' &= -\sqrt{6} (\Gamma -1) x z^2 \,.
  \label{059}
\end{align}
Note that the last term in both the equations for $x$ and $y$ diverges as $z\rightarrow 1$. This is expected since $z\rightarrow 1$ corresponds to $\lambda\rightarrow+\infty$. In order to remove these infinities we can multiply the right hand sides of (\ref{058})--(\ref{059}) by $(1-z)$. This operation allows us to study the properties of the $z=1$ plane and does not change the qualitative dynamical features of the system in the other regions of the phase space\footnote{Strictly speaking this operation is not mathematically well-defined. However in this case we are just removing the divergent terms on the $z=1$ plane leaving the rest of the phase space basically invariant since, in general, for any dynamical system $\mathbf{x'=f(x)}$ the new dynamical system constructed as $\mathbf{x}'=\xi(\mathbf{x})\mathbf{f(x)}$ for a positive defined function $\xi(\mathbf{x})>0$ will present the same critical points with the same stability properties. We could have equally kept Eqs.~(\ref{058})--(\ref{059}) and studied the dynamics on the $z=1$ plane only considering the diverging terms and neglecting all the others.}, since $(1-z)$ is always positive for $0\leq z<1$. After this little trick we obtain
\begin{align}
  x' &= \frac{1}{2} (1-z)\left\{3 (1-w) x^3-3 x \left[w \left(y^2-1\right)+y^2+1\right]\right\}+\frac{\sqrt{3}}{\sqrt{2}} y^2 z \,,\label{eq:x_power_law_V_compact}\\
  y' &= -\frac{1}{2} y\, (1-z)\left[3 (w-1) x^2+3 (w+1) \left(y^2-1\right)\right]-\frac{\sqrt{3}}{\sqrt{2}}\, x y z \,,\label{eq:y_power_law_V_compact}\\
  z' &= -\sqrt{6} (\Gamma -1)(1-z) x z^2 \label{eq:z_power_law_V_compact} \,,
\end{align}
which is now regular at $z=1$.

Note that this system presents the invariant submanifolds $y=0$ and $z=0$. This implies that a true global attractor for the system will have to have both $y=0$ and $z=0$.  Since we are only considering $y>0$, we can have  a ``semi-global'' attractor (i.e.~a global attractor for $y>0$ orbits) if a fixed point has $z=0$.

\begin{table}[t]
\begin{center}
\begin{tabular}{|c|c|c|c|c|c|c|c|c|}
\hline
Point & $x$ & $y$ & $z$ & Existence & $w_{\rm eff}$ & Accel. & $\Omega_\phi$ & $w_\phi$  \\
\hline
$O_z$ & $0$ & $0$ & Any & $\forall\, w,\alpha$ & $w$ & No & $0$ & - \\
$A_\pm$ & $\pm 1$ & $0$ & $0$ & $\forall\, w,\alpha$ & $1$ & No & $1$ & $1$ \\
$B_x$ & Any & 0 & 1 & $\forall\, w,\alpha$ & $w+x^2(1-w)$ & No & $x^2$ & 1 \\
$C$ & 0 & 1 & 0 & $\forall\, w,\alpha$ & $-1$ & Yes & 1 & $-1$ \\
\hline
\end{tabular}
\caption{Critical points of the system (\ref{eq:x_power_law_V_compact})--(\ref{eq:z_power_law_V_compact}) with existence and physical properties.}
\label{tab:power_law_CP_physics}
\end{center}
\end{table}
\begin{table}[t]
\begin{center}
\begin{tabular}{|c|c|c|c|}
\hline
Point & Eigenvalues & Hyperbolicity & Stability \\
\hline
\multirow{2}*{$O_z$} & \multirow{2}*{$\left\{0,-\frac{3}{2} (w\pm 1) (z-1)\right\}$} & \multirow{2}*{Non-hyperbolic} & \multirow{2}*{Saddle} \\ & & & \\
\multirow{2}*{$A_+$} & \multirow{2}*{$\{0,3,3(1- w)\}$} & \multirow{2}*{Non-hyperbolic} & Saddle if $\alpha>0$ ($\Gamma>1$) \\  & & & Unstable if $\alpha<0$ ($\Gamma<1$)\\ & & & \\
\multirow{2}*{$A_-$} & \multirow{2}*{$\{0,3,3(1- w)\}$} & \multirow{2}*{Non-hyperbolic} & Unstable if $\alpha>0$ ($\Gamma>1$) \\ & & & Saddle if $\alpha<0$ ($\Gamma<1$)  \\ & & & \\
\multirow{4}*{$B_x$} & \multirow{4}*{$\left\{0,-\frac{\sqrt{3}}{\sqrt{2}} x,\sqrt{6} (\Gamma -1) x\right\}$} & \multirow{4}*{Non-hyperbolic} & Saddle if $x>0$ and $\alpha>0$ ($\Gamma>1$) \\ & & & Stable if $x>0$ and $\alpha<0$ ($\Gamma<1$) \\ & & & Saddle if $x<0$ and $\alpha>0$ ($\Gamma>1$) \\ & & & Unstable if $x<0$ and $\alpha<0$ ($\Gamma<1$) \\ & & & \\
\multirow{2}*{$C$} & \multirow{2}*{$\{0,-3,-3 (1+w)\}$} & \multirow{2}*{Non-hyperbolic} & Stable if $\alpha>0$ ($\Gamma>1$) \\ & & & Saddle if $\alpha<0$ ($\Gamma<1$)\\
\hline
\end{tabular}
\caption{Critical points of the system (\ref{eq:x_power_law_V_compact})--(\ref{eq:z_power_law_V_compact}) with stability properties.}
\label{tab:power_law_CP_stability}
\end{center}
\end{table}

\subsubsection{Critical points and phenomenology}

We are now ready to discuss the critical points of the system (\ref{eq:x_power_law_V_compact})--(\ref{eq:z_power_law_V_compact}) whose existence and phenomenological properties have been listed in Tab.~\ref{tab:power_law_CP_physics}, while Tab.~\ref{tab:power_law_CP_stability} shows their stability properties.

\begin{itemize}
\item {\it Points~$O_z$}. The $z$-axis is a critical line. This means that all the points with $x=y=0$ are critical points whose existence does not depend on the theoretical parameters $w$ and $\alpha$. Being both $x$ and $y$ equal to zero, the effective EoS coincides with the matter EoS and the relative energy density of the scalar field vanishes leaving $w_\phi$ undetermined. Since these are not isolated critical points we expect that at least one eigenvalue of the Jacobian vanishes. This is indeed the case, as one can note from Tab.~\ref{tab:power_law_CP_stability}, meaning that these points are non-hyperbolic and that linear stability theory cannot be used. Moreover, since there is only one vanishing eigenvalue, the critical line of these points corresponds to the $z$-axis and the centre manifold theorem cannot apply. To determine the stability properties, from Tab.~\ref{tab:power_law_CP_stability} we can see that the non vanishing eigenvalues of Points~$O_z$ are given by $-\frac{3}{2} (w\pm 1) (z-1)$. Since $0\leq z\leq 1$ and for physically acceptable matter fluids $0\leq w\leq 1/3$, we obtain that one of these eigenvalues is always positive while the other is always negative. The critical lines of Points~$O_z$ is thus generally unstable, and we can also conclude that it will act as a saddle line since the non-zero eigenvalues have opposite sign. As we will see, this behaviour will in fact be confirmed by numerical computations.

\item {\it Points~$A_\pm$}. The two points at $(\pm 1,0,0)$ are again the scalar field kinetic dominated solutions that we already encountered for the exponential potential (Sec.~\ref{sec:exp_potential}). They exist for all values of $w$ and $\alpha$ and their phenomenological properties remain the same with $w_{\rm eff}=w_\phi=1$ and $\Omega_\phi=1$, which means they represent stiff-fluid dominated solutions. From Tab.~\ref{tab:power_law_CP_stability} we see that one eigenvalue of the Jacobian is zero implying that Points~$A_\pm$ are isolated non-hyperbolic critical points. Since the remaining eigenvalues are both always positive, we can conclude that Points~$A_\pm$ are both asymptotically unstable. However in order to understand whether they can be saddle points or past attractors we must study the flow along the centre manifold which in this case coincides with the $z$-direction, i.e.~with the centre subspace. The flow restricted to the $z$-direction passing through the Points~$A_\pm$ is given by $z'=\mp\sqrt{6}(1-z)z^2(\Gamma-1)$. Since $0\leq z\leq 1$ the stability is determined by the parameter $\alpha$ through $\Gamma$. If $\Gamma>1$ ($\alpha>0$) then Point~$A_+$ is a saddle point while Point~$A_-$ is asymptotically stable in the past. On the other hand, if $\Gamma<1$ ($\alpha<0$) then Point~$A_+$ is a past attractor and Point~$A_-$ is a saddle point. This has been summarised in Tab.~\ref{tab:power_law_CP_stability}.

\item {\it Point~$B_x$}. The straight line connecting Points~$(\pm 1,0,1)$ is another critical line. Since all critical points at infinity, i.e.~on the $z=1$ plane, are points belonging to this line, Points~$B_x$ completely characterise the asymptotic behaviour of trajectories as $\lambda\rightarrow+\infty$. The effective EoS at these points will depend on the scalar field relative energy $\Omega_\phi=x^2$ as $w_{\rm eff}=w+x^2(1-w)$. However, since the potential energy of quintessence vanishes ($y=0$) the scalar field EoS can only be determined by its kinetic part and thus $w_{\phi}=1$ with no possible acceleration for the universe ($0\leq w_{\rm eff}\leq 1$). As we can see in Tab.~\ref{tab:power_law_CP_stability} there is only one zero eigenvalue meaning that the centre manifold for these points is nothing but the critical line itself. The only exception is when $x=0$ where all the eigenvalues vanish. This is the point where the two critical lines $O_z$ and $B_x$ intersect and, as one can again understand from Tab.~\ref{tab:power_law_CP_stability}, also all the eigenvalues of Points~$O_z$ vanish at this point, i.e.~at $z=1$. Using numerical evaluations\footnote{The stability of these types of points should be treated with the so called ``blow-up techniques'' \citep{dumortier1991}, here however we will not present the full analysis, referring the reader to the literature of these advanced techniques.} we will see that this point will in fact recall the features of a centre (see e.g.~\citet{Tamanini:2014nvd}). In general the stability of the critical line $B_x$ will depend on both the parameter $\alpha$ (through $\Gamma$) and the value of the coordinate $x$. If $\alpha>0$ ($\Gamma>1$) the non vanishing eigenvalues have opposite sign no matter the value of $x$ and we can conclude that Points~$B_x$ are saddle points in this case. On the other hand if $\alpha<0$ ($\Gamma<1$) the non zero eigenvalues have the same sign: positive if $x<0$ and negative if $x>0$. As we well know linear stability theory fails in these cases, however since the centre manifold corresponds to the critical line $B_x$ (i.e.~it is flat), we can expect that orbits near $B_x$ are perpendicularly attracted or repelled according to the sign of the non vanishing eigenvalues. Points~$B_x$ will then be attractive if $x>0$ and repulsive if $x<0$. As we will see with numerical techniques, this is indeed the right stability behaviour.

\item {\it Point~$C$}. The final critical point of the system (\ref{eq:x_power_law_V_compact})--(\ref{eq:z_power_law_V_compact}) is the scalar field dominated point at $(0,1,0)$. This corresponds to nothing but a cosmological constant-like dominated solution where the universe undergoes a de Sitter accelerated expansion. In fact Point~$C$ is dominated by the potential energy of the scalar field which is constant due to the vanishing of its kinetic energy ($x^2=0$). As shown in Tab.~\ref{tab:power_law_CP_stability} one of the eigenvalues of the Jacobian vanishes at Point~$C$ implying that this point is an isolated non-hyperbolic critical point. Since the remaining eigenvalues are both negative, to determine the stability we must either find a Lyapunov function or apply the centre manifold theorem. In what follows we will show explicitly how this can be done.
\end{itemize}

\subsubsection{Stability of a non-hyperbolic critical point: Lyapunov and centre manifold approaches}

Due to the presence of non-hyperbolic critical points, one has to employ methods other than linear stability theory to understand the stability of those points. We will now show that point $C$ is asymptotically stable if $\Gamma>1$.
To do so, let us begin with the following candidate Lyapunov function
\begin{align}
  V = x^8 + (y-1)^6 + c_1 z^4 + c_2 x^2 z^4 \,,
\end{align}
where $c_1$ and $c_2$ are two constants which will be suitably chosen. This function is differentiable and positive definite near the point $(0,1,0)$. Let us now compute the crucial quantity $V'$ which becomes
\begin{align}
  V' = 8 x^7 x' + 6(y-1)^5 y' + 4 c_1 z^3 z' + c_2 (2x x' z^4 + 4 x^2 z^3 z') \,,
\end{align}
where $x'$,$y'$,$z'$ are taken from (\ref{eq:y_power_law_V_compact}). The resulting expression is quite involved and its negativity near the critical point is not guaranteed. However, let us introduce spherical polar coordinates, similar to Sec.~\ref{sec:Lyapunovexample}, $x=r \sin\theta \cos\phi$, $y=1+r\sin\theta\sin\phi$, $z=r\cos\theta$. Then, perform an expansion in the radial coordinate which yields
\begin{multline}
  V' =
  \Bigl(
  \sqrt{6}\sin\theta\cos^5\negmedspace\theta \cos\phi\, (c_2-4 c_1 (\Gamma -1)) -
  6 c_2 \sin^2\negmedspace\theta \cos^4\negmedspace\theta \cos^2\negmedspace\phi \\-
  18 (w+1) \sin^6\negmedspace\theta\sin^6\negmedspace\phi
  \Bigr) r^6 + \mathcal{O}(r^7) \,.
\end{multline}
We observe that only the first term can change its sign due to the odd powers of the trigonometric functions appearing. We can eliminate these terms by setting $c_2 = 4 c_1 (\Gamma -1)$. This choice is only permissible if $\Gamma > 1$, otherwise, our function $V$ would no longer be a Lyapunov function. This gives
\begin{align}
  V' =
  6 \sin^2\negmedspace\theta \Bigl(
  -4(\Gamma -1) c_1 \cos^4\negmedspace\theta \cos^2\negmedspace\phi -
  3(w+1) \sin^4\negmedspace\theta\sin^6\negmedspace\phi
  \Bigr) r^6 + \mathcal{O}(r^7) \,.
\end{align}
We can now set $c_1=1$ for simplicity. Therefore, as long as $w>-1$ and $\Gamma > 1$, we have found a suitable Lyapunov function which satisfies $V' \leq 0$. Note that the zero is possible along the direction where $\theta=0$, this is the $z$-direction or polar direction. This implies that point $C$ is stable but not asymptotically stable (see Theorem~\ref{lyapunovtheorem} in Sec.~\ref{sec:lyapunov}). Since the above function also satisfies $V(x,y,z) \rightarrow \infty$ as $\|(x,y,z)\| \rightarrow \infty$, we can also state that this point is globally stable.

The centre manifold approach, which we introduced in Sec.~\ref{SecCMT}, allows us moreover to determine the shape of the centre manifold. Here we sketch the computations leaving the details for the reader (see Sec.~\ref{sec:centre_manifold_example} for an explicit example).
First we must find the eigenvectors of the Jacobian at Point~$C$ which are $(1,0,0)$, $(0,1,0)$ and $(1/\sqrt{6},0,1)$ with the last one corresponding to the vanishing eigenvalues. Since the eigenvectors are not all aligned with the orthonormal axis, we need to rewrite the dynamical system with respect to the basis given by the eigenvectors themselves. The corresponding change of coordinates $\mathbf{x\mapsto \tilde x}$ is given by $x=\tilde{x}+\tilde{z}/\sqrt{6}$, $y=\tilde{y}$ and $z=\tilde{z}$.
Using these new coordinates the autonomous system of equations \eqref{eq:x_power_law_V_compact}--\eqref{eq:z_power_law_V_compact} can now be rewritten in the form provided by Eqs.~\eqref{cmexvec}, and the centre manifold methods described in Sec.~\ref{SecCMT} can be applied.
With a power-law ansatz for the centre manifold function $\mathbf{h}=(h_x,h_y)$ we obtain, up to the third order in $\tilde{z}$
\begin{align}
  h_x(\tilde{z}) &= \frac{\tilde{z}^2}{\sqrt{6}}+\frac{\sqrt{6}}{18} \left(\Gamma +2 \right) \tilde{z}^3+\mathcal{O}(\tilde{z}^4) \,,\label{060}\\
  h_y(\tilde{z}) &= -\frac{\tilde{z}^2}{12}-\frac{\tilde{z}^3}{6} +\mathcal{O}(\tilde{z}^4) \,.\label{061}
\end{align}
These two functions determine the shape of the one dimensional centre manifold of Point~$C$. Thanks to the theorems presented in Sec.~\ref{SecCMT}, at the smallest order in $\tilde{z}$ the dynamics along this centre manifold is given by
\begin{align}
\tilde{z}' = - \left(\Gamma-1\right) \tilde{z}^3 + \mathcal{O}(\tilde{z}^4) \,.
\label{066}
\end{align}
We can thus conclude that Point~$C$ is stable if $\Gamma>1$ ($\alpha>0$) and unstable (saddle point) if $\Gamma<1$ ($\alpha<0$). As we will see, numerical evaluations will not only confirm this stability behaviour but will also show us how well the centre manifold (\ref{060})--(\ref{061}) computed with approximation methods matches the actual one.

\subsubsection{Inverse power-law potentials: tracking solutions}

All the critical points of the system \eqref{eq:x_power_law_V_compact}--\eqref{eq:z_power_law_V_compact} are non-hyperbolic (cf.~Tab.~\ref{tab:power_law_CP_stability}) and their stability properties have not been as easy to find as in the exponential potential case of Sec.~\ref{sec:exp_potential}.
The picture we obtained from the analysis above is that there are two possible regimes for the power-law potential (\ref{def:inv_power_law_pot}) depending on the value of $\alpha$ being positive (inverse power-law) or negative (direct power-law), which corresponds to $\Gamma$ being bigger or smaller than one. Tab.~\ref{tab:power_law_CP_stability} suggests that the only future attractor of the phase space is Point~$C$ if $\Gamma>1$ and Points~$B_x$ (with $x>0$) if $\Gamma<1$. As we are now going to show using numerical plotting of the phase space, this is indeed the behaviour of the phase space. In what follows we will focus on the matter EoS value $w=0$, but the results will not change for other values inside the physically meaningful interval $0\leq w\leq 1/3$.
Here we mainly focus on the $\Gamma>1$ ($\alpha>0$) case, and then we will briefly discuss the $\Gamma<1$ ($\alpha<0$) case in Sec.~\ref{sec:quintessence_direct_power_law}.

\begin{figure}
\centering
\includegraphics[width=\columnwidth]{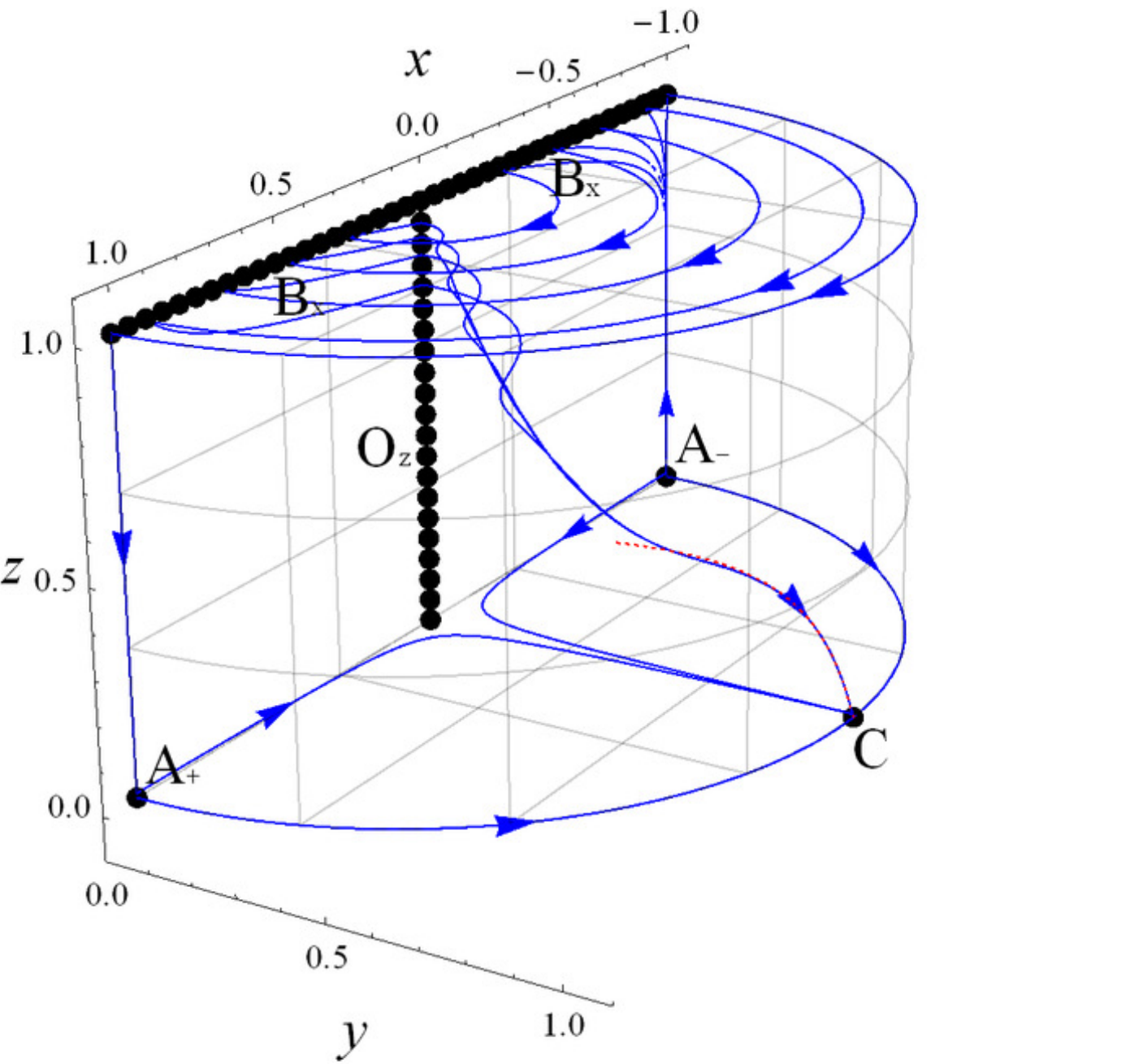}
\caption{Phase space of quintessence with inverse power-law potential corresponding to the dynamical system (\ref{eq:x_power_law_V_compact})--(\ref{eq:z_power_law_V_compact}). The values $w=0$ and $\alpha=10$ ($\Gamma=1.1$) have been chosen. The black/thick points denote critical points with the $x=y=0$ line and $z=1$ plus $y=0$ line being critical lines. The late time attractor is the dark energy dominated Point~$C$ and orbits approaching this point are first attracted by its centre manifold approximated by the red/dashed line. In this plot the tracking behaviour of solutions moving from higher to lower values of $z$ (i.e.~of $\lambda$) is particularly clear.}
\label{fig:power_law_3D_plot}
\end{figure}

In Fig.~\ref{fig:power_law_3D_plot} the phase space for the value $\alpha=10$, corresponding to $\Gamma=1.1$, has been plotted. The late time attractor is the dark energy dominated Point~$C$, while the  past attractor is Point~$A_-$. The red/dashed line denotes the centre manifold of Point~$C$ which in this plot has been approximated up to the 7th order in $\tilde{z}$. The interesting phenomenological applications of quintessence with a power-law potential are all summarised in Fig.~\ref{fig:power_law_3D_plot}, and we will deal with them after having completely clarified the dynamics of the phase space. Numerical examples will mainly be provided for $\Gamma=1.1$ ($\alpha=10$), though different values do not alter the qualitative dynamical features (as long as $\Gamma>1$).

As one can realise from the pictures, the centre manifolds of Points~$A_\pm$ are both linear along the $z$ directions. The numerical results thus confirms that these centre manifolds coincide with their corresponding centre subspaces. Furthermore the centre manifold of Point~$A_-$ is repulsive while the one of Point~$A_+$ is attractive. This implies that while Point~$A_-$ is a past attractor, Point~$A_+$ is a saddle point, as denoted also by the solutions plotted in Fig.~\ref{fig:power_law_3D_plot}.

The dynamical system (\ref{eq:x_power_law_V_compact})--(\ref{eq:z_power_law_V_compact}) restricted to the $z=1$ plane can be analytically solved by the following solution \citep{Ng:2001hs}
\begin{align}
  x(\eta) &= A \tanh\left[\frac{\sqrt{3}}{\sqrt{2}}(\eta-\eta_0)\right] \,,
  \label{062}\\
  y(\eta) &= A\, \mbox{sech}\left[\frac{\sqrt{3}}{\sqrt{2}}(\eta-\eta_0)\right] \,,
  \label{063}
\end{align}
where $A$ and $\eta_0$ are two constants. The flow on the $z=1$ plane is thus composed by circular orbits ($x^2 + y^2 = A^2$) and Point~$(0,0,1)$ effectively acts as a centre, i.e.~a critical point where all the eigenvalues of the Jacobian matrix have vanishing real part.

Going back to Fig.~\ref{fig:power_law_3D_plot} we see that every trajectory escaping to the $z=1$ plane, after completing the circular tour from negative to positive $x$, is then attracted by Points~$O_z$. More interestingly though we see that for almost all these trajectories there is a late time convergence towards a single orbit which asymptotically approaches the centre manifold of Point~$C$. The matter to dark energy transition can thus be easily described by one of these orbits which always experiences a finite period of matter domination before the universe becomes dark energy dominated. This convergence behaviour during the matter to dark energy transition is phenomenologically important since it can help in solving the fine tuning problems we encountered with the $\Lambda$CDM model and with the exponential potential \citep{Zlatev:1998tr,Steinhardt:1999nw}. We now focus our attention on this transition from Points~$O_z$ to Point~$C$.

\begin{figure}
\centering
\includegraphics[width=\columnwidth]{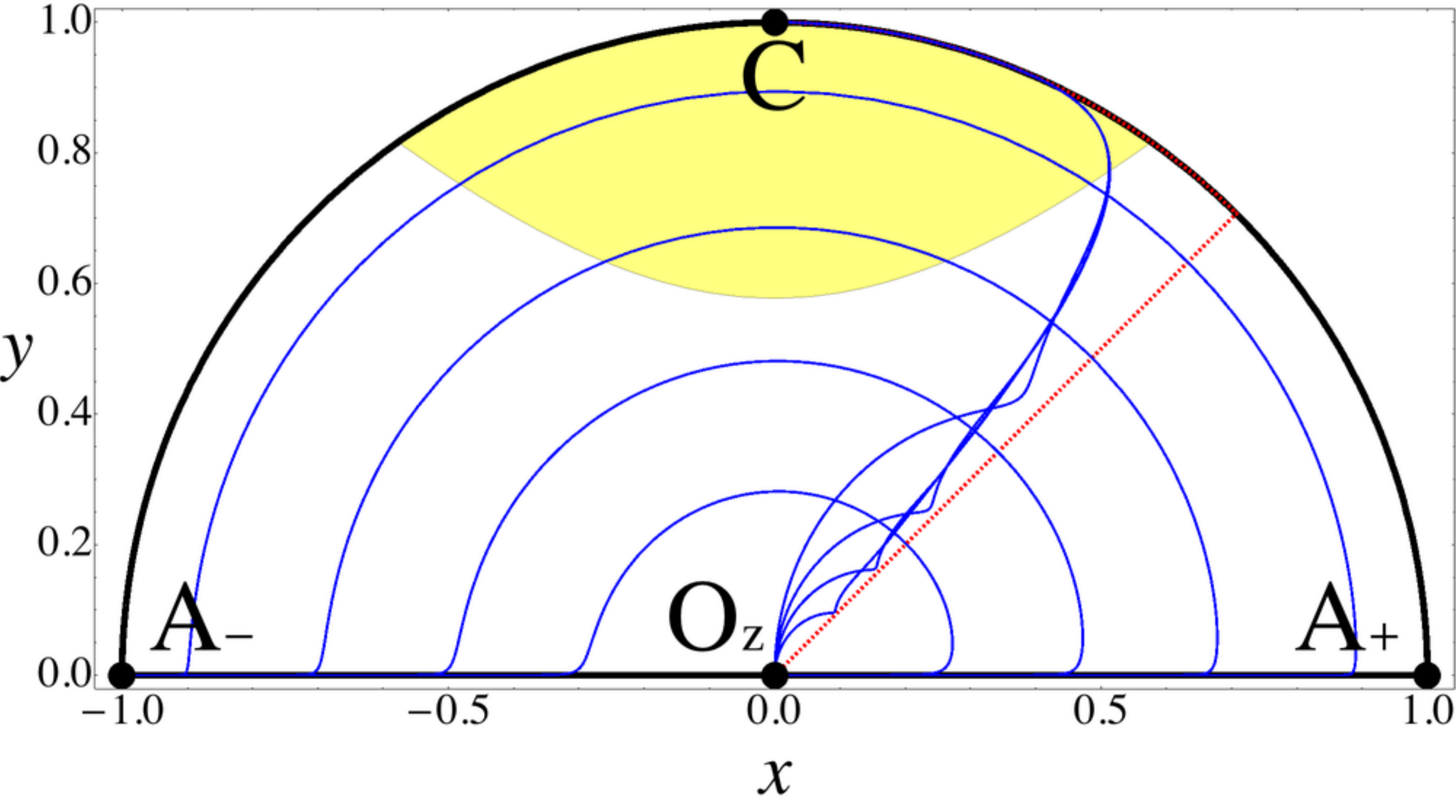}
\caption{Projection onto the $(x,y)$-plane of solutions of the system (\ref{eq:x_power_law_V_compact})--(\ref{eq:z_power_law_V_compact}) with $\Gamma=1.1$ ($\alpha=10$) and different initial conditions (i.e.~of the trajectories in Fig.~\ref{fig:power_law_3D_plot}.). The tracking behaviour is characterised by the orbits following the red/dashed line representing the future attractor (scaling and dark energy solutions) of the exponential potential case of Sec.~\ref{sec:exp_potential} for different values of $\lambda$. Orbits whose initial circular motion on the $z=1$ plane is closer to Points~$O_z$ eventually join the tracking behaviour more rapidly. The yellow/shaded region denotes the projected part of the phase space where the universe is accelerating.
}
\label{fig:power_law_projection_0}
\end{figure}

\begin{figure}
\centering
\includegraphics[width=\columnwidth]{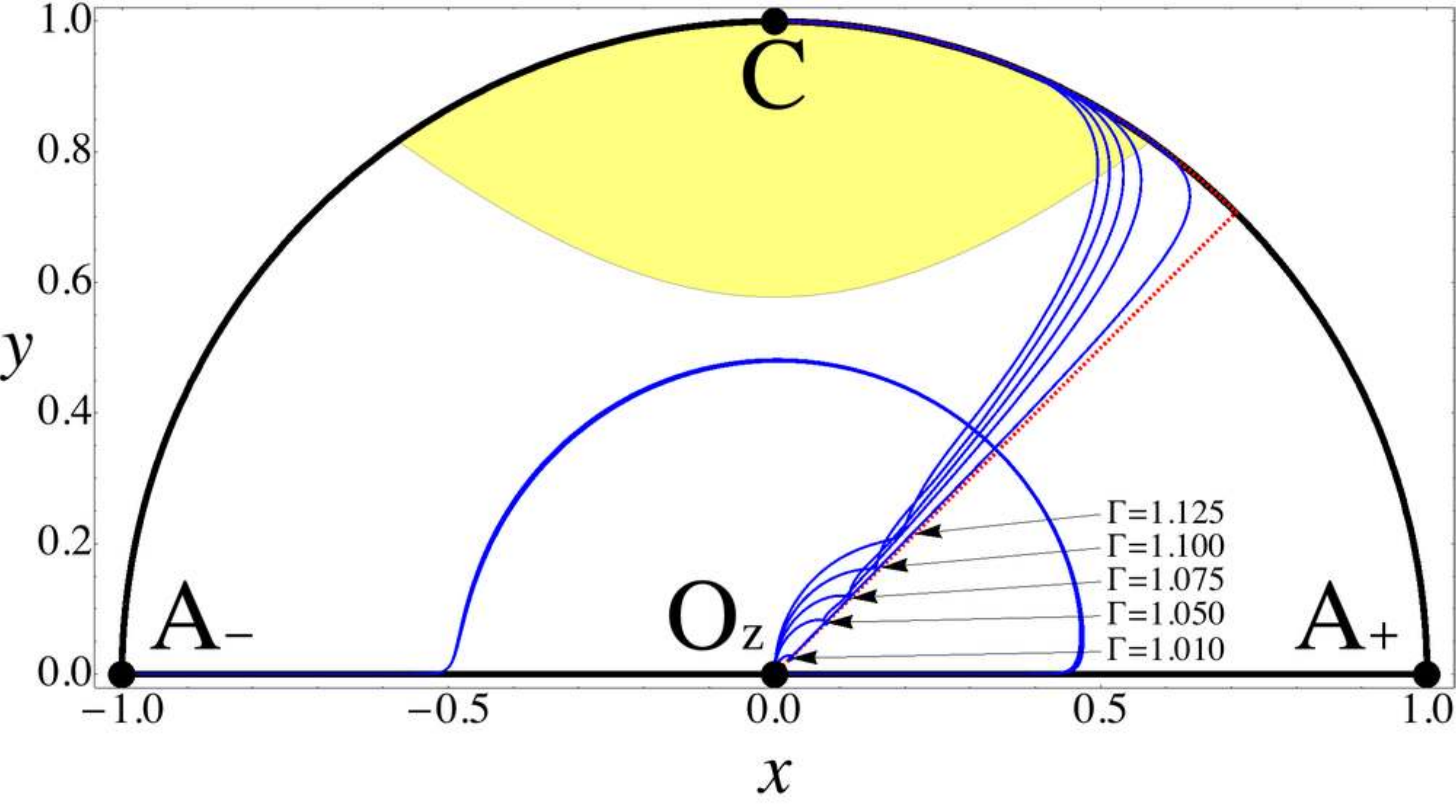}
\caption{Projection onto the $(x,y)$-plane of solutions of the system (\ref{eq:x_power_law_V_compact})--(\ref{eq:z_power_law_V_compact}) with the same initial conditions but corresponding to different values of $\Gamma$ (i.e.~of $\alpha$). The tracking behaviour is well represented by the orbits shadowing the red/dashed line, which has the same meaning as in Fig.~\ref{fig:power_law_projection_0}. The closer $\Gamma$ is to unity, the faster and more efficiently the solutions join the tracking behaviour. The yellow/shaded region denotes the projected part of the phase space where the universe is accelerating.
}
\label{fig:power_law_projection}
\end{figure}

To better visualise what happens during this period, two projections on the $(x,y)$-plane have been drawn\footnote{See \citet{UrenaLopez:2011ur} where similar projections were considered.}. In Fig.~\ref{fig:power_law_projection_0} trajectories with different initial conditions in the case $\Gamma=1.1$ are plotted, while in Fig.~\ref{fig:power_law_projection} orbits with the same initial conditions but corresponding to different values of $\Gamma$ are presented. From these two pictures one can understand that as the solutions leave the matter dominated saddle Point~$O_z$, they are first attracted by and eventually follow the red/dashed line shown in both figures. This represents the position of the attractor solution in the exponential potential case of Sec.~\ref{sec:exp_potential} for all possible values of $\lambda$. The line connecting the origin to the unit circle represents the scaling solutions where the scalar field EoS matches the matter EoS, while the remaining red/dashed line on the circumference stands for the late time attracting dark energy dominating solution of the exponential potential case. The dynamics of orbits in Figs.~\ref{fig:power_law_projection_0} and \ref{fig:power_law_projection} can then be understood in terms of a tracking evolution along the positions where the scaling solutions would appear. Because of this behaviour these solutions are known as {\it tracking solutions}.


Tracking solutions are phenomenologically interesting since they allow the scalar field to follow a matter EoS for a finite period of time and then to switch to the dark energy dominated solutions. The term ``tracking'' refers to the ability of the scalar field to approximately follow the matter evolution in such a way that this approximation eventually fails at late time and a cosmological constant-like EoS is attained with the universe undergoing an asymptotic de Sitter expansion. Note that while the trajectories follows the scaling solutions, the energy density of the scalar field increases with the kinetic energy roughly remaining proportional to the potential energy. This is a behaviour which the exponential potential scaling solutions follow exactly as $\lambda$ changes. Moreover since the orbits are attracted by the positions where the scaling solutions would appear, the points on the red/dashed line of Figs.~\ref{fig:power_law_projection_0} and \ref{fig:power_law_projection} are sometimes called {\it instantaneous critical points}, though mathematically they are not critical points.

As shown in Fig.~\ref{fig:power_law_projection_0} the closer to Points~$O_z$ the orbits take their tour on the $z=1$ plane, the faster they join the tracking behaviour. In fact the solution which is first attracted by the red/dashed line corresponds to the one drawing the smaller circle, while the solution which struggles the most to join the tracking behaviour corresponds to the bigger circle. A similar situation happens for different values of $\Gamma$ as presented in Fig.~\ref{fig:power_law_projection}. The closer $\Gamma$ is to one, the faster and more efficiently the trajectories attain the tracking nature and the more  they remain near the red/dashed line. Models with larger $\alpha$ will lead to a better tracking behaviour as can be understood comparing the $\Gamma=1.01$ ($\alpha=100$) with the $\Gamma=1.125$ ($\alpha=8$) trajectories in Fig.~\ref{fig:power_law_projection}. The condition $\Gamma\simeq 1$ is generally known to be necessary for the achievement of a tracking solution also in models of quintessence with a dynamically changing $\Gamma$ \citep{Steinhardt:1999nw}. If $\Gamma$ is effectively (but not equal to) one the orbits would accurately follow the red/dashed line in Figs.~\ref{fig:power_law_projection_0} and \ref{fig:power_law_projection}. In this situation the dynamics would describe an effective transition from a scaling solution to a dark energy dominated universe, which we could not obtain from the exponential potential case in Sec.~\ref{sec:exp_potential}. In other words the dynamics would equal the exponential potential one with a variable $\lambda$, and the relative energy of the scalar field would start dominating the universe without spoiling the matter-like evolution. We also notice that as $\Gamma$ becomes closer to one, the instantaneous critical points effectively act as attracting spirals \citep{Ng:2001hs}. This can be seen in the $\Gamma=1.05$ and $\Gamma=1.01$ trajectories of Fig.~\ref{fig:power_law_projection} where a small spiralled attraction is achieved before the tracking behaviour becomes increasingly less powerful.

\begin{figure}
\centering
\includegraphics[width=\columnwidth]{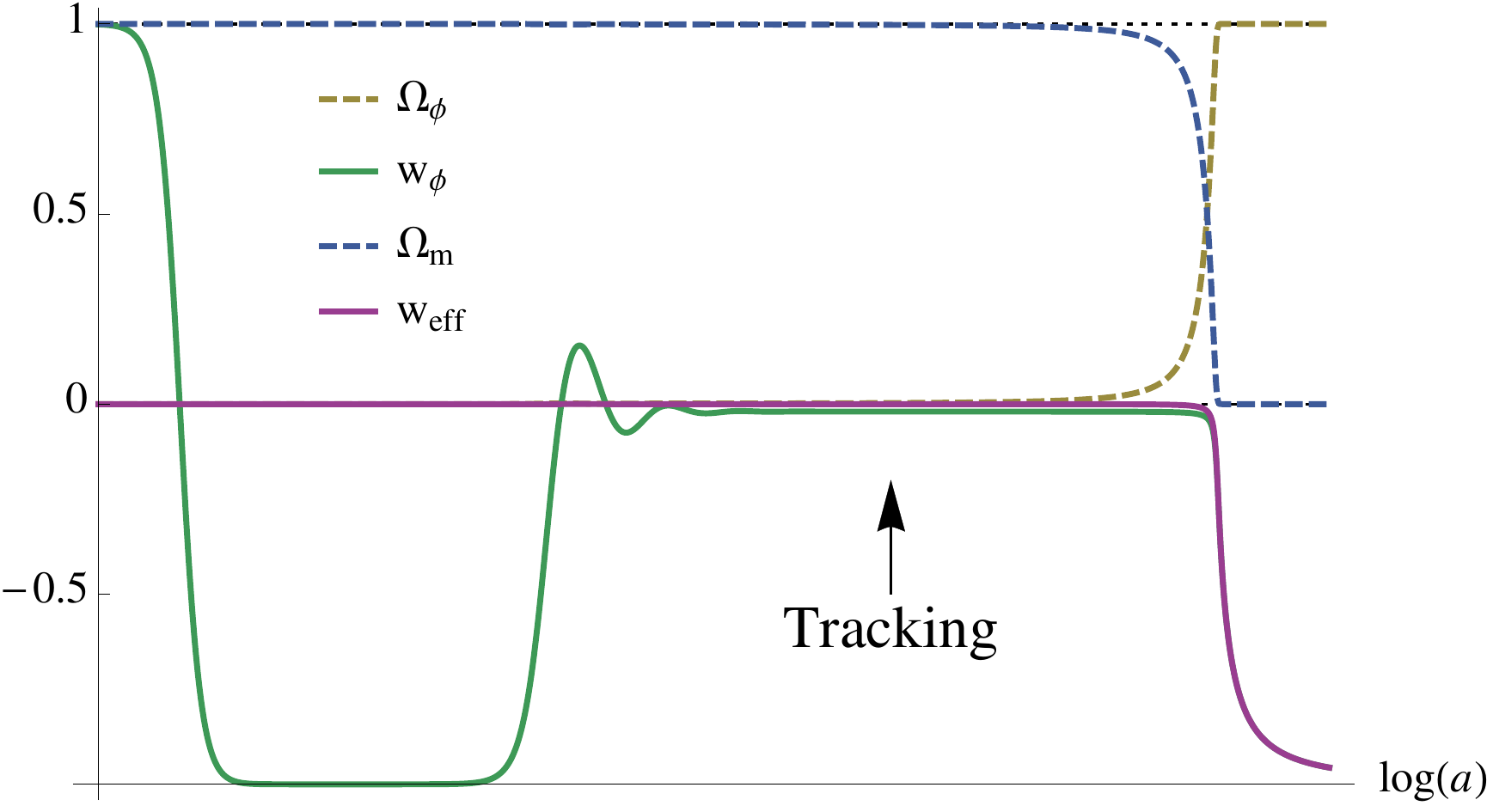}
\caption{Late time evolution of the physically relevant quantities during the matter to dark energy transition of quintessence with inverse power-law potential (\ref{def:inv_power_law_pot}) and $\alpha=100$ ($\Gamma=1.01$). Note the tracking and frozen behaviours of the scalar field (here $w=0$).}
\label{fig:power_law_late_time_physics}
\end{figure}

In order to better understand the dynamics of the scalar field during the matter to dark energy transition, in Fig.~\ref{fig:power_law_late_time_physics} the evolution of the phenomenologically interesting quantities has been plotted for the best solution of Fig.~\ref{fig:power_law_projection}, the one with $\Gamma=1.01$ ($\alpha=100$) (a smaller value of $\alpha$ could be equivalently chosen considering different initial conditions). The tracking behaviour is particularly in evidence as the quintessence EoS shadows the matter EoS before converging towards the $-1$ value of dark energy. Before the tracking regime we note that the scalar field EoS assumes also the cosmological constant value $-1$, though, being its relative energy negligible in comparison to the matter one during that period, this has no influence on the effective evolution of the universe ($w_{\rm eff}=w$). This phase of the scalar field is known as the {\it frozen field} epoch since the energy density of the quintessence field remains constant as the universe expands \citep{Zlatev:1998tr,Steinhardt:1999nw}. Note also the oscillating behaviour as the field approaches the tracking regimes. This is due to the spirally attractive nature of the instantaneous critical points, whose effects are stronger for larger values of $\alpha$ ($\Gamma$ closer to one). At the end of the tracking phase the EoS parameter of the scalar field drops from $w$ to $-1$. The dynamical relation that holds between $w_\phi$ and $\Omega_\phi$ during this phase predicts late time phenomenological signatures whose observation could in principle distinguish between quintessence and the cosmological constant \citep{Zlatev:1998tr,Gong:2014dia}.

To conclude the analysis on the $\Gamma>1$ case we discuss the problem of fine tuning and initial conditions. As we mentioned before trajectories entering the tracking regime can in principle solve this problem. One has to be careful that tracking solutions do not solve the cosmic coincidence problem as one can immediately realise from Fig.~\ref{fig:power_law_late_time_physics} where the matter to dark energy transition happens again at the present cosmological time for no apparent reason. The inverse power-law potential helps in solving another coincidence problem, namely the fine tuning problem of the initial conditions. Recall that the exponential potential quintessence model was able to describe the matter to dark energy transition only for very special initial conditions: the ones allowing for a sufficiently long period of matter domination. In the inverse power-law potential instead almost every orbit which first passes near the $z=1$ plane, i.e.~for which $\lambda\gg 1$ at early times, will eventually reach the tracking regime and then describe a late time transition to dark energy domination, a result known as the {\it tracker theorem} \citep{Zlatev:1998tr,Steinhardt:1999nw,UrenaLopez:2011ur}. This situation is relatively insensitive to the initial conditions, but how general are the solutions passing near the $z=1$ plane? In other words, why should the evolution of phenomenologically relevant trajectories reach the $z=1$ plane after leaving the past attractor Point~$A_-$? To answer this question we will make two arguments: a mathematical one and a physical one.

First we recall that the unstable centre manifold of Point~$A_-$ coincides with its centre subspace which is notably parallel to the $z$ axis. Every trajectory leaving the past attractor is immediately attracted by this centre manifold towards the increasing $z$ direction. Considering then a random solution escaping Point~$A_-$, it is highly probable that such a solution will travel along its centre manifold as only very particular initial conditions will force this orbit not to reach high values of $z$. This happens exactly because the unstable centre manifold of Point~$A_-$ is linear in the $z$ direction. The situation is similar to the one we encountered for the future attractor Point~$C$ where all the orbits approaching the critical point are first captured by its stable centre manifold. In a similar way, if we let time travel backwards, all the orbits approaching the past attractor are first attracted by its centre manifold with the majority of them joining at high values of $z$ near the $z=1$ plane. This implies that given random initial conditions near the past attractor Point~$A_-$ it is highly probable to find a solution passing sufficiently near the $z=1$ plane and thus describing the tracking behaviour at late time.

The second argument is more physical and regards the value of the amplitude of the quintessence field $\phi$. Recalling the definitions of $\lambda$ (\ref{def:lambda_dynamical}) and of the Ratra-Peebles potential (\ref{def:inv_power_law_pot}) we have
\begin{align}
	\lambda = -\frac{1}{\kappa}\frac{V_{,\phi}}{ V} = \frac{\alpha}{\kappa \phi} \,, \quad\mbox{i.e.}\quad z=\frac{1}{1+\kappa \phi/\alpha} \,, \label{064}
\end{align}
which implies that the amplitude of the scalar field $\phi$ is inversely related to $\lambda$. If we make the assumption that $\kappa\phi\ll \alpha$ at some time in the past, we automatically select trajectories near the $z=1$ plane where $\lambda\rightarrow+\infty$. A small scalar field amplitude in the early universe is phenomenologically favoured since it avoids possible problems that can arise with large field excursions and the possibility of the effective field description breaking down in such a regime. In fact it allows the effects of the scalar field to be negligible when other fields, for example the inflaton, have to drive the universe evolution. From a phenomenological perspective it is thus natural to require the condition $\kappa\phi\ll \alpha$ at early times, meaning that the physically acceptable solutions in Fig.~\ref{fig:power_law_3D_plot} will be the ones approaching the $z=1$ plane and converging to the tracking behaviour at late time.

\begin{figure}
\centering
\includegraphics[width=\columnwidth]{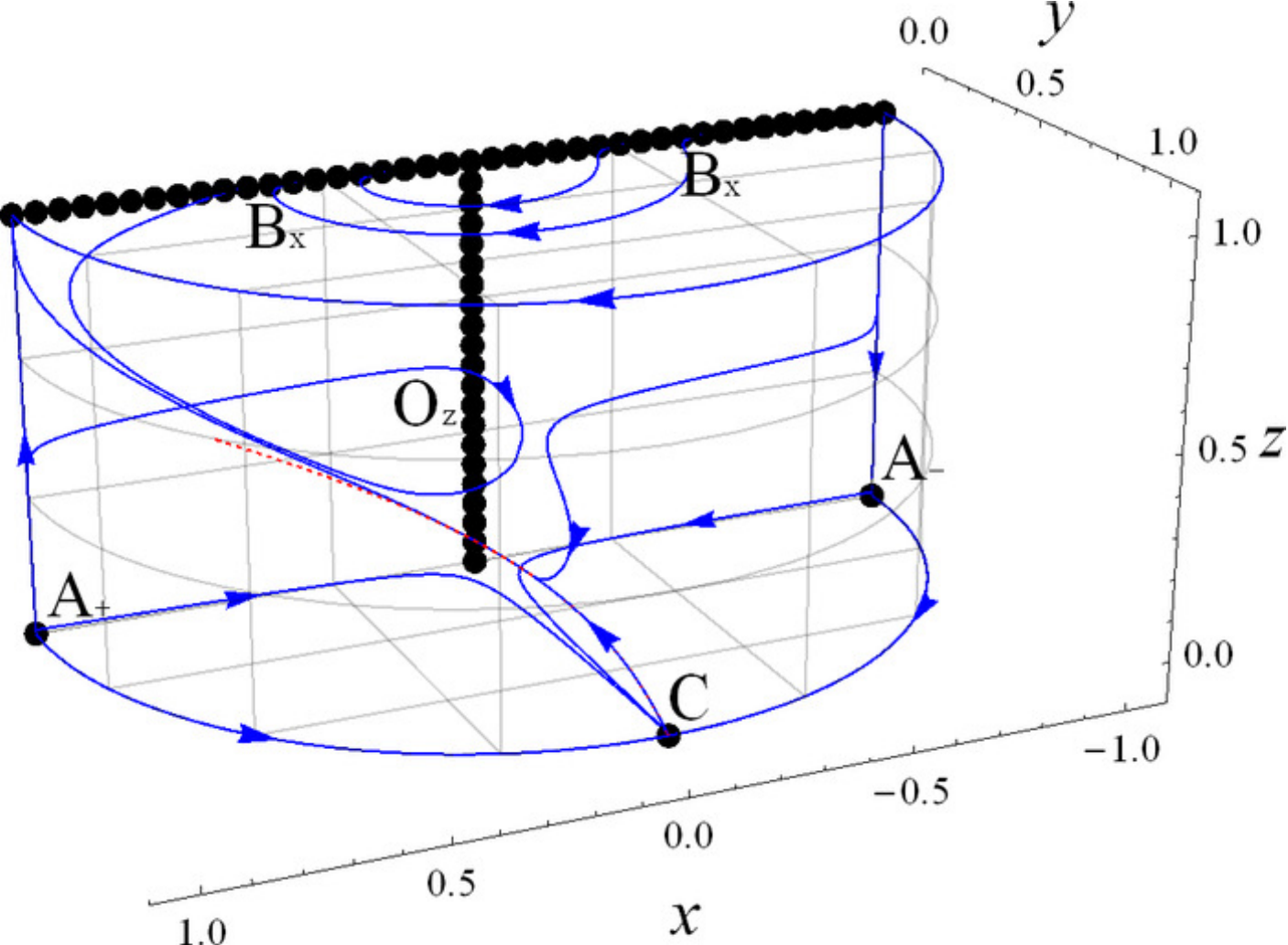}
\caption{Phase space of quintessence for a direct power-law potential corresponding to the dynamical system (\ref{eq:x_power_law_V_compact})--(\ref{eq:z_power_law_V_compact}). The values $w=0$ and $\alpha=-10$ ($\Gamma=0.9$) have been chosen. Again black/thick points denote critical points with the lines $B_x$ and $O_z$ being critical lines. The late time attractors are Points~$B_x$ for $x>0$, while the past attractors are either Point~$A_+$ or Points~$B_x$ for $x<0$. Point~$C$ is now a saddle point attracting every solution near the $z=0$ plane and repelling them along its centre manifold (thawing behaviour), here denoted up to 7th order in $\tilde{z}$ by the red/dashed line.}
\label{fig:power_law_3D_plot_inverted}
\end{figure}

\subsubsection{Direct power-law potentials}
\label{sec:quintessence_direct_power_law}

We now turn our discussion to the direct power-law potential case: $\alpha<0$ ($\Gamma<1$).
The dynamics of this model has been extensively studied by \citet{Alho:2015cza}, who investigated the global behaviour of the phase space using different variables, but similarly considering compactification and centre manifold analyses.
In Fig.~\ref{fig:power_law_3D_plot_inverted} the phase space for the values $\alpha=-10$ ($\Gamma=0.9$) and $w=0$ has been drawn. We will first briefly outline the dynamical properties of the phase space and then derive its phenomenological implications.

From Fig.~\ref{fig:power_law_3D_plot_inverted} it seems that the future attractors are Points~$B_x$ for $x>0$, while the past attractors are either Point~$A_+$ or Points~$B_x$ for $x<0$. This is the result we obtained before with analytical methods (see Tab.~\ref{tab:power_law_CP_stability}). Trajectories on the $z=1$ plane follow the same behaviour we discussed for the inverse power-law potential. In fact the dynamics on this plane does not depend on $\alpha$, as one can see from the solutions \eqref{062}--\eqref{063}. Trajectories on the $z=0$ plane are instead attracted by Point~$C$ and then repelled along its centre manifold. They eventually reach Points~$B_x$ (for $x>0$) which act as an attractor.

We can now discuss the physical implications of this model. The future possible attractors are Points~$B_x$ for $x>0$ which, as shown in Tab.~\ref{tab:power_law_CP_physics}, never describe an accelerating universe. It seems thus that a direct power-law potential ($\alpha<0$) is not suited for characterising the late time transition from decelerated to accelerated expansion. However Point~$C$ is now a saddle point which attracts solutions on the $z=0$ plane. For some of these orbits, the ones passing close to the origin, a matter to dark energy transition can be identified by the heteroclinic orbit connecting Point~$O_z$ ($z=0$) to Point~$C$. This is nothing but an effective cosmological constant solution since, the dynamics on the $z=0$ plane is given by a constant quintessence potential ($z=0$, i.e.~$\lambda=0$, implies $V_{,\phi}=0$ by definition \eqref{def:lambda_dynamical}). The final state for these trajectories is not represented by Point~$C$ but eventually they always escape towards Point~$B_x$ ($x>0$). Scalar field models of dark energy with an initial small value of $z$ (i.e.~of $\lambda$) are known as {\it thawing models} (see e.g.~\citet{Clemson:2008ua} and \citet{Pantazis:2016} for a discussion about thawing and freezing dark energy models). For these solutions $\lambda$ is an increasing function of time and the accelerated cosmological constant phase (Point~$C$) never represents the final state of the universe. Although they constitute other possible models of dark energy, they cannot solve the fine tuning problem, as the tracking solutions do, since special initial conditions are required for a sufficiently long matter dominated era.

\subsection{Dynamics with other scalar field potentials}
\label{sec:other_potentials}

In this section we deal with more complicated potentials that arise in quintessence scenarios and study them within a unified framework. To begin, we follow the approach first introduced by \citet{Zhou:2007xp}, and developed by others\footnote{Some authors call this approach the {\it method of $f$-devisers} \citep{Escobar:2013js}.} \citep{Fang:2008fw,Matos:2009hf,UrenaLopez:2011ur}, where different forms of the potential $V(\phi)$ can be translated to different forms of the function $\Gamma(\lambda)$. As we mentioned in Sec.~\ref{sec:quintessence}, if the function $\lambda(\phi)$, as defined by (\ref{def:lambda_dynamical}), is invertible then one can consider $\Gamma(\phi)$ as a function of $\lambda$ and close Eqs.~(\ref{eq:x_can_scalar_field_gen_V}), (\ref{eq:y_can_scalar_field_gen_V}) and (\ref{eq:lambda}) to an autonomous dynamical system, namely
\begin{align}
	x'&=-\frac{3}{2} \left[2x+(w-1) x^3+ x (w+1) \left( y^2-1\right)-\frac{\sqrt{2}}{\sqrt{3}} \lambda  y^2\right]\,, \label{eq:x_other_V}\\
	y'&=-\frac{3}{2} y \left[(w-1) x^2+(w+1) \left(y^2-1\right)+\frac{\sqrt{2}}{\sqrt{3}} \lambda  x\right]\,,\label{eq:y_other_V}\\
	\lambda' &= -\sqrt{6}f(\lambda)x \,. \label{eq:lambda_other_V}
\end{align}
where we have defined
\begin{align}
	f(\lambda)=\lambda^2[\Gamma(\lambda)-1] \,.
	\label{067}
\end{align}
Instead of considering a specific function $\Gamma(\lambda)$, i.e.~a specific potential $V(\phi)$, in what follows we will try to obtain as much information as possible from the system (\ref{eq:x_other_V})--(\ref{eq:lambda_other_V}) leaving $\Gamma$ as an arbitrary function of $\lambda$. We will not assume any particular symmetry for the function $\Gamma$, meaning that the only symmetry of the system (\ref{eq:x_other_V})--(\ref{eq:lambda_other_V}) will be the $y\mapsto-y$ reflection\footnote{Remember that the $(x,\lambda)\mapsto(-x,-\lambda)$ symmetry holds only if $\Gamma(\lambda)=\Gamma(-\lambda)$.}. The (physical) phase space under consideration is thus the infinite positive $y$ half unit cylinder stretching from $\lambda\rightarrow-\infty$ to $\lambda\rightarrow+\infty$.

First we need to find the possible critical points of the system (\ref{eq:x_other_V})--(\ref{eq:lambda_other_V}). We start looking at Eq.~(\ref{eq:lambda_other_V}) whose left hand side can be zero either if $x=0$ or $f(\lambda)=0$. As can be seen in Tab.~\ref{tab:other_CP_physics}, in the first case ($x=0$) we find critical points either if $y=0$ (Points~$O_\lambda$) or if $y=1$ and $\lambda=0$ (Point~$D$). As long as $\Gamma$ can be written as a function of $\lambda$ (and $f(0)$ is finite), these two critical points are independent of the quintessence model under investigation, i.e.~they are critical points for all possible potentials $V(\phi)$.
The second possibility ($f(\lambda)=0$) can be realised for more than one value of $\lambda$. If $\lambda_*$ is such that $f(\lambda_*)=0$, i.e.~$\lambda_*$ is a zero of the function $f(\lambda)$, then the remaining two Eqs.~(\ref{eq:x_other_V}) and (\ref{eq:y_other_V}) describe exactly the exponential potential system of Sec.~\ref{sec:exp_potential} and thus we will find again the same critical points of Tab.~\ref{tab:exp_CP_physics} with $\lambda=\lambda_*$. Apart from the points where $x=y=0$, which are always critical points, we find that there are up to four (depending on the value of $\lambda_*$) critical points for every zero of the function $f(\lambda)$ (Points~$A_\pm^*$, $B^*$ and $C^*$). All these points have been listed in Tab.~\ref{tab:other_CP_physics} with their phenomenological properties.

\begin{table}
\begin{center}
\begin{tabular}{|c|c|c|c|c|c|c|c|}
\hline
Point & $x$ & $y$ & $\lambda$ & Existence & $w_{\rm eff}$ & Accel. & $\Omega_\phi$ \\
\hline
\multirow{2}*{$O_\lambda$} & \multirow{2}*{0} & \multirow{2}*{0} & \multirow{2}*{Any} & \multirow{2}*{Always} & \multirow{2}*{$w$} & \multirow{2}*{No} & \multirow{2}*{0} \\ & & & & & & & \\
\multirow{2}*{$A^*_\pm$} & \multirow{2}*{$\pm 1$} & \multirow{2}*{0} & \multirow{2}*{$\lambda_*$} & \multirow{2}*{$\forall\;\lambda_*$} & \multirow{2}*{1} & \multirow{2}*{No} & \multirow{2}*{1} \\ & & & & & & & \\
\multirow{2}*{$B^*$} & \multirow{2}*{$\frac{\sqrt{3}}{\sqrt{2}}\frac{1+w}{\lambda_*}$} & \multirow{2}*{$\sqrt{\frac{3(1-w^2)}{2\lambda_*^2}}$} & \multirow{2}*{$\lambda_*$} & \multirow{2}*{$\lambda_*^2\geq 3(1+w)$} & \multirow{2}*{$w$} & \multirow{2}*{No} & \multirow{2}*{$\frac{3(1+w)}{\lambda_*^2}$} \\ & & & & & & & \\
\multirow{2}*{$C^*$} & \multirow{2}*{$\lambda_*/\sqrt{6}$} & \multirow{2}*{$\sqrt{1-\frac{\lambda_*^2}{6}}$} & \multirow{2}*{$\lambda_*$} & \multirow{2}*{$\lambda_*^2<6$} & \multirow{2}*{$\frac{\lambda_*^2}{3}-1$} & \multirow{2}*{$\lambda_*^2<2$} & \multirow{2}*{1} \\ & & & & & & & \\
\multirow{2}*{$D$} & \multirow{2}*{0} & \multirow{2}*{1} & \multirow{2}*{0} & \multirow{2}*{Always} & \multirow{2}*{-1} & \multirow{2}*{Yes} & \multirow{2}*{1} \\ & & & & & & & \\
\hline
\end{tabular}
\caption{Critical points of the system (\ref{eq:x_other_V})--(\ref{eq:lambda_other_V}) with existence and physical properties. $\lambda_*$ is any zero of the function $f(\lambda)$ given in (\ref{067}).}
\label{tab:other_CP_physics}
\end{center}
\end{table}

For every node $\lambda_*$ of the function $f(\lambda)$ the number of critical points to add in the phase space depends on the value of $\lambda_*$ itself. If $\lambda_*=0$ the only two critical points to add are Points~$A_\pm^*$ since in this case Point~$C^*$ coincides with Point~$D$. If $0<\lambda_*^2<3(1+w)$ we add three critical points: Points~$A_\pm^*$ and Point~$C^*$. If $3(1+w)\leq\lambda_*^2< 6$ we add all four critical points: Points~$A_\pm^*$, Point~$B^*$ and Point~$C^*$. Finally if $\lambda_*^2\geq 6$ we add only Points~$A_\pm$ and Point~$B^*$. In general the number of critical points in the phase space will depend on the quintessence potential through the function $f(\lambda)$.
Note that the analysis we have performed is valid only if the function $f(\lambda)$ is finite for all possible values of $\lambda$. If at some $\lambda_\infty$ we have $f(\lambda_\infty)=\pm\infty$, the dynamical system (\ref{eq:x_other_V})--(\ref{eq:lambda_other_V}) is no longer differentiable and, in order to apply dynamical systems techniques, a change of variables must first be taken into account. In this case it is not guaranteed that the critical points will follow the scheme outlined in Tab.~\ref{tab:other_CP_physics}.
Furthermore we have not considered critical points at infinity since these will depend on the specific quintessence models through the function $\Gamma(\lambda)$. In general we can state that if $\Gamma(\lambda)$ is finite as $\lambda\rightarrow\pm\infty$, the only critical points at infinity will be given by the critical line at $x=0$, exactly as it happens in the power-law case of Sec.~\ref{sec:power_law_potential}. However if the function $\Gamma(\lambda)$ diverges as $\lambda\rightarrow\pm\infty$ the dynamics at infinity could be more complicated and each model has to be analysed separately.

\begin{table}
\begin{center}
\begin{tabular}{|c|c|c|c|}
\hline
P & Eigenvalues & Stability \\
\hline
\multirow{2}*{$O_\lambda$} & \multirow{2}*{$\{0,\frac{3}{2} (w\pm 1)\}$} & \multirow{2}*{Saddle} \\ & & \\
\multirow{3}*{$A_-^*$} & \multirow{3}*{$\{3 - 3 w\,,\, 3 + \frac{\sqrt{3}}{\sqrt{2}} \lambda_*,\,\sqrt{6}\lambda_*^2 \Gamma'_*\}$} & \multirow{2}*{Unstable if $\lambda_*>-\sqrt{6}$ and $\Gamma'_*>0$} \\ & & \multirow{2}*{Saddle if $\lambda_*<-\sqrt{6}$ or $\Gamma'_*<0$} \\ & & \\
\multirow{2}*{$A_+^*$} & \multirow{2}*{$\{3 - 3 w\,,\, 3 - \frac{\sqrt{3}}{\sqrt{2}} \lambda_*,\,-\sqrt{6}\lambda_*^2 \Gamma'_*\}$} & Unstable if $\lambda_*<\sqrt{6}$ and $\Gamma'_*<0$ \\ & & Saddle if $\lambda_*>\sqrt{6}$ or $\Gamma'_*>0$ \\
\multirow{3}*{$B^*$} & \multirow{3}*{$\{\frac{3}{4\lambda_*}\left[(w-1)\lambda_*\pm\Delta\right],\,-3(w+1)\lambda_*\Gamma'_*\}$} & \multirow{2}*{Stable if $\lambda_*\Gamma'_*>0$} \\ & & \multirow{2}*{Saddle if $\lambda_*\Gamma'_*<0$} \\ & & \\
\multirow{2}*{$C^*$} & \multirow{2}*{$\{\frac{\lambda_*^2}{2}-3\,,\,\lambda_* ^2-3w-3,\,-\lambda_*^3\Gamma'_*\}$} & Stable if $\lambda_*^2<3(1+w)$ and $\lambda_*\Gamma'_*>0$ \\ & & Saddle if $3(1+w)\leq\lambda^2<6$ or $\lambda_*\Gamma'_*<0$ \\
\multirow{3}*{$D$} & \multirow{3}*{$\{-3(w+1),\,-\frac{3}{2}\left(1\pm\sqrt{1-4f(0)/3}\right)\}$} & \multirow{2}*{Stable if $f(0)>0$} \\ & & \multirow{2}*{Saddle if $f(0)<0$} \\ & & \\
\hline
\end{tabular}
\caption{Stability properties for the critical points of the system (\ref{eq:x_other_V})--(\ref{eq:lambda_other_V}). Here $\Gamma'_*$ is the derivative of $\Gamma(\lambda)$ evaluated at $\lambda_*$, $f(\lambda)$ is given by Eq.~(\ref{067}) and $\Delta=\sqrt{(w-1)[(7+9w)\lambda_*^2-24(w+1)^2]}$.}
\label{tab:other_CP_stability}
\end{center}
\end{table}

The linear stability of the critical points has been summarised in Tab.~\ref{tab:other_CP_stability}. In general this depends on the different values of $\lambda_*$ and on the form of the function $\Gamma(\lambda)$. Points on the critical line $O_\lambda$ are always saddle points since the two non vanishing eigenvalues have opposite sign (recall that $0\leq w\leq 1/3$). Points~$A_\pm^*$ are never stable points and represent saddle or unstable points depending on the sign of $\Gamma'_*$ (the derivative of $\Gamma(\lambda)$ with respect to $\lambda$ evaluated at $\lambda_*$) and on the sign of $\lambda_*^2 -6$. Whenever Point~$B^*$ exists, i.e.~when $\lambda_*^2\geq 3(w+1)$, it represents a stable point if $\lambda_*\Gamma'_*>0$ and a saddle point if $\lambda_*\Gamma'_*<0$. Point~$C^*$ is a stable point if both $\lambda_*^2< 3(w+1)$ and $\lambda_*\Gamma'_*>0$, while it is a saddle point if either $3(1+w)\leq\lambda^2<6$ or $\lambda_*\Gamma'_*<0$. Note that in the case $\lambda_*\Gamma'_*=0$, corresponding to $f'(0)=0$, Points~$A_\pm^*$, $B^*$ and $C^*$ all become non-hyperbolic and the linear theory fails to address their stability nature. This is the case of the power-law potential of Sec.~\ref{sec:power_law_potential}. Finally Point~$D$ is a stable or a saddle point depending on $f(0)$ being positive or negative respectively. If $f(0)=0$ one of the eigenvalues vanishes and the point becomes non-hyperbolic. To address the stability in this case one can rely on the centre manifold theorem and perform a similar computation to the one we considered for Point~$C$ of Sec.~\ref{sec:power_law_potential}, arriving at the following results. If $f(0)=0$ then Point~$D$ is stable\footnote{Note that this result is in agreement with the appendix of \citet{Fang:2008fw}, although the case $f(0)\neq 0$ was not considered in that work.} if $\Gamma(0)>1$ and a saddle if $\Gamma(0)<1$. If also $\Gamma(0)=0$, the stability will depend on the sign of $\Gamma'(0)$.

The critical points for a general quintessence model where $\Gamma$ can be written as a function of $\lambda$ are repetitions of the critical points one finds with the exponential and power-law potentials. From Tab.~\ref{tab:other_CP_physics} we can see that Points~$A_\pm^*$, $B^*$ and $C^*$ have the same phenomenological properties as Points~$A_\pm$, $B$ and $C$ of the exponential case in Sec.~\ref{sec:exp_potential}, while Points~$O_\lambda$ and $D$ correspond to the critical line $O_z$ and Point~$C$ of the power-law potential of Sec.~\ref{sec:power_law_potential}. Of course there can be several of these points in the phase space depending on the number and values of the nodes of the function $f(\lambda)$, namely $\lambda_*$. The resulting phase space dynamics can be highly complicated and the stability of these points now depends on the properties of the function $\Gamma(\lambda)$. Multiple late time attractors can be present, as is evident from Tab.~\ref{tab:other_CP_stability}. If none of these critical points constitutes a future attractor, periodic orbits could appear in the phase space or, as in the power-law potential case with $\alpha<0$, the attractor can be a critical point at infinity.

There are only two points which are relevant for dark energy phenomenology: Point~$C^*$ and Point~$D$. The first one represents an accelerating universe only if $\lambda_*<2$, while the second one characterises a cosmological constant-like dominated universe where a de Sitter expansion is guaranteed. They are important especially when representing future attractors. Note that Point~$D$ appears in every quintessence model and if $f(0)>0$ it can always constitute a possible late time dark energy dominated solution. Moreover the matter dominated Points~$O_\lambda$ are saddle points for every possible potential. One can thus expect to find a late time matter to dark energy transition independently of the quintessence model chosen. Of course this will in general depend on initial conditions, but for some of these models scaling solutions can also be achieved whenever Point~$B^*$ appears in the phase space. If this point (or one of them if more than one are present) happens to be a saddle, then the matter to dark energy transition can be described by an heteroclinic orbit connecting Point~$B^*$ with Point~$D$ (or Point~$C^*$ for a different $\lambda_*$ with $\lambda_*^2<2$). In this case the fine tuning problem of initial conditions can be avoided if the basin of attraction of Point~$B^*$ is sufficiently large. This situation is similar to the tracking regime we encountered for the inverse power-law case of Sec.~\ref{sec:power_law_potential}, although in that case the scaling solutions were only instantaneous critical points. To determine which potentials are capable of yielding scaling solutions is thus an important issue in quintessence models \citep{Nunes:2000yc,Copeland:2004qe}.

\begin{table}
\begin{center}
\begin{tabular}{|c|c|c|}
\hline
$V(\phi)$ & $\Gamma(\lambda)-1$ & References \\
\hline
\multirow{2}*{$V_1e^{\alpha\phi}+V_2e^{\beta\phi}$} & \multirow{2}*{$-\left(\alpha+\lambda\right)\left(\beta+\lambda\right)/\lambda^2$} & \multirow{2}*{\citet{Jarv:2004uk,Li:2005ay}} \\ & & \\
$V_0\cos^2(\alpha\phi)$ or & \multirow{2}*{$-\frac{1}{2}-\frac{\alpha^2}{\lambda^2}$} & \citet{Ng:2000di} \\
$\frac{V_0}{2}\left[1\pm\cos(2\alpha\phi)\right]$ & & \citet{UrenaLopez:2011ur,Gong:2014dia} \\
\multirow{2}*{$\exp\left[\alpha\exp\left(\beta\phi\right)\right]$} & \multirow{2}*{$-\beta/\lambda$} & \multirow{2}*{\citet{Ng:2001hs}} \\ & & \\
\multirow{2}*{$V_0\phi^{-n} e^{\alpha \phi}$} & \multirow{2}*{$(1+\alpha/\lambda)^2/n$} & \multirow{2}*{\citet{Ng:2001hs}} \\ & & \\
\multirow{3}*{$V_0 \sinh^n(\sigma\phi)$ or} & \multirow{4}*{$-\frac{1}{n}+\frac{n\sigma^2}{\lambda^2}$} & \citet{Fang:2008fw} \\
	\multirow{3}*{$V_0 \cosh^n(\sigma\phi)$} & & \citet{Roy:2013wqa} \\ & & \citet{Garcia-Salcedo:2015ora} \\ & & \citet{Paliathanasis:2015gga} \\
\multirow{2}*{$V_0 \left[\cosh(\sigma\phi)-1\right]+\Lambda$} & \multirow{2}*{Eq.~(\ref{068})} & \multirow{2}*{\citet{Matos:2009hf}} \\ & & \\
\multirow{2}*{$V_0 e^{\alpha\phi(\phi+\beta)/2}$} & \multirow{2}*{$\alpha/\lambda^2$} & \multirow{2}*{\citet{Fang:2008fw}} \\ & & \\
\multirow{2}*{$V_0 e^{\alpha/\phi}$} & \multirow{2}*{$1/\sqrt{\alpha\lambda}$} & \multirow{2}*{\citet{Fang:2008fw}} \\ & & \\
\multirow{2}*{$V_0\left(\eta+e^{-\alpha\phi}\right)^{-\beta}$} & \multirow{2}*{$\frac{1}{\beta}+\frac{\alpha}{\lambda}$} & \multirow{2}*{\citet{Zhou:2007xp,Fang:2008fw}} \\ & & \\
\multirow{2}*{$V_0 e^{\alpha \phi} (1 + \beta \phi)$} & \multirow{2}*{$- \frac{(\alpha + \lambda)^2}{\lambda^2}$} & \multirow{2}*{\citet{Clemson:2008ua}} \\ & & \\
\hline
\end{tabular}
\caption{Quintessence potentials for which $\Gamma$, as defined in Eq.~\eqref{def:Gamma}, can be related to $\lambda$ and reference to the relevant dynamical system literature. Here we have taken $\kappa = 1$ for simplicity and all quantities except $\phi$ are constant parameters.}
\label{tab:other_potentials}
\end{center}
\end{table}

Now that we have gained useful information on the general behaviour of quintessence models leading to a dynamical $\Gamma$, we can spend some words on a few specific models. In Tab.~\ref{tab:other_potentials} we have listed quintessence potentials considered in the literature for which the relation $\Gamma(\lambda)$ can be determined. Some of them are motivated by well known high energy phenomenology, others by the simple relation they provide for $\Gamma(\lambda)$. Quintessence models with a pseudo-Nambu-Goldstone-boson (PNGB) potential given by cos-like expressions (see Tab.~\ref{tab:other_potentials}) are amongst the most studied with dynamical systems techniques \citep{Ng:2000di,UrenaLopez:2011ur,Gong:2014dia}. They represent thawing dark energy models with an intermediate inflationary phase and a future complicated behaviour which strongly depends on the potential parameters but never leads to accelerated expansion. Other well studied potentials within this approach are cosh-like potentials \citep{Matos:2009hf,Kiselev:2006jy} which can be justified by string theory phenomenology and are employed in unified dark matter models. For example \citet{Matos:2009hf} considered the potential
\begin{align}
	V(\phi)=V_0\left[\cosh(\sigma\phi)-1\right]+\Lambda \,,
\end{align}
corresponding to
\begin{align}
	\Gamma(\lambda)-1 = \pm \left(\frac{1}{\lambda^2}-1\right) \frac{\sqrt{1+\alpha(\alpha-2)\lambda^2}}{\alpha-1\pm\sqrt{1+\alpha(\alpha-2)\lambda^2}} \,,
	\label{068}
\end{align}
where $\alpha=\Lambda/V_0$ and the plus/minus sign represents different branches of the solution. The interesting phenomenological properties of these models can be found in the intermediate behaviour where the scalar field acts as a dark matter candidate. The late time solution is given by a de Sitter expansion produced by a cosmological constant added to the potential.
Finally one of the most studied potentials \citep{Barreiro:1999zs,Jarv:2004uk,Li:2005ay} is the so-called double exponential potential
\begin{align}
	V(\phi) = V_1e^{\alpha\phi}+V_2e^{\beta\phi} \,,
\end{align}
with $V_1$, $V_2$, $\alpha$ and $\beta$ all constant. It represents a straightforward generalisation of the exponential potential case of Sec.~\ref{sec:exp_potential} and it is interesting under a phenomenological point of view since, depending on the values of the parameters, scaling and dark energy dominated solutions can appear together in the phase space. Although a simple $\Gamma(\lambda)$ relation arises in the double exponential potential case (see Tab.~\ref{tab:other_potentials}), for this particular model it is easier to employ EN-like variables defined as $y^2=V_1e^{\alpha\phi}/(3H^2)$ and $z^2=V_2e^{\beta\phi}/(3H^2)$ \citep{Li:2005ay}. With these variables the physical phase space becomes automatically compact due to the Friedmann constraint which would read $x^2+y^2+z^2\leq 1$.

Quintessence models with potentials characterised by hyper-geometric functions are usually associated to unified dark matter (UDM) scenarios, and have been largely explored with dynamical systems techniques as one can realise from Tab.~\ref{tab:other_potentials}. Some authors have also studied these models for non-flat cosmology finding interesting mathematical features. \cite{LukesGerakopoulos:2008rr} investigated closed universes in a UDM model with a potential of the form of $V(\phi)=V_0 \cosh^2(p\phi) + \Lambda$. They noticed that models which are close to being spatially flat ($k\rightarrow 0^{+}$) exhibit a chaotic behaviour after a long time, whereas pure spatially flat models ($k=0$) do not. Subsequently \cite{Acquaviva:2016fxs} reinterpreted the same analysis in a more general setting by introducing new dimensionless variables allowing for the compactification of the phase space.

Let us now discuss what results one can derive making only minimal assumptions on the scalar field potentials. In this respect we highlight the work by \cite{Alho:2015ila} who analysed the general features and asymptotic properties of quintessence models with a general positive monotonic potential. Splitting their analysis into two cases: bounded and unbounded $\lambda$, they looked at the monotonicity of observable variables, and examined the structure of the system on the different boundaries of the phase space, finding that $\lambda$ must be globally and asymptotically bounded for observational viability.

To conclude this section we recall that quintessence potentials for which the function $\Gamma(\lambda)$ cannot be obtained analytically must be analysed using another approach. In these cases different variables than the EN ones might represent a better choice to characterise the dynamics of the system (e.g.~see \citet{Miritzis:2003ym,Hao:2003aa,Faraoni:2012bf}). However, as shown in Tab.~\ref{tab:other_potentials}, for almost every quintessence potential considered as a viable model in the literature the relation $\Gamma(\lambda)$ can easily be obtained. This provides a unified framework to analyse the isotropic\footnote{Anisotropic spacetimes for canonical scalar fields with complicated potentials have been studied by \citet{Fadragas:2013ina}.} background dynamics of canonical scalar field models of dark energy and only highly complicated potentials will fail to fit into such a scheme. Note also that every potential in Tab.~\ref{tab:other_potentials} yields a function $\Gamma(\lambda)$ which is finite as $\lambda\rightarrow\pm\infty$. As we mentioned before, this implies that for all these models the dynamics at infinity will be identical to the one described in Sec.~\ref{sec:power_law_potential} for the power-law potential.

\subsection{Multiple scalar field models}
\label{sec:multi_scalar_fields}

Theoretically, it is possible that dark energy is composed of more than one scalar field. One can study a model with $N$ minimally coupled canonical scalar fields by considering the Lagrangian
\begin{align}
  \mathcal{L}_{\phi_1,...\,,\phi_N} = -\sum_{i=1}^N \partial\phi_i^2 - V(\phi_1,...\,,\phi_N) \,,
  \label{071a}
\end{align}
where $V(\phi_1,...\,,\phi_N)$ is a general potential depending on all scalar fields.
The Friedmann and acceleration equations obtained from the Einstein field equations sourced by the scalar field (\ref{071a}) plus a standard matter fluid component are
\begin{align}
\frac{3 H^2}{\kappa^2} &= \rho + \left(\sum_i \frac{\dot\phi_i^2}{2}\right) + V \,,\label{073}\\
\frac{1}{\kappa^2}\left(2\dot H+3H^2\right) &= -p - \left(\sum_i \frac{\dot\phi_i^2}{2}\right) + V \,,
\end{align}
while the Klein-Gordon equation of the $i$th scalar field is
\begin{align}
\ddot\phi_i +3H\dot\phi_i +\frac{\partial V}{\partial\phi_i} = 0\,, \qquad i=1,\ldots,N\,.
\label{072}
\end{align}

\subsubsection{Non-interacting potentials: assisted quintessence}

A well-known model of this type is known as {\it assisted quintessence}, being a late time adaptation of the {\it assisted inflationary model} \citep{Liddle:1998jc, Copeland:1999cs, Malik:1998gy}. The multifield potential is in this case given by
\begin{align}
  V(\phi_1,...\,,\phi_N) = \sum_i V_i(\phi_i) = \sum_i V_{0i} e^{-\lambda_i\kappa\phi_i} \,,
  \label{074}
\end{align}
where $\lambda_i$ and $V_{0i}$ are $N$ parameters. The potential (\ref{074}) is the sum of $N$ individual exponential potential terms for each single scalar field. Knowing the dynamics of the single quintessence model with an exponential potential, it is clear that the multifield potential (\ref{074}) is the simplest choice one can make to result in a reasonably simple dynamical system. In analogy to the single field case (\ref{def:ENvars}), one defines $2N$ normalised variables as
\begin{align}
  x_i = \frac{\kappa\dot\phi_i}{\sqrt{6}H} \,,\quad\mbox{and}\quad y_i=\frac{\kappa \sqrt{V_i}}{\sqrt{3}H} \,,
\end{align}
which reduces the cosmological equations to the $2N$-dimensional dynamical system represented by
\begin{align}
  x_i' &= -3x_i +\frac{\sqrt{3}}{\sqrt{2}}\lambda_i y_i^2 +\frac{3}{2}x_i \sum_j\left[x_j^2+(w+1)\left(1-x_j^2-y_j^2\right) \right] \,,\label{075}\\
  y_i' &= -\frac{\sqrt{3}}{\sqrt{2}}\lambda_i x_iy_i +\frac{3}{2} y_i \sum_j\left[x_j^2+(w+1)\left(1-x_j^2-y_j^2\right) \right] \,,\label{076}
\end{align}
subject to the Friedmann constraint
\begin{align}
  \Omega_{\rm m}+\sum_ix_i^2+\sum_iy_i^2 = 1 \,.
\end{align}
where as before $w=p/\rho$. The most important feature in the assisted inflationary model \citep{Liddle:1998jc}, is that late time accelerated expansion can be achieved even if the individual exponential potentials are not flat enough. In fact, from Sec.~\ref{sec:exp_potential} we know that for the single field exponential potential a future dark energy dominated attractor can only be obtained if $\lambda^2<2$. The late time dynamics of the assisted quintessence model (\ref{074}) can however be mapped into one of a single scalar field $\tilde\phi$ with potential $V(\tilde\phi)= V_0\exp(-\tilde\lambda\kappa\tilde\phi)$ where
\begin{align}
  \frac{1}{\tilde\lambda^2}=\sum_{i}\frac{1}{\lambda_i^2} \,.
  \label{077}
\end{align}
It is now clear how assisted quintessence works. Even if each single exponential potential $V_i$ is such that $\lambda_i^2>2$, according to Eq.~(\ref{077}) the average $\tilde\lambda$ determining the dynamics at late times can satisfy $\tilde\lambda^2<2$. This is an interesting result especially because steep potentials seem to be more natural in high energy physics phenomenology.

The assisted inflationary and quintessence models have been analysed using dynamical systems techniques by several authors. \citet{Coley:1999mj} studied in detail the model with two and three scalar fields, while \cite{Li:2017xpa} delivered a general analysis for two scalar fields with arbitrary non-interacting potentials following the approach we presented in Sec.~\ref{sec:other_potentials}.
\citet{Guo:2003eu} considered the case where some of the single exponential potentials $V_i$ can be negative. \citet{Huey:2001ah} generalised the exponential potentials with a temperature (background) dependent coupling and delivered a general analysis about tracking solutions in such models. \Citet{Karthauser:2006ix} showed that scaling solutions for the uncoupled (non-interacting scalar fields) potential $V=\sum_i V_i$ appear only if the $V_i$s are all exponential and for other kinds of potentials a coupling, such as the one motivated by string theory phenomenology that they considered, is needed. The dynamics of multiple quintessence fields with an interaction to multiple matter fluids, has been investigated by \cite{Amendola:2014kwa}, with similar results to the non-interacting case. Finally \citet{Kim:2005ne} reviewed the assisted quintessence scenario and proved that no assisted behaviour arises if the $V_i$s are all of the power-law type.

\subsubsection{Interacting potentials}

At this point we turn our attention to another well studied possibility of multi-scalar quintessence. This time we will consider the multiplicative (or cross-coupling) multifield exponential potential defined as \citep{Copeland:1999cs}
\begin{align}
  V(\phi_1,...\,,\phi_N) = \sum_i V_i(\phi_1,...\,,\phi_N) = \sum_i \Lambda_i \exp\sum_j \lambda_{ij} \phi_j \,,
  \label{080}
\end{align}
where $\lambda_{ij}$ are now $N\times N$ parameters and $\Lambda_i$ are $N$ constants. This model is sometimes called {\it generalised assisted quintessence} or inflation, depending if late or early time applications are respectively considered. Note that now the scalar fields are interacting with each other. As before, the resulting Einstein field equations and Klein-Gordon equation can be cast into dynamical systems form by introducing $2N$ variables.

The dynamical system based on (\ref{080}) has been studied by \citet{Collinucci:2004iw} where detailed examples with two and three scalar fields were provided and useful de Sitter future attractors were found. \citet{Hartong:2006rt} showed that such accelerated solutions can be obtained also with some negative $V_i$s and then generalised the analysis to the case of non vanishing spatial curvature. A simpler model where all the $\Lambda_i$s are equal and there are only $N$ parameters $\lambda_i$ have been considered by \citet{vandenHoogen:2000cf} and \citet{Guo:2003rs}. In this case we can reduce the dimensionality of the system to $N+1$ equations defining a single new variable for the potential. Moreover if $\sum_i\lambda_i^2<3(w+1)$ the late time attractor is always given by a dark energy inflationary solutions \citep{Guo:2003rs}.

To conclude the section we mention few examples considering dynamical systems arising from more complicated multifield potentials. \citet{Zhai:2005ub} analysed two scalar fields with a rather complicated potential and a kinetic coupling motivated by quintessential inflation phenomenology. They showed that within this model de Sitter acceleration can be obtained both at early and late times. A kinetic coupling between the scalar fields is also present in the analysis of \citet{vandeBruck:2009gp} where the multifield potentials (\ref{074}) and (\ref{080}) are considered in such a framework. Finally \citet{Marsh:2012nm} studied dynamical systems in the context of string theory phenomenology where a complicated two-field potential for moduli and axion fields appears leading to a future evolution of the universe containing multiple epochs of accelerated expansion.

%% file: chapters/05_noncanonical/noncanonical.tex
\section{Dark energy from non-canonical scalar fields}
\label{chap:noncanonicalscalarfields}

In Sec.~\ref{chap:scalarfields} we considered dark energy arising from the potential energy contribution of a canonical scalar field. We now turn our attention to the case of non-canonical scalar fields. As we will see, these models are non-trivial because in general non-canonical scalar fields suffer from theoretical issues which do not appear in the canonical case, for example the existence of ghosts and unphysical solutions. However the generic form of such contributions can be easily motivated from high energy phenomenology and they can provide a bridge between cosmological observations and high energy physics. For this reason they have generated much interest in the dark energy and modified gravity literature (for reviews see e.g.~\citet{Copeland:2006wr,Clifton:2011jh}).

\subsection{Phantom dark energy}
\label{sec:phantom}

The first non-canonical scalar field model we study is mathematically the simplest one to deal with. Its Lagrangian is almost the same as the canonical one (\ref{def:can_scalar_field_lagrangian}) with only the {\it sign} of the kinetic term being opposite to that of quintessence. Explicitly we have
\begin{equation}
	\mathcal{L}_\phi = +\frac{1}{2}\partial\phi^2 - V(\phi) \,,
	\label{def:phantom_lagrangian}
\end{equation}
where $V(\phi)$ is a self-interacting potential.

The scalar field defined by Lagrangian (\ref{def:phantom_lagrangian}) is known as a {\it phantom field} since its EoS is capable of reaching values $w_\phi<-1$, which lie in the so-called {\it phantom regime} \citep{Caldwell:1999ew,Caldwell:2003vq}. A dark energy model able to produce an EoS with values below $-1$ is interesting from a phenomenological point of view since the phantom regime is slightly favoured by astronomical observations, though not with statistical significance. The present value for the dark energy EoS parameter is measured to be $-1.006\pm0.045$ from the Planck 2015 results \citep{Ade:2015xua} and whilst the $\Lambda$CDM model with the constant value $w_\Lambda=-1$ still fits the observational results, phantom dark energy with $w_{\rm de}<-1$ could in principle better accommodate the data. Recall that the EoS of quintessence is constrained to the interval $[-1,1]$; see Eq.~(\ref{eq:EoSPhi}). This implies that a canonical scalar field cannot account for a dark energy EoS in the phantom regime. If future observations indicate $w_{\rm de} < -1$, then we will need to fully understand the implications of phantom like dark energy in order to determine whether they are an acceptable explanation for dark energy or whether they have their own pathological aspects.

Unfortunately leaving the canonical paradigm means also the appearance of new theoretical problems (see e.g.~\citet{Carroll:2003st,Cline:2003gs,Clifton:2011jh}). The most evident issue in the case of the phantom field (\ref{def:phantom_lagrangian}) is the introduction of negative energies. Flipping the sign of the kinetic energy inevitably leads to a total energy of the scalar field which is no longer bounded from below. From a quantum perspective this implies the appearance of {\it ghosts} (modes violating unitarity) in the theory, while from a classical point of view solutions of the equations of motion are no longer stable under small perturbations and the dominant energy condition is violated \citep{Carroll:2003st,Cline:2003gs}. Given those serious concerns we will consider the phantom scalar field as a simple phenomenological model interpreting it as an emergent phenomenon not to be trusted at the deepest fundamental level.

A general cosmological fluid with an EoS in the phantom regime leads to a future singularity known as the big rip. To see this, if we go back to Eq.~(\ref{036}) and assume $w = w_{\rm de} <-1$, once we substitute this into the Friedmann equation (\ref{eq:Friedmann}), instead of obtaining Eq.~(\ref{eq:mattersolw}), the expanding solution for the scale factor will be
\begin{equation}
	a(t)\propto (t_0-t)^{\frac{2}{3(w_{\rm de}+1)}} \,,
	\label{081}
\end{equation}
where $t_0$ is some time in the future. Note that the exponent of $t_0-t$ in Eq.~(\ref{081}) is negative for $w_{\rm de}<-1$ and thus $a(t)$ is indeed expanding as $t$ increases. The interesting feature of this solution is that there is a set of initial condition such that at some future time $t=t_0$ the scale factor diverges. This implies that at some time in the future the expansion will become so fast that everything in the universe will be ripped apart. This future singularity is known as the {\it big rip} \citep{Caldwell:2003vq} and always happens in universes which are perpetually phantom dominated.

At this point we start analysing the dynamics of the phantom scalar field (\ref{def:phantom_lagrangian}).
Assuming a flat FLRW metric, the cosmological equations arising from Lagrangian (\ref{def:phantom_lagrangian}) in the presence of a matter fluid with\footnote{Here again $0 \leq w \leq 1/3$, not to be confused with the discussion above where $w = w_{\rm de} <-1$.} $p= w \rho$ are
\begin{align}
3H^2 &= \kappa^2 \left(\rho-\frac{1}{2}\dot\phi^2+V\right) \,,\label{eq:Friedmann_Phi_phantom}\\
2\dot H+3H^2 &= \kappa^2 \left(-w\rho+\frac{1}{2}\dot\phi^2+V\right) \,,\label{eq:acc_Phi_phantom} \\
\ddot\phi+3H\dot\phi -V_{,\phi} &= 0 \,,\label{eq:KG_FRW_phantom}
\end{align}
where as before a dot means differentiation with respect to the coordinate time $t$. Note the opposite sign with respect to Eqs.~(\ref{eq:FriedmannPhi})--(\ref{eq:KGFRW}) in all the terms where a derivative of $\phi$ appears. The EoS of the scalar field is now given by
\begin{equation}
	w_\phi = \frac{\frac{1}{2}\dot\phi^2+V(\phi)}{\frac{1}{2}\dot\phi^2-V(\phi)} \,,
	\label{eq:EoS_phantom}
\end{equation}
while its energy density is
\begin{equation}
	\rho_\phi = -\frac{1}{2}\dot\phi^2 + V(\phi) \,,
\end{equation}
and is clearly negative whenever the kinetic energy is larger than the potential energy. Moreover, assuming a positive scalar field potential, when the kinetic energy equals the potential energy the EoS (\ref{eq:EoS_phantom}) diverges. This can be taken as a first warning that the theoretical pathologies mentioned above can yield non-physical behaviour.

Since the cosmological equations (\ref{eq:Friedmann_Phi_phantom})--(\ref{eq:KG_FRW_phantom}) are almost the same as the canonical scalar field ones, the use of the expansion normalised variables (\ref{def:ENvars}) will again be of great advantage. Defining
\begin{equation}
	x=\frac{\kappa\dot\phi}{\sqrt{6}H} \,, \quad\quad y=\frac{\kappa\sqrt{V}}{\sqrt{3}H} \,, \quad\quad \lambda = -\frac{V_{,\phi}}{\kappa V} \,,
\end{equation}
the cosmological equations (\ref{eq:Friedmann_Phi_phantom})--(\ref{eq:KG_FRW_phantom}) can be rewritten as
\begin{align}
	x' &= \frac{1}{2} \left\{3 (w-1) x^3-3 x \left[w \left(y^2-1\right)+y^2+1\right]-\sqrt{6} \lambda  y^2 \right\} \,,\label{eq:x_phantom}\\
	y' &= -\frac{1}{2} y \left[-3 (w-1) x^2+3 (w+1) \left(y^2-1\right)+\sqrt{6} \lambda  x\right] \,,\label{eq:y_phantom}\\
	\lambda' &= -\sqrt{6} (\Gamma-1) x \lambda^2 \,.\label{eq:lambda_phantom}
\end{align}
The Friedmann constraint is
\begin{equation}
	\Omega_\phi= -x^2+y^2 = 1-\Omega_{\rm m} \leq 1 \,,
	\label{eq:Friedmann_constr_phantom}
\end{equation}
where $\Omega_{\rm m}$ (the relative energy density of matter) is assumed to be positive and
\begin{equation}
	\Gamma = \frac{V V_{,\phi\phi}}{V_{,\phi}^2} \,.
\end{equation}

As in the quintessence scenario, Eqs.~(\ref{eq:x_phantom})--(\ref{eq:lambda_phantom}) do not form an autonomous system of equations unless $\Gamma$ can be written as a function of $\lambda$ in which case they represent a 3D autonomous dynamical system. A similar analysis for arbitrary potentials as the one conducted in Sec.~\ref{sec:other_potentials} could be performed here for the phantom field and interesting potentials, such as the power-law one of Sec.~\ref{sec:power_law_potential}, could be studied. However we will focus on the exponential case where Eqs.~(\ref{eq:x_phantom})--(\ref{eq:y_phantom}) constitute a 2D autonomous system and $\lambda$ becomes a constant.
As well as being the simplest, it is also the most studied case in the literature \citep{Hao:2003ww,UrenaLopez:2005zd}. We will now review it and in doing so we will also consider the full global analysis of the model including its behaviour at infinity.

In what follows we will assume
\begin{equation}
	V(\phi) = V_0 e^{- \lambda \kappa \phi} \,,
	\label{eq:exp_V_phantom}
\end{equation}
where $V_0>0$ is a positive constant and $\lambda$ a parameter. In this case
eqs.~(\ref{eq:x_phantom})--(\ref{eq:Friedmann_constr_phantom}) reduce to a 2D dynamical system in the variables $x$ and $y$. The dynamical analysis of the phantom scalar field with the variables $(w_\phi,\Omega_\phi)$ has been considered by \citet{Fang:2014bta}.

The Friedmann constraint (\ref{eq:Friedmann_constr_phantom}) now fails to close the (physical) phase space to a compact set. The forbidden non-physical regions lie above and below the hyperbolae $y=\pm\sqrt{1+x^2}$ with the physical phase space constrained between the two. The system (\ref{eq:x_phantom})--(\ref{eq:y_phantom}) is again invariant under the transformation $y\mapsto -y$, meaning that the dynamics in the negative $y$ half-plane will be a reflection of the one in the positive $y$ half-plane. Having set $V>0$, we can assume that the relevant dynamics is contained in the upper half-plane~$y>0$. Note also that the $(x,\lambda)\mapsto (-x,-\lambda)$ symmetry holds in the dynamical system (\ref{eq:x_phantom})--(\ref{eq:y_phantom}). As in the canonical case we will thus only need to consider positive $\lambda$ values since negative values will lead to the same dynamics after a reflection in the $y$-axis. Note that the scalar field EoS (\ref{eq:EoS_phantom}) can now be rewritten as
\begin{equation}
	w_\phi = \frac{x^2+y^2}{x^2-y^2} \,,
	\label{eq:EoS_phantom_xy}
\end{equation}
which diverges whenever $x^2=y^2$.

\begin{table}
\begin{center}
\begin{tabular}{|c|c|c|c|c|c|c|c|}
\hline
Point & $x$ & $y$ & Existence & $w_{\rm eff}$ & Accel. & $\Omega_\phi$ & Stability \\
\hline
\multirow{2}*{$O$} & \multirow{2}*{0} & \multirow{2}*{0} & \multirow{2}*{$\forall\;\lambda,w$} & \multirow{2}*{$w$} & \multirow{2}*{No} & \multirow{2}*{0} & \multirow{2}*{Saddle} \\ & & & & & & & \\
\multirow{2}*{$C$} & \multirow{2}*{$-\lambda/\sqrt{6}$} & \multirow{2}*{$\sqrt{1+\frac{\lambda^2}{6}}$} & \multirow{2}*{$\forall\;\lambda,w$} & \multirow{2}*{$-1- \lambda^2/3$} & \multirow{2}*{Yes} & \multirow{2}*{1} & \multirow{2}*{Stable} \\ & & & & & & & \\
\hline
\end{tabular}
\caption{Critical points of the system (\ref{eq:x_phantom})--(\ref{eq:y_phantom}) with the exponential potential \eqref{eq:exp_V_phantom}. Existence, physical and stability properties of the critical points are also reported.}
\label{tab:phantom_exp_CP_physics}
\end{center}
\end{table}

The critical points of the system (\ref{eq:x_phantom})--(\ref{eq:y_phantom}) with the exponential potential \eqref{eq:exp_V_phantom} are listed, together with their existence, physical and stability properties, in Tab.~\ref{tab:phantom_exp_CP_physics}. There are only two finite critical points of this system:
\begin{itemize}
	\item {\it Point~$O$}. The origin of the phase space is again a critical point representing a matter dominated universe: $\Omega_{\rm m}=1$ and $w_{\rm eff}=w$. Its existence is independent of the values of $w$ and $\lambda$ and, as in the case of quintessence models, it always acts as a saddle point, attracting trajectories along the $x$-axis and repelling them towards the $y$-axis (as one can note by checking the eigenvectors of the Jacobian matrix).
	\item {\it Point~C}. The only non-trivial critical point appearing in the phase space is the scalar field dominated ($\Omega_\phi=1$) Point~$C$ (see Tab.~\ref{tab:phantom_exp_CP_physics} for the coordinates). It exists for all values of $w$ and $\lambda$ and, being scalar field dominated, it lies on the upper hyperbola $y=\sqrt{1+x^2}$. The effective EoS at this point matches the scalar field EoS and takes the value $w_{\rm eff}=w_\phi=-1- \lambda^2/3$, which is in the phantom regime for every value of $\lambda$ different from zero.
	As $\lambda$ increases from zero to higher values Point~$C$ moves away from point~$(0,1)$ along the upper hyperbola.
	Finally Point~$C$ is always a stable point and as we will see below it always represents the future attractor for orbits in the $y>0$ half plane.
\end{itemize}

Although there are only two critical points, these are the ones required to describe a late time transition from matter to phantom domination. In fact a heteroclinic orbit connecting Point~$O$ to Point~$C$ would represent such phenomenological behaviour. Note that as we increase the value of $\lambda$ (i.e.~the potential steepens), we see that Point~$C$ moves further into the phantom regime with correspondingly lower values of $w_{\rm eff}$. In order to characterise the small deviation allowed by the observational data, only small values of $\lambda$ should be considered in this model ($\lambda^2\simeq 3/10$ for $w_\phi = w_{\rm de}\simeq -1.1$).

\begin{figure}[t]
\centering
\includegraphics[width=\columnwidth]{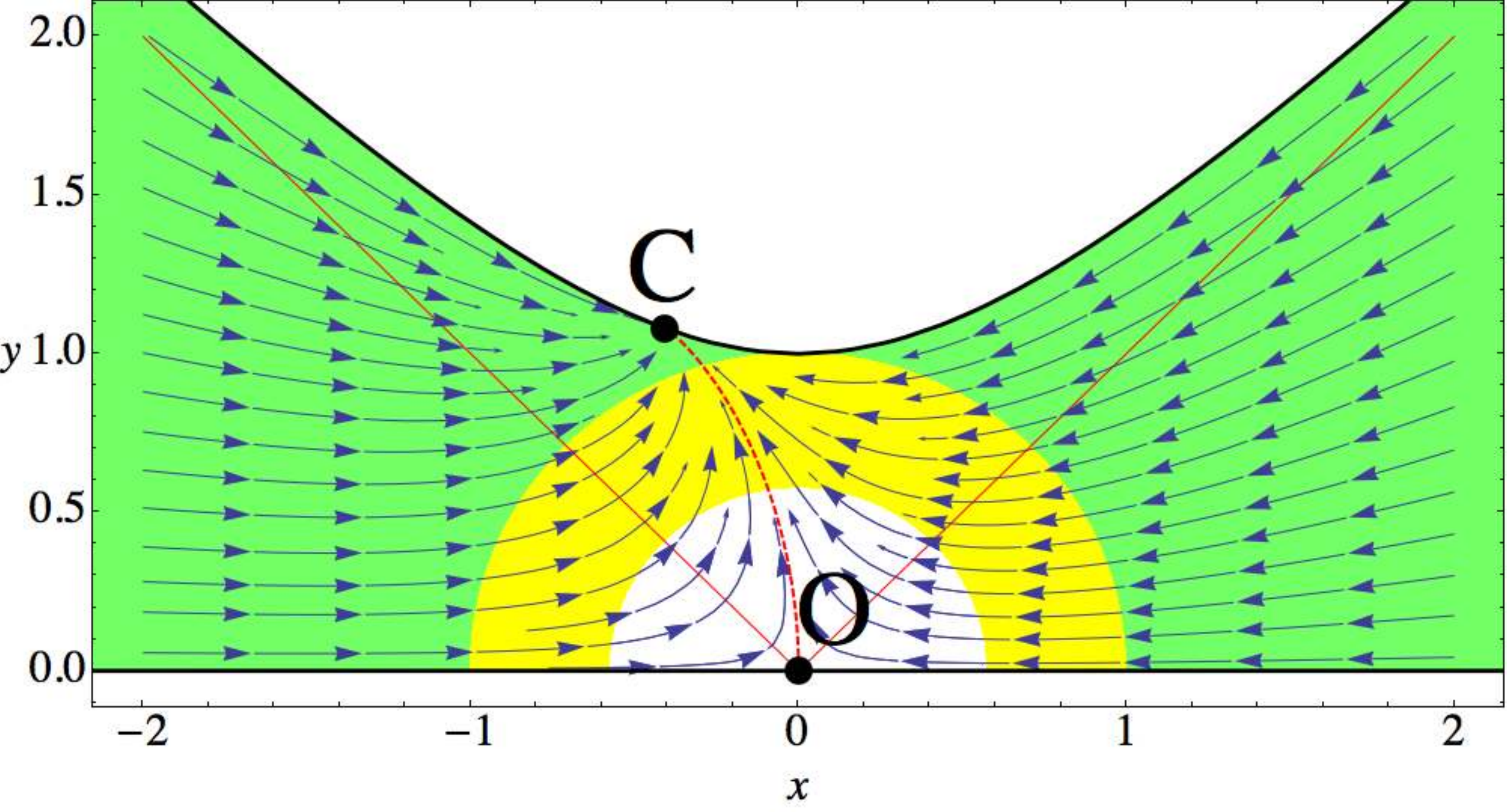}
\caption{Phase space portrait near the origin of the dynamical system (\ref{eq:x_phantom})--(\ref{eq:y_phantom}) with the values $w=0$ and $\lambda=1$. Point~$C$ represents a phantom dominated point, while the dashed/red line denotes the heteroclinic orbit connecting Point~$O$ to Point~$C$ and characterising the matter to phantom transition. The external green/shaded region shows where the universe is phantom dominated ($w_{\rm eff}<-1$), while the internal yellow/shaded region shows where the universe undergoes a standard accelerated expansion ($-1<w_{\rm eff}<-1/3$). The two red solid lines correspond to $y=\pm x$ where the scalar field EoS \eqref{eq:EoS_phantom} diverges.}
\label{fig:phantom_exp_local}
\end{figure}

We can now have a look at the phase space portrait near the origin. This has been plotted in Fig.~\ref{fig:phantom_exp_local} for the values\footnote{Different values of $w$ in the physically meaningful region $0\leq w \leq 1/3$ do not alter the qualitative analysis that follows.} $w=0$ and $\lambda=1$. The qualitative behaviour of the flow in the phase space does not change for different values of $\lambda$. Point~$C$ will always constitute a future attractor moving along the $y=\sqrt{1+x^2}$ hyperbola as the value of $\lambda$ changes, while Point~$O$ will always be a saddle point. The hyperbola $y=\sqrt{1+x^2}$ divides the physically allowed region ($\Omega_{\rm m}>0$) of the phase space to the non-physical one ($\Omega_{\rm m}<0$) above itself. The external green/shaded region in Fig.~\ref{fig:phantom_exp_local} represents the area of the phase space where the universe is phantom dominated ($w_{\rm eff}<-1$), while the internal yellow/shaded region shows where the universe undergoes standard accelerated expansion ($-1<w_{\rm eff}<-1/3$). Note that outside the unit disk only phantom behaviour is possible. The phase space for the phantom scalar field is not compact and trajectories can extend to infinity. From Fig.~\ref{fig:phantom_exp_local} it is clear that the past attractors of the phase space must be represented by points at infinity.

\begin{figure}
\centering
\includegraphics[width=\columnwidth]{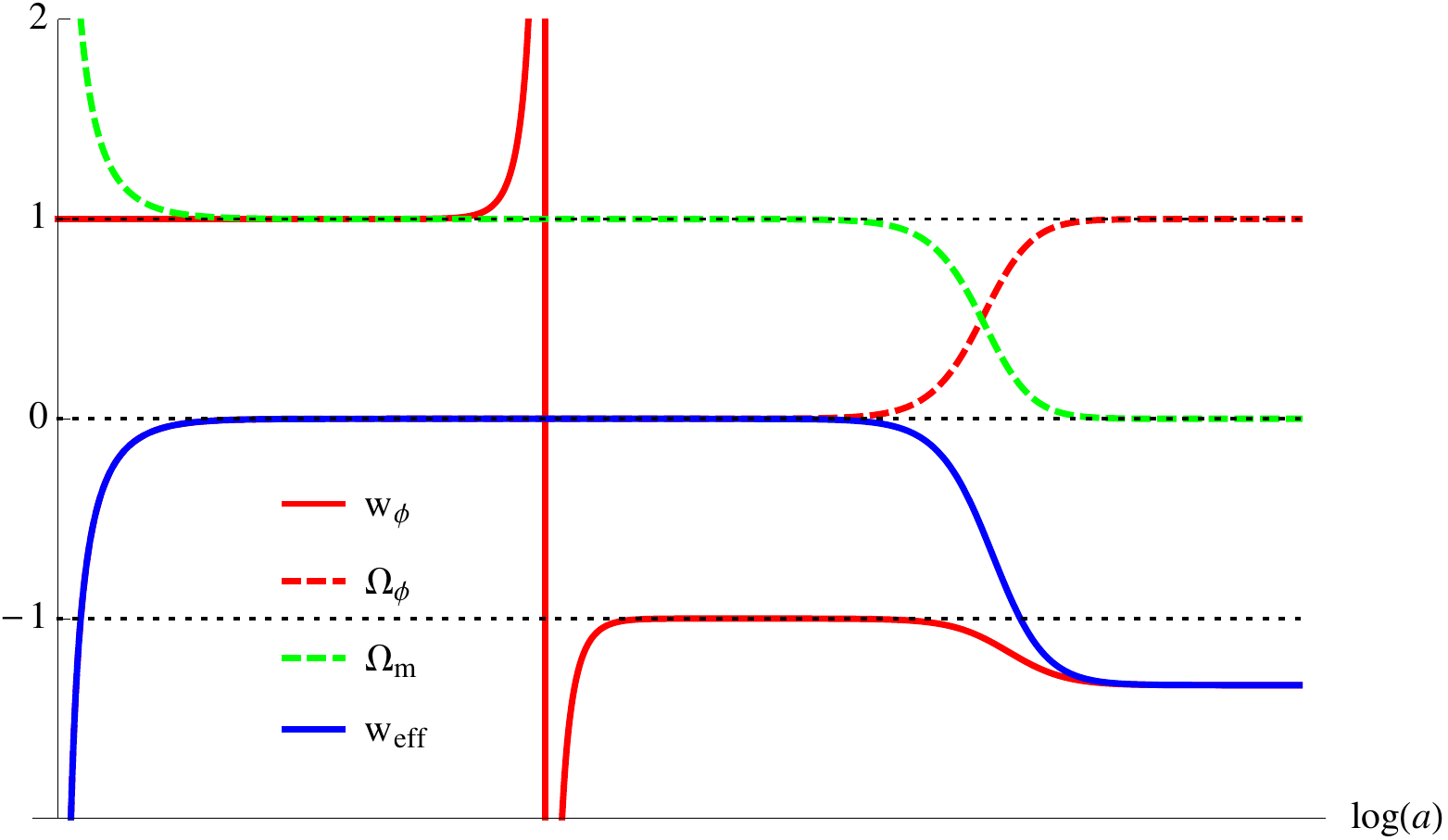}
\caption{Evolution of the phenomenological quantities for an orbit shadowing the matter to phantom transition in the phantom dark energy model.}
\label{fig:phantom_exp_physics}
\end{figure}

The whole late time dynamics of the phantom scalar field model is completely self-explained by Fig.~\ref{fig:phantom_exp_local}. The red/dashed line denotes the heteroclinic orbit connecting the matter dominated Point~$O$ with the phantom dominated Point~$C$, and trajectories aiming at characterising the late time transition from matter to phantom acceleration must shadow this orbit. In Fig.~\ref{fig:phantom_exp_physics} the evolutions of the energy densities of matter ($\Omega_{\rm m}$) and the phantom field ($\Omega_\phi$) together with the effective EoS ($w_{\rm eff}$) and the scalar field EoS ($w_\phi$) have been plotted for a trajectory shadowing the heteroclinic orbit between Point~$O$ and Point~$C$ and coming from $x\rightarrow+\infty$ as $\eta\rightarrow-\infty$ (recall $\eta = \log(a)$). The late time matter to phantom transition is well explained by the effective EoS dropping from zero to a value below $-1$ exactly when the scalar field starts dominating. Right before the transition the scalar field acts as a negligible cosmological constant with the value $w_\phi=-1$. Nevertheless, although the late time behaviour could well characterise the observed universe, the scalar field experiences non-physical features during its early time evolution.

First at very early times the scalar field energy density diverges to $-\infty$, while the matter energy density compensates this situation diverging at $+\infty$. This non-physical behaviour is independent of the trajectory coming from positive or negative values of $x$ as $\eta\rightarrow-\infty$ since effectively $\Omega_\phi\simeq-x^2$ at early times. This implies that the scalar field will always present a negative energy in the very early universe. Its EoS during this period is $w_\phi\simeq 1$ describing a kinetic dominated stiff fluid, while the effective EoS diverges $\to -\infty$ in the past, but it stabilises around the matter value as soon as the scalar field energy density approaches zero.

At some point well into the matter dominated era the scalar field EoS becomes discontinuous, heading towards $+\infty$ and suddenly emerging at $-\infty$. This always happens when $x^2=y^2$, as noticed before in Eq.~(\ref{eq:EoS_phantom_xy}), and no orbit in the phase space can avoid the discontinuity\footnote{It appears when a solution crosses one of the lines $y=\pm x$. Note that before the discontinuity the scalar field EoS parameter is one, while after it the EoS approaches the value $w_\phi=-1$. From Fig.~\ref{fig:phantom_exp_local} it is clear that every trajectory, a part from the scalar field dominated solutions on the upper hyperbola, must eventually cross one of these lines.}. Although for orbits shadowing the matter to phantom transition this discontinuity happens extremely close to Point~$O$ where the scalar field energy density is negligible, the fact that a singularity appears in the dark energy EoS represents the direct effect of the theoretical problems mentioned above. It is telling us that a phantom scalar field model of dark energy should be trusted for phenomenological applications only after this discontinuity has occurred. Everything that comes before should be ignored, assuming that some other physical mechanisms come into play and that the effective description of dark energy as a phantom scalar field ceases. This also applies to the very early unphysical behaviour which for the same reason should be neglected. If the model is seriously considered only after the non-physical behaviours have happened, then it can effectively describe the late time expansion of the universe with a distinctive signature on the dark energy EoS which could be measured by forthcoming observations.

In the final part of this section we will determine the behaviour of the flow at infinity. As we mentioned above, and as it is clear from Fig.~\ref{fig:phantom_exp_local}, the phase space of the system (\ref{eq:x_phantom})--(\ref{eq:y_phantom}) is not compact. In order to analyse the flow at infinity we must employ the techniques developed in Sec.~\ref{sec:poincarecomp}.

The first step is to determine critical points at infinity using Eq.~(\ref{020}). The polynomial terms of higher order in Eqs.~(\ref{eq:x_phantom}) and (\ref{eq:y_phantom}) are of the third order in $x$ and $y$. If we consider only these higher terms, then Eq.~(\ref{020}) vanishes identically $G_4 = 0$. This means that we must consider the next to leading order to check for critical points at infinity. The second order terms in $x$ and $y$ yield
\begin{align}
  G_3 = -\frac{\sqrt{6}}{2}\lambda \cos(2\theta)\sin\theta = 0 \,,
\end{align}
which gives the (physical) solutions $\theta=0,\pi/4,3\pi/4,\pi$ corresponding to Points~$A_\pm$ and $B_\pm$. Moreover, the function $G_3$ is negative in the intervals $[0,\pi/4]$ and $[3\pi/4,\pi]$, implying that the flow is clockwise near the unit circle between $B_-$ and $A_+$, and between $A_-$ and $B_+$, and counter-clockwise otherwise.

\begin{figure}
\centering
\includegraphics[width=\columnwidth]{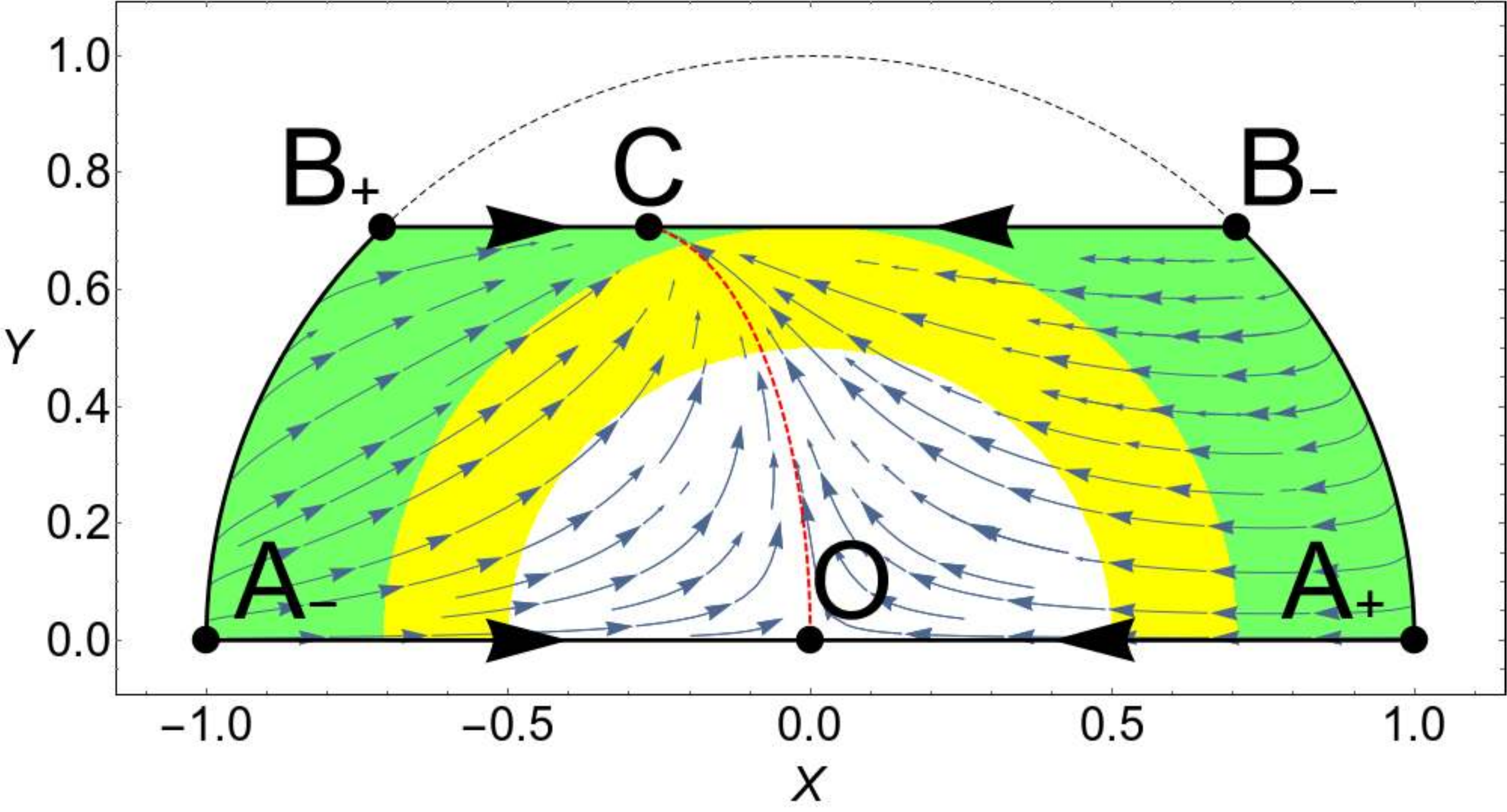}
\caption{Global phase space portrait of the system (\ref{eq:x_phantom})--(\ref{eq:y_phantom}).}
\label{fig:phantom_exp_global}
\end{figure}

This behaviour is confirmed by numerical computation. In Fig.~\ref{fig:phantom_exp_global} the global phase space for the dynamical system (\ref{eq:x_phantom})--(\ref{eq:y_phantom}) has been drawn. The phase space has been compactified using the projection onto the Poincar\'{e} sphere given by Eq.~(\ref{eq:Poincare_sphere}). Points on the unit circle correspond to infinity, and we correctly find that the physical phase space contains 4 such critical points, $A_\pm$ and $B_\pm$. These points represent the possible past attractors of the system which can be split into two invariant sets: trajectories on the right of the heteroclinic orbit connecting Point~$O$ with Point~$C$ have Point~$B_-$ as their past attractor, while trajectories on the left have Point~$A_-$ as their past attractors.

The constant line $Y=1/\sqrt{2}$ corresponds to the upper hyperbola $y=\sqrt{1+x^2}$ in the non-compact phase space. The region of the phase space above this line, but still inside the unit disk (dashed line in Fig.~\ref{fig:phantom_exp_global}) due to the compactification, stands for the physically forbidden region where the matter energy density is negative ($\Omega_{\rm m}<0$). We must neglect this region in our analysis.

In conclusion the phantom scalar field model of dark energy we have examined can be considered as a viable description of late time cosmology capable of predicting a dark energy EoS in the phantom regime ($w_{\rm de}<-1$). If future observations will constrain the dark energy EoS to lie in the phantom regime, then a phantom scalar field with a negative kinetic energy is certainly the simplest model which can account for such behaviour. Its dynamics is extremely simple to analyse with only two finite critical points appearing in the phase space and no qualitative dependence on the theoretical parameters. The 2D phase space is mathematically simple to handle with the dynamical systems tools we developed in Sec.~\ref{sec:dynamicalsystems}, including the behaviour at infinity.

However, the phantom scalar field with an exponential potential and matter with an EoS parameter $0\leq w\leq 1/3$ has its drawbacks. First of all its early time phenomenology presents singularities which clearly indicates the non-viability of the model during this period. These infinities are the direct consequence of more fundamental problems connected with negative energies appearing in the theory. Nevertheless the model should be considered as an emergent phenomenon at late times and assumed not to hold at early times where different mechanisms should come into play to cure the pathologies.
Moreover the model does not solve the cosmic coincidence problem and suffers from the fine tuning of initial conditions, exactly as quintessence with an exponential potential does; see Sec.~\ref{sec:exp_potential}. In fact the only solutions which well characterise the transition from dark matter to dark energy domination are the ones shadowing the heteroclinic orbit between Points~$O$ and $C$, which of course require special initial conditions.

Different choices of potential or matter fluid lead, of course, to different results. For example, \citet{UrenaLopez:2005zd} extended the analysis to matter fluids outside the $[0,1/3]$ interval in which situation scaling solutions appear but at the cost of unusual matter fluids. \citet{Li:2003ft} instead generalised the phantom dynamics to multiple scalar fields with an $O(N)$ symmetry and found the lower bound $w_{\rm de}>-3$ for phantom stable attractor solutions.
It might also be the case that the choice of other potentials leads to the solution of some of these problems, as we saw in the quintessence case of Sec.~\ref{chap:scalarfields}.  \citet{Hao:2003th} discussed the dynamics of the phantom field for a general potential, looking in particular at tracking, de Sitter and big rip solutions.
Again \citet{Hao:2003aa} showed that de Sitter-like solutions always arise as late time attractors if the phantom scalar field potential has a non-vanishing maximum.
\citet{Nojiri:2005sx} instead classified the possible singularities appearing in general phantom dark energy models (including interactions with the matter fluid) and studied them with dynamical systems methods.

Finally for investigations where the phantom field is coupled to the matter sector, we refer the reader to Sec.~\ref{sec:coupled_noncanonical_scalars}.

\subsection{Quintom models of dark energy}
\label{sec:quintom}

We learned in Sec.~\ref{chap:scalarfields} that the EoS of quintessence must satisfy $w_{\rm de}\geq -1$, while in Sec.~\ref{sec:phantom} we saw that for a phenomenologically acceptable phantom model, (i.e.~valid only after the early time discontinuities in its EoS have taken place), the phantom scalar field EoS is constrained to be $w_{\rm de}<-1$. There is no way to cross this phantom barrier (i.e.~the cosmological constant value $w_{\rm de}=-1$) with a single canonical or phantom scalar field. An interesting model which allows for such a crossing to take place was proposed by \citet{Feng:2004ad}. This scenario of dark energy gives rise to the EoS larger than -1 in the past and less than -1 today, satisfying current observations. It can be achieved with more general non-canonical scalar fields, but the simplest model is represented by the quintom Lagrangian made up of two scalar fields, one canonical field $\phi$ and one phantom field $\sigma$:
\begin{equation}
  \mathcal{L}_{\rm quintom} = -\frac{1}{2}\partial\phi^2 +\frac{1}{2}\partial\sigma^2 - V(\phi,\sigma) \,,
  \label{def:quintom_Lagrangian}
\end{equation}
where $V(\phi,\sigma)$  is a general potential for both the scalar fields. Note the opposite sign of the kinetic terms implying that $\phi$ is a canonical scalar field and $\sigma$ is a phantom scalar field. The model~(\ref{def:quintom_Lagrangian}) has been dubbed {\it quintom dark energy} from the fusion of the words quintessence and phantom.

In this section we will deal with the simplest quintom models without delivering a complete dynamical systems analysis. Detailed references to the literature for works studying quintom dark energy with dynamical systems methods will be provided. The reader interested in the theory and phenomenology of these models can refer to the extensive review by \citet{Cai:2009zp}.
Similarly to the multi-field quintessence models we encountered in Sec.~\ref{sec:multi_scalar_fields}, we can distinguish between interacting and non-interacting potentials for the two scalar fields $\phi$ and $\sigma$ of the quintom scenario. We will deal first with the non-interacting case and then discuss interacting models.

\subsubsection{Non-interacting potentials}
\label{sec:quintom-non-interacting}

The potential $V(\phi,\sigma)$ for non-interacting scalar fields can be generally written as
\begin{equation}
	V(\phi,\sigma) = V_1(\phi) +V_2(\sigma) \,,
	\label{093}
\end{equation}
where $V_1(\phi)$ and $V_2(\sigma)$ are arbitrary self-interacting potentials for the two scalar fields $\phi$ and $\sigma$ respectively. In this situation we have two separated scalar fields which can be treated exactly as if they were single models.
The cosmological equations are given by the Friedmann and acceleration equations
\begin{align}
	3H^2 &= \kappa^2\left[\rho + \frac{1}{2}\dot\phi^2 + V_1(\phi) -\frac{1}{2}\dot\sigma^2 +V_2(\sigma)\right] \,,\label{082}\\
	 2\dot H +3H^2  &= -\kappa^2\left[w \rho +\frac{1}{2}\dot\phi^2 - V_1(\phi) -\frac{1}{2}\dot\sigma^2 -V_2(\sigma)\right] \,,
\end{align}
and by the Klein-Gordon equations
\begin{align}
	\ddot\phi +3H \dot\phi + \frac{\partial V_1(\phi)}{\partial\phi} &=0 \,,\\
	\ddot\sigma +3H \dot\sigma - \frac{\partial V_2(\sigma)}{\partial\sigma} &=0 \,. \label{083}
\end{align}
In order to recast them into a dynamical system we define EN variables as (cf.~with definition~\eqref{def:ENvars})
\begin{align}
x_\phi = \frac{\kappa \dot\phi}{\sqrt{6} H} \,,\qquad y_\phi = \frac{\kappa \sqrt{V_1}}{\sqrt{3}H} \,,\qquad \lambda_\phi = -\frac{1}{\kappa V_1} \frac{\partial V_1}{\partial\phi} \,,\\
x_\sigma = \frac{\kappa \dot\sigma}{\sqrt{6} H} \,,\qquad y_\sigma = \frac{\kappa \sqrt{V_2}}{\sqrt{3}H} \,,\qquad \lambda_\sigma = -\frac{1}{\kappa V_2} \frac{\partial V_2}{\partial\sigma} \,.
\label{088}
\end{align}
To find the dynamical system governing the cosmological evolution, we proceed in the same way as with the quintessence and phantom field. Eventually Eqs.~(\ref{082})--(\ref{083}) yield
\begin{align}
x_\phi ' &= \frac{3}{2} \left\{x_{\phi } \left[(w-1) (x_{\sigma }^2+1)-(w+1) (y_{\phi }^2+y_{\sigma }^2)\right]- (w-1) x_{\phi }^3+\frac{\sqrt{3}}{\sqrt{2}} \lambda_{\phi } y_{\phi }^2\right\} ,\label{084}\\
y_\phi ' &= -\frac{ y_{\phi }}{2} \left[3 (w-1) (x_{\phi }^2-x_{\sigma }^2)+3 (w+1) \left(y_{\sigma }^2+y_{\phi }^2-1\right)+\sqrt{6} \lambda _{\phi} x_{\phi }\right] \,,\\
x_\sigma '&= \frac{3}{2} \left\{(w-1) x_{\sigma }^3-x_{\sigma } \left[(1-w)(1- x_{\phi }^2)+(w+1) (y_{\phi }^2+y_\sigma^2)\right]-\frac{\sqrt{2}}{\sqrt{3}} \lambda_{\sigma } y_{\sigma }^2\right\} ,\\
y_\sigma ' &= -\frac{ y_{\sigma }}{2} \left[3 (w-1) (x_{\phi }^2-x_{\sigma }^2) +3 (w+1) \left(y_{\sigma }^2+y_{\phi }^2-1\right)+\sqrt{6} \lambda _{\sigma } x_{\sigma }\right] \,,\label{085} \\
\lambda_\phi ' &= -\sqrt{6} (\Gamma_\phi-1) x\lambda_\phi^2 \,, \\
\lambda_\sigma ' &= -\sqrt{6} (\Gamma_\sigma-1) x\lambda_\sigma^2 \,, \label{086}
\end{align}
where
\begin{equation}
\Gamma_\phi = V_1\frac{V_{1,\phi\phi}}{V_{1,\phi}^2} \,,\qquad \Gamma_\sigma = V_2\frac{V_{2,\sigma\sigma}}{V_{2,\sigma}^2} \,,
\end{equation}
and the Friedmann constraint
\begin{equation}
-x_{\sigma }^2+x_{\phi }^2+y_{\sigma }^2+y_{\phi }^2 =1-\Omega_{\rm m} \leq 1 \,,
\label{eq:H3}
\end{equation}
holds.
Note that the presence of the negative $x_{\sigma }^2$ term in (\ref{eq:H3}) means that the Friedmann constraint can be thought of as making the inside of an hyperboloid $\mathbb{H}^3$. If $\Gamma_\phi$ and $\Gamma_\sigma$ can be written as functions of $\lambda_\phi$ and $\lambda_\sigma$ respectively, then Eqs.~(\ref{084})--(\ref{086}) constitute a 6D dynamical system and a similar analysis of the one we performed in Sec.~\ref{sec:other_potentials} can be conducted. This has indeed been done in detail by \citet{Leon:2012vt} where critical points and the stability analysis have been studied and also the specific potential $V(\phi,\sigma)=A\sinh^2(\alpha\phi)+B\cosh^2(\beta\sigma)$ has been considered as an example.

In what follows however we will briefly discuss the exponential potential case where
\begin{equation}
V_1(\phi)= A\, e^{-\lambda_\phi\kappa\phi} \quad\mbox{and}\quad V_2(\sigma)= B\, e^{-\lambda_\sigma\kappa\sigma} \,,
\label{087}
\end{equation}
with $A$, $B$, $\lambda_\phi$, $\lambda_\sigma$ all constant. This represents the simplest situation where Eqs.~(\ref{084})--(\ref{085}) form an autonomous 4D dynamical system. The exponential potentials (\ref{087}) have been analysed by \citet{Guo:2004fq} who found quintessence dominated solutions, phantom future attractors and scaling solutions (see also \citet{Cai:2009zp} for details). We will not list here all the critical points with their properties, since these are identical to the exponential potential case, except for the fact that now some of them can appear simultaneously in the phase space.
We will limit our discussion to the interesting late time phenomenology arising from this model.

Using the variables (\ref{088}) the energy densities and EoS of the scalar fields can be written as
\begin{align}
\Omega_\phi = x_{\phi }^2+y_{\phi }^2 \,, \qquad w_\phi=\frac{x_{\phi }^2-y_{\phi }^2}{x_{\phi }^2+y_{\phi }^2} \,,\label{089}\\
\Omega_\sigma = -x_{\sigma }^2+y_{\sigma }^2 \,, \qquad w_\sigma=\frac{-x_{\sigma }^2-y_{\sigma }^2}{y_{\sigma }^2-x_{\sigma }^2} \,,
\end{align}
while the energy density and EoS of dark energy, given by the added contributions of $\phi$ and $\sigma$, are
\begin{align}
\Omega_{\rm de} = -x_{\sigma }^2+x_{\phi }^2+y_{\sigma }^2+y_{\phi }^2 \,, \qquad w_{\rm de}= \frac{-x_{\sigma }^2+x_{\phi }^2-y_{\sigma }^2-y_{\phi }^2}{-x_{\sigma }^2+x_{\phi }^2+y_{\sigma }^2+y_{\phi }^2} \,.
\end{align}
The effective EoS is now given by
\begin{equation}
w_{\rm eff} = w \left(x_{\sigma }^2-x_{\phi }^2-y_{\sigma }^2-y_{\phi }^2+1\right)-x_{\sigma }^2+x_{\phi }^2-y_{\sigma }^2-y_{\phi }^2 \,,
\label{090}
\end{equation}
where one can see the separate contributions of the phantom and quintessence fields respectively.

\begin{figure}
\centering
\includegraphics[width=\columnwidth]{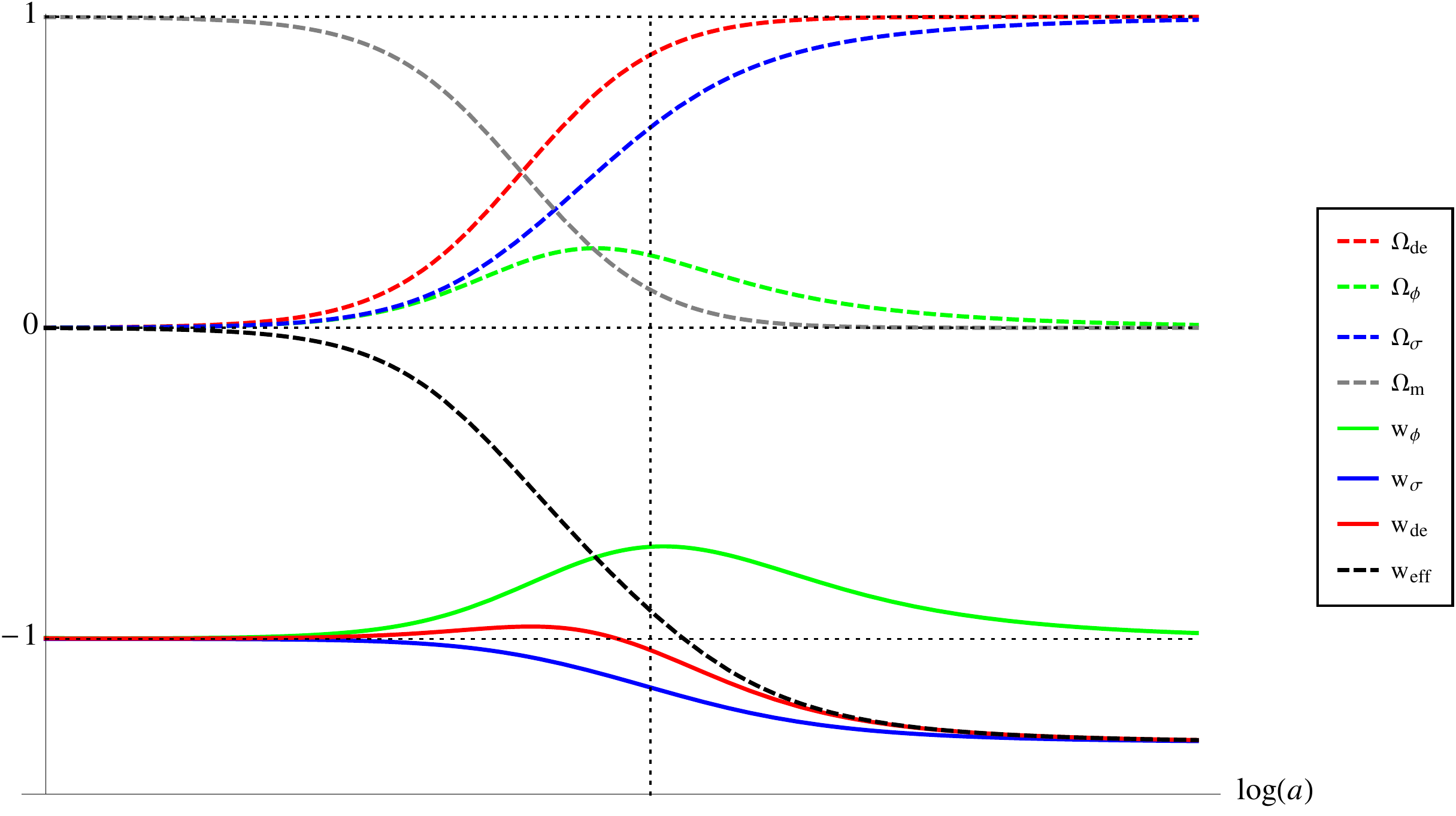}
\caption{Late time evolution for the phenomenological quantities of the uncoupled quintom model with exponential potentials. The values $\lambda_\phi=2$, $\lambda_\sigma=1$ and $w=0$, together with suitable initial conditions, have been chosen.}
\label{fig:quintom_uncoupled}
\end{figure}

In Fig.~\ref{fig:quintom_uncoupled} the late time evolution of the phenomenological quantities (\ref{089})--(\ref{090}) for the quintom uncoupled model with exponential potentials (\ref{087}) have been plotted. The initial conditions have been chosen to highlight the characteristics of the quintom paradigm while the values $\lambda_\phi=2$, $\lambda_\sigma=1$ and $w=0$ have been considered. Only the late time evolution is shown when the singularities associated with the phantom field have already taken place and the effective field approach applies (see Sec.~\ref{sec:phantom}). In the example of Fig.~\ref{fig:quintom_uncoupled} the phantom field asymptotically dominates in the future, with the quintessence field never completely dominating during its evolution. The matter to dark energy transition begins with both the scalar field energies growing from negligible values to an order of magnitude comparable with the matter energy density. However before dark energy dominates, the quintessence energy decreases again while the phantom energy begins to dominate.

The interesting features happen in the EoS parameters during the matter to dark energy transition (solid lines below zero in Fig.~\ref{fig:quintom_uncoupled}). As one can see from Fig.~\ref{fig:quintom_uncoupled}, $w_\phi$ (green line - fourth one from the bottom at present time, i.e.~at the vertical dotted line) always lies above the phantom barrier, while $w_\sigma$ (blue line - first one from the bottom) always stays below $-1$. Their effective contribution however, denoted by $w_{\rm de}$ (red line - second one from the bottom), is able to cross the phantom barrier being greater than $-1$ before the matter to dark energy transition and below $-1$ thereafter. The EoS of dark energy is thus above $-1$ in the past and below $-1$ at both the present and future times. Note that the effective EoS is still in the quintessence region at the present time (denoted by the vertical dotted line in Fig.~\ref{fig:quintom_uncoupled}) while the dark energy EoS is effectively in the phantom regime. The average evolution of the universe is not influenced by the change in the nature of dark energy. Indeed the transition from matter to phantom domination for the effective EoS (black dashed line in Fig.~\ref{fig:quintom_uncoupled}) happens similarly as in the single phantom scalar field case of Sec.~\ref{sec:phantom}. This changes somehow for different choices of the model parameters or initial conditions, but if today the effective EoS and the dark energy EoS are constrained to be above and below $-1$ respectively, then the simplest qualitative evolution of the universe is provided by the one depicted in Fig.~\ref{fig:quintom_uncoupled}.

The situation described in Fig.~\ref{fig:quintom_uncoupled} is the one slightly favoured by observational data, though not at a statistically significant level. The quintom model with uncoupled exponential potentials is thus able to provide a dynamical crossing of the phantom barrier unifying the properties of quintessence and phantom dark energy. If future observations will constrain the dark energy EoS to be below $-1$ today but above $-1$ in the past, then the quintom scenario of Fig.~\ref{fig:quintom_uncoupled} is arguably the simplest framework where such a situation can arise, although it fails to address the fundamental problems associated with phantom fields.

\subsubsection{Interacting potentials}
\label{sec:quintom-interacting}

We now turn our attention to quintom models with a potential coupling the two scalar fields. As in the case of multiple quintessence fields (Sec.~\ref{sec:multi_scalar_fields}), the simplest and most studied potential is again of the exponential type, namely
\begin{equation}
V(\phi,\sigma) = V_0\, e^{-\lambda_\phi\kappa\phi-\lambda_\sigma\kappa\sigma} \,.
\label{092}
\end{equation}
Using a dynamical systems approach, \citet{Lazkoz:2006pa} obtained tracking, phantom and quintessence solutions.

The cosmological equations for the general coupled case are
\begin{align}
3H^2 &= \kappa^2\left[\rho + \frac{1}{2}\dot\phi^2 -\frac{1}{2}\dot\sigma^2 +V(\phi,\sigma)\right]\,,\label{094}\\
	2\dot H +3H^2  &=-\kappa^2\left[ w \rho +\frac{1}{2}\dot\phi^2 -\frac{1}{2}\dot\sigma^2 -V(\phi,\sigma)\right] \,,
\end{align}
and
\begin{align}
	\ddot\phi +3H \dot\phi + \frac{\partial V}{\partial\phi} &=0 \,,\\
	\ddot\sigma +3H \dot\sigma - \frac{\partial V}{\partial\sigma} &=0 \,.
	\label{095}
\end{align}
The coupled potential (\ref{092}) is mathematically simpler to analyse than the uncoupled one (\ref{093}), because it only requires one EN variable for the potential energy rather than two. This is a similar situation to the one we encountered in Sec.~\ref{sec:multi_scalar_fields} where for $N$ multiple interacting quintessence fields with a coupling exponential potential only one EN variable was employed instead of $N$. In fact defining the EN variables
\begin{align}
x_\phi = \frac{\kappa \dot\phi}{\sqrt{6} H} \,,\qquad
x_\sigma = \frac{\kappa \dot\sigma}{\sqrt{6} H} \,,\qquad y = \frac{\kappa \sqrt{V}}{\sqrt{3}H}  \,,
\end{align}
Eqs.~(\ref{094})--(\ref{095}) can be recast into the following 3D autonomous dynamical system
\begin{align}
x_\phi ' &= \frac{1}{2} \left\{3 x_{\phi } \left[(w-1) x_{\sigma }^2-(w+1) y^2+w-1\right]-3 (w-1) x_{\phi }^3+\sqrt{6} y^2 \lambda _{\phi }\right\} \,,\\
x_\sigma '&= \frac{1}{2} \left\{3 (w-1) x_{\sigma }^3-3 x_{\sigma } \left[(w-1) x_{\phi }^2+(w+1) y^2-w+1\right]-\sqrt{6} y^2 \lambda _{\sigma }\right\} \,,\\
y'&= -\frac{1}{2} y \left[3 (w-1) (x_{\phi }^2-x_{\sigma }^2)+3 (w+1) \left(y^2-1\right)+\sqrt{6} \lambda _{\sigma } x_{\sigma }+\sqrt{6} \lambda _{\phi} x_{\phi }\right] \,,
\end{align}
where the Friedmann constraint
\begin{equation}
x_\phi^2 -x_\sigma^2 +y^2 = 1 - \Omega_{\rm m} \leq 1 \,,
\end{equation}
must hold. The fact that the dynamical system of this model is three dimensional means that suitable plots of the phase space can be drawn, as for example the ones presented by \citet{Lazkoz:2006pa}. However we will focus again on the late time phenomenology leaving the dynamical systems details to the mentioned references.

\begin{figure}
\centering
\includegraphics[width=\columnwidth]{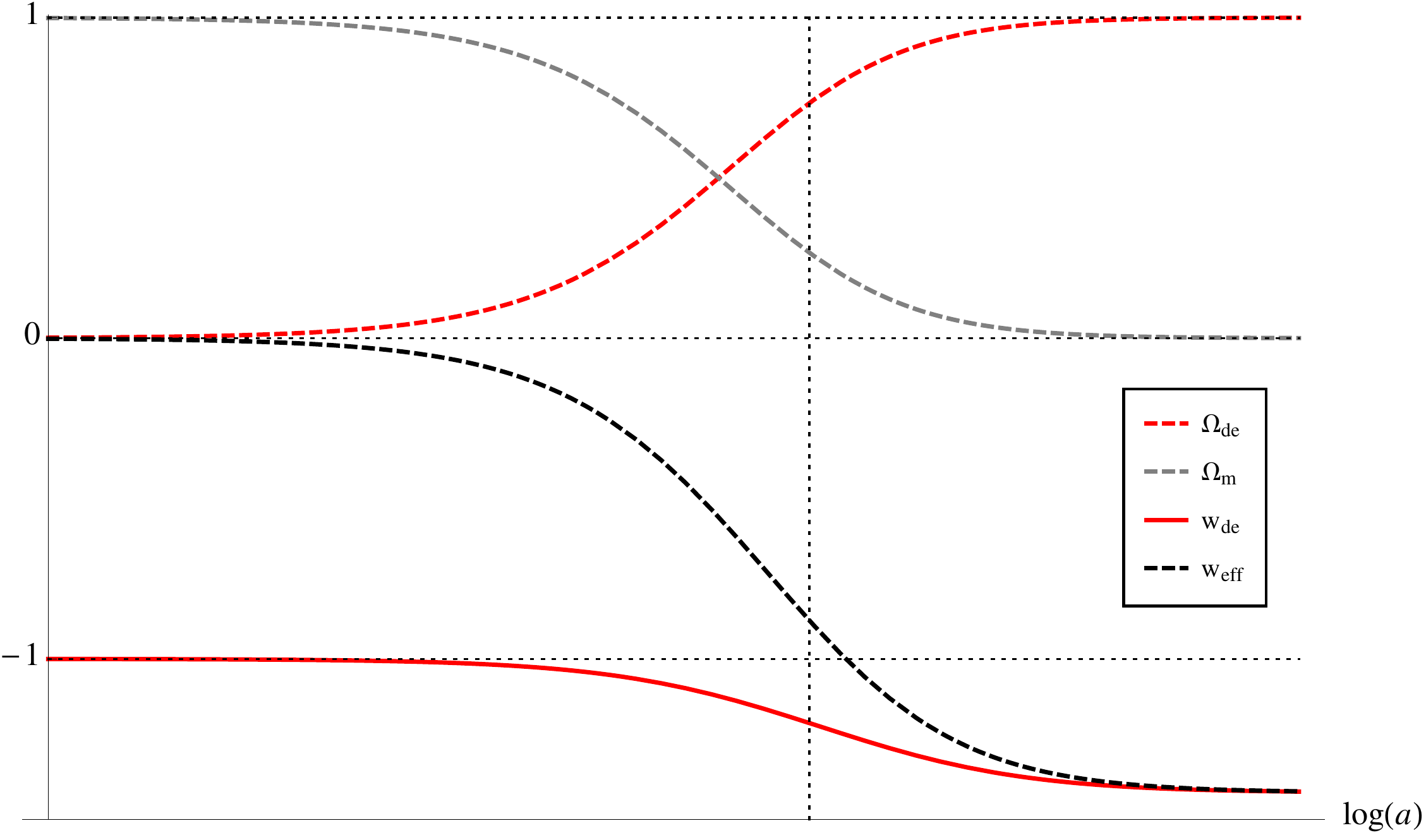}
\caption{Late time evolution for the phenomenological quantities of the coupled quintom model with exponential potentials (\ref{092}). The values $\lambda_\phi=1$, $\lambda_\sigma=1.5$ and $w=0$, together with suitable initial conditions, have been chosen for the plot.}
\label{fig:Quintom_coupled}
\end{figure}

The energy density and EoS of dark energy, given by the mixed contributions of $\phi$ and $\sigma$, are now given by
\begin{align}
\Omega_{\rm de} = -x_{\sigma }^2+x_{\phi }^2+y^2 \,, \qquad w_{\rm de}= \frac{-x_{\sigma }^2+x_{\phi }^2-y^2}{-x_{\sigma }^2+x_{\phi }^2 +y^2} \,,
\end{align}
while the effective EoS is
\begin{equation}
w_{\rm eff} = w \left(x_{\sigma }^2-x_{\phi }^2- y^2 +1\right)-x_{\sigma }^2+x_{\phi }^2-y^2 \,.
\end{equation}
In Fig.~\ref{fig:Quintom_coupled} the late time evolution of these quantities have been plotted for the values $\lambda_\phi=1$, $\lambda_\sigma=1.5$ and $w=0$. The initial conditions have again been chosen in order to highlight the qualities of this quintom model.
For the coupled case dark energy acts as a cosmological constant during the matter dominated epoch eventually switching to phantom values during the deceleration to acceleration phase. This is clear in Fig.~\ref{fig:Quintom_coupled} where the dark energy EoS (red solid line) is constantly $-1$ in the past before taking values below $-1$. Note that also the coupled exponential potential quintom model manages to fit the best astronomical data having both $w_{\rm eff}>-1$ and $w_{\rm de}<-1$ today (vertical dotted line in Fig.~\ref{fig:Quintom_coupled}).
This represents thus another model of dynamical crossing of the phantom barrier, even though the dark energy EoS in the past is basically $-1$, and thus the dark energy EoS is never higher than $-1$.

Unfortunately neither the coupled nor the uncoupled quintom models with exponential potentials seem to solve the cosmic coincidence problem and the fine tuning of initial conditions. The matter to dark energy transition happens again at the present time without an apparent reason, while highly special initial conditions are required to fit the observations. It might be possible that employing potentials beyond the exponential case will lead to a solution of these problems, at least partially. However, for the case of arbitrary decoupled potentials beyond the exponential case, the cosmic coincidence problem still exists as discussed by \citet{Leon:2012vt}.
They included both radiation and dark matter, investigated the phase-space structure of the quintom dark energy paradigm in the framework of a spatially flat and homogeneous universe with arbitrary decoupled potentials, and found certain general conditions under which the phantom-dominated solution is a late time attractor which generalised previous results found for the case of an exponential potential.

For the case of arbitrary coupled potentials beyond the exponential case, this problem remains open. \citet{Zhang:2005eg} worked with the potential $V(\phi,\sigma)=A\exp(-\lambda_\phi\kappa\phi)+B\exp(-\lambda_\sigma\kappa\sigma)+C\exp(-\lambda_\phi\kappa\phi/2)\exp(-\lambda_\sigma\kappa\sigma/2)$ where the EN variables (\ref{088}) can be used with the interacting term giving rise to contributions proportional to $y_\phi y_\sigma$. They also proposed a quintom model with mass varying neutrinos instead of the quintessence field. \citet{Setare:2008si} generalised the quintom exponential case to multiple scalar fields with an $O(N)$ symmetry showing that in this case the phantom dominated solutions always represents the future attractor of the system. In general however it might be possible that different variables are better suited for the analysis of more complicated potentials. This is indeed the approach considered by some authors such as \citet{Lazkoz:2007mx} and \citet{Setare:2008dw}.
Finally some models have been proposed with a kinetic coupling between the scalar fields. \citet{Saridakis:2009jq} introduced the coupling $\partial_\mu\phi\partial^\mu\sigma$, while \citet{Wei:2005fq} and \citet{Alimohammadi:2006qi} analysed the so-called {\it hessence dark energy} scenario with the Lagrangian $\mathcal{L}=-\partial\phi^2/2+\phi^2\partial\sigma^2/2-V(\phi)$.

\subsection{Tachyons and DBI scalar fields}
\label{sec:tachyons}

{\it Tachyons} are particles predicted by {\it string theory} in its low-energy effective field theory description \citep{Green:1987sp,Mazumdar:2001mm,Sen:2002an,Sen:2002in,Sen:2002nu}.
Applications of tachyonic scalar fields to late time cosmology were considered soon after they arose from high-energy physics \citep{Padmanabhan:2002cp,Gibbons:2002md, Gibbons:2003gb,Bagla:2002yn,Gorini:2003wa}.
They can be defined by the {\it Dirac-Born-Infeld (DBI) Lagrangian}\footnote{Sometimes the tachyonic Lagrangian $\mathcal{L}_{\rm tachyons} =-V(\phi)\sqrt{-\det(g_{\mu\nu}+\partial_\mu\phi\partial_\nu\phi)}$ is assumed instead of (\ref{def:tachyon_Lagrangian}); see e.g.~\citet{Copeland:2004hq}. Nevertheless the cosmological equations (\ref{eq:Friedmann_tachyon})--(\ref{eq:KG_tachyon}) can be equally derived from both the Lagrangians.}
\begin{equation}
\mathcal{L}_{\rm tachyons} = V(\phi) \sqrt{1+\partial\phi^2} \,,
\label{def:tachyon_Lagrangian}
\end{equation}
where again $\partial\phi^2=g^{\mu\nu}\partial_\mu\phi\partial_\nu\phi$ and $V$ is a general function of $\phi$ which is generally called the scalar field potential, though it does not correspond to the potential energy. Note that in order for the Lagrangian to be mathematically consistent, i.e.~to be real, the assumption $1+\partial\phi^2\geq 0$ must be made a priori. Also the dimensionality of the scalar field, i.e.~its physical units, is now taken in order to render $\partial\phi^2$ dimensionless.

The flat FLRW cosmological equations derived from the Lagrangian (\ref{def:tachyon_Lagrangian}) minimally coupled to general relativity, together with a standard matter fluid component, are
\begin{align}
\frac{3H^2}{\kappa^2} &= \rho + \frac{V}{\sqrt{1-\dot\phi^2}}  \,,\label{eq:Friedmann_tachyon}\\
\frac{2\dot H}{\kappa^2} &= -(w+1)\rho - \frac{\dot\phi^2V}{\sqrt{1-\dot\phi^2}} \,,
\end{align}
and the scalar field equation
\begin{equation}
\frac{\ddot\phi}{1-\dot\phi^2} +3H\dot\phi +\frac{V_{,\phi}}{V} =0 \,,
\label{eq:KG_tachyon}
\end{equation}
where again $V_{,\phi}$ denotes the derivative of $V$ with respect to $\phi$. The assumption $1+\partial\phi^2\geq 0$ now translates into $\dot\phi^2\leq 1$ which implies the consistency of Eqs.~(\ref{eq:Friedmann_tachyon})--(\ref{eq:KG_tachyon}). The scalar field energy density and pressure can be written as
\begin{equation}
\rho_\phi = \frac{V}{\sqrt{1-\dot\phi^2}} \quad\mbox{and}\quad p_\phi = -V\sqrt{1-\dot\phi^2} \,,
\end{equation}
and its EoS becomes
\begin{equation}
w_\phi = \frac{p_\phi}{\rho_\phi} = -1 +\dot\phi^2 \,.
\end{equation}
Note that whenever the scalar field kinetic energy vanishes the EoS takes the cosmological constant value $-1$.

In order to convert Eqs.~(\ref{eq:Friedmann_tachyon})--(\ref{eq:KG_tachyon}) into a dynamical system, we need to define suitable dimensionless variables. Following \citet{Copeland:2004hq} we introduce the variables
\begin{equation}
x = \dot\phi \quad\mbox{and}\quad y = \frac{\kappa \sqrt{V}}{\sqrt{3}H} \,.
\label{def:nonENvars}
\end{equation}
Note that $x$ is dimensionless due to the non-standard units of $\phi$ (mass$^{-1}$).
The cosmological equations (\ref{eq:Friedmann_tachyon})--(\ref{eq:KG_tachyon}) can now be rewritten as
\begin{align}
x' &= \left(x^2-1\right) \left(3 x-\sqrt{3} \lambda  y\right) \,,\label{eq:x_tachyon}\\
y' &= -\frac{1}{2} y \left[\frac{3 y^2 \left(w-x^2+1\right)}{\sqrt{1-x^2}}-3 (w+1)+\sqrt{3} \lambda  x y\right] \,,\label{eq:y_tachyon}\\
\lambda '&= \sqrt{3}\left(\Gamma-\frac{3}{2}\right) xy\lambda^2 \,,\label{eq:lambda_tachyon}
\end{align}
where we have defined
\begin{equation}
\lambda = -\frac{V_{,\phi}}{\kappa V^{3/2}} \quad\mbox{and}\quad \Gamma = \frac{VV_{,\phi\phi}}{V_{,\phi}^2} \,,
\label{104}
\end{equation}
and the Friedmann constraint
\begin{equation}
\frac{y^2}{\sqrt{1-x^2}} = 1-\Omega_{\rm m} \leq 1 \,,
\label{eq:tachyon_Friedmann_const}
\end{equation}
must be satisfied.
Eqs.~(\ref{eq:x_tachyon})--(\ref{eq:lambda_tachyon}) are consistent in the range $x^2\leq 1$ which follow from the constraint $\dot\phi^2\leq 1$. The limit $x\rightarrow \pm 1$ must be handled with care but, as we will see, only the points~$(x,y)=(\pm 1,0)$ will be part of the phase space due to the Friedmann constraint (\ref{eq:tachyon_Friedmann_const}).

Note the difference in the definition of $\lambda$ with respect to the canonical case (\ref{def:lambda_dynamical}) where $V$ appeared linearly in the denominator. The function $V(\phi)$ corresponding to a constant $\lambda$ is the inverse square potential
\begin{equation}
V(\phi)=\frac{M^2}{\phi^2} \,,
\label{099}
\end{equation}
where $M$ is a constant with units of mass which relates to $\lambda$ as $M=2/(\kappa\lambda)$. The simplest case of tachyonic dark energy is thus characterised by the inverse square potential and not by the exponential potential as in canonical quintessence.

Eqs.~(\ref{eq:x_tachyon})--(\ref{eq:lambda_tachyon}) do not represent an autonomous system due to the appearance of $\Gamma$ which is still a function of $\phi$. However, exactly as in the quintessence models (Sec.~\ref{sec:other_potentials}), both $\lambda$ and $\Gamma$ are functions of $\phi$ implying that for suitable $\lambda(\phi)$, the quantity $\Gamma$ can be written as a function of $\lambda$, namely $\Gamma(\lambda)$. In this case Eqs.~(\ref{eq:x_tachyon})--(\ref{eq:lambda_tachyon}) close to a 3D autonomous dynamical system and a general analysis similar to the one we considered for quintessence in Sec.~\ref{sec:quintessence} can be performed \citep{Fang:2010zze}. We will however focus on the $\Gamma=3/2$ case corresponding to the potential (\ref{099}) and to a constant $\lambda$.

\subsubsection{Tachyons with an inverse square potential}

For the inverse square potential (\ref{099}), Eqs.~(\ref{eq:x_tachyon})--(\ref{eq:y_tachyon}) constitute an autonomous 2D dynamical system which has been studied by \citet{Copeland:2004hq} (for a similar dynamical analysis see also \citet{Aguirregabiria:2004xd}). The variables $x$ and $y$ must satisfy the Friedmann constraint (\ref{eq:tachyon_Friedmann_const}) which renders the phase space compact. Moreover the system (\ref{eq:x_tachyon})--(\ref{eq:y_tachyon}) is odd-parity invariant, i.e.~it is invariant under the mapping $(x,y)\mapsto (-x,-y)$. This implies that only half of the $(x,y)$-space needs to be analysed and we will choose the positive $y$ half plane for convenience.

\begin{figure}
\centering
\includegraphics[width=\columnwidth]{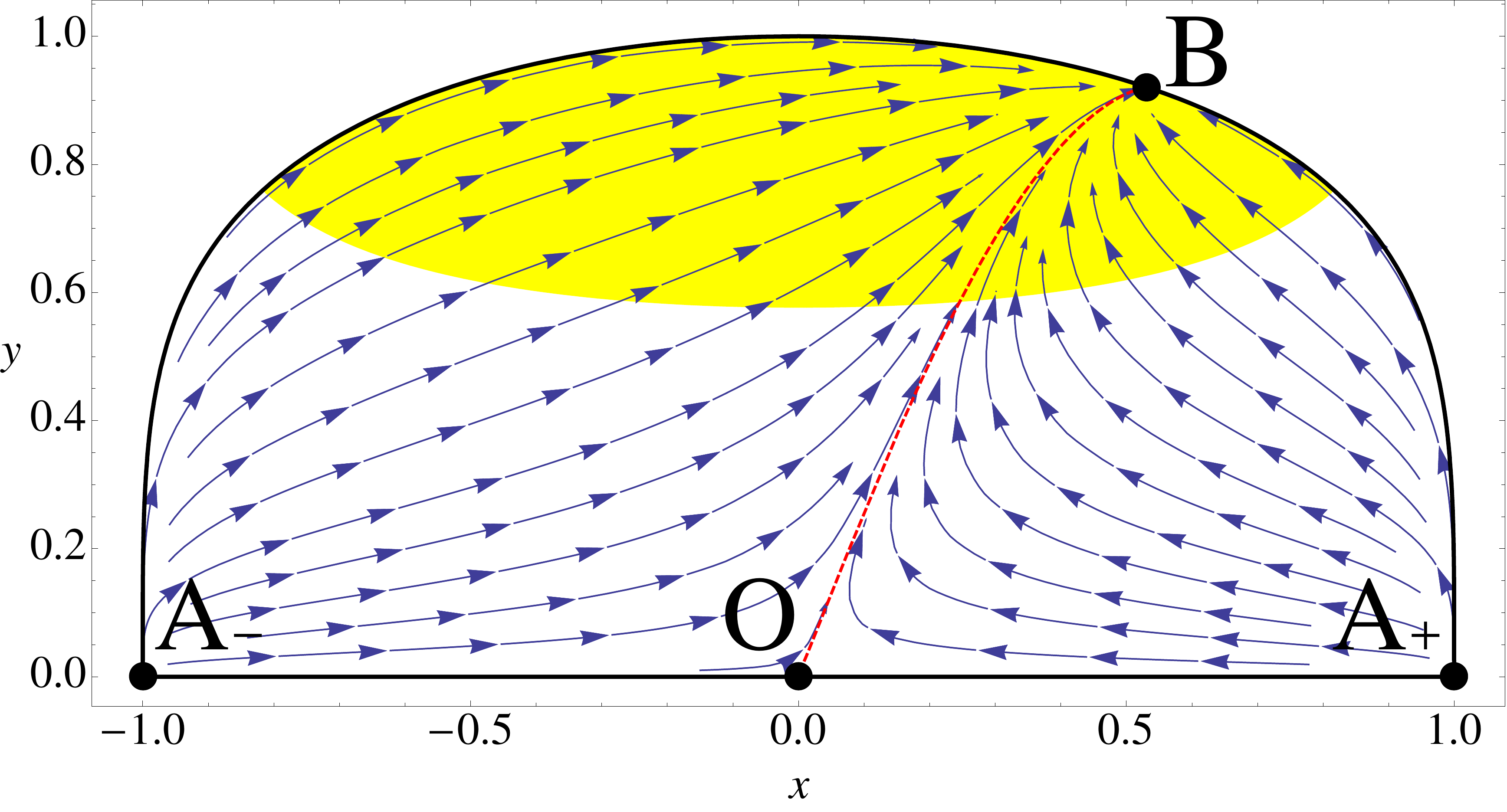}
\caption{Phase space of the dynamical system (\ref{eq:x_tachyon})--(\ref{eq:y_tachyon}) with the values $\lambda=1$ and $w=0$. The yellow/shaded region denotes the area of the phase space where the universe undergoes accelerated expansion.}
\label{fig:tachyon_phase_space}
\end{figure}

Another symmetry of the dynamical system (\ref{eq:x_tachyon})--(\ref{eq:y_tachyon}) is represented by the mapping $(x,\lambda)\mapsto (-x,-\lambda)$ which is the same appearing in canonical quintessence. It implies that only positive values of $\lambda$ need to be considered since negative values would lead to the same dynamics after a reflection over the $y$-axis.
From all this information we learn that the phase space for the tachyonic potential (\ref{099}) is compact and constrained in the region $-1\leq x\leq 1$ and $0\leq y\leq 1$. Its exact shape depends on the Friedmann constraint and can be seen in Fig.~\ref{fig:tachyon_phase_space}. Note that only at the points~$(\pm 1,0)$ of the phase space the dynamical system (\ref{eq:x_tachyon})--(\ref{eq:y_tachyon}) is undetermined.

The relative energy density and EoS of the tachyonic scalar field can be written as
\begin{equation}
\Omega_\phi = \frac{y^2}{\sqrt{1-x^2}} \quad\mbox{and}\quad w_\phi = -1+x^2 \,,\\
\label{102}
\end{equation}
while the effective EoS of the universe is
\begin{equation}
w_{\rm eff} = w \left(1-\frac{y^2}{\sqrt{1-x^2}}\right)- y^2\sqrt{1-x^2} \,.
\label{103}
\end{equation}
Note that the tachyonic EoS is constrained in the interval $-1\leq w_\phi \leq 0$ since $x^2\leq 1$.

\begin{table}
\begin{center}
\begin{tabular}{|c|c|c|c|c|c|c|c|}
\hline
Point & $x$ & $y$ & Existence & $w_{\rm eff}$ & $w_\phi$ & $\Omega_\phi$ & Stability \\
\hline
\multirow{2}*{$O$} & \multirow{2}*{0} & \multirow{2}*{0} & \multirow{2}*{$\forall\;\lambda,w$} & \multirow{2}*{$w$} & \multirow{2}*{$-1$} & \multirow{2}*{0} & \multirow{2}*{Saddle} \\ & & & & & & & \\
\multirow{2}*{$A_\pm$} & \multirow{2}*{$\pm 1$} & \multirow{2}*{0} & \multirow{2}*{$\forall\;\lambda,w$} & \multirow{2}*{$0$} & \multirow{2}*{0} & \multirow{2}*{--} & \multirow{2}*{Unstable} \\ & & & & & & & \\
\multirow{2}*{$B$} & \multirow{2}*{$y_B\,\lambda/\sqrt{3}$} & \multirow{2}*{$y_B$} & \multirow{2}*{$\forall\;\lambda,w$} & \multirow{2}*{$-1 +y_B^2\,\lambda^2/3$} & \multirow{2}*{$-1 +y_B^2\,\lambda^2/3$} & \multirow{2}*{1} & \multirow{2}*{Stable} \\ & & & & & & & \\
\hline
\end{tabular}
\caption{Critical points of the system (\ref{eq:x_tachyon})--(\ref{eq:y_tachyon}) with existence, physical and stability properties. The coordinate $y_B$ is given in Eq.~(\ref{100}).}
\label{tab:tachyon_CP}
\end{center}
\end{table}

The critical points of the system (\ref{eq:x_tachyon})--(\ref{eq:y_tachyon}) are listed, together with their phenomenological and stability properties, in Tab.~\ref{tab:tachyon_CP}. There are four critical points:
\begin{itemize}
\item {\it Point~$O$}. The origin of the phase space is again a matter dominated point ($\Omega_{\rm m}=1$ and $w_{\rm eff}=w$) where the scalar field energy vanishes ($\Omega_\phi=0$). It exists for every value of the parameters and always represents a saddle point attracting orbits along the $x$-axis and repelling them towards the $y$-axis. The tachyon EoS at this point takes the value $-1$ meaning that the negligible scalar field freezes and acts as a cosmological constant.
\item {\it Points~$A_\pm$}. The two points~$(\pm 1,0)$ are always present in the phase space and always represent the past attractors being them the only unstable nodes. Interestingly the tachyonic EoS vanishes at these point implying that the scalar field acts as pressure-less matter (dust) in the early universe. Points~$A_\pm$ are not formally part of the phase space since the dynamical system (\ref{eq:x_tachyon})--(\ref{eq:y_tachyon}) is singular for $x=\pm 1$. Nevertheless they effectively act as critical points, though care must be taken in dealing with them knowing that standard dynamical systems techniques cannot apply. Their properties can only be derived studying the limit of the flow in their neighbourhood. For example the scalar field energy density (\ref{102}) is undetermined and only its limit as solutions approach Point~$A_\pm$ can be evaluated\footnote{Although Points~$A_\pm$ formally lie on the $\Omega_\phi=1$ line, implying scalar field domination, at $x=\pm 1$ the dynamical system is singular and thus only the limit as trajectories approach these point is mathematically correct. In fact, as we will see, different values of $\Omega_\phi$ are obtained approaching Points~$A_\pm$ from different directions.}.
\item {\it Point~$B$}. The last point in the phase space has coordinates $(x,y)=(y_B\,\lambda/\sqrt{3},y_B)$ where
\begin{equation}
y_B = \left(\frac{\sqrt{\lambda^4+36}-\lambda^2}{6}\right)^{1/2} \,.
\label{100}
\end{equation}
It is a scalar field dominated point ($\Omega_\phi=1$) where the tachyonic EoS assumes the value $w_\phi = w_{\rm eff} = -1+y_B^2\,\lambda^2/3$. It always represents the future attractor of the phase space being the only stable point and appearing for every value of the parameters $w$ and $\lambda$. Point~$B$ denotes a late time dark energy dominated solution if $\lambda^2<2\sqrt{3}$ in which case $w_{\rm eff}<-1/3$ at this point.
\end{itemize}
If $w$ is allowed to take also negative values, then another critical point describing a matter scaling solution point can appear in the phase space \citep{Aguirregabiria:2004xd,Copeland:2004hq}.
We do not discuss this solution as we are only considering values within the range $0\leq w\leq 1/3$.

Considering the phase space, it is clear from the properties of the critical points, that for all admissible values of the parameters $\lambda$ and $w$ the qualitative behaviour of the phase space will be the same. In Fig.~\ref{fig:tachyon_phase_space} the phase space for the values $w=0$ and $\lambda=1$ has been plotted. The future attractor is Point~$B$ which describes a dark energy dominated solution whenever it falls inside the yellow/shaded region ($\lambda^2<2\sqrt{3}$), denoting the area of the phase space where the universe undergoes acceleration. All the orbits are heteroclinic solutions connecting Points~$A_\pm$, the past attractors, to Point~$B$, except for the heteroclinic orbit between Point~$O$ and Point~$B$ (red/dashed line) and the ones between Points~$A_\pm$ and Point~$O$. The heteroclinic orbit connecting Point~$O$ to Point~$B$ divides the phase space into two invariant sets with past attractors Point~$A_+$ and Point~$A_-$ respectively.

The phase space depicted in Fig.~\ref{fig:tachyon_phase_space} can be employed to characterise a late time matter to a dark energy transition. Every orbit shadowing the heteroclinic solution connecting Point~$O$ to Point~$B$ will indeed describe a matter dominated era followed by eternal accelerated expansion. The early time behaviour is complicated to derive since only the limit to Points~$A_\pm$, the past attractors, makes sense mathematically. In any case since at Points~$A_\pm$ the tachyonic EoS vanishes, the scalar field will always behave as non-relativistic matter at early times. The effective EoS of the universe will then be constrained in the range $0\leq w_{\rm eff}\leq 1/3$ as the orbits approach Points~$A_\pm$ in the past, with the precise value depending on the matter EoS parameter $w$ and on the relative energy density of the tachyon. On the extremes if $\Omega_\phi=0$ then $w_{\rm eff}=w$, while if $\Omega_\phi=1$ then $w_{\rm eff}=0$. Of course if $w=0$ then $w_{\rm eff}=0$ irrespective of the tachyonic energy $\Omega_\phi$ as the solutions approach Points~$A_\pm$.

\begin{figure}
\centering
\includegraphics[width=\columnwidth]{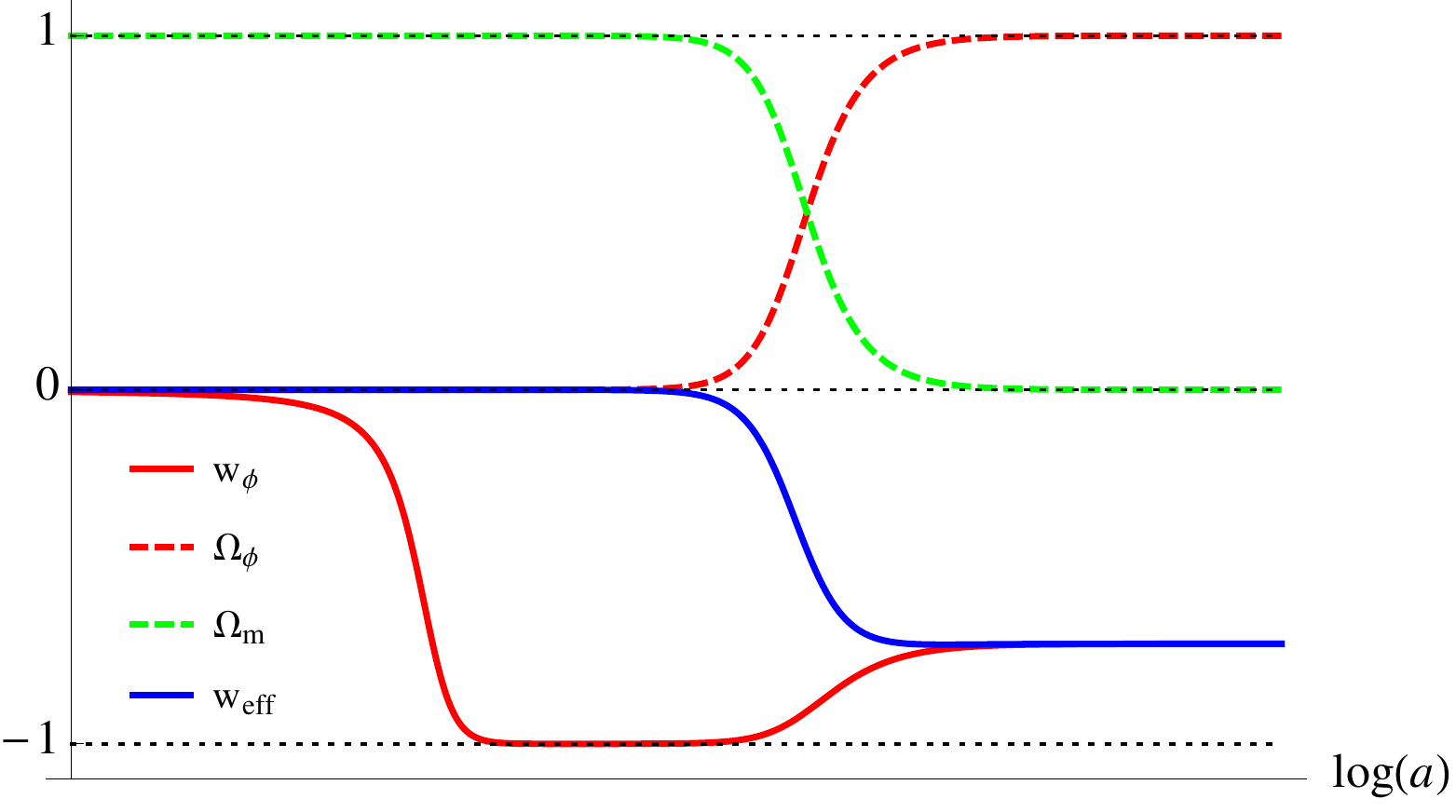}
\caption{Evolution of the phenomenological quantities (\ref{102}) and (\ref{103}) for a solution shadowing the heteroclinic sequence $A_-\rightarrow O\rightarrow B$ in Fig.~(\ref{fig:tachyon_phase_space}).}
\label{fig:tachyon_w_plot_1}
\end{figure}
\begin{figure}
\centering
\includegraphics[width=\columnwidth]{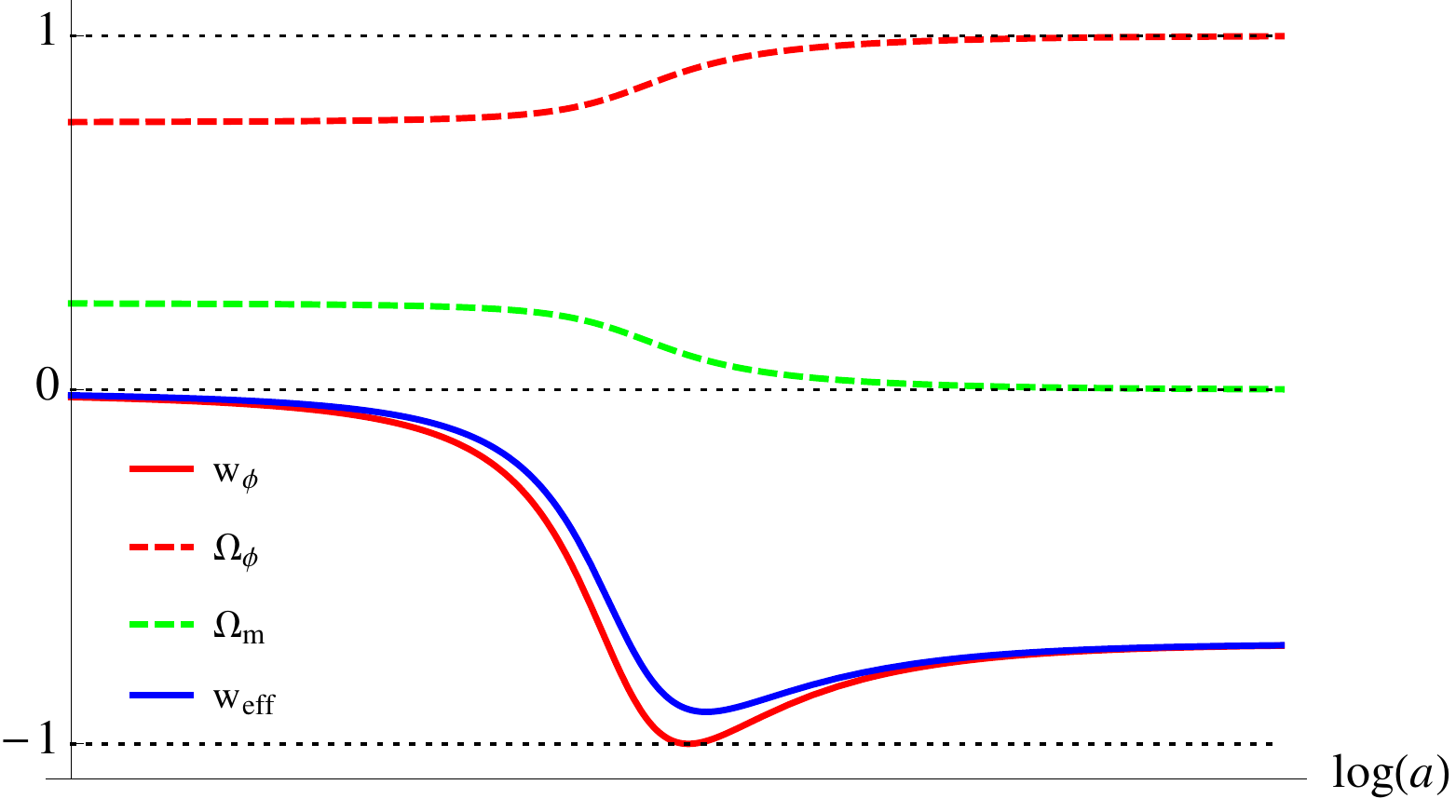}
\caption{Evolution of the phenomenological quantities (\ref{102}) and (\ref{103}) for a solution sufficiently close to, but not completely shadowing, the heteroclinic orbit connecting Point~$A_-$ to Point~$B$ in Fig.~(\ref{fig:tachyon_phase_space}).}
\label{fig:tachyon_w_plot_2}
\end{figure}

We now provide two examples of how the phenomenological quantities (\ref{102}) and (\ref{103}) evolve for two different trajectories in the phase space of Fig.~\ref{fig:tachyon_phase_space}. In Fig.~\ref{fig:tachyon_w_plot_1} these quantities have been plotted for a trajectory shadowing the heteroclinic sequence $A_-\rightarrow O\rightarrow B$, while in Fig.~\ref{fig:tachyon_w_plot_2} they have been plotted for a solution passing along, but not completely shadowing, the heteroclinic orbit connecting Point~$A_-$ to Point~$B$.
Both of them represent a late time matter to dark energy transition as can be seen in the dynamics of $w_{\rm eff}$ (blue/solid line) which is zero at early times and below $-1/3$ at late times\footnote{Phantom behaviour cannot be obtained in this model.}.
The dark tachyonic EoS however behaves slightly differently in the two cases. In Fig.~\ref{fig:tachyon_w_plot_1} it starts from zero (Point~$A_-$), decreases to $-1$ during matter domination (Point~$O$) and finally reaches the dark energy regime ($w_\phi\simeq -0.72$) at late times (Point~$B$). In Fig.~\ref{fig:tachyon_w_plot_2} $w_\phi$ starts again from zero (Point~$A_-$) but then almost immediately reaches the dark energy value (Point~$B$) falling for a brief moment towards $-1$ as the solution follows the boundary of the phase space in its transition from Point~$A_-$ to Point~$B$.

The most important differences between Figs.~\ref{fig:tachyon_w_plot_1} and \ref{fig:tachyon_w_plot_2} arise however in the evolution of the energy densities of matter and dark energy. In Fig.~\ref{fig:tachyon_w_plot_1} the late time behaviour, when dark energy comes to dominate, is the expected one. At early times matter always dominates even in the neighbourhood of Point~$A_-$, since when $w_\phi\rightarrow 0$ one still find $\Omega_{\rm m}=1$. This implies that as we approach Point~$A_-$ in the past following the $y$-axis we find $\Omega_\phi\rightarrow 0$, as it is also evident first taking the limit $y\rightarrow 0$ of $\Omega_\phi$ and then the $x\rightarrow\pm 1$ one.
A completely different situation emerges from Fig.~\ref{fig:tachyon_w_plot_2} where $\Omega_\phi$ and $\Omega_{\rm m}$ are of the same order of magnitude in the neighbourhood of Point~$A_-$. In fact the more an orbit shadows the boundary of the phase space in approaching Point~$A_-$, the more dark energy dominates with $\Omega_\phi=1$ exactly on the boundary. This is the reason why we chose a trajectory which does not follow exactly the heteroclinic orbit connecting Point~$A_-$ to Point~$B$. Such a solution would have given complete dark energy domination in the past as it approached Point~$A_-$, while from Fig.~\ref{fig:tachyon_w_plot_2} it is clear that a scaling behaviour is possible for orbits emerging from Point~$A_-$.

Although the evolution of the universe is practically the same at early times since $w_{\rm eff}=0$ for every orbit approaching Points~$A_\pm$, the domination by matter or tachyons will depend on the direction of the trajectory escaping from Points~$A_\pm$.
Note that in this tachyonic dark energy model matter domination is achieved at early times no matter the initial conditions one chooses, implying that the fine tuning problem of initial conditions does not arise in this framework. Moreover since the tachyonic field behaves as pressure-less matter at early times, and can dominate as in Fig.~\ref{fig:tachyon_w_plot_2}, this model can also be used as a dark matter unified theory, where both dark matter and dark energy are described by a single scalar field \citep{Padmanabhan:2002sh}.

\subsubsection{Other tachyonic models}

The inverse square potential (\ref{099}) constitutes the simplest tachyonic model to study with dynamical systems techniques, but applications to other potentials have also been considered in the literature. \citet{Fang:2010zze} performed a general analysis of the dynamical system (\ref{eq:x_tachyon})--(\ref{eq:lambda_tachyon}) for a general function $\Gamma(\lambda)$. The same approach, with a redefinition of the tachyonic field, has been also used by \citet{Quiros:2009mz} who studied the exponential, inverse power-law and $V(\phi)=V_0[\sinh(\lambda\phi)]^{-\alpha}$ potentials as examples. The inverse power-law potential, where $\Gamma$ as given by (\ref{104}) is constant, together with more general potentials have also been considered by \citet{Copeland:2004hq}, while the exponential potential has been analysed also by \citet{Guo:2003zf}. In all these models future stable accelerated attractors can be easily obtained.
\citet{Hao:2003aa} and \citet{Chingangbam:2004xe} proved that these dark energy late time solutions always arises when the corresponding potentials admits a minimum.
\citet{Li:2010eua} performed the dynamical analysis for tachyons in spatially curved FLRW universes focusing in particular on scaling solutions and \citet{Fang:2014bta} rewrote the dynamical system (\ref{eq:x_tachyon})--(\ref{eq:lambda_tachyon}) in the variables $(\Omega_\phi,w_\phi,\lambda)$ which can be directly compared with astronomical observations.

Extended tachyonic models have also appeared in the dynamical systems literature.
\citet{Gumjudpai:2005ry}, \citet{Farajollahi:2011jr} and again \citet{Farajollahi:2011ym} added a coupling between the tachyon field and the matter sector obtaining new scaling solutions (cf.~Sec.~\ref{sec:coupled_noncanonical_scalars}). \Citet{Guo:2004dta} expanded the system to multiple tachyons with inverse square potentials showing that the dynamics is qualitatively similar to the single field scenario. \Citet{delaMacorra:2007mc} studied the dynamical system of a tachyon scalar field $T$ and potential $V(T)$ coupled to a canonically normalised scalar field $\phi$ with an arbitrary interaction term $B(T, \phi)$ in the presence of a barotropic fluid. The results show that, the effective EoS for the tachyon field changes due to the interaction with the scalar field, and it is possible for a tachyon field to redshift as matter in the absence of an interaction term $B$ and as radiation when $B$ is turned on.

For phantom tachyons, where the sign of the kinetic term $\partial\phi^2$ is the opposite of the one appearing in (\ref{def:tachyon_Lagrangian}), the condition for late time phantom stable domination is that $V(\phi)$ have a maximum \citep{Hao:2003aa}. Phantom tachyons have also been studied by \citet{Fang:2010zze,Gumjudpai:2005ry,Fang:2014bta}. A quintom tachyonic model, where a standard tachyonic field is coupled to a phantom tachyon, has been proposed by \citet{Shi:2008df} as an alternative solution for a dynamical crossing of the phantom barrier.

\subsection{Generalised DBI scalar field models}
\label{sec:generalised-dbi}

In this section we turn our attention to the case of non-canonical scalar field models defined by extensions of the DBI Lagrangian \eqref{def:tachyon_Lagrangian}. These models are usually motivated by D-brane (higher-dimensional theories) phenomenology (see e.g.~\citet{Silverstein:2003hf}) and can be defined by the scalar field {\it generalised DBI Lagrangian}
\begin{equation}
	\mathcal{L}_{\rm DBI} = \frac{1}{f(\phi)}\left(\sqrt{1+f(\phi)\partial\phi^2}-1\right)-V(\phi) \,,
	\label{def:generalised_DBI_Lagrangian}
\end{equation}
where $f(\phi)$ and $V(\phi)$ are arbitrary functions of the scalar field $\phi$. The cosmological equations deriving from the Lagrangian (\ref{def:generalised_DBI_Lagrangian}) are
\begin{align}
\frac{3H^2}{\kappa^2} = \rho + \frac{\gamma^2}{\gamma+1}\dot\phi^2 + V(\phi) \,,\label{105}\\
\ddot\phi + \frac{3H}{\gamma^2}\dot\phi + \frac{V_{,\phi}}{\gamma^3} + \frac{f_{,\phi}}{2f} \frac{(\gamma+2)(\gamma-1)}{\gamma(\gamma+1)} \dot\phi^2 = 0 \,,\\
\dot\rho +3H (1+w)\rho = 0 \,,\label{106}
\end{align}
where we have defined
\begin{equation}
	\gamma = \frac{1}{\sqrt{1-f(\phi)\dot\phi^2}} \,.
\end{equation}
Introducing the variables \citep{Copeland:2010jt}
\begin{equation}
	x=\frac{\gamma\kappa\dot\phi}{\sqrt{3(\gamma+1)H}} \,, \qquad y= \frac{\kappa \sqrt{V}}{\sqrt{3}H} \,,\qquad \tilde\gamma = \frac{1}{\gamma} \,,\label{106a}
\end{equation}
which in the limit $\gamma\rightarrow 1$ reduce to the canonical EN variables originally defined in (\ref{def:ENvars}), Eqs.~(\ref{105})--(\ref{106}) can be conveniently rewritten in a dynamical system formulation as
\begin{align}
x' &= \frac{1}{2}\sqrt{3\tilde\gamma(1+\tilde\gamma)}\lambda y^2 + \frac{3}{2}x\left[(1+\tilde\gamma)(x^2-1)+(1+w)(1-x^2-y^2)\right] \,,\label{107}\\
y'&= -\frac{1}{2} \sqrt{3\tilde\gamma(1+\tilde\gamma)}\lambda xy + \frac{3}{2} y \left[(1+\tilde\gamma)x^2+(1+w)(1-x^2-y^2)\right] \,,\\
\tilde\gamma' &= \frac{\tilde\gamma(1-\tilde\gamma^2)}{\sqrt{1+\tilde\gamma}} \left[3\sqrt{1+\tilde\gamma} +\sqrt{3\tilde\gamma} \frac{1}{x}(\mu x^2 -\lambda y^2)\right] \,,\\
\lambda' &= -\sqrt{3\tilde\gamma(1+\tilde\gamma)}\lambda^2 (\Gamma-1)x \,,\\
\mu' &= -\sqrt{3\tilde\gamma(1+\tilde\gamma)}\mu^2 (\Xi-1)x \,,\label{108}
\end{align}
where
\begin{equation}
	\lambda = -\frac{V_{,\phi}}{\kappa V} \,,\qquad \mu= -\frac{f_{,\phi}}{\kappa V} \,,
\end{equation}
and
\begin{equation}
	\Gamma = \frac{V \,V_{,\phi \phi}}{V_{,\phi}^2} \,, \qquad \Xi= \frac{f\, f_{,\phi \phi}}{f_{,\phi}^2}\,.
\end{equation}

The Friedmann constraint (\ref{105}) imposes the condition
\begin{equation}
	x^2 + y^2 = 1-\Omega_{\rm m} \leq 1 \,,
\end{equation}
which is the same arising in canonical quintessence.

The energy density and EoS of the scalar field are given by
\begin{equation}
	\Omega_\phi = x^2+y^2 \quad\mbox{and}\quad w_\phi = \frac{\tilde\gamma x^2-y^2}{x^2+y^2} \,,
\end{equation}
while the effective EoS is
\begin{equation}
	w_{\rm eff} = \tilde\gamma x^2 - y^2 + w (1-x^2-y^2) \,.
\end{equation}
Note how in this model the phenomenological quantities depend also on the new variable $\tilde\gamma$, which is in general constrained as $0\leq \tilde\gamma \leq 1$.

In analogy with other simpler models, if the variables $\Gamma$ and $\Xi$ can be written as functions of $\lambda$ and $\mu$ respectively, then Eqs.~(\ref{107})--(\ref{108}) would represent a 5D autonomous system. The simplest case is again determined by $V$ and $f$ being exponential functions where $\lambda$ and $\mu$ become constant and the system reduces to three dimensions. If instead $f$ and $V$ are of the power-law type, then the quantity $\Gamma$ and $\Xi$ are constant and one obtains the simplest 5D system. These cases have been studied in detail by \citet{Copeland:2010jt} where the corresponding dark energy attractors and scaling solutions were properly analysed.
\citet{Guo:2008sz} considered the case $f\propto \phi^{-4}$ and $V\propto \phi^2$ which can be naturally justified by D-brane phenomenology. They found scaling solutions for negative (non-physical) values of $w$ in analogy to the simpler tachyonic model. A general analysis of future attractor solutions for the generalised DBI model (\ref{def:generalised_DBI_Lagrangian}) and different form of the functions $f$ and $V$, has been performed by \citet{Ahn:2009hu,Ahn:2009xd} showing that if $V(\phi)$ has a minimum then there is always a dark energy late time solution. \citet{Panpanich:2017nft} studied the cosmological dynamics of the D-BIonic model proposed by \citet{Burrage:2014uwa} and of a Dirac-Born-Infeld model in which the scalar field is coupled to a matter fluid for different types of couplings. They found a new class of analytic scaling solutions yielding accelerated expansion for the case of an exponential potential and exponential couplings to matter. In particular they showed that in these coupled models, because the D-BIonic theory possesses a screening mechanism, they can solve the coincidence as well as the dark energy problem, as they obtain observationally consistent energy densities for the matter and dark energy components.

Finally \citet{Fang:2004qj} studied the dynamics of the phantom case of the Lagrangian (\ref{def:generalised_DBI_Lagrangian}) where the sign of $\partial\phi^2$ is inverted. They found stable phantom late time attractors whenever $V$ admits a positive maximum and then considered an example with potential $V(\phi)= V_0 (1+\phi/\phi_0)\exp(-\phi/\phi_0)$ for a constant $\phi_0$. \citet{Kaeonikhom:2012xr} instead added a coupling between the generalised DBI field and the matter sector, while \citet{Gumjudpai:2009uy} extended the Lagrangian (\ref{def:generalised_DBI_Lagrangian}) to $\mathcal{L}_{\rm DBI} = W(\phi)/f(\phi)\sqrt{1+f(\phi)\partial\phi^2}-1/f(\phi)-V(\phi)$.

\subsection{$k$-essence}
\label{sec:k-essence}

In this section we review dynamical systems applications to more general scalar field models characterised by Lagrangians with general higher-order derivative terms. In order to simplify the following equations we will denote the scalar field kinetic term with
\begin{equation}
	X = -\frac{1}{2}\partial\phi^2 = - \frac{1}{2} \partial_\mu\phi\partial^\mu\phi \,.
	\label{def:X}
\end{equation}
Note that for a homogeneous scalar field $(\phi=\phi(t))$ in a FLRW metric (\ref{eq:FRWmetric}) the kinetic energy (\ref{def:X}) is always positive reducing to
\begin{equation}
	X = \frac{1}{2} \dot\phi^2 \,.
\end{equation}

The broad class of models we discuss here is known as {\it $k$-essence} from kinetic generalisation of quintessence. Its Lagrangian is an arbitrary function $P$ of the kinetic term and the scalar field $\phi$ and can be written as
\begin{equation}
	\mathcal{L}_{k\rm-essence} = P(X,\phi) \,.
	\label{def:k-essence_Lagrangian}
\end{equation}
It clearly contains as subclasses the canonical, phantom and tachyonic scalar fields.
The new cosmological features of these models will mainly come from the contribution of the higher-order terms.
The cosmological equations following from the scalar field Lagrangian (\ref{def:k-essence_Lagrangian}), again considering also a standard matter fluid sector, are
\begin{align}
	\frac{3 H^2}{\kappa^2} &= \rho + 2X\frac{\partial P}{\partial X} -P \,,\label{eq:Friedmann_k-essence}\\
	\frac{1}{\kappa^2}\left(2\dot H+3H^2\right) &= -p -P \,,\label{eq:acc_k-essence}
\end{align}
together with the scalar field equation
\begin{align}
	\left(\frac{\partial P}{\partial X}+2X\frac{\partial^2 P}{\partial X^2}\right) \ddot\phi +3H \frac{\partial P}{\partial X}\dot\phi +2X \frac{\partial^2 P}{\partial X\partial\phi} -\frac{\partial P}{\partial \phi} = 0 \,.
	\label{eq:KG_k-essence}
\end{align}
Note that the function $P$ plays the role of the scalar field pressure $p_\phi$ in Eq.~(\ref{eq:acc_k-essence}), which is why the letter $P$ has been used\footnote{The common convention in the literature is to use the lower case letter $p$ for the Lagrangian (\ref{def:k-essence_Lagrangian}). We will however use the upper case $P$ in order to not create confusion with the matter pressure.}. From Eq.~(\ref{eq:Friedmann_k-essence}) instead the scalar field energy density can be obtained as
\begin{equation}
	\rho_\phi = 2X\frac{\partial P}{\partial X} -P \,,
	\label{def:k-essence_energy}
\end{equation}
which reduces to the usual $\rho_\phi = X+V$ in the canonical case.
The EoS for the $k$-essence scalar field can be written as
\begin{equation}
	w_\phi = \frac{P}{\rho_\phi} = P \left(2X\frac{\partial P}{\partial X} -P\right)^{-1} \,,
	\label{eq:w_kess}
\end{equation}
with the allowed range and possible singularities strongly dependent on the function $P(X,\phi)$.

As mentioned in Sec.~\ref{sec:phantom}, a non-canonical scalar field Lagrangian will in general introduce theoretical problems at both the quantum and classical levels. In order to avoid these instabilities, at least at the classical level, the $k$-essence Lagrangian (\ref{def:k-essence_Lagrangian}) must satisfy some consistency conditions, which come from the requirement that solutions of the theory must be stable under small perturbations. These conditions can be translated into two constraints over the energy density and {\it speed of sound} of the scalar field which are required to be positive. In quantitative terms the constraints
\begin{equation}
	\rho_\phi \geq 0 \quad\mbox{and}\quad c_s^2 \geq 0 \,,
\end{equation}
where the scalar field energy density $\rho_\phi$ is given by Eq.~(\ref{def:k-essence_energy}) and the speed of sound of adiabatic perturbations is defined as
\begin{equation}
	c_s^2 = \frac{\partial P}{\partial X}/\frac{\partial \rho_\phi}{\partial X} \,,
\end{equation}
must hold (see \citet{Garriga:1999vw} for more details).

The $k$-essence scalar field was first considered as a model of dark energy by \citet{Chiba:1999ka} and \citet{ArmendarizPicon:2000dh,ArmendarizPicon:2000ah} who found that tracking behaviour is possible in these general models and late time accelerated solutions are easy to obtain.
Eqs.~(\ref{eq:Friedmann_k-essence})--(\ref{eq:KG_k-essence}) are however too complicated to directly study with dynamical systems techniques due to the unknown dependence upon the arbitrary function $P(X,\phi)$. For an in depth dynamical analysis some assumptions to reduce the general function $P$ must first be taken into account.
Through the literature several different forms of the $k$-essence Lagrangian have been considered, however for dynamical systems applications we can broadly divide them into three subclasses: $P = X G(X/V(\phi))$, $P(X,\phi)= K(\phi)\tilde{P}(X)$ and $P=F(X)-V(\phi)$.
In what follows we review these models under the dynamical system perspective.

\subsubsection{Models with Lagrangian $\mathcal{L} = X G(X/V(\phi))$}

The first subclass of $k$-essence Lagrangian we will discuss considers the function
\begin{equation}
 	P(X,\phi) = X G(Y) \quad\mbox{with}\quad Y = \frac{X}{V(\phi)} \,,
 	\label{096}
 \end{equation}
where $G$ is an arbitrary function, $\lambda$ a constant and the ``potential'' is usually considered of the exponential type $V(\phi) \propto e^{\lambda \kappa \phi}$, for which the EN variables (\ref{def:ENvars}) are well suited since the argument of the function $G$ becomes proportional to $x^2/y^2$. Note that for $G(Y) = 1-1/Y$ the Lagrangian \eqref{096} reduces to the canonical one $\mathcal{L}_\phi = X - V$. The reason why the Lagrangian (\ref{096}) is important is its connection with scaling solutions. More precisely \citet{Piazza:2004df} (see also \citet{Tsujikawa:2004dp,Gong:2006sp}) proved that scaling solutions appear in $k$-essence models only if a Lagrangian of the type (\ref{096}) is assumed with $V(\phi) \propto e^{-\lambda \kappa \phi}$. As discussed in Sec.~\ref{sec:exp_potential}, scaling solutions are of great relevance for the cosmic coincidence problem since stable accelerated scaling solutions could completely solve this issue.

\citet{Tsujikawa:2006mw} showed that for the general class (\ref{096}) with $V(\phi) \propto e^{-\lambda \kappa \phi}$, even in the presence of a coupling to the matter sector, the dark energy late time solutions are always unstable in the presence of a scaling solution unless they are of the phantom type. Subsequently \citet{Amendola:2006qi} generalised this analysis including also a matter to dark energy coupling dependent on $\phi$ and found that the Lagrangian (\ref{096}) with an exponential $V(\phi)$ can be generalised to $P=Q^2(\phi) X G(X Q^2(\phi) e^{\lambda \kappa \phi})$ where $Q(\phi)$ is the coupling function (cf.~Sec.~\ref{sec:coupled_noncanonical_scalars}). They also performed a dynamical systems analysis finding critical points corresponding to scaling, quintessence and phantom dominated solutions. Remarkably they showed that a dynamical sequence with one early time matter scaling solution and one late time dark energy scaling solution never occurs in such models.

An interesting and simple $k$-essence model within the class \eqref{096} is defined by the function
\begin{equation}
	G(Y)=1 - \frac{1}{Y} + c_1\, Y \,,
	\label{eq:kinetic_corrections}
\end{equation}
with $c_1$ is a constant. This model provides square kinetic corrections to the canonical scalar field Lagrangian which can be justified by high energy theory phenomenology. For the exponential case $V(\phi) \propto e^{-\lambda \kappa \phi}$ its dynamics has first been explored by \citet{Piazza:2004df}, but a complete dynamical systems analysis has been presented by \citet{Tamanini:2014mpa}. This has been subsequently extended by \citet{Dutta:2016bbs} to scalar field potentials $V(\phi)$ beyond the exponential type. The late time dynamics of this system are similar to that of the canonical quintessence case with the appearance of dark energy and scaling solutions. However the early time dynamics is completely different from the canonical one. In particular the scalar field kinetic dominated solutions no longer appear in the phase space of this model. The early time behaviour is now characterised by a matter dominated solution, which is in better agreement with a radiation or dark matter dominated epoch as required by observations. The model can thus be used to describe a universe where dark energy becomes important only at late times while dark matter dominates at early times.

\citet{Tamanini:2014mpa} also investigated the model
\begin{equation}
	G(Y) = 1 - \frac{1}{Y} + \frac{c_2}{\sqrt{Y}} \,,
\end{equation}
for some constant $c_2$. The background dynamics of this model presents a richer phenomenology compared to the canonical case. In this scenario the early time behaviour is similar to the canonical case, although super-stiff ($w_{\rm eff} >1$) transient regions are always present in the phase space. The main differences appear in the late time evolution where phantom dominated solutions, dynamical crossings of the phantom barrier and new scaling solutions emerge from the dynamics. This model can thus be used to describe a late time dark energy dominated universe capable of dynamically crossing the phantom barrier. Such scenario is similar to the quintom paradigm (see Sec.~\ref{sec:quintom}) and slightly favoured by astronomical observations. The drawbacks of such model arise at the level of perturbations where instabilities of the scalar field always appears (see \citet{Vikman:2005} for a detailed discussion on crossing the phantom barrier).

Finally some authors have also considered theories of multiple non-canonical scalar fields within this $k$-essence class. \citet{Tsujikawa:2006mw} and \citet{Ohashi:2009xw} proved that for multiple $k$-essence scalar fields, each one with a Lagrangian $P_i=X_i G(X_i e^{\lambda_i \kappa \phi_i})$, assisted behaviour is possible; see Sec.~\ref{sec:multi_scalar_fields}. \citet{Chiba:2014sda} studied scaling solutions in the same model with interactions to the matter sector, showing that accelerated scaling solutions can be obtained and argued that the cosmic coincidence problem could be avoided.

\subsubsection{Models with Lagrangian $\mathcal{L} = V(\phi) F(X)$}

Whereas the Lagrangian (\ref{096}) is well known to give rise to scaling solutions, in the second subclass of $k$-essence we consider, tracking solutions are the dominant feature. The function $P(X,\phi)$ for this category can be generally written as the factorised product
\begin{equation}
	P(X,\phi)= V(\phi) F(X) \,,
	\label{097}
\end{equation}
where $V(\phi)$ and $F(X)$ are arbitrary functions of $\phi$ and $X$ respectively. Note that this class of theories includes the tachyonic Lagrangian discussed in Sec.~\ref{sec:tachyons}.

Lagrangians of this kind have been proposed in the first models of $k$-essence. \citet{Chiba:1999ka} (see also \citet{Yang:2010vv,Fang:2014qga}) considered square kinetic corrections as $P(X,\phi)=V(\phi)(-X+X^2)$ (with $V$ usually of the power-law type) and showed that tracking behaviour, as well as quintessence and phantom attractors, arise in such models. Then \citet{ArmendarizPicon:2000dh,ArmendarizPicon:2000ah} showed that tracking solutions naturally appear for the models (\ref{097}) during the radiation dominated era, with the scalar field EoS reducing to $-1$ (cosmological constant) during the matter dominated era and driving the late time accelerated expansion as effective quintessence. Although these tracking solutions could solve the fine tuning problem of initial conditions, \citet{Malquarti:2003hn} showed that the basin of attraction of the $k$-essence tracking solutions is smaller than the quintessence one, implying that canonical scalar field models are still better candidates to address this issue.

Models of $k$-essence of the type (\ref{097}) plus a self-interacting potential for the scalar field have also been studied. \citet{Malquarti:2003nn} demonstrated that for such models, which include (\ref{097}), if the scalar field is approximately constant during some period of its cosmic evolution, then the theory effectively reduces to quintessence for that period of time.
\citet{Piazza:2004df} (see also \citet{Gumjudpai:2005ry}) studied Lagrangians of the type $P=-X+A e^{\lambda_1 \kappa \phi}X^2-Be^{-\lambda_2 \kappa \phi}$ which can be motivated by string theory phenomenology. They also considered a coupling between the matter fluid and the scalar field and performed a dynamical analysis of the subclass $\lambda_1=\lambda_2=\lambda$, which corresponds to the model defined in Eq.~\eqref{eq:kinetic_corrections}.

We now present a general discussion on the $k$-essence model \eqref{097} from a dynamical system perspective. For this type of scalar fields we define the following variables
\begin{align}
  x=\dot\phi\,, \quad
  y=\frac{\kappa\sqrt{V}}{\sqrt3 H}\,, \quad
  \lambda=-\frac{V'}{\kappa V^{3/2}}\,, \quad
  \Gamma=\frac{V V''}{V'^2}\,,
\label{eqs26}
\end{align}
which are the same dynamical variables we employed to analyse tachyonic models in Sec.~\ref{sec:tachyons} (see Eqs.~\eqref{def:nonENvars} and \eqref{104}).

Using the variables \eqref{eqs26}, the cosmological equations derived from the $k$-essence model \eqref{097} together with a standard matter fluid, can be recast into the following autonomous system
\begin{align}
  x'&= \frac{-\sqrt{3}}{F_X+2XF_{XX}}[\sqrt{3}xF_X+y\lambda(F-2X F_X)]\,,
  \label{eqs27} \\
  y'&= \frac{y}{2}[-\sqrt{3}\lambda x y-6 w XF_X y^2+3(w+1) F y^2+3(w+1)]\,,
  \label{eqs28} \\
  \lambda '&= -\sqrt{3} x y\lambda^2\left(\Gamma-\frac{3}{2}\right)\,,
  \label{eqs29}
\end{align}
where $F_X$ and $F_{XX}$ are respectively the first and second derivatives of $F$ with respect to $X$. The right hand side of Eq.~\eqref{eqs29} vanishes for the inverse square potential, or more generally for the potential $V(\phi)=(\frac{1}{2}\kappa \lambda \phi-c_1)^{-2}$, which yields $\Gamma = 3/2$. This is analogous to the tachyonic scalar field (cf.~Sec.~\ref{sec:tachyons}) and in contrast with the case of quintessence where the dynamical equations for the variable $\Gamma$ (which was defined differently) vanishes for the exponential potential (see Sec.~\ref{sec:exp_potential}).

The relative energy density of the scalar field $\Omega_{\phi}$ and its EoS $w_\phi$ can be written as
\begin{align}
  \Omega_{\phi} &= \left(2XF_X -F\right)y^2=\left(x\frac{dF}{dx}-F\right)y^2\,,
  \label{eqs33} \\
  w_{\phi} &= \frac{F}{\Omega_\phi} y^2 = -1 + \frac{xy^2}{\Omega_{\phi}} \frac{dF}{dx} =\left(x\frac{dF}{dx}-F\right)^{-1} F\,,
  \label{eqs34}
\end{align}
where the function $F$ has been written in terms of $x$ thanks to the relation $2X = \dot\phi^2 = x^2$.
Using (\ref{eqs33}) and (\ref{eqs34}), we can rewrite the above dynamical system directly using the observable quantities $(\Omega_{\phi}, \gamma_{\phi}, \lambda)$, where $\gamma_\phi$ is defined as $\gamma_\phi = w_\phi +1$.
This change of variables yields
\begin{align}
  \Omega_{\phi}'&= 3(w+1-\gamma_{\phi})\Omega_{\phi}(1-\Omega_{\phi})\,,
  \label{eqs35} \\
  \gamma_{\phi}'&= \sqrt{3}\left(\sqrt{3}\gamma_{\phi}-\lambda x y\right)
  \left(\gamma_{\phi}-1-\frac{1}{2\Xi+1}\right) \,,
  \label{eqs36} \\
  \lambda'&= -\sqrt{3}x y \lambda^2\left(\Gamma-\frac{3}{2}\right) \,,
  \label{eqs37}
\end{align}
where we denoted
\begin{equation}
	\Xi = \frac{X F_{XX}}{F_X} = \frac{1}{2} \left[ x \frac{d^2F}{dx^2} \left(\frac{dF}{dx}\right) ^{-1} -1 \right] \,.
\end{equation}
Note that in Eqs.~\eqref{eqs35}--\eqref{eqs37} the remaining variables $x$ and $y$ should be intended as functions of $\Omega_\phi$ and $\gamma_\phi$.

If the function $F(X)$ takes the form of $F(X)=\sqrt{1-2\varsigma X}=\sqrt{1-\varsigma x^2}$ (where $\varsigma= 1$ stands for standard tachyons and $\varsigma= -1$ stands for phantom tachyons), then $x=\sqrt{\varsigma \gamma_{\phi}}$, $y=\sqrt{\Omega_{\phi}}(1-\gamma_{\phi})^{\frac{1}{4}}$ and $1+\frac{1}{2\Xi+1}=2-\gamma_{\phi}$.
In this case we obtain the tachyonic Eqs.~(\ref{eq:x_tachyon})--(\ref{eq:lambda_tachyon}) directly from Eqs.~(\ref{eqs27})--(\ref{eqs29}).
Similarly we can rewrite Eqs.~(\ref{eq:x_tachyon})--(\ref{eq:lambda_tachyon}) from the variables $(x, y, \lambda)$ directly to the observable quantities $(\Omega_{\phi},\gamma_{\phi}, \lambda)$ as
\begin{align}
  \label{eqs23}
  \Omega_{\phi}' &= 3(w+1-\gamma_{\phi})\Omega_{\phi}(1-\Omega_{\phi})\,, \\
  \label{eqs24}
  \gamma_{\phi}' &= 2(1-\gamma_{\phi})\left[\lambda \sqrt{3\varsigma \gamma_{\phi}\Omega_{\phi}}(1-\gamma_{\phi})^{\frac{1}{4}}-3\gamma_{\phi}\right]\,, \\
  \label{eqs25}
  \lambda' &= -\sqrt{3\varsigma \gamma_{\phi}\Omega_{\phi}}(1-\gamma_{\phi})^{\frac{1}{4}}\lambda^2\left(\Gamma-\frac{3}{2}\right) \,.
\end{align}
On the other hand if the Lagrangian of $k$-essence takes the form of $P(\phi, X)=V(\phi)(-X+X^2)$, i.e.~$F(X) = - X +X^2$, which has been proposed as a kinetically driven quintessence model \citep{Chiba:1999ka}, one obtains the following dynamical system
\begin{align}
  \label{eqsadd6}
 \Omega_{\phi}'&= 3(w+1-\gamma_{\phi})\Omega_{\phi}(1-\Omega_{\phi}) \,, \\
  \label{eqsadd7}
  \gamma_{\phi}'&= \frac{(\gamma_{\phi}-2)(3\gamma_{\phi}-4)}{3\gamma_{\phi}-8}\left(3\gamma_{\phi}-\lambda\sqrt{3(4-3\gamma_{\phi})\Omega_{\phi}}\right) \,, \\
  \label{eqsadd8}
  \lambda'&= -\lambda^2\sqrt{3(4-3\gamma_{\phi})\Omega_{\phi}}\left(\Gamma-\frac{3}{2}\right) \,.
\end{align}

We can now emphasise the usefulness of the dynamical variables $(\Omega_{\phi},\gamma_{\phi}, \lambda)$ for non-canonical scalar fields, including also models beyond the class of $k$-essence models defined by the Lagrangian \eqref{097}.
For a general scalar field it is in fact quite convenient to work with the observable quantities $(\gamma_{\phi}, \Omega_{\phi})$ rather than the variables $(x, y)$. The reasons are listed in the following \citep{Fang:2014bta}.
Firstly, $(\gamma_{\phi}, \Omega_{\phi})$ are directly related to observable quantities which characterise some properties of dark energy. Analysing the system based on $(\gamma_{\phi},\Omega_{\phi})$, we can immediately visualise how the EoS of dark energy $w_{\phi}$ and the density parameter $\Omega_{\phi}$ evolve in time. Secondly, the form of autonomous systems for $x$ and $y$ are completely different for different models within the $k$-essence class defined by Eq.~\eqref{097}, i.e.~for different functions $F(X)$, as one can realise looking at Eqs.~(\ref{eqs27})--(\ref{eqs29}).
Using the variables $(\gamma_{\phi},\Omega_{\phi})$ instead the dynamical equation for $\Omega_{\phi}'$ retains always the same expression, namely
\begin{equation}
  \Omega_{\phi}' =3(w+1-\gamma_{\phi})\Omega_{\phi}(1-\Omega_{\phi}) \,,
  \label{eqs47}
\end{equation}
regardless of the actual form of the function $F(X)$.
This results can actually be generalised to all non-coupled dark energy models as it is easy to demonstrate starting from the equations
\begin{equation}
	3 H^2=\kappa^2(\rho+\rho_{\rm de}) \,, \quad \dot{\rho}_{\rm de}+3H \rho_{\rm de} \gamma_{\rm de} = 0 \,,\quad \dot\rho + 3H \rho (w+1) =0 \,.
	\label{eq:got}
\end{equation}
In the present case clearly $\Omega_{\rm de} = \Omega_\phi$ and $\gamma_{\rm de} = \gamma_\phi$.
Nevertheless any model of dark energy satisfying Eqs.~(\ref{eq:got}), no matter how complicated are the expressions for $\Omega_{\rm de}$ and $\gamma_{\rm de}$, will yield Eq.~\eqref{eqs47}.
For this reason the variables $(\gamma_{\phi},\Omega_{\phi})$ are particularly suited for dark energy models where arbitrary functions appear, as for example $k$-essence.

\begin{table}
\begin{center}
\begin{tabular}{|c|c|c|}
\hline
 $\Omega_{\phi}$ & $\gamma_{\phi}(=w_\phi+1)$ & Properties  \\
\hline
\multirow{3}*{$0$} &\multirow{3}*{Determined by $\gamma_{\phi}'=0$} & \multirow{2}*{Matter dominated,} \\ & & \multirow{2}*{no accelerating expansion} \\ & & \\
\multirow{3}*{$1$} & \multirow{3}*{Determined by $\gamma_{\phi}'=0$} & Dark energy dominated,\\ &  & possible accelerating expansion, \\ & &  no solution to cosmic coincidence problem \\
\multirow{4}*{Determined by $\gamma_{\phi}'=0$} & \multirow{4}*{$w+1$}  & \multirow{2}*{Scaling solution,} \\
& &  \multirow{2}*{behaving as matter,}\\
& & \multirow{2}*{no accelerating expansion}\\
& & \\
\hline
\end{tabular}
\caption{
Possible critical points allowed by Eq.~\eqref{eqs47}.
}
\label{tab:kessence varibles}
\end{center}
\end{table}

We can conclude from  Eq.~(\ref{eqs47}) that there are only three possible critical points for $\Omega_{\phi}$ allowed in these models, namely $\Omega_{\phi}=0$, $\Omega_{\phi}=1$ and the case $\gamma_{\phi}= w+1$ (i.e.~$w_{\phi}=w$) where the actual value of $\Omega_{\phi}$ is determined by solving the remaining equations in the dynamical system (see Tab.~\ref{tab:kessence varibles}).
The cases $\Omega_{\phi}=0$ and $\Omega_{\phi}=1$ are completely opposite solutions, respectively corresponding to universes completely dominated by the scalar field and by the matter fluid. In the remaining case, we could in principle obtain any value of $\Omega_{\phi}$ under the constraint $0<\Omega_{\phi}<1$. However, for this scaling solution, the EoS of dark energy $w_{\phi}$ coincides with the EoS of the matter fluid $w$, so there cannot be any accelerated expansion. Since the observations suggest that we are living in an accelerated expanding universe with $\Omega_{\phi}\sim0.7$, none of these three critical points correspond to the present observed universe, and thus the cosmic coincidence problem cannot be solved with such models. As we will see in Sec.~\ref{chap:IDE}, this scenario can instead be attained in models of dark energy interacting with the matter sector.

\subsubsection{Models with Lagrangian $\mathcal{L} = F(X)-V(\phi)$}

Another interesting subclass of $k$-essence models is represented by the Lagrangian
\begin{equation}
	P=F(X)-V(\phi) \,,
	\label{098}
\end{equation}
where a non-canonical kinetic term appears together with a standard self-interacting potential for the scalar field. As shown in \citet{DeSantiago:2012nk} (see also \citet{De-Santiago:2014apa}), who also considered bouncing solutions\footnote{Solutions where the universe passes from expansion to contraction (or the other way around).}, a detailed dynamical systems analysis can be performed introducing the new variables
\begin{equation}
	x = \frac{\kappa}{\sqrt{3}H}\sqrt{2X\,F_X-F} \quad\mbox{and}\quad y = \frac{\kappa \sqrt{V}}{\sqrt{3}H} \,,
	\label{eq:Santiagovars}
\end{equation}
where $F_X$ denotes the derivative of $F$ with respect to $X$. Note that for the canonical case $F=X$ these reduce to the EN variables (\ref{def:ENvars}). Besides late time accelerated solutions, \citet{DeSantiago:2012nk} found also that scaling solutions are possible for some special functions $F$ and $V$ of the model (\ref{098}), even if this cannot be written in the form (\ref{096}).
Within this $k$-essence class \citet{Graham:2014hva} considered a model with a varying fine structure constant $\alpha$. Subcases of a ghost condensate and dilatonic ghost condensate model with the Lagrangian $P=-X+\frac{e^{\lambda\phi}X^2}{M^4}-V(\phi)$, were considered in detail, before more general models were analysed. The dynamics reveal that models of this kind are generally unacceptable as the variation of $\alpha$ is found to be too fast to be compatible with terrestrial constraints.

The variables \eqref{eq:Santiagovars} are well defined only for $2XF_X-F>0$. In order to encompass all possible models within the $k$-essence class defined by Eq.~\eqref{098}, we can redefine and extend them as \citep{DeSantiago:2012nk}
\begin{align}
  \label{eqs38}
  x=\frac{\kappa}{\sqrt{3}H} \sqrt{(2XF_X-F)\varsigma}\,, \quad
  y = \frac{\kappa \sqrt{V}}{\sqrt{3}H}\,, \quad
  \sigma = -\frac{1}{\kappa \sqrt{3(2XF_X-F)\varsigma}} \frac{d \log V}{dt} \,,
\end{align}
where $\varsigma=1$ for $2XF_X-F>0$ and $\varsigma=-1$ for $2XF_X-F<0$.

Using the variables\eqref{eqs38}, from the cosmological equations sourced by the scalar field \eqref{098} plus a matter fluid, we can obtain the following dynamical system
\begin{align}
 x'&= \frac{3}{2}[\sigma \varsigma y^2-x(w_k+1)]+\frac{3}{2}x[(1+w)(1-y^2)+(w_k-w)\varsigma x^2]\,,
  \label{eqs39} \\
 y'&= -\frac{3}{2}y \left[\sigma x- (1+w)(1-y^2)+(w_k-w)\varsigma x^2 \right]\,,
  \label{eqs40} \\
  \sigma' &= -3 \sigma^2 x(\Gamma-1)+\frac{3\sigma[2\Xi(w_k+1)+w_k-1]}{2(2\Xi+1)(w_k+1)}(w_k+1-\frac{\sigma y^2}{\varsigma x})\,,
  \label{eqs41}
\end{align}
where
\begin{align}
  \label{eqs42}
  w_k =\frac{F}{2XF_X-F} = \gamma_k-1 \,, \quad
  w_{\phi} = \frac{p_{\phi}}{\rho_{\phi}}=\frac{w_k\varsigma x^2-y^2}{\varsigma x^2+y^2} \,,\quad
  \Omega_{\phi} = \varsigma x^2+y^2\,, \quad \Xi=\frac{XF_{XX}}{F_X} \,.
\end{align}
The dynamical system (\ref{eqs39})--(\ref{eqs41}) can be rewritten using the variables $(\Omega_{\phi}, \gamma_{\phi}, \sigma)$, directly related to observable quantities.
The results of this change of variables is
\begin{align}
 \Omega_{\phi}'&= 3(w+1-\gamma_{\phi})\Omega_{\phi}(1-\Omega_{\phi}) \,,
  \label{eqs44} \\
 \gamma_{\phi}'&= 3\left[\varsigma\gamma_{\phi}\left(\gamma_{\phi}-1-\frac{1}{2\Xi+1}\right)+2\sigma\sqrt{\frac{\varsigma\Omega_{\phi}\gamma_{\phi}}{\gamma_k}}\frac{(\gamma_k-\gamma_{\phi})(\Xi+1)}{(2\Xi+1)\gamma_k}\right] \,,
  \label{eqs45} \\
  \sigma'&= -3\sigma\left\{\sigma(\Gamma-1)\sqrt{\frac{\varsigma\Omega_{\phi}\gamma_{\phi}}{\gamma_k}}+\left(\frac{1}{2\Xi+1}-\frac{\gamma_k}{2}\right)\left[\sigma\sqrt{\frac{\varsigma\Omega_{\phi}\gamma_{\phi}}{\gamma_k}}\left(\frac{1}{\gamma_k}-\frac{1}{\gamma_{\phi}}\right)+1\right]\right\} \,.
  \label{eqs46}
\end{align}
where we recall that $\gamma_\phi = 1 + w_\phi$. These equations are quite complicated, however they reduce to the canonical quintessence case discussed in Sec.~\ref{sec:quintessence} and to the phantom dark energy case discussed in Sec.~\ref{sec:phantom} for $F(X)=\varsigma X$. If we chose the potential to vanish ($V(\phi)=0$) then $P=F(X)$, which is the so-called purely kinetic united model \citep{Scherrer:2004PRL,De-Santiago:2014apa}. For the simple case $P=F(X)=A_1\sqrt{X}-A_2 X^{\beta}$ with $A_1$, $A_2$ being constants, and $\beta\neq -1$. \citep{Scherrer:2004PRL,Chimento:2004}, the complicated dynamical system (\ref{eqs44})--(\ref{eqs46}) simplifies to
\begin{align}
  \label{eqsadd16}
 \Omega_{\phi}'&= 3(w+1-\gamma_{\phi})\Omega_{\phi}(1-\Omega_{\phi}) \,, \\
  \label{eqsadd17}
  \gamma_{\phi}'&= 3\gamma_{\phi}\left(\frac{2\beta-1}{2\beta} \gamma_{\phi}-1\right) \,.
\end{align}
These equations can be solved analytically as Eq.~\eqref{eqsadd17} decouples from Eq.~\eqref{eqsadd16}. One can thus obtain the following exact solution for $\gamma_{\phi}$ and $\Omega_{\phi}$
\begin{align}
  \gamma_{\phi}(\eta) = \frac{1}{c_5 e^{3\eta}+1/\alpha} \,, \quad
  \Omega_{\phi}(\eta) = \frac{1}{c_6(c_5\alpha e^{3\eta}+1)^{-\alpha}e^{3(\alpha-w-1)\eta}+1} \,,
\label{eqsadd18}
\end{align}
where $c_5$ and $c_6$ are constants, and $\alpha=2\beta/(2\beta-1)$. At very early times, i.e.~when $\eta\rightarrow -\infty$, we get $\gamma_{\phi}\approx \alpha=2\beta/(2\beta-1)$. In this case the scalar field will mimic the evolution of matter with zero pressure ($\gamma_\phi = 1$) whenever $\beta\gg 1$. At late times instead, i.e.~when $\eta\rightarrow +\infty$, $\gamma_{\phi}\approx 0$ so the scalar field behaves as the cosmological constant. Moreover, as long as $\alpha>0$, at late times we also have $\Omega_{\phi}\approx 1$. In this scenario the purely kinetic united model, with the Lagrangian without a potential term, provides a de Sitter expansion at late times and can account for dark matter at early times. Finally \cite{Li:2016grl} considered the case of $F(X) = X^{\alpha}$ and used a dynamical systems approach to derive the general conditions for the existence of stable scaling solutions, for both power-law and exponential potentials when the expansion is dominated by a background barotropic fluid.

\subsection{Higher-order scalar fields beyond $k$-essence: Galileons and Horndeski theories}

We finally turn our discussion from $k$-essence models to other higher-order scalar field models.
Some of them are motivated by high energy physics, some by phenomenological insights and some even by theoretical issues such as the avoidance of ghosts in the higher order scalar field terms.

\citet{Gao:2009me} considered the direct insertion of a kinetic term in the energy-momentum tensor of the scalar field, completely bypassing the Lagrangian set up. They used this model for a unified approach to both dark matter and dark energy with the scalar field behaving as matter (dust) at early times.

\citet{Leon:2012mt} delivered a detailed dynamical systems analysis for generalised {\it Galileon cosmologies} where the higher order terms of the scalar field satisfy the Galilean symmetry $\phi\mapsto \phi+c$ and $\partial\phi\mapsto\partial\phi+b_\mu$ with $c$ and $b_\mu$ constant. They showed that in this model the higher order contributions do not influence the late time cosmological dynamics where the evolution is governed by an effective canonical scalar field. \citet{DeArcia:2015ztd} considered the Galileon scalar field when no additional matter fields are present, and found that the asymptotic dynamics in the vacuum strongly depart from standard quintessence.

In Appendix C of their paper, \citet{Zumalac:2013} presented a dynamical system analysis, yielding all the fixed points of a specific type of DBI Galileon theory analysed in the Einstein frame. Subsequently \cite{Sakstein:2015jca} studied the phase space of the same model with particular emphasis on the role of the disformal transformation.  They found that these theories are in strong tension with dark energy observations, giving an EoS of the dark energy equal to $-3$.

\citet{horndeski1974second} provided the most general scalar-tensor theory which guarantees second order equations of motion (see also \citet{Deffayet:2011gz}). More recently this was extended to include models beyond Horndeski by \citet{Gleyzes:2014dya} and \citet{Langlois:2015cwa} which could consistently accommodate terms of higher order in derivatives in the equations of motion. Working with Horndeski's original action, it provided the basic ingredients required for a novel self tuning mechanism to alleviate the famous cosmological constant problem as shown by \citet{Charmousis:2011bf}. The self tuning model consists of four geometric terms in the action, with each term containing a free potential function of the scalar field; the four together being labelled as the Fab-Four. \citet{Copeland:2012qf} developed a dynamical systems approach to begin the task of deriving the cosmology associated with the Fab-Four Lagrangian. Performing a phase plane analysis of the system a number of fixed points for the system were obtained, with new solutions emerging from the trade-off between the various potentials. As well as obtaining inflationary solutions they also found conventional radiation/matter-like solutions, but in regimes where the energy density is dominated by a cosmological constant, and where we do not have any explicit forms of radiation or matter. In a similar vein a class of Horndeski models has been explored via Dynamical Systems and shown to lead to a spatially flat de Sitter vacuum fixed by the theory itself by \cite{Martin-Moruno:2015bda, Martin-Moruno:2015kaa}. Finally \citet{Gomes:2013ema,Gomes:2015dhl} studied scaling solutions for general higher order Lagrangians of the {\it Horndeski} type, where despite introducing higher order terms for the scalar field the equations of motion remain of the second order (no ghosts). They found matter dominated solutions followed by an accelerating scaling solution, sequence which could not be obtained with simpler scalar fields. This particular late time evolution can in principle solve the cosmic coincidence problem and provide a viable cosmic history for the observed universe.

To summarise higher-order scalar fields beyond the $k$-essence paradigm are interesting since they can provide a dynamical evolution mixing features of different canonical and non-canonical scalar fields. For example late time tracking, scaling, quintessence and phantom behaviours are possible with viable attempts at solving the fine tuning and cosmic coincidence problems. Unfortunately the rather complicated equations of motion arising in these theories prevent simple applications of dynamical systems theories and rather involved analysis are required, often dealing with non-compact high-dimensional autonomous systems.

We conclude this section by emphasising the importance of linking the models we are discussing with observations. The recent remarkable result announced by the LIGO/VIRGO collaboration of the detection of gravitational waves from a neutron star-neutron star merger (GW170817) \citep{TheLIGOScientific:2017qsa} and the simultaneous measurement of an optical counterpart (gamma-ray burst GRB 170817A) \citep{GBM:2017lvd} has had a dramatic impact on a wide class of models including the Horndeski models described in this section. This is because the close arrival time of the gravitational and electromagnetic waves limits the difference in speed of photons and gravitons to be less than about one part in $10^{15}$ which is considerably smaller than the differences predicted in a class of Horndeski models (see \citet{Lombriser:2015sxa} and \citet{Lombriser:2016yzn} for earlier work describing the potential impact on Horndeski models of this form of multi-messenger astronomy). This class of models therefore is ruled out as being cosmologically interesting; for details see \cite{Creminelli:2017sry,Baker:2017hug,Wang:2017rpx,Sakstein:2017xjx,Ezquiaga:2017ekz}. Nevertheless these results do not rule out the wide class of beyond Horndeski models (\cite{Gleyzes:2014dya,Langlois:2015cwa}) which have more free parameters associated with the extra terms present.

%% file: chapters/06_coupled/coupled.tex
\section{Interacting dark energy models}
\label{chap:IDE}

This section is devoted to the study of dark energy interacting with dark matter through a non-gravitational coupling. The presentation will of course be focused on the dynamical properties of these models, but the reader interested in their phenomenology and observational implications can refer to \citet{Copeland:2006wr,Timothy:2013iia,Bolotin:2013jpa,Valiviita:2015dfa,Wang:2016lxa}.

\subsection{Introducing a dark sector coupling: accelerating scaling solutions}
\label{sub:introducing_a_dark_sector_coupling}

As we mentioned at the end of Sec.~\ref{sec:FRW}, the different components that source the right-hand side of the Einstein field equations can in principle interact with one another.
In particular this is the case for dark energy (DE) which is allowed to interact with (dark) matter (DM) as a particular example of Eq.~(\ref{032}), namely
\begin{align}
  \nabla^\mu T_{\mu\nu}^{\rm (m)} = -Q_\nu \,, \quad\mbox{and}\quad 
  \nabla^\mu T_{\mu\nu}^{\rm (de)} = Q_\nu \,,
  \label{069}
\end{align}
where $T_{\mu\nu}^{\rm (m)}$ is the matter energy-momentum tensor, $T_{\mu\nu}^{\rm (de)}$ is the DE energy-momentum tensor and $Q_\nu$ is the {\it interaction vector} which determines the coupling between the dark energy and matter fluids.

In this section we will assume $T_{\mu\nu}^{\rm (m)}$ describes both DM and baryonic matter in a unifying manner, in line with the usual approach in the dynamical system literature. This choice can be justified considering that at cosmological scales DM constitutes the majority of the dust component sourcing the Einstein equations, as reported by observations \citep{Ade:2015xua}. Thus the contribution of baryonic matter can be ignored in a first approximation. Furthermore the observational constraints on a possible coupling between dark energy and dark matter are not so restrictive at cosmological scales (for some recent works see e.g.~\citet{Wang:2016lxa}), although stringent bounds can be derived from Solar System tests \citep{Will:2014xja} constraining an eventual fifth force arising from a dark sector interaction \citep{Carroll:1998zi}. Nevertheless several ways of avoiding these stringent local constraints have been proposed, generally involving some kind of screening mechanism \citep{Khoury:2003aq,Khoury:2003rn,Gubser:2004uf,Hinterbichler:2010es} where the fifth force hides its presence at Solar System distances (for a review of screening mechanisms see \citet{Clifton:2011jh,Joyce:2014kja}). The result is that for well designed models, DE can be allowed to interact with the whole non-relativistic matter sector on cosmological scales, as long as both local and cosmological observations are satisfied.

In what follows we will describe the matter sector, including both DM and baryonic matter, as a simple perfect fluid with $p=w\rho$ and we will focus on the value $w=w_{\rm m}=0$ to describe the late-time cosmic evolution (we will thus also refer to matter and DM interchangeably). We will not however consider a possible interaction between DE and radiation  and the weak coupling between baryons and radiation is neglected as usual. This will allow us to cover the majority of cosmological applications where dark energy interacts with the matter sector.

From a dynamical system perspective, the main consequence of introducing a coupling between DE and DM is the appearance of \textit{accelerating scaling solution}, namely scaling solutions (cf.~Sec.~\ref{sec:exp_potential}) where $w_{\rm eff} < -1/3$. This type of solution is phenomenologically important since it can be employed to solve the cosmic coincidence problem (see Sec.~\ref{sec:cosmoconstpbm}). In fact if the late time cosmological attractor is characterised by $\Omega_{\rm m} \simeq 0.3$, $\Omega_{\rm de} \simeq 0.7$ and $w_{\rm eff} \simeq - 0.7$, which can indeed be realised by an accelerating scaling solution, then the observed state of the universe is actually its final state, and the cosmic coincidence problem would be solved, or at least strongly alleviated.

In the following sections we will discuss several examples of accelerating scaling solutions found in different interacting DE models. However, since this is a general feature associated to the dark sector coupling, we will briefly demonstrate here why these particular solutions cannot be obtained in non-interacting models and how a coupling between DM and DE can produce them. The proof follows the argument mentioned in Sec.~\ref{sec:k-essence}. Consider non-interacting cosmological equations sourced by matter and DE, specifically
\begin{align}
	3H^2 = \kappa^2 (\rho_{\rm m} + \rho_{\rm de}) \,, \quad \dot\rho_{\rm m} + 3H \rho_{\rm m} (w_{\rm m}+1) = 0 \,, \quad \dot\rho_{\rm de} + 3H \rho_{\rm de} (w_{\rm de}+1) = 0 \,.
	\label{eq:p01}
\end{align}
From these equations one can easily derive a dynamical equation for $\Omega_{\rm de} = \kappa^2 \rho_{\rm de}/(3H^2)$, the relative energy density of DE, as
\begin{align}
	\Omega_{\rm de}' = 3 \Omega_{\rm de} (1 -\Omega_{\rm de}) (w_{\rm m} - w_{\rm de}) \,,
\end{align}
where as usual a prime denotes differentiation with respect to $\eta = \log a$.
It is clear from this equation that critical points are allowed only for the values $\Omega_{\rm de} = 0$, $\Omega_{\rm de} = 1$ and if $w_{\rm de} = w_{\rm m}$.
The first two are nothing but the DM and DE dominated solutions, respectively.
The third one is indeed a scaling solution where the EoS of DE must be constant and equal to the matter EoS.
The actual value of $\Omega_{\rm de}$, constrained within the range $0 < \Omega_{\rm de} < 1$, will then be determined by the remaining dynamical equations.
If there is no interaction between DM and DE, then the only possible scaling solution will never describe an accelerating universe since it will be always characterised by a universe with $w_{\rm eff} \equiv w_{\rm de}\Omega_{\rm de} + w_{\rm m} \Omega_{\rm m} = w_{\rm m}$, and thus always be decelerating.
This proof is valid for all non-interacting DE models for which Eqs.~\eqref{eq:p01} hold, no matter how complicated the expressions for $\rho_{\rm de}$ and $w_{\rm de}$ might be.
This includes higher-order scalar field models such as $k$-essence (see Sec.~\ref{sec:k-essence}) and even some modified gravity models (cf.~Sec.~\ref{chap:modifiedgrav}).

The situation changes if a coupling between DM and DE is added.
In this case Eqs.~\eqref{eq:p01} generalises to
\begin{align}
	3H^2 = \kappa^2 (\rho_{\rm m} + \rho_{\rm de}) \,, \quad \dot\rho_{\rm m} + 3H \rho_{\rm m} (w_{\rm m}+1) = -Q \,, \quad \dot\rho_{\rm de} + 3H \rho_{\rm de} (w_{\rm de}+1) = Q \,,
	\label{eq:p03}
\end{align}
where the last two equations follow directly from the 00-components of Eqs.~\eqref{069} in a FLRW universe, with $Q$ being the 0-component of the interacting vector $Q_\mu$.
The dynamical equation for $\Omega_{\rm de}$ now reads
\begin{align}
	\Omega_{\rm de}' = 3 \Omega_{\rm de} (1 -\Omega_{\rm de}) (w_{\rm m} - w_{\rm de}) + q \,,
\end{align}
where we have defined the dimensionless quantity $q$ (not to be confused with the deceleration parameter)
\begin{align}
	q = \frac{\kappa^2 Q}{3 H^3} \,.
\end{align}
It is now clear that, no matter what the actual dynamical expression of $q$ is, a scaling solution (for which $\Omega_{\rm de}$ is neither one nor zero) can be found whenever
\begin{align}
	w_{\rm de} = w_{\rm m} + \frac{q}{\Omega_{\rm de} (1 -\Omega_{\rm de})} \,.
\end{align}
The effective EoS of the universe at this critical point will read
\begin{align}
	w_{\rm eff} = w_{\rm m} + \frac{q}{1-\Omega_{\rm de}} \,,
\end{align}
which, provided a sufficiently negative $q$, will indeed describe an accelerating universe. A negative $q$ corresponds to a transfer of energy from dark matter to dark energy. This is expected as dark energy dominates cosmology at late times. For example for $w_{\rm m}=0$ and $\Omega_{\rm de} = 0.7$ one needs $q < -0.1$. The dark sector interaction thus allows for a solution of the cosmic coincidence problem through the appearance of accelerating scaling solutions.

Of course the details of the actual accelerating scaling solution will depend on the specific model of DE and coupling function $Q$.
In what follows we will present the dynamics of different interacting DE models, showing in detail how accelerating scaling solutions can be found.

\subsection{Coupled perfect fluids}
\label{sec:coupled_fluids}

In this section we assume that every matter component sourcing the Einstein field equations can be described by a perfect fluid energy-momentum tensor at cosmological distances.
The analysis will be similar to the one presented in Sec.~\ref{sec:LCDM_dynam}, but here an interaction between DE and the matter sector will be considered.

Given these premises, the Friedmann constraint can be written as
\begin{align}
	3 H^2 = \kappa^2 \left( \rho_{\rm m} + \rho_{\rm de} + \rho_{\rm r} \right) \,,
	\label{eq:Friedmann_IDE_fluids}
\end{align}
where $\rho_{\rm m}, \rho_{\rm de}$ and $\rho_{\rm r}$ are the energy densities of matter, DE and radiation, respectively. The radiation, matter and dark energy conservation equations become
\begin{align}
	\dot\rho_{\rm r} +  4H \rho_{\rm r} &= 0 \label{eq:baryons} \,,\\
	\dot\rho_{\rm m} + 3H \rho_{\rm m} &= -Q \,, \label{eq:DM_fluid_Q}\\
	\dot\rho_{\rm de} + 3H (\rho_{\rm de}+p_{\rm de}) &= Q \,, \label{eq:DE_fluid_Q}
\end{align}
where $Q$ denotes the energy exchange between DE and the matter sector, while radiation is separately conserved. As mentioned above, with respect to \eqref{069} one can define $Q$ as the time component of the interaction vector $Q_\mu$. An over-dot denotes as usual differentiation with respect to $t$, while the DE pressure is related to its energy density by a constant equation of state: $p_{\rm de} = w_{\rm de} \rho_{\rm de}$. Note that now the choice $w_{\rm de}=-1$ does not automatically imply $\dot\rho_{\rm de}=0$, as it happens when DE does not interact. The case $w_{\rm de}=-1$ with $Q\neq 0$ is sometimes referred to as the interacting vacuum energy scenario. The sign of $Q$ determines the direction of the energy transfer: if $Q>0$ the matter fluid is giving energy to DE, while if $Q<0$ it is DE which is releasing energy into the matter sector.

Eqs.~\eqref{eq:Friedmann_IDE_fluids}--\eqref{eq:DE_fluid_Q} can be recast into an autonomous system of equations defining the following dimensionless variables
\begin{align}
  x = \frac{\kappa^2 \rho_{\rm m}}{3 H^2} \,, \quad y = \frac{\kappa^2 \rho_{\rm de}}{3 H^2} \,, \quad \sigma = \frac{\kappa^2 \rho_{\rm r}}{3 H^2} \,,
  \label{eq:vars_IDE_fluids}
\end{align}
together with
\begin{align}
  q = \frac{\kappa^2 Q}{3 H^3} \,.
  \label{eq:def_q_fluids}
\end{align}
Note that throughout this section the variables \eqref{eq:vars_IDE_fluids} are similar to the ones defined in Eq.~\eqref{eq:defxyOmega} and used for the dynamical system analysis of $\Lambda$CDM in Sec.~\ref{sec:LCDM_dynam}.

However when an interaction between DE and DM is present, assuming the energy density $\rho_{\rm de}$ and $\rho_{\rm m}$ to be positive excludes possible viable solutions in the phase space; see e.g.~\citet{Quartin:2008px}. For this reason in what follows we will not assume $\rho_{\rm de}>0$ and $\rho_{\rm m}>0$, although ultimately the physically viable trajectories in the phase space should satisfy these conditions.
This positivity issue might be related to the ambiguity in the definition of the energy density variables for multiple interacting fluids in General Relativity \citep{Tamanini:2015iia}. Mathematically this implies that the physical phase space will no longer be compact, as is the case for $\Lambda$CDM. Although $\rho_{\rm m}$ and $\rho_{\rm de}$ will no longer be restricted to be positive, in what follows we will nevertheless assume that $\rho_{\rm r}\geq 0$ ($\sigma \geq 0$) since no interaction is considered for radiation.

The Friedmann equation \eqref{eq:Friedmann_IDE_fluids} written in terms of the variables \eqref{eq:vars_IDE_fluids} yields again the constraint
\begin{align}
	x + y + \sigma = 1 \,,\label{Friedmann-mixed}
\end{align}
which can be used to eliminate $\sigma$ in favour of $x$ and $y$ in the following equations.
Moreover since $\sigma \geq 0$, we have the constraint
\begin{align}
	x + y \leq 1 \,,
\end{align}
which reduces the physical phase space for this interacting model, without making it compact although we are no longer assuming $x\geq 0$ or $y\geq 0$, as we did in Sec.~\ref{sec:LCDM_dynam}. Using the variables \eqref{eq:vars_IDE_fluids}, and assuming $q$ is a function only of $x$ and $y$, the cosmological evolution is described by the following two-dimensional dynamical system
\begin{align}
  x' &= -x \left(-3 w_{\rm de} y+x+y-1 \right) - q \,, \label{eq:IDE_fluid_x} \\
  y' &= -y \left(-3 w_{\rm de} (y-1)+x+y-1 \right) + q \,, \label{eq:IDE_fluid_y}
\end{align}
where as usual a prime denotes differentiation with respect to $\eta = \log a$. The effective EoS becomes
\begin{align}
	w_{\rm eff} = \frac{p_{\rm tot}}{\rho_{\rm tot}} = \frac{1}{3} \left[1-x+(3 w_{\rm de}-1) y\right] \,,
\end{align}
where $p_{\rm tot} = w_{\rm de} \rho_{\rm de} + \rho_{\rm r}/3$ and $\rho_{\rm tot} = \rho_{\rm m} + \rho_{\rm de} + \rho_{\rm r}$.
One can immediately notice that $x=\sigma=0$ implies $w_{\rm eff} = w_{\rm de}$ (DE domination), $y=\sigma=0$ implies $w_{\rm eff} = 0$ (DM domination) and $x=y=0$ implies $w_{\rm eff} = 1/3$ (radiation domination).

\begin{table}
\begin{center}
\begin{tabular}{|c|c|c|}
\hline
$Q$ & $q$ & References \\
\hline
\multirow{2}*{$\eta\,H^\lambda\rho_{\rm m}^\alpha\rho_{\rm de}^\beta$} & \multirow{2}*{--} & \multirow{2}*{\citet{Nunes:2000ka}} \\
& & \\
\multirow{4}*{$H (\alpha \rho_{\rm m} + \beta \rho_{\rm de})$} & \multirow{4}*{$\alpha x + \beta y$} & \citet{Olivares:2007rt}\\
& & \citet{CalderaCabral:2008bx} \\
& & \citet{Quartin:2008px} \\
& & \citet{Li:2010ju}\\
\multirow{2}*{$\alpha \rho_{\rm m} + \beta \rho_{\rm de}$} & \multirow{2}*{--} & \multirow{2}*{\citet{CalderaCabral:2008bx}} \\
& & \\
$\beta H \rho_{\rm de}^\alpha \rho_{\rm m}^{1-\alpha}$ & $\beta x (y/x)^\alpha$
& \citet{Chen:2008ca} \\
\multirow{2}*{$3H (\lambda + \alpha \rho_{\rm m} + \beta \rho_{\rm de})$} & \multirow{2}*{--} & \multirow{2}*{\citet{Quercellini:2008vh}} \\
& & \\
$\lambda \rho_{\rm m}$ & -- & \citet{Li:2010ju} \\
\multirow{2}*{$3H \eta (\rho_{\rm m} + \rho_{\rm de})^\lambda \rho_{\rm m}^\alpha \rho_{\rm de}^\beta $} & \multirow{2}*{--} & \multirow{2}*{\citet{Arevalo:2011hh}} \\
& & \\
\multirow{1}*{$ \lambda \rho_{\rm m} \rho_{\rm de} / H $} & \multirow{1}*{$3 \lambda x y$} & \citet{Perez:2013zya} \\
& & \multirow{2}*{\citet{Szydlowski:2016epq}} \\
$ \lambda\, a^{-3} $ & -- & \multirow{2}*{\citet{Haba:2016swv}} \\
& & \\
\hline
\end{tabular}
\caption{Perfect fluid couplings considered in the literature of cosmological dynamical systems. In the cases where $q$, as defined in Eq.~(\ref{eq:def_q_fluids}), can be written in terms of the variables $x$ and $y$, we have provided the relation.
Here we have assumed $\kappa=1$, and $\alpha$, $\beta$, $\lambda$, $\eta$ are arbitrary constants of suitable dimensions.
}
\label{tab:coupled_fluid_models}
\end{center}
\end{table}

At this point one must make a choice regarding the dark sector coupling, i.e.~one must specify the time dependence of $Q$. Since $Q$ denotes the energy exchanged between DE and DM, from a physical perspective it is natural to assume that it depends only on other relevant dynamical quantities, namely $\rho_{\rm m}$, $\rho_{\rm de}$, $\rho_{\rm r}$ and $a$ (and possibly derivatives of these).
Several choices have been considered in the literature; see Tab.~\ref{tab:coupled_fluid_models}.
For example \citet{Nunes:2000ka} considered the very general coupling $Q \propto H^\lambda \rho_{\rm m}^\alpha \rho_{\rm de}^\beta$, with $\lambda$, $\alpha$ and $\beta$ constants satisfying $\lambda + 2 (\alpha + \beta -1) = 0$, showing that not only future DE attractors and scaling solutions can be obtained, but also that oscillatory behaviour is possible, where the universe periodically undergoes a period of accelerated expansion.

For a dynamical system analysis the most convenient choices of $Q$ are the ones leading to $q$ being function of $x$ and $y$ only, since in such cases the system of equations \eqref{eq:IDE_fluid_x}--\eqref{eq:IDE_fluid_y} remains autonomous. An interesting application of this case is the coupling $Q \propto \rho_{\rm m} \rho_{\rm de} / H$ which, as shown by \citet{Perez:2013zya}, can lead to a generalised Lokta-Volterra system and consequently to a dynamics similar to the predator-prey scenario. A similar scenario might be obtained by coupling radiation to the dark fluids, leading to endless alternate periods of acceleration and deceleration expansion \citep{Fay:2015sjt}. On the other hand, if $q$ cannot be written as a function of $x$ and $y$, then further variables must be introduced. An example is given by \citet{Szydlowski:2015rga} who studied an extension of the $\Lambda$CDM model with a varying cosmological constant $\Lambda(t) = \Lambda_{\rm bare} + \alpha^2 / t^2$ that can be interpreted as an interacting model between dark matter and dark energy with $Q=-d\Lambda(t)/dt$.

\begin{table}
\begin{center}
\begin{tabular}{|c|c|c|c|c|c|c|}
\hline  
Point & $x$ & $y$ & $w_{\rm eff}$ & Eigenvalues & Stability \\
\hline & & & & & \\
\multirow{4}*{$O$} & \multirow{4}*{0} & \multirow{4}*{0} & \multirow{4}*{1/3} & \multirow{4}*{$\{ 1, 1 - 3 w_{\rm de} + 3\beta  \}$} & Saddle if  \\ & & & & & $w_{\rm de} > \beta+1/3$ \\
& & & & & Unstable if  \\ & & & & & $w_{\rm de} < \beta+1/3$ \\ & & & & &\\ 
\multirow{2}*{$A$} & \multirow{2}*{1} & \multirow{2}*{0} & \multirow{2}*{0} & \multirow{2}*{$\{ -1, -3 w_{\rm de} + 3\beta \}$} & Stable if $w_{\rm de}>\beta$ \\ & & & & & Saddle if $w_{\rm de}< \beta$ \\ & & & & & \\
\multirow{6}*{$B$} & \multirow{6}*{$\beta/w_{\rm de}$} & \multirow{6}*{$1-\beta/w_{\rm de}$} & \multirow{6}*{$w_{\rm de}-\beta$} & \multirow{6}*{$\{ 3 w_{\rm de} - 3\beta, -1 + 3 w_{\rm de} - 3\beta \}$} & \multirow{1}*{Stable if $w_{\rm de}< \beta$} \\ & & & & & Saddle if \\ & & & & & $\beta<w_{\rm de}< \beta+1/3$ \\ & & & & & Unstable if  \\& & & & & $w_{\rm de}> \beta+1/3$ \\ & & & & & \\
\hline
\end{tabular}
\caption{Critical points of the system \eqref{eq:IDE_fluid_x}--\eqref{eq:IDE_fluid_y} with physical and stability properties for the case where $Q$ is given by Eq.~(\ref{eq:fluid_coupling_example}).}
\label{tab:IDE}
\end{center}
\end{table}

In the remaining part of this section we will consider a specific dark sector coupling chosen in order to show the interesting features that might arise in DE-DM interacting models, with a simple dynamical system investigation.
Let us then assume
\begin{align}
	Q = 3 \beta H \rho_{\rm de} \,, \quad{\rm implying}\quad q = 3 \beta y \,,
	\label{eq:fluid_coupling_example}
\end{align}
where we have introduced a dimensionless constant $\beta$. It is easy to check that in this case Eqs.~\eqref{eq:IDE_fluid_x}--\eqref{eq:IDE_fluid_y} presents the invariant submanifold $y=0$. This line divides the physical phase space in a sector in which the dark energy component has a positive ($y>0$) or negative ($y<0$) energy density (see the definition \eqref{eq:vars_IDE_fluids}). 

The phase space consists of only three finite critical points listed in Tab.~\ref{tab:IDE}. The possible finite stable critical points are Point~$A$, a matter dominated solution, or Point~$B$ which describes an accelerated expansion if $w_{\rm eff} = w_{\rm de} - \beta < -1/3$. Of these only Point~$A$ can be a global attractor, whereas Point~$B$ can be an attractor only for models in which $\rho_{\rm de}>0$. 

\begin{figure}
\centering
\includegraphics[width=.6\textwidth]{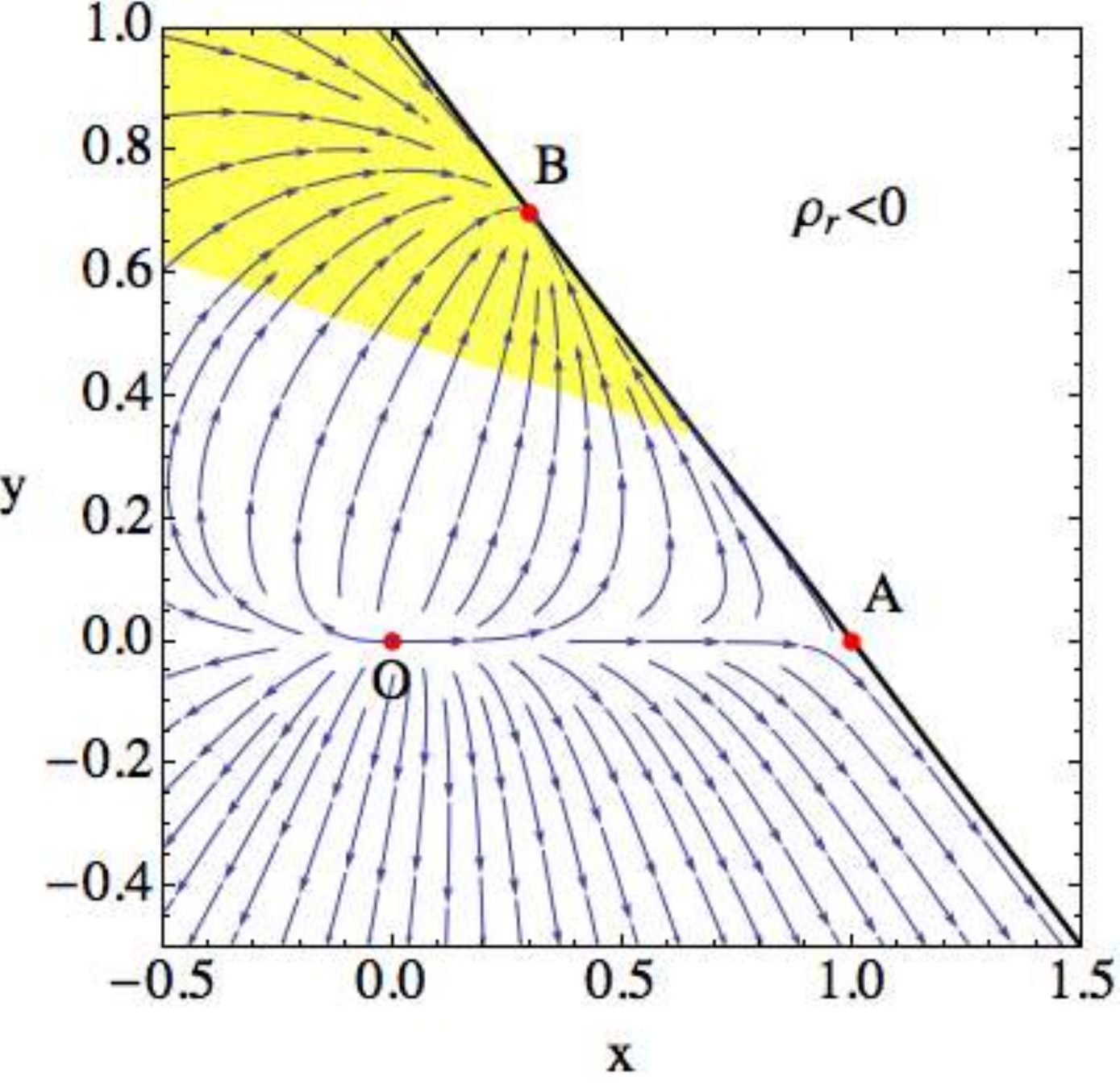}
\caption{Phase space portrait of the dynamical system (\ref{eq:IDE_fluid_x})--(\ref{eq:IDE_fluid_y}) and (\ref{eq:fluid_coupling_example}) with the values $w_{\rm de}=-1$ and $\beta=-0.3$. The phase space above the $x+y=1$ line is not viable since $\rho_{\rm r}<0$ there, while the yellow-shaded region identifies accelerated expansion. Point~$B$ represents an accelerating scaling solution.
}
\label{fig:coupled_fluids_phase_space}
\end{figure}

From Tab.~\ref{tab:IDE} we see that Points~$O$ and $A$ are the usual radiation and matter dominated critical points, appearing also in the standard $\Lambda$CDM dynamics. Point~$B$ represents a different solution compared to the $\Lambda$CDM case, where instead of the usual cosmological constant dominated solution with $x=0$, $y=1$ and $w_{\rm eff} = -1$, we find now a scaling solution depending on the interacting parameter $\beta$. Note that for $\beta=0$ and $w_{\rm de}=-1$ we recover the cosmological constant solution. If $w_{\rm de}-\beta<-1/3$ Point~$B$ is a stable attractor and describes an accelerated scaling solution which might be used to solve the cosmic coincidence problem. In this scenario in fact the final state of the universe corresponds to what we observe, namely comparable energy densities of DE and DM, thereby solving the cosmic coincidence problem. More specifically the orbit connecting Points~$O\rightarrow A\rightarrow B$ (or more correctly passing sufficiently close to them) will correctly describe the dynamics of the universe from a radiation dominated phase, through a matter dominated phase and finally to a scaling accelerated solution representing the present observational situation with $y = \Omega_{\rm de} \simeq 0.7$, $x = \Omega_{\rm dm} \simeq 0.3$ and $w_{\rm eff} \simeq -0.7$. This situation can be achieved for example choosing the parameters $w_{\rm de} = -1$ (vacuum DE) and $\beta = 0.3$ as shown in the phase space depicted in Fig.~\ref{fig:coupled_fluids_phase_space}. Note that for the physical heteroclinic orbit $O\rightarrow A\rightarrow B$ we always have $\rho_{\rm m}>0$ and $\rho_{\rm de}>0$, so the problem of negative energy densities is avoided in such a scenario.

We stress that the coupling term \eqref{eq:fluid_coupling_example} is not the only one that can be used to find solutions able to alleviate the cosmic coincidence problem. In fact other interactions between dark sector fluids yield accelerating scaling solutions, a simple example is $Q \propto H \rho_{\rm m}$ \citep{Chimento:2003iea}.

\subsection{Coupled quintessence}
\label{sec:coupled_quintessence}

In this section we will assume that DE is generally described by a scalar field, namely quintessence (cf.~Sec.~\ref{chap:scalarfields}). The matter sector will again be parametrised as a perfect fluid describing both baryonic and dark matter. In what follows we will neglect the contribution of radiation focusing only on the dynamics at late times. For the sake of generality we will however describe the matter sector as a general matter fluid with EoS parameter $w \geq 0$, which correctly recovers non-relativistic matter only in the limit $w=0$. The dynamics of quintessence with similar assumptions was first analysed by \citet{Amendola:1999qq,Amendola:1999er,Zimdahl:2001ar}.

On a FLRW background the Friedmann constraint becomes
\begin{align}
  3 H^2 = \kappa^2 \left( \rho + \rho_{\phi} \right) \,,
  \label{eq:Friedmann_IDE_quintessence}
\end{align}
while the matter conservation Eqs.~(\ref{069}) now read
\begin{align}
  \dot\rho + 3H \rho (w+1) &= -Q \,, \label{eq:matter_fluid_Q}\\
  \dot\rho_\phi + 3H (\rho_\phi+p_\phi) &= Q \,. \label{eq:scalar_field_fluid_Q}
\end{align}
Here $Q$ is again the time component of $Q_\mu$ (cf.~Eqs.~\eqref{069}), $\rho$ is the matter fluid energy density and the scalar field energy density $\rho_\phi$ and pressure $p_\phi$ are given by (\ref{def:energy_scalar_field}) and (\ref{def:pressure_scalar_field}), respectively.
Note that Eq.~(\ref{eq:scalar_field_fluid_Q}) is equivalent to the Klein-Gordon equation (\ref{eq:KGFRW}) with a non-vanishing source on the right hand side, namely
\begin{align}
  \ddot\phi +3H\dot\phi + V_{,\phi} = \frac{Q}{\dot\phi} \,.
  \label{eq:KG_eq_FRW_Q}
\end{align}
The cosmological equations governing the evolution of the universe are now Eqs.~\eqref{eq:Friedmann_IDE_quintessence}--\eqref{eq:scalar_field_fluid_Q}, or equivalently the Friedmann equation (\ref{eq:FriedmannPhi}), the acceleration equation (\ref{eq:accPhi}) and the Klein-Gordon equation (\ref{eq:KG_eq_FRW_Q}).

In order to recast these equations into a dynamical system we consider again the EN variables (\ref{def:ENvars}) and (\ref{def:lambda_dynamical}), in terms of which we obtain
\begin{align}
	x' &= \frac{1}{2} \left\{3 (1-w) x^3-3 x \left[w \left(y^2-1\right)+y^2+1\right]+\sqrt{6} \lambda  y^2\right\} +q \,,\label{eq:x_Q}\\
	y' &= -\frac{1}{2} y \left[3 (w-1) x^2+3 (w+1) \left(y^2-1\right)+\sqrt{6} \lambda  x\right] \,,\label{eq:y_Q}\\
	\lambda' &= -\sqrt{6} x\lambda ^2 (\Gamma-1) \,,\label{eq:lambda_Q}
\end{align}
where we have now defined
\begin{align}
	q = \frac{\kappa Q}{\sqrt{6}H^2\dot\phi} \,.
	\label{def:q}
\end{align}
Note that only Eq.~(\ref{eq:x_Q}) is modified by the interaction, while Eqs.~(\ref{eq:y_Q}) and (\ref{eq:lambda_Q}) are the same as Eqs.~(\ref{eq:y_can_scalar_field_gen_V}) and (\ref{eq:lambda}). As long as one chooses a quintessence model leading to a well defined $\Gamma(\lambda)$ relation (see Sec.~\ref{sec:other_potentials}), the only unknown quantity in Eqs.~(\ref{eq:x_Q})--(\ref{eq:lambda_Q}) is $q$ itself. If $q$ can be written in terms of the variables $x$, $y$ and $\lambda$, then Eqs.~(\ref{eq:x_Q})--(\ref{eq:lambda_Q}) will constitute an autonomous dynamical system which in general will be three dimensional unless an exponential potential is assumed for the scalar field, in which case the system becomes 2D. If instead $q$ cannot be expressed in terms of the other dynamical variables, then a further dynamical variable must be introduced increasing in this way the dimensionality of the system (see for example~\citet{Boehmer:2008av}).

\begin{table}
\begin{center}
\small{\begin{tabular}{|c|c|c|}
\hline
$Q$ & $q$ & References \\
\hline
\multirow{13}*{$\beta \rho \dot\phi$} & \multirow{13}*{$\frac{\sqrt{3}}{\sqrt{2}}\beta(1-x^2-y^2)$} & \\
	& & \citet{Amendola:1999qq,Amendola:1999er} \\
	& & \citet{Billyard:2000bh} \\
	& & \citet{Holden:1999hm} \\
	& & \citet{Amendola:2000uh} \\
	& & \citet{TocchiniValentini:2001ty} \\
	& & \citet{Gumjudpai:2005ry} \\
	& & \citet{Gonzalez:2006cj} \\
	& & \citet{Boehmer:2008av} \\
	& & \citet{Cicoli:2012tz} \\
	& & \citet{Tzanni:2014eja}\\
	& & \citet{Singh:2015rqa} \\
	& & \citet{Landim:2016gpz} \\
	& & \\
\multirow{3}*{$\beta H \rho$} & \multirow{3}*{$\frac{\beta}{2}(1-x^2-y^2)/x$} & \citet{Billyard:2000bh} \\
	& & \citet{Boehmer:2008av} \\
	& & \citet{Chen:2008pz} \\
\multirow{2}*{$\beta \rho \dot\phi \phi/a^4$} & \multirow{2}*{--} & \multirow{2}*{\citet{Liu:2005wga}} \\
	& & \\
\multirow{2}*{$\beta\dot\phi^2$} & \multirow{2}*{--} & \citet{Mimoso:2005bv} \\ & & \citet{Zhang:2012zz} \\
\multirow{2}*{$\beta H\dot\phi^2$} & \multirow{2}*{$\beta x$} & \multirow{2}*{\citet{Mimoso:2005bv}} \\ & & \\
\multirow{2}*{$\beta \rho$} & \multirow{2}*{--} & \multirow{2}*{\citet{Boehmer:2008av}} \\ & & \\
\multirow{2}*{$\beta \rho^2/H$} & \multirow{2}*{--} & \multirow{2}*{\citet{Chen:2008pz}} \\ & & \\
\multirow{2}*{$\beta \rho \dot\phi^2/H$} & \multirow{2}*{$3 \beta x (1-x^2-y^2)$} & \multirow{2}*{\citet{Chen:2008pz}} \\ & & \\
\multirow{2}*{$\alpha \rho_\phi^2+\beta \rho^2 +\gamma \rho \rho_\phi$} & \multirow{2}*{--} & \multirow{2}*{\citet{Boehmer:2009tk}} \\ & & \\
\multirow{4}*{$\beta \rho f(\phi) \dot\phi$} & \multirow{4}*{--} & \citet{LopezHonorez:2010ij} \\
 & & \citet{Morris:2013hua} \\ & & \citet{Hossain:2014xha} \\ & & \citet{Roy:2014hsa} \\
$\eta \,(\dot\rho_i+3 \beta H \rho_i)$ & \multirow{3}*{--} & \multirow{3}*{\citet{Wei:2010fz}} \\ with $\eta=-(1+\dot{H}/H^2)$ & & \\ and $\rho_i=\rho,\,\rho_\phi,\,\rho+\rho_\phi$ & & \\
\multirow{2}*{$\beta H (\rho_\phi- \rho)$} & \multirow{2}*{$\beta(x^2+y^2-1/2)/x$} & \multirow{2}*{\citet{Zhang:2014zfa}} \\ & & \\
\multirow{2}*{$\alpha \dot\rho_\phi + \beta \dot\rho$} & \multirow{2}*{--} & \multirow{2}*{\citet{Shahalam:2015sja}} \\ & & \\
\multirow{2}*{$\alpha (\rho + \rho_\phi) \dot\phi$} & \multirow{2}*{$\alpha / \sqrt{6}$} & \multirow{2}*{\citet{Landim:2016gpz}} \\ & & \\
\hline
\end{tabular}}
\caption{Quintessence to dark matter couplings considered in the dynamical system literature. When $q$, as defined in Eq.~(\ref{def:q}), can be written in terms of the variables $x$ and $y$, we have provided that relationship. Here we have assumed $\kappa=1$ and $\alpha$, $\beta$, $\gamma$ are arbitrary constants.}
\label{tab:coupled_quintessence_models}
\end{center}
\end{table}

Tab.~\ref{tab:coupled_quintessence_models} lists a number of approaches using several different couplings $Q$. In all of these cases the scalar field potential is taken to be exponential. Whenever possible the dimensionless quantity $q$ in terms of $x$ and $y$ has been provided. For the other cases such a relation cannot be found and a new variable must be introduced thereby increasing the dynamical system to a three dimensional one\footnote{In most situations such a variable is defined as $z=H_0/(H+H_0)$ with $H_0$ a constant (see e.g.~\citet{Boehmer:2008av,Boehmer:2009tk}).}.
For more complicated couplings or potentials, it might be more convenient and transparent to replace the EN variables with new ones (see e.g.~\citet{Leon:2008de,Fadragas:2014mra} and also \citet{vandeBruck:2016jgg} for the case of a quintessence field interacting with multiple matter fluids).

The coupling $Q=\beta\kappa \rho \dot\phi$ with $\beta$ a dimensionless constant is probably the most studied through the dynamical systems literature as the references in Tab.~\ref{tab:coupled_quintessence_models} confirm. This particular interaction comes from Brans-Dicke theory (once it has been rephrased into the Einstein frame), but it can also hold for more general non-minimally coupled gravitational theories (see e.g.~\citet{Amendola:1999er,Holden:1999hm}). It has also been used in models of multifield quintessence coupled to several matter fluids \citep{Amendola:2014kwa}, with similar results to those presented in Sec.~\ref{sec:multi_scalar_fields}, and it has even been generalised to introduce a coupling to DM inhomogeneities \citep{Marra:2015iwa}.

In what follows we will briefly review some of the features of this model as an example. This will show us how the interaction between dark energy and matter can affect the dynamics of the system. The reader interested in dynamical systems studies for this model can refer to the references provided in Tab.~\ref{tab:coupled_quintessence_models}.

For the coupling $Q=\beta\kappa \rho \dot\phi$ we can easily find from \eqref{def:q} that
\begin{align}
	q= \frac{\sqrt{3}}{\sqrt{2}}\beta \left(1-x^2-y^2\right) \,.
	\label{070}
\end{align}
This implies that for this model there is no need to introduce new variables in the dynamical system. For the sake of simplicity in our example we will also consider an exponential potential for which $\lambda$ is a constant in Eqs.~(\ref{eq:x_Q}) and (\ref{eq:y_Q}), while Eq.~(\ref{eq:lambda_Q}) is automatically satisfied. Eqs.~(\ref{eq:x_Q}) and (\ref{eq:y_Q}) constitute then an autonomous system and the phase space will be nothing but the upper half unit disk in the $(x,y)$-plane, i.e.~the same as the uncoupled case of Sec.~\ref{sec:exp_potential}. Note that the presence of the coupling breaks the $(x,\lambda)\mapsto (-x,-\lambda)$ symmetry. However in this particular case this can be restored considering also a reflection of the parameter $\beta$: $(x,\lambda,\beta)\mapsto (-x,-\lambda,-\beta)$. This implies that the phase space dynamics for opposite values of $\lambda$ will again be the same after a reflection over $x$ and opposite values of $\beta$ are considered. In other words, to analyse the whole dynamics of the system it suffices to consider positive values of $\lambda$, though both positive and negative values for $\beta$ must be taken into account\footnote{Of course we could consider only positive values of $\beta$ if both negative and positive values of $\lambda$ had been taken into account instead.}.

\begin{table}
\begin{center}
\begin{tabular}{|c|c|c|c|c|c|c|}
\hline
Point & $x$ & $y$ & Existence & $w_{\rm eff}$ & Accel. \\
\hline
\multirow{2}*{$O_\beta$} & \multirow{2}*{$\frac{\sqrt{3}}{\sqrt{2}}\frac{\beta}{1-w}$} & \multirow{2}*{0} & \multirow{2}*{$\beta^2<\frac{3}{2}(w-1)^2$} & \multirow{2}*{$w+\frac{2\beta^3}{3(1-w)}$} & \multirow{2}*{No} \\ & & & & & \\
\multirow{2}*{$A_\pm$} & \multirow{2}*{$\pm 1$} & \multirow{2}*{0} & \multirow{2}*{$\forall\;\lambda,w$} & \multirow{2}*{1} & \multirow{2}*{No} \\ & & & & & \\
\multirow{2}*{$B$} & \multirow{2}*{$\frac{\sqrt{3}}{\sqrt{2}}\frac{1+w}{\lambda-\beta}$} & \multirow{2}*{$\sqrt{\frac{3(1-w^2)+2\beta(\beta-\lambda)}{2(\beta-\lambda)^2}}$} & \multirow{2}*{Fig.~\ref{fig:coupled_quintessence_stability}} & \multirow{2}*{$\frac{w\lambda+\beta}{\lambda-\beta}$} & \multirow{2}*{Fig.~\ref{fig:coupled_quintessence_stability}} \\ & & & & & \\
\multirow{2}*{$C$} & \multirow{2}*{$\frac{\lambda}{\sqrt{6}}$} & \multirow{2}*{$\sqrt{1-\frac{\lambda^2}{6}}$} & \multirow{2}*{$\lambda^2<6$}  & \multirow{2}*{$\frac{\lambda^2}{3}-1$} & \multirow{2}*{$\lambda^2<2$} \\ & & & & & \\
\hline
\end{tabular}
\caption{The critical points of the dynamical system (\ref{eq:x_Q})--(\ref{eq:y_Q}) with the coupling (\ref{070}), including their existence and physical properties.}
\label{tab:coupled_CP_physics}
\end{center}
\end{table}

\begin{figure}[t]
\centering
\includegraphics[width=.7\columnwidth]{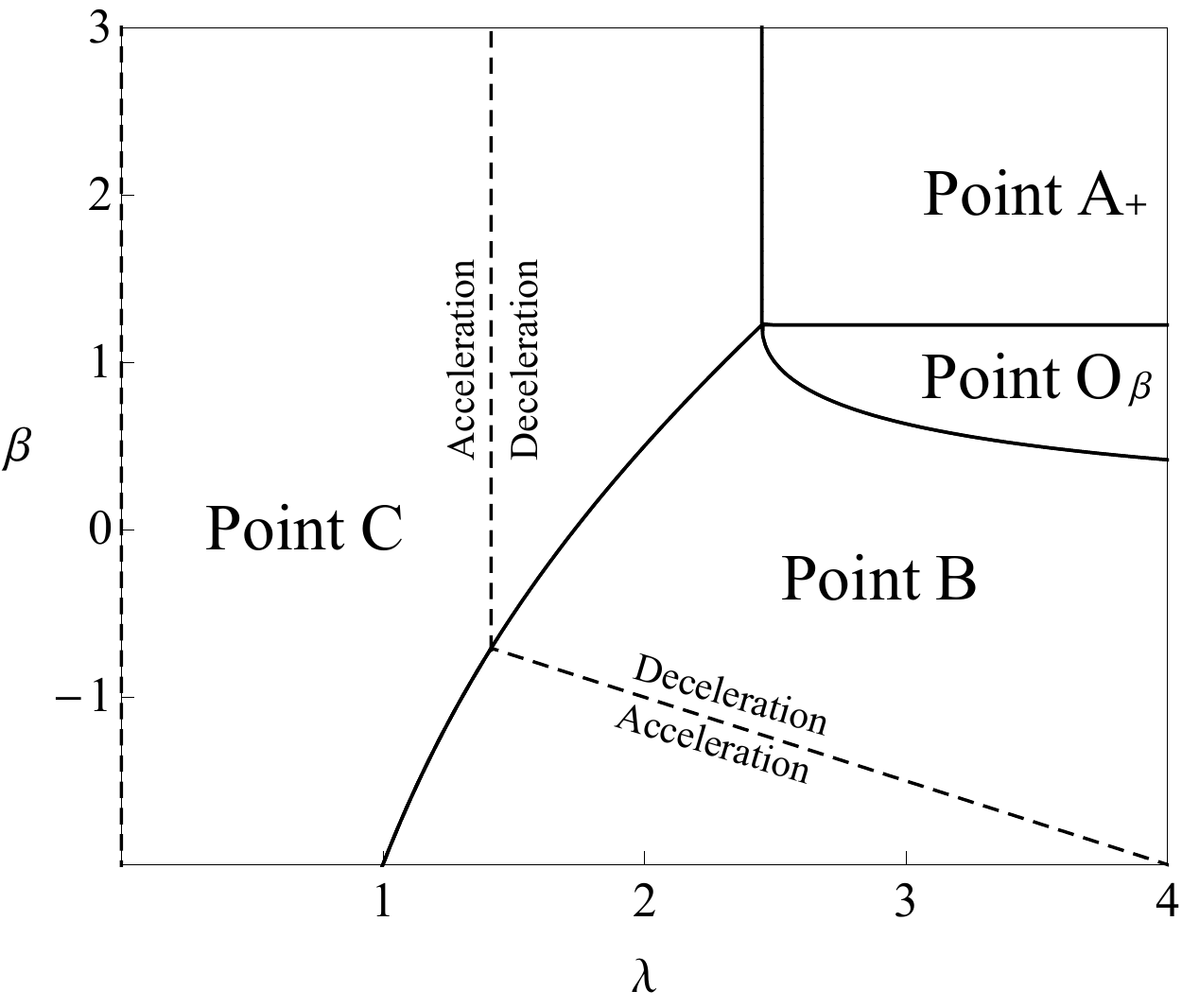}
\caption{Future attractors in the parameter space of the dynamical system (\ref{eq:x_Q})--(\ref{eq:y_Q}) with $w=0$ and the coupling (\ref{070}) (cf.~Fig.~1 by \citet{TocchiniValentini:2001ty}). The dashed line delimits the region where the universe undergoes accelerated expansion at the corresponding stable critical point.}
\label{fig:coupled_quintessence_stability}
\end{figure}

The critical points of the dynamical system (\ref{eq:x_Q})--(\ref{eq:y_Q}) with the coupling (\ref{070}) are listed in Tab.~\ref{tab:coupled_CP_physics}. Points~$A_\pm$ and $C$ are exactly in the same position and with the same phenomenological properties of their corresponding points in the uncoupled case (cf.~Tab.~\ref{tab:exp_CP_physics}). This does not come as a surprise because the coupling (\ref{070}) vanishes when the scalar field dominates, i.e.~when $x^2+y^2=1$. The remaining two points instead are changed by the interaction. The origin is no longer a critical point and Point~$O_\beta$ now lies on the $x$-axis. It is no longer a matter dominated point and its effective EoS will now depend on the parameter $\beta$, though no acceleration is possible since for this point~$w\leq w_{\rm eff}\leq 1$. The scaling solution described by Point~$B$ is also affected by the coupling. Its position and properties now depend on $\beta$, although we always have $\Omega_\phi\propto\Omega_{\rm m}$ which determines the nature of the scaling solution. Interestingly the effective EoS parameter at this point reads
\begin{align}
	w_{\rm eff} = \frac{w\lambda+\beta}{\lambda-\beta} \,,
\end{align}
which implies that Point~$B$ characterises an accelerating universe in some region of the $(\lambda,\beta)$ parameter space, as shown in Fig.~\ref{fig:coupled_quintessence_stability}.
We will not deliver a detailed stability analysis as we did in Sec.~\ref{sec:exp_potential} for each critical point of Tab.~\ref{tab:coupled_CP_physics}. However in Fig.~\ref{fig:coupled_quintessence_stability} a self contained explanation on the possible future attractors of the system depending on the values of $\lambda$ and $\beta$ is provided. Note that with a non-vanishing coupling  Points~$O_\beta$ and $A_\pm$ can also represent future attractors. These situations are however not interesting for dark energy phenomenology since no accelerated expansion can be obtained at these points.

\begin{figure}
\centering
\includegraphics[width=.7\columnwidth]{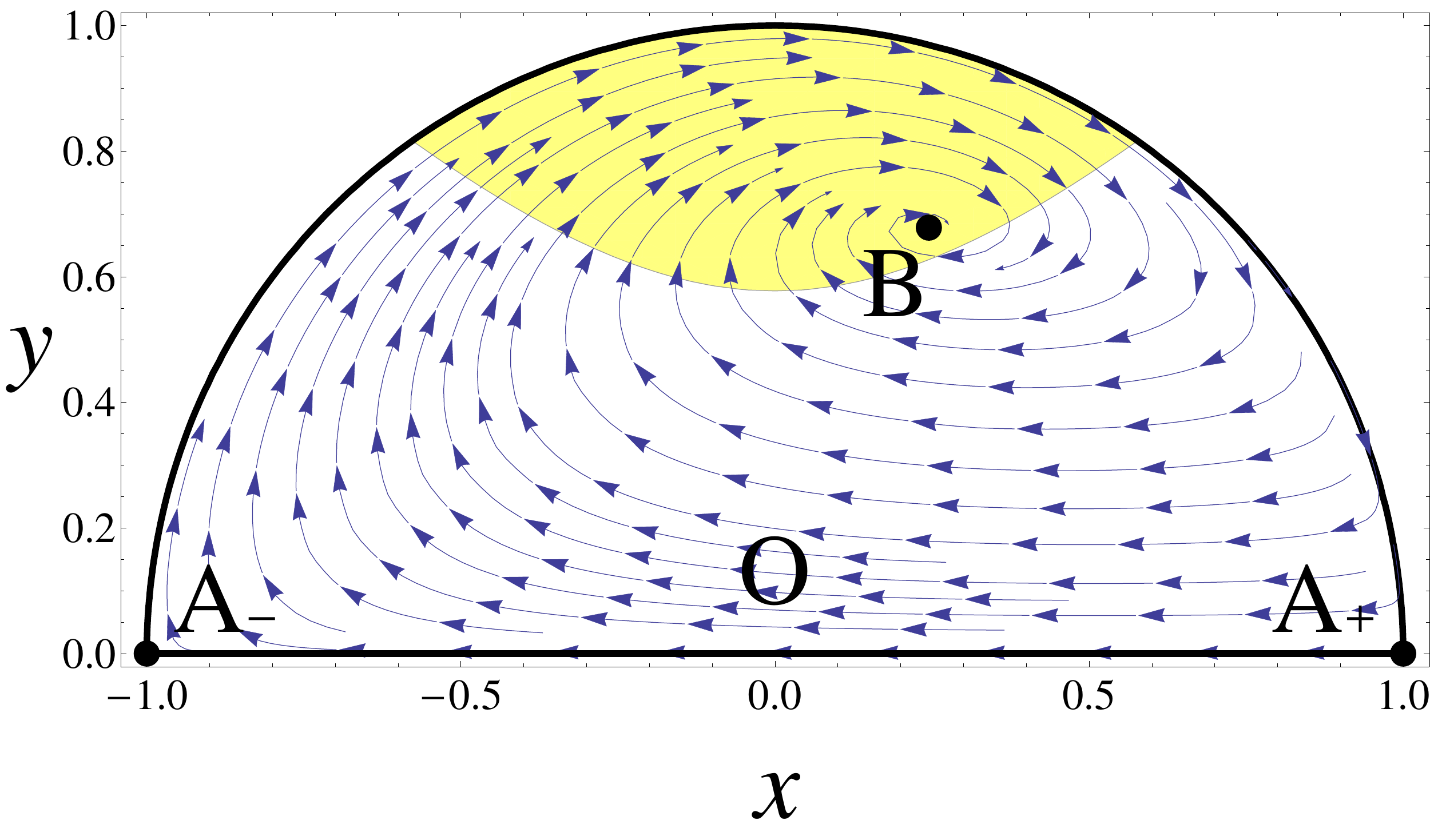}
\caption{Phase space portrait of the dynamical system (\ref{eq:x_Q})--(\ref{eq:y_Q}) and the coupling (\ref{070}) with the values $w=0$, $\lambda=3$, $\beta=-2$. The scaling solution of Point~$B$ represents in this case an accelerating universe since it lies inside the yellow/shaded region.}
\label{fig:coupled_quintessence_phase_space}
\end{figure}

The most interesting feature of this model, and usually of all the coupled quintessence models (cf.~Sec.~\ref{sub:introducing_a_dark_sector_coupling}), is that the scaling solution of Point~$B$ can now give an accelerating universe, as shown for example in the phase space portrait of Fig.~\ref{fig:coupled_quintessence_phase_space}. This could in principle solve the cosmic coincidence problem since an everlasting expanding solution with $\Omega_\phi\simeq 0.7$ can now be achieved. Unfortunately whenever Point~$B$ describes an accelerating solution Point~$O_\beta$ does not appear in the phase space. This implies the absence of a matter dominated saddle point, as shown in Fig.~\ref{fig:coupled_quintessence_phase_space}, and thus the expansion history of the universe cannot be fully described. Note also that unless $\beta$ is very small Point~$O_\beta$ does not describe a matter dominated solution at all.
Although it fails to account for a matter dominated epoch, this example provides useful insights to build a suitable dark energy model capable of solving the cosmic coincidence problem. In fact it is easy to realise a slightly more complicated model where a matter dominated solution or a scaling solution with $w_{\rm eff}=w$ appear; for example with a double exponential potential \citep{Tzanni:2014eja}. It is expected that more complicated models could yield both a tracking regime and an accelerating scaling solution. In such a situation both the fine tuning problem of initial conditions and the cosmic coincidence problem would be solved within a single theoretical dark energy model.

\subsection{Coupled non-canonical scalar fields}
\label{sec:coupled_noncanonical_scalars}

In this section we deal with non-canonical scalar field models where a coupling with the matter sector is present. In general all models discussed in Sec.~\ref{chap:noncanonicalscalarfields} can be generalised by adding a coupling in the dark sector in exactly the same way as with quintessence; cf.~Sec.~\ref{sec:coupled_quintessence}. In this section we will not present any detailed dynamical system analysis, but we will limit the discussion in order to only expose the main results of a few specific models and to give references to the literature.

We start by considering phantom scalar fields coupled to the matter sector; cf.~Sec.~\ref{sec:phantom}. Several authors have performed dynamical systems investigations for this kind of model, mostly using some of the couplings listed in Tab.~\ref{tab:coupled_quintessence_models} and mainly considering an exponential potential for the scalar phantom field \citep{Guo:2004xx,Gumjudpai:2005ry,Chen:2008ft,Wei:2010fz}. The coupling term most frequently analysed for this models is $Q \propto \rho \dot\phi$, whose results have been also summarised in the extended review on DE by \citet{Copeland:2006wr}, showing that scaling solutions cannot be stable as in the quintessence case. \citet{Leon:2009dt} studied a model where the mass of dark matter particles depends on the phantom field for both exponential and power-law potentials, and later \citet{Boehmer:2011tp} applied centre manifold techniques to the same model. \citet{Shahalam:2017fqt} investigated different derivative interactions of a phantom field with a dark matter component. They chose three different derivative couplings (linear and combination of linear), and their dynamical analysis showed that neither coupling can alleviate the coincidence problem. The general consensus arising from these investigations appears to be that the interaction between the phantom scalar field and the matter sector cannot solve the cosmic coincidence problem, as it happens in the case of quintessence (see Sec.~\ref{sec:coupled_quintessence}), because of the lack of accelerating scaling solutions constituting late time attractors.

The situation is different for more general $k$-essence models; cf.~Sec.~\ref{sec:k-essence}.
\citet{Amendola:2006qi} found the most general scalar field Lagrangian admitting scaling solutions with a matter coupling depending on $\phi$: $P(X,\phi) = Q^2(\phi) X f(X Q^2(\phi) e^{\lambda \kappa \phi})$ where $Q(\phi)$ is the coupling function and $f$ an arbitrary function. They also performed a dynamical systems analysis finding critical points corresponding to scaling, quintessence and phantom dominated solutions. Remarkably they showed that a dynamical sequence with one early time matter scaling solution and one late time dark energy scaling solution never occurs in such models, generalising in this way the results found in Sec.~\ref{sec:coupled_quintessence} for quintessence. Later a similar analysis was performed by \citet{Gomes:2013ema,Gomes:2015dhl} in order to find the most general Horndeski Lagrangian admitting scaling solution. Specific $k$-essence models have also been considered in the dynamical system literature. For example \citet{Das:2014yoa} studied the scalar fields defined by the Lagrangian $P = X^2 - V(\phi)$ with a coupling $Q \propto H \dot\phi^4.$

Tachyonic models (see Sec.~\ref{sec:tachyons}) have also been coupled to the matter sector. \citet{Gumjudpai:2005ry}, \citet{Farajollahi:2011jr} and again \citet{Farajollahi:2011ym} showed that scaling solutions can be found for the coupled tachyon field. \citet{Landim:2015poa} studied the dynamics of tachyonic DE coupled to DM including also the contribution of radiation and showed that the sequence of eras radiation domination $\rightarrow$ DM domination $\rightarrow$ DE domination can be attained. \citet{Mahata:2015lja} considered instead a coupled DBI scalar field, while \citet{Kaeonikhom:2012xr} investigated the dynamics of a model involving a generalised DBI field interacting with the matter sector (cf.~Sec.~\ref{sec:generalised-dbi}).

Some authors have also considered coupled theories of multiple non-canonical scalar fields $\phi_i$ with higher-order terms. \citet{Chiba:2014sda} studied scaling solutions for the general Lagrangian $P_i = X_i G(X_i e^{\lambda_i \kappa \phi_i})$ with interactions to the matter sector, showing that accelerated scaling solutions can be obtained and that the cosmic coincidence problem could be avoided. A complex scalar field coupled to the matter sector, for a quintessence, phantom and tachyonic Lagrangian, has been studied by \citet{Landim:2015uda}, who showed that the observed DM to DE transition can be obtained in all these models.

Finally the dynamics of an interacting DE model where DM undergoes microscopic diffusion into a non-canonical scalar field which generalises the cosmological constant, has been investigated by \citet{Alho:2014ola}, while \citet{Nozari:2016ilx} considered a scalar field interacting with the matter sector which also presents non-minimal gravitational coupling given by higher order kinetic terms.

\subsection{Scalar-fluid models}
\label{sec:scalar_fluid_models}

In almost all models considered so far in this section the coupling between DE and DM is introduced phenomenologically at the level of the field equations, irrespectively of whether DE is described as a perfect fluid (Sec.~\ref{sec:coupled_fluids}) or as a scalar field (Secs.~\ref{sec:coupled_quintessence} and \ref{sec:coupled_noncanonical_scalars}). Exceptions are provided by a few scalar field couplings which can be derived either by specifying the fundamental form of the matter Lagrangian, e.g.~in terms of spinor matter fields or non-relativistic point masses, or by conformal transformations in scalar-tensor theories, sometimes motivated by high energy phenomenology. An example is the coupling $Q \propto \rho \dot\phi$ which appears in the Einstein frame representation of scalar-tensor theories \citep{Amendola:1999er}. However apart from these few well justified couplings, every other interaction between DE and DM is phenomenological in nature, without a fundamental Lagrangian description behind it.

The only way to consistently define an interaction between DE and DM at the Lagrangian level is to describe both dark components with a sufficiently general Lagrangian description allowing for an arbitrary coupling between the DE and DM degrees of freedom. This can be achieved by so-called \textit{Scalar-Fluid theories}, where a scalar field DE is coupled to a matter sector defined by a fluid Lagrangian formalism. Scalar fluid theories were independently introduced by \citet{Pourtsidou:2013nha} and \citet{Boehmer:2015kta,Boehmer:2015sha} using two different formalisms for the matter fluid Lagrangian. We will focus on the second approach since a few dynamical systems investigations have been performed for it in the context of cosmology \citep{Boehmer:2015kta,Boehmer:2015sha,Tamanini:2016klr,Dutta:2017kch}, while there are none for the first approach.

The action for Scalar-Fluid theories can generally be written as \citep{Boehmer:2015kta,Boehmer:2015sha,Koivisto:2015qua}
\begin{align}
\mathcal{L}_{\rm tot} = \mathcal{L}_{\rm grav} +\mathcal{L}_{\rm m}+ \mathcal{L}_\phi+ \mathcal{L}_{\rm int} \,,
\label{SF:010}
\end{align}
where $\mathcal{L}_{\rm grav}$ is the gravitational Einstein-Hilbert Lagrangian, $\mathcal{L}_\phi$ is the scalar field Lagrangian, $\mathcal{L}_{\rm m}$ is the Lagrangian for the matter sector and $\mathcal{L}_{\rm int}$ defines the interaction between the scalar field and the matter fields.
In what follows we will assume that $\mathcal{L}_\phi$ describes a canonical scalar field (cf.~Eq.~\eqref{def:can_scalar_field_lagrangian}), although more general Lagrangians could in principle also be considered. The Lagrangian for the matter fluid can be written within Brown's formalism \citep{Brown:1992kc} as
\begin{align}
\mathcal{L}_{\rm m} = -\sqrt{-g}\,\rho(n,s) + J^\mu\left(\varphi_{,\mu}+s\theta_{,\mu}+\beta_A\alpha^A_{,\mu}\right) \,,
\label{SF:001}
\end{align}
where $g$ is the determinant of the metric tensor $g_{\mu\nu}$ and $\rho$ is the energy density of the matter fluid. The function $\rho(n,s)$ is prescribed as a function of $n$, the particle number density, and $s$, the entropy density per particle. $\varphi$, $\theta$ and $\beta_A$ are all Lagrange multipliers with $A$ taking the values $1,2,3$ and $\alpha_A$ are the Lagrangian coordinates of the fluid. The vector-density particle number flux $J^\mu$ is related to $n$ as
\begin{align}
J^\mu=\sqrt{-g}\,n\,u^\mu\,, \qquad |J|=\sqrt{-g_{\mu\nu}J^\mu J^\nu}\,, \qquad n=\frac{|J|}{\sqrt{-g}} \,,
\label{SF:056}
\end{align}
where $u^\mu$ is the fluid 4-velocity satisfying $u_\mu u^\mu=-1$.
The independent dynamical variables which have to be considered in the variation of the Lagrangian (\ref{SF:001}) are $g^{\mu\nu}$, $J^\mu$, $s$, $\varphi$, $\theta$, $\beta_A$ and $\alpha^A$.
The reader interested in more details regarding this relativistic fluid formalism can find them in \citet{Brown:1992kc}.

The important issue in Scalar-Fluid theories is that one can now define the Lagrangian coupling term between DM and DE in terms of the scalar field and the fluid's degrees of freedom defining the matter Lagrangian \eqref{SF:001}.
There are two very general interacting terms proposed so far in the literature:
\begin{align}
  \mathcal{L}_{\rm int} &= -\sqrt{-g}\, f_a(n,s,\phi) \,; &\mbox{(algebraic coupling)}  \\
  \mathcal{L}_{\rm int} &= f_d(n,s,\phi) J^\mu \partial_\mu\phi \,; &\mbox{(derivative coupling)} \label{eq:deriv-coupling}
\end{align}
where $f_a$ and $f_d$ are two arbitrary functions of $n$, $s$ and $\phi$. The first one does not present any derivative of the scalar field and it generalises well known couplings commonly used for screening light scalar degrees of freedom at Solar system scales \citep{Brax:2015fcf}. The second interacting terms constitutes instead a new approach  where one can couple the gradient of the scalar field to the velocity of the matter fluid. In a cosmological context these two different couplings lead to the following interacting terms in the (background) matter conservation equations (cf.~Eqs.~\eqref{eq:matter_fluid_Q}--\eqref{eq:KG_eq_FRW_Q}) \citep{Koivisto:2015qua}:
\begin{align}
  Q &= \frac{\partial f_a}{\partial\phi} \dot\phi \,; &\mbox{(algebraic coupling)} \label{SF:alg_Q} \\
  Q &= -3H n^2 \frac{\partial f_d}{\partial n} \dot\phi \,; &\mbox{(derivative coupling)} \label{SF:der_Q}
\end{align}
Recalling that $f_a$ is a function of $n$ and that for non-relativistic fluids $\rho \propto n$, one can realise how the coupling \eqref{SF:alg_Q} extends the scalar-tensor coupling where $Q \propto \rho \dot\phi$. The derivative interaction \eqref{SF:der_Q} represents instead a new way of coupling the scalar field to the matter sector. In particular the Hubble rate $H$, which is usually considered in the phenomenological choices of $Q$ for dimensional reason, now appears naturally from a local Lagrangian description.

From a cosmological point of view Scalar-Fluid theories are interesting since they can give rise to new and different phenomenology at both early and late times. For example for both algebraic and derivative couplings one can find accelerated scaling solutions, DM to DE transitions, phantom domination and transient periods of super-acceleration ($\dot{H}>0$), as well as early time inflationary solutions \citep{Boehmer:2015kta,Boehmer:2015sha,Tamanini:2016klr,Dutta:2017kch}. As an example the simplest derivative coupling, $f_d\propto 1/n$, where the gradient of the scalar field is linearly coupled to the matter fluid four-velocity, yields an early time matter saddle point attracting all early universe trajectories ($H\gg H_0$). This particular solution can be used to solve the problem of fine tuning of initial conditions, since all trajectories at early times will be forced to go through a matter dominated phase before ending in a DE dominated attractor \citep{Boehmer:2015sha}.

In general Scalar-Fluid theories represent a recently introduced new paradigm to couple a scalar field to the matter sector. Their properties and applications are still to be fully investigated and it might well be that future analysis will unveil some new interesting phenomenology for cosmology. For example see the recent works by \citet{Skordis:2015yra} and \citet{Pourtsidou:2016ico} where the possible observable implications of coupled quintessence models with pure momentum exchange interactions (the equivalent of (\ref{eq:deriv-coupling})) are discussed.

We conclude this section with a word of caution. The analysis we have presented here is a classical one, based on classical fields. Things can turn unpleasant when quantum corrections are taken into account as recently emphasised by \citet{DAmico:2016jbm} and \citet{Marsh:2016ynw}. Proposals that are based on the assumption that dark matter is made up of heavy particles with masses which are very sensitive to the value of dark energy turn out to be strongly constrained. The problem arises because quintessence-generated long-range forces and radiative stability of the quintessence potential together require that the dark matter and dark energy are completely decoupled. This problem can be alleviated if the dark energy fields are suitably screened in high matter density regions or if the dark energy and a fraction of dark matter are very light axions. These can then have significant mixings which are radiatively stable. Such axion like models can naturally occur in multi-axion realisations of monodromies, and as is shown by \citet{DAmico:2016jbm}, the mixings can lead to interesting signatures which are observable and are within current cosmological limits.

%% file: chapters/07_higherfields/higherfields.tex
\section{Non-scalar field models}
\label{sec:non-scalar_models}

In this section we will discuss dynamical systems techniques applied to dark energy models involving fields other than just the usual scalar field. The literature regarding dark energy models going beyond the scalar field paradigm is quite vast (for reviews see \citet{Copeland:2006wr}, \citet{Clifton:2011jh} and \citet{Joyce:2014kja}), though not as extensive as work dedicated to scalar fields. We will focus on models which have been analysed with dynamical systems techniques\footnote{It is worth stressing at this point that in making the choice of the models we present, we are, not by any means, advocating that these models are physically motivated or well justified. We are considering them only as examples of the  variety of ways in which dynamical system techniques can be applied to understand the dynamics of dark energy models. The same considerations apply to the models presented in the next chapter.}. These models tend to be motivated by either particle physics or phenomenological applications and consist of higher spin fields or alternative fluid models, for instance. Gravity is assumed to be described by general relativity, dark energy models based on alternative theories of gravity will be the subject of Sec.~\ref{chap:modifiedgrav}. The majority of this section will consist of brief discussions with references to the original literature and a more detailed dynamical systems analysis will be confined to one specific model.

As we mentioned in Sec.~\ref{chap:Cosmology} the particle physics approach to dark energy requires the introduction of new matter degrees of freedom needed to drive the accelerated expansion of the universe. These degrees of freedom are usually associated with new particles yet to be discovered. The simplest case is represented by a scalar particle, i.e.~a scalar field, and it is the case we have considered so far in Secs.~\ref{chap:scalarfields} - \ref{chap:IDE}. Of course, the major part of the literature on the subject considers scalar fields because they are both simple to handle mathematically and able to give a low-energy effective field description of high-energy theories. Moreover the simplest way to describe unknown degrees of freedom in field theory is usually through scalar fields, unless these degrees of freedom are somehow related, for example through a (gauge) symmetry. It is thus natural to first characterise (dynamical) dark energy as a scalar field, and only if such a description fails the experimental tests seek more complicated solutions.

It might be the case however that different particle physics models of dark energy predict distinctive observational signatures or new phenomenological insights with respect to scalar field cosmology. Studying their theory and dynamics is thus important not only to build alternative routes to solve the dark energy mystery, but also to drive future experiments towards possible signals that may differentiate between various dark energy models.

In (quantum) field theory a scalar particle is defined by having spin-0 and thus by being invariant under local and global Lorentz transformations. The first natural extension of a scalar field is represented by particles with non-zero spin, such as spinors (spin-$1/2$ particles like electrons) and vectors fields (spin-$1$ particles like photons), or tensor fields (spin-$2$ particles like the graviton).

\subsection{Spinor fields}
\label{sec:spinors}

Spinor fields are less frequently employed to model dark energy than vector fields. Standard spinors, or Dirac spinors, describe the standard model fermions and are usually not used to drive the early-time or late-time accelerated expansion of the Universe. One of the few examples which explain this behaviour goes back to \cite{Ribas:2005vr}, where cosmological models with acceleration were studied with Dirac fields. Nonetheless, spinor field models are useful as models of dark matter beyond the standard model of particle physics since they are better suited to characterise non-relativistic matter.

However, there exists a class of spinor models which have been considered as an alternative model of dark matter and dark energy, the so-called ELKO spinors introduced in \citet{Ahluwalia:2004sz,Ahluwalia:2004ab}. They are non-standard spinors, according to the Lounesto general classification, ELKO spinors belong to the class of flag-pole spinors. They have mass dimension one and obey the relationship $(CPT)^2=-\mathbb{I}$. These models are able to provide interesting phenomenological features in both early and late time cosmology. These spinor fields can only couple directly with gravity which renders them naturally invisible to radiation and hence can be naturally viewed as dark spinors.

An extensive review on the theory and cosmological applications of ELKO spinors has been compiled by \citet{Boehmer:2010ma}. Dark energy models of ELKO spinors have been studied with dynamical systems methods by \citet{Wei:2010ad} and \citet{Sadjadi:2011uu}, who found that within this framework a solution of the fine tuning problem of initial conditions is difficult to obtain. Nevertheless \citet{Basak:2012sn} claimed that an alleviation to the problem can be achieved, though \citet{Pereira:2014wta} criticised the result showing that no isolated critical points appear at early times. They also found that phantom behaviour is possible in such models. \cite{Pereira:2014pqa} reviews the cosmology of ELKO spinors, and considers an interaction between the ELKO spinors and dark matter. Such a scenario is able to give rise to the late time acceleration of the universe via dark matter particles into the ELKO field.  Along similar lines \citet{S.:2014dja} considered a new dynamical system approach assuming an ELKO field interacting with dark matter. They were able to find stable fixed points which are able to alleviate the cosmological coincidence problem.

It should be added that the original ELKO construction suffered some shortcomings which were subsequently addressed, in particular issues with respect to local Lorentz invariance. It is fair to say that no consensus has been reached with respect to the physical viability of these non-standard spinors.
However, the model can be rigorously derived from an action principle and hence can be seen as an interesting toy model in the quest to improve our understanding of dark matter and dark energy.

In the following we will review the work of \citet{Wei:2010ad} in some detail, where ELKO spinors are the matter field responsible for the acceleration of the universe. The action used for this work is given by
\begin{align}
  S=\int d^{4}x\sqrt{-g}\Big[g^{\mu\nu}\nabla_{(\mu}\stackrel{\neg}{\lambda}\nabla_{\nu)}\lambda-V(\stackrel{\neg}{\lambda}\lambda)\Big]\,,
\end{align}
where $\nabla_{\mu}$ is the covariant derivative and $\stackrel{\neg}{\lambda}$ and $\lambda$ are the Elko dual spinor and the Elko spinor respectively. Let us now study flat FLRW cosmology and  assume that the spinor fields are homogeneous giving us \citep{Wei:2010ad}
\begin{eqnarray}
  \lambda_{\{-,+\}}=\phi(t)\frac{\xi}{\sqrt{2}}\,,\quad \lambda_{\{+,-\}}=\phi(t)\frac{\zeta}{\sqrt{2}}\,,
\end{eqnarray}
where $\xi$ and $\zeta$ are constant spinors satisfying $\stackrel{\neg}{\zeta}\zeta=\,\stackrel{\neg}{\xi}\xi=2$ and $\phi(t)$ is a homogeneous and real scalar field, first introduced in \citet{Boehmer:2007dh}. The notation $\lambda_{\{\mp,\pm\}}$ stands for the two possible eigenspinors of the charge conjugation operator (for details see \citet{Ahluwalia:2004sz,Ahluwalia:2004ab}). For this theory, the flat FLRW equations can be written as follows
\begin{align}
  3H^2 &= \kappa^2(\rho_{\phi}+\rho)\,,\label{HfCosm1}\\
  2\dot{H} &= -\kappa^2(\rho_{\phi}+\rho+p_{\phi}+p)\label{HfCosm2}\,,
\end{align}
where $\rho$ and $p$ are the energy density and pressure of the cosmic (dark) matter fluid and we have defined
\begin{align}
  \rho_{\phi} &= \frac{1}{2}\dot{\phi}^2+V(\phi)+\frac{3}{8}H^2\phi^2\,,\\
  p_{\phi} &= \frac{1}{2}\dot{\phi}^2-V(\phi)-\frac{3}{8}H^2\phi^2-\frac{1}{4}\dot{H}\phi^2-\frac{1}{2}H\phi\dot{\phi}\,.
\end{align}
Now, we assume that there is an interaction term $Q$ between the spinor dark energy and the matter, yielding the conservation equations
\begin{align}\label{ConsHf}
  \dot{\rho}_{\phi}+3H(\rho_{\phi}+p_{\phi})=-Q\,, \quad \dot{\rho}+3H(\rho+p)=Q\,.
\end{align}
Note that the total fluid described by $\rho_{\rm tot}=\rho_{\phi}+\rho$ and $p_{\rm tot}=p_{\phi}+p$ satisfies the usual conservation equation $\dot{\rho}_{\rm tot}+3H(\rho_{\rm tot}+p_{\rm tot})=0$. In principle the coupling $Q$ can be an arbitrary function. However, in order to use dynamical systems techniques, we will always choose $Q$ so that it can be expressed in terms of the dynamical variables (cf.~Sec.~\ref{chap:IDE}). Now, let us assume a linear EoS $p=w\rho$ and define the following dimensionless variables
\begin{align}\label{Ch7var}
  x = \frac{\kappa \dot{\phi}}{\sqrt{6}H}\,,\quad
  y = \frac{\kappa \sqrt{V}}{\sqrt{3}H}\,,\quad
  z = \frac{\kappa \phi}{2\sqrt{2}}\,,\quad
  u = \frac{\kappa \sqrt{\rho}}{\sqrt{3}H}\,.
\end{align}

The dynamical system for this model can be written as
\begin{align}
  x'&=(s-3)x+\frac{\sqrt{3}}{2}z-\sqrt{\frac{3}{2}} \frac{y^2  }{\kappa} \frac{ V_{,\phi}}{V}- q\,, \\
  y'&=sy+\frac{x}{\sqrt{2}H}\frac{ V_{,\phi}}{\sqrt{V}}\,, \\
  z'&=\frac{\sqrt{3}}{2}x\,,\\
  u'&=\Big[s-\frac{3}{2}(w+1)\Big]u+\frac{x}{u} q\,,
\end{align}
where $s=[3x^2-\sqrt{3}x z+\frac{3}{2}(w+1) u^2](1-z^2)^{-1}$, and $q$ depends on the chosen interaction term. We note that for the above equations to become a well-defined dynamical system, one needs to find an explicit form of $V_{,\phi}/\sqrt{V}$ in terms of the other variables. Following \citet{Wei:2010ad}, the coupling term is given by
\begin{align}
  q = \displaystyle\frac{\kappa Q}{\sqrt{6}H^3\phi'}\,.
\end{align}
Note that the case $z=1$ is excluded from the analysis since it represents the trivial case where $\phi$ is a constant. This dynamical system satisfies the Friedmann constraint $x^2+y^2+z^2+u^2=1$.
Geometrically speaking the phase space is confined to the surface of a 3-sphere.
Therefore, one can reduce this 4-dimensional system to a 3-dimensional one by eliminating one of the four variables.
In what follows we choose to eliminate $u$.
Consequently, we can re-write the dynamical system as
\begin{align}
  x' &= -3x+\frac{x\left[2 \sqrt{3} z x-3 (1+w)  \left(-z^2-x^2-y^2+1\right)-6 x^2\right]}{2 \left(z^2-1\right)}+\frac{\sqrt{3}}{2}  z-\sqrt{\frac{3}{2}} \frac{y^2  }{\kappa} \frac{ V_{,\phi}}{V}-q\,,
  \label{spinor1}\\
  y' &= \frac{y \left[2 \sqrt{3} z x-3 (1+w)  \left(-z^2-x^2-y^2+1\right)-6 x^2\right]}{2 \left(z^2-1\right)}+\sqrt{\frac{3}{2}}\frac{ x y }{\kappa }\frac{ V_{,\phi}}{V}\,,
  \label{spinor2}\\
  z' &= \frac{\sqrt{3}}{2}x\,.
  \label{spinor3}
\end{align} 
We assume the energy potential to be positive, the phase space is constrained as
\begin{align}
  x^2 + y^2 + z^2 \leq 1\,, \quad
  y\geq 0\,,
\end{align}
which represents a half of a unit sphere.
Let us now consider the exponential potential
\begin{align}
  V(\phi) = V_{0} e^{-\lambda \kappa \phi}\,,
\end{align}
where $V_{0}$ and $\lambda $ are positive constants. In \citet{Wei:2010ad} it was shown that in the cases $Q=0$ and $Q=\alpha \kappa \rho\dot{\phi}$ there are two critical points, neither of which is a stable attractor. However, for the second case, there is one critical point which is a scaling solution. Moreover, there are no critical points (and hence no attractors) for the interactions $Q=3\beta \rho_{\rm tot}$ and $Q=3\alpha H \rho$, respectively where $\beta$ and $\alpha$ are constants and $\rho_{\rm tot}=\rho+\rho_{\phi}$. We now consider a new coupling
\begin{align}
  q=\alpha z^2=\alpha \frac{\kappa^2 \phi^2}{8}\,,
  \label{Qspinor}
\end{align}
where $\alpha$ is a constant.
This coupling corresponds to an interacting energy given by
\begin{equation}
  Q \propto \phi^2 \rho_{\rm tot} \dot\phi \,.
\end{equation}
In this case, there are four critical points which are shown in Tab.~\ref{tab:spinor1} where we have defined the constants
\begin{align}
\beta_1=\frac{2 \left(2 \alpha +\sqrt{6} \lambda \right)}{\sqrt{3}}\,,\quad \beta_2= \frac{\sqrt{8 \sqrt{6} \alpha  \lambda +24 \lambda ^2+3}}{\sqrt{3}}\,.
\label{eq:beta_consts}
\end{align}

\begin{table}[!htb]
  \begin{center}
   \begin{tabular}{|c|c|c|c|c|}
	\hline
	Point & $x$ & $y$ & $z$ & Existence  \\
	\hline & & & & \\
	$O$ & 0 & 0 & 0 & $\forall \ \beta_1, \beta_2$  \\ & & & & \\
	\multirow{3}{*}{	$A_{-}$} & 	\multirow{3}{*}{0} & 	\multirow{3}{*}{$\displaystyle\frac{\sqrt{\beta_{1}^2-(\beta_{2}-1)^2}}{\beta_{1}}$} & 	\multirow{3}{*}{$\displaystyle\frac{1-\beta_{2}}{\beta_{1}}$} &   $\beta_{2}<1\land (\beta_{1}+1\leq \beta_{2}\lor \beta_{1}+\beta_{2}\geq 1)$\\
	& & & & or \ $\beta_{2}=1$ \\
		& & & & or \ $\beta_{2}>1\land (\beta_{1}+\beta_{2}\leq 1\lor \beta_{1}+1\geq \beta_{2})$ \\ & & & & \\
			\multirow{3}{*}{$A_{+}$ }& 	\multirow{3}{*}{0} & 	\multirow{3}{*}{$\displaystyle\frac{\sqrt{\beta_{1}^2-(\beta_{2}+1)^2}}{\beta_{1}}$} &   	\multirow{3}{*}{$\displaystyle\frac{1+\beta_{2}}{\beta_{1}}$}
			& $\beta_{2}<-1\land \beta_{1}\geq -\beta_{2}-1$ \\
				& & & & or \ $\beta_2=-1$  \\
			& & & & or \ $\beta_{2}>-1\land \beta_{1}\geq \beta_{2}+1$ \\	& & & & \\
				\multirow{2}{*}{$B$} & 	\multirow{2}{*}{0} & 	\multirow{2}{*}{0} & 	\multirow{2}{*}{$\displaystyle\frac{2 \beta_{1}}{\beta_{1}^2-\beta_{2}^2+1}$} &  	\multirow{2}{*}{$\beta_1^3+\beta_{1}<\beta_{1}\beta_{2}^2\lor \frac{1-\beta_{2}^2}{\beta_{1}}+\beta_{1}\geq 2$} \\ & & & & \\ & & & & \\
			\hline
   \end{tabular}
  \end{center}
  \caption{Critical points of the dynamical system (\ref{spinor1})--(\ref{spinor3}) along with the conditions for existence of the point. The two constants $\beta_1$ and $\beta_2$ are defined in Eq.~\eqref{eq:beta_consts}.
  }
  \label{tab:spinor1}
\end{table} 

We have introduced these new constants to simplify the analysis of the critical points. For the same reason, we have also discarded the limiting case where $\beta_1=0$ (or $\alpha=-\sqrt{6}\lambda/2$). Tab.~\ref{tab:spinor2} shows the effective state parameter, acceleration and the stability of the critical points of the system.

\begin{table}
  \begin{center}
    \begin{tabular}{|c|c|c|c|c|c|}
      \hline
      Point & $w_{\rm eff}$ & Acceleration & $\Omega_{\rm m}$ & $\Omega_{\phi}$ & Stability  \\
      \hline & & & & &\\
      $O$ & $w$ &  No & 1 & 0 & Saddle point \\& & & & &\\

      $A_{\pm}$ & $-1$ &  Always & 0 & 1 & See Fig.~\ref{fig:spinor3} \\  & & & & &\\

      \multirow{2}{*}{$B$}  & 	\multirow{2}{*}{$w$} & 	\multirow{2}{*}{No}& \multirow{2}{*}{$1-\frac{4 \beta_1^2}{\left(\beta_1^2-\beta_2^2+1\right)^2}$} & \multirow{2}{*}{$\frac{4 \beta_1^2}{\left(\beta_1^2-\beta_2^2+1\right)^2}$} & Stable if $w<-1$  \\
      & & & & & Unstable if $w \geq \frac{1}{3} \left(2 \sqrt{3}+3\right)$\\& &  & & & \\ \hline
    \end{tabular} 
  \end{center}
  \caption{Effective state parameter and acceleration of the critical points given in Tab.~\ref{tab:spinor1} for the dynamical system of the model (\ref{spinor1})--(\ref{spinor3}).
  }
  \label{tab:spinor2}
\end{table}

\begin{figure}[!htb]
\begin{center}
  \includegraphics[width=0.47\textwidth]{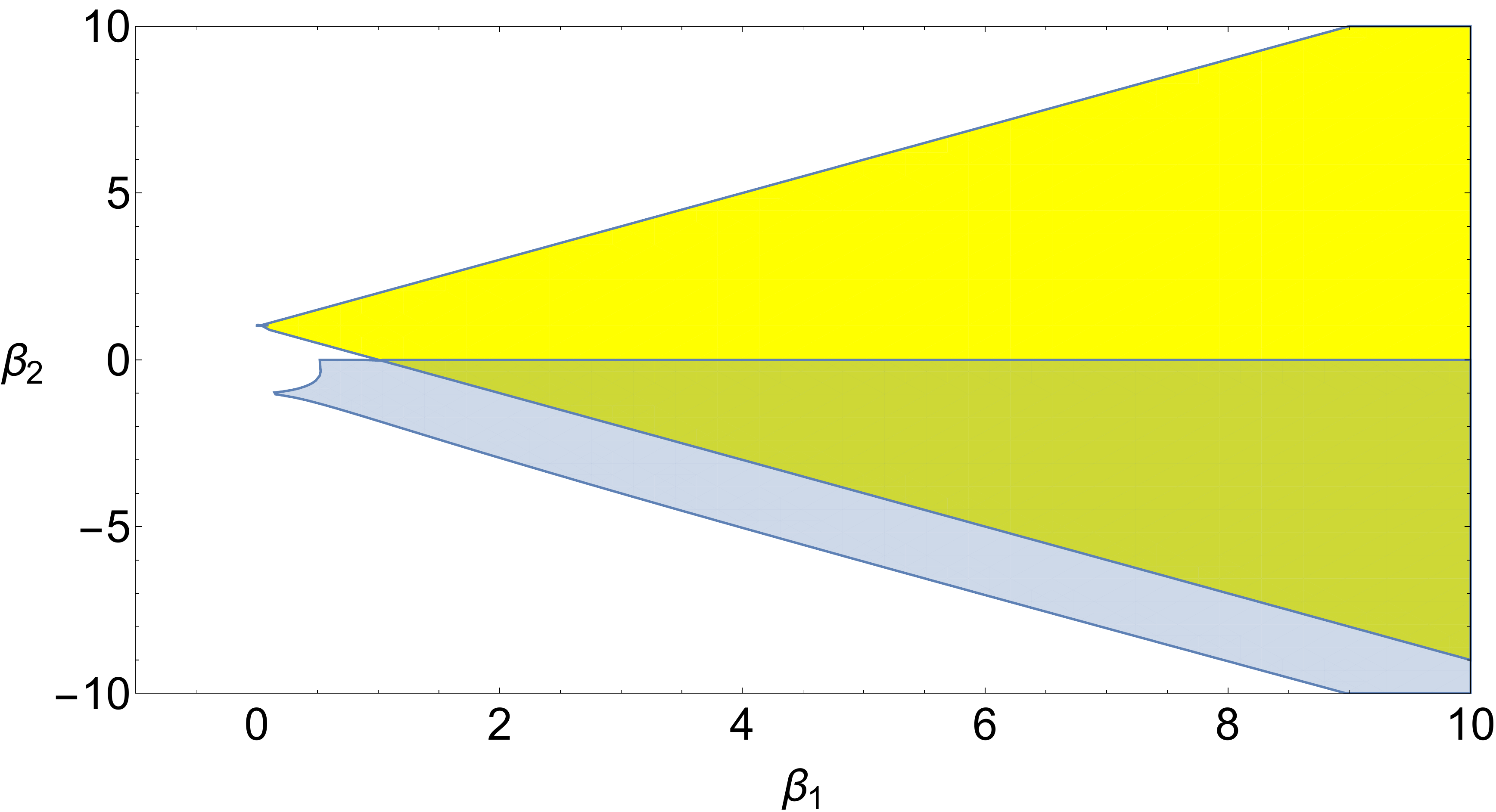}
  \hfill
  \includegraphics[width=0.47\textwidth]{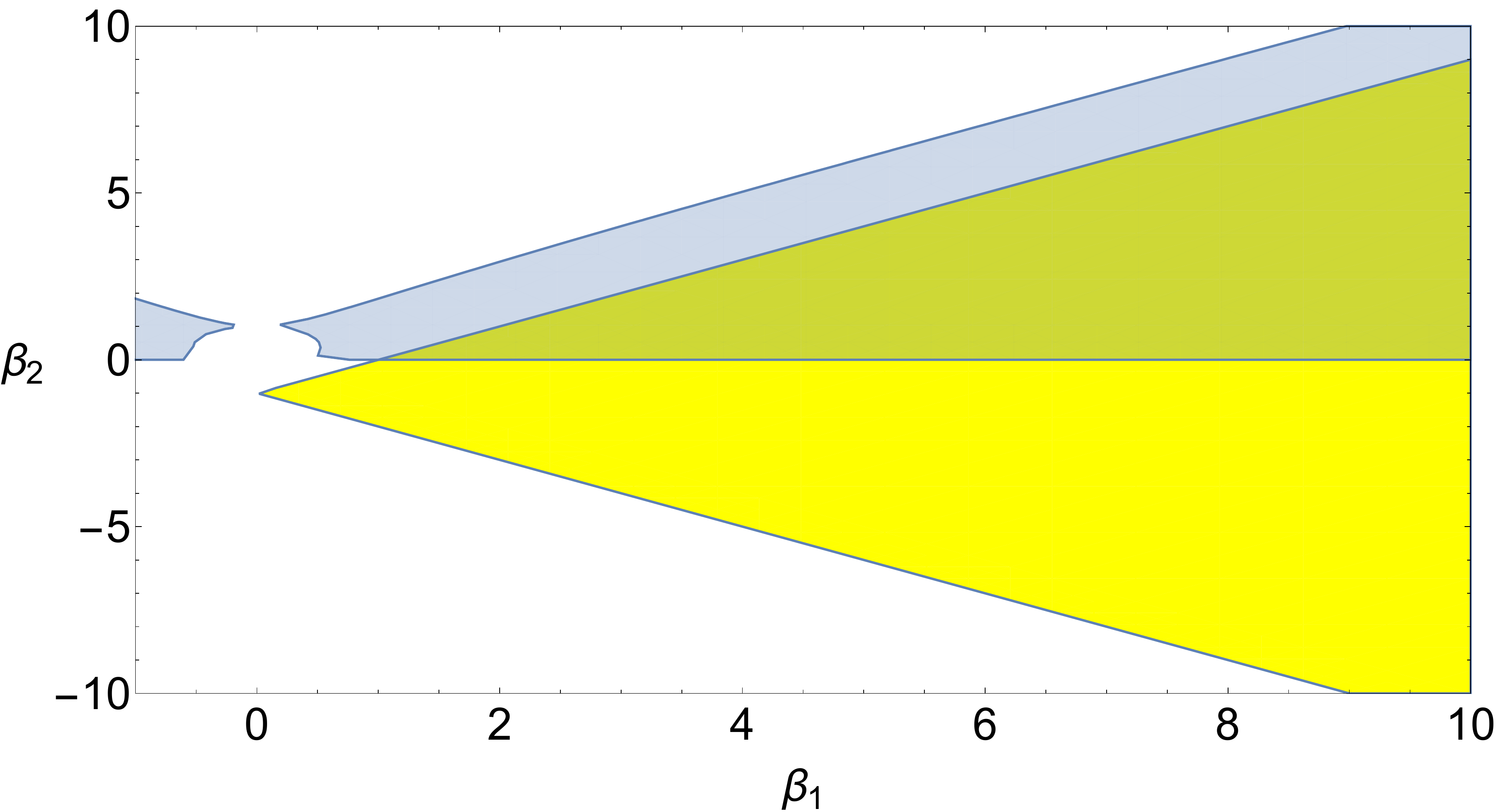}
\end{center}
\caption{Stability regions for the points $A_{\pm}$ given by the model (\ref{spinor1})-(\ref{spinor3}) with $w=0$ (dust). The figure on the left (right) shows the stability region for the point $A_{-}$ ($A_{+}$). The grey region represents the region where all the eigenvalues of the Jacobian evaluated at the respective point are negative. The yellow region represents the region where the point exists and hence the green region (superposition of the two) represents the region where the point is stable.
}
\label{fig:spinor3}
\end{figure}

The stability regions for the points $A_{\pm}$ are depicted in Fig.~\ref{fig:spinor3} in the case where $w=0$. Note that for the dust case, if $A_{+}$ is stable then $A_{-}$ becomes either unstable or a saddle point.

Let us now take a specific model where all four critical points exist and moreover the point $A_{+}$ is stable (see Tab.~\ref{tab:spinor1} and Fig.~\ref{fig:spinor3}). To do that, let us consider $\lambda=1,w=0$ and $\alpha=2$, or in terms of our new constants $\beta_1=2 (\sqrt{6}+4)/\sqrt{3}$ and $\beta_2=\sqrt{9+16\sqrt{2/3}}$. For this particular choice of parameters, the points $O$ and $B$ have eingenvalues
\begin{align}
  O:\quad\left\{-\frac{1}{4} \left(3+\sqrt{21}\right),\frac{3}{2},-\frac{1}{4} \left(3-\sqrt{21}\right)\right\} \,,
\end{align}
and
\begin{align}
  B:\quad\left\{\frac{3}{2},-\frac{1}{4} \left(3-i \sqrt{3}\right),-\frac{1}{4} \left(3+i \sqrt{3}\right)\right\} \,,
\end{align}
respectively. Thus, these two points are unstable. The eingenvalues of the other two critical points $A_{\pm}$ are approximately,
\begin{align}
A_{-}:\quad{}&\Big\{0.84, -3.42+0.87 i, -3.42-0.87 i\Big\}\,,\\
A_{+}:\quad{}&\Big\{-1.02,-2.48+2.02 i,-2.48-2.024 i\Big\}\,.
\end{align}
Hence, for the chosen parameters $A_{+}$ is a stable attractor which represents a late time accelerating solution with effective equation of state $w_{\rm eff} = -1$.

\begin{figure}
  \begin{center}
    \includegraphics[width=0.9\textwidth]{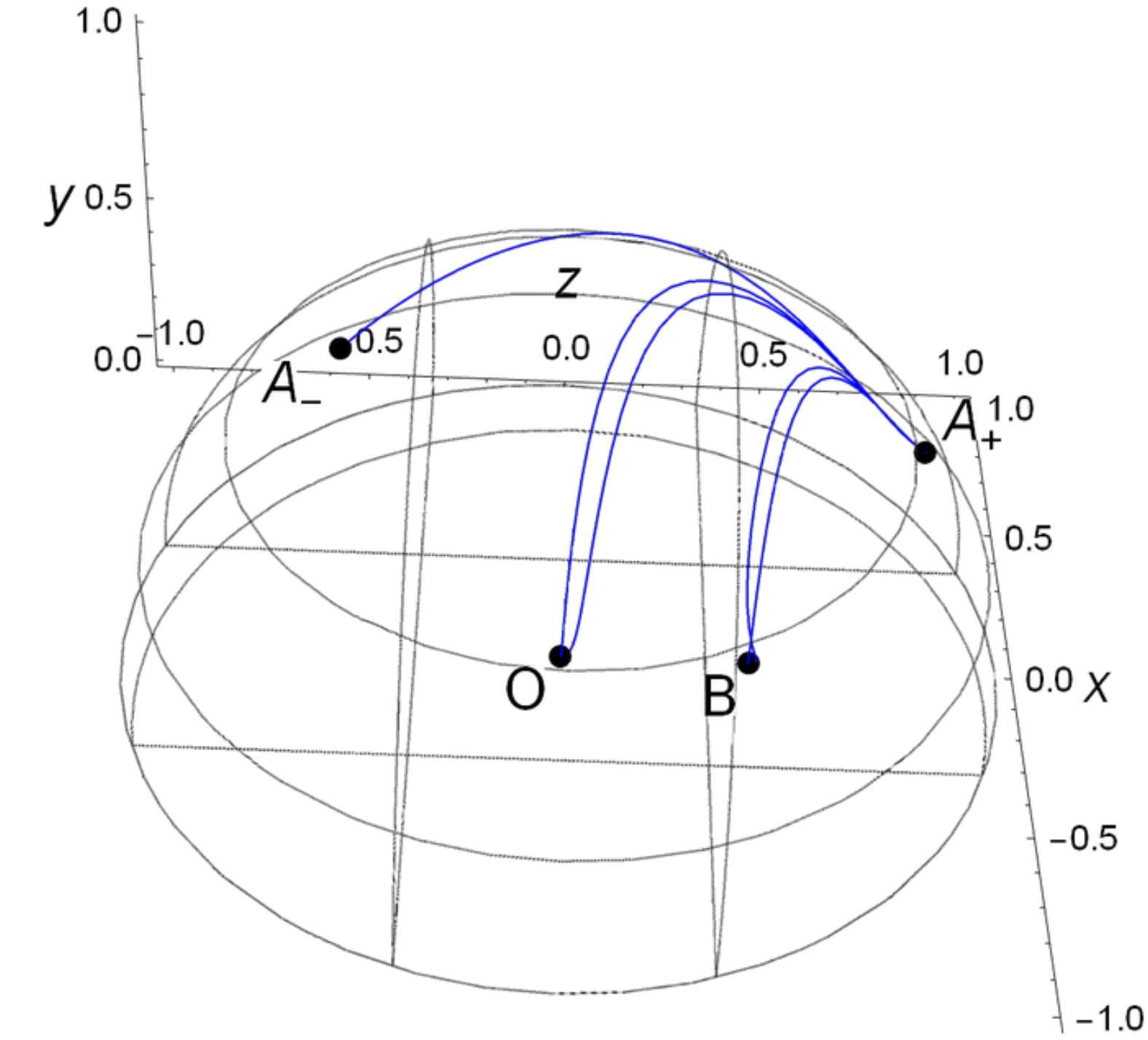}
  \end{center}
  \caption{Phase space portrait of the dynamical system (\ref{spinor1})-(\ref{spinor3}) and the coupling (\ref{Qspinor}) with the values $w=0$, $\lambda=1$ and $\alpha=2$. For these values, $A_+$ is a stable attractor. }
  \label{fig:spinor1}
\end{figure}

\begin{figure}
  \begin{center}
    \includegraphics[width=0.9\textwidth]{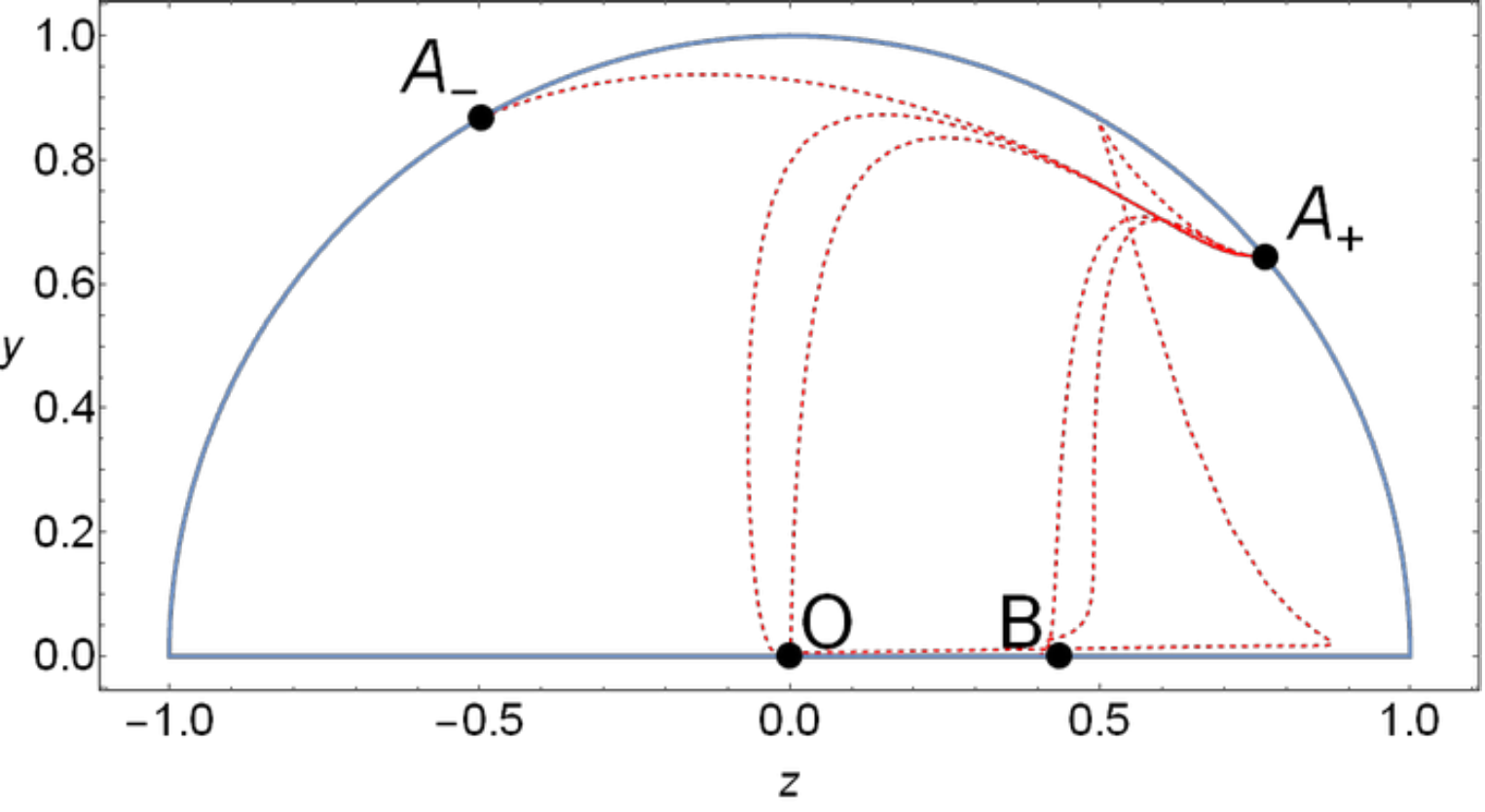}
  \end{center}
  \caption{2D Phase space portrait in the $yz$ plane of the dynamical system (\ref{spinor1})-(\ref{spinor3}) and the coupling (\ref{Qspinor}) with the values $w=0$, $\lambda=1$ and $\alpha=2$. For these values, $A_+$ is a spiral stable attractor.
  }
  \label{fig:spinor2}
\end{figure}

Figs.~\ref{fig:spinor1} and \ref{fig:spinor2} show the behaviour of the dynamical system for this specific model, and we see that it displays some interesting features. Trajectories starting near the point $A_{-}$ represent a universe with early time accelerated expansion and late-time accelerated expansion at $A_{+}$. On the other hand, trajectories starting near $O$ and $B$ correspond to matter dominated states which evolve to become dark energy dominated.

\subsection{Vector fields}

Vector fields arise in the standard model of particle physics not only to describe the electro-magnetic interaction, but also as mediators of the nuclear forces. At the centimetre scale the electro-magnetic field is sufficiently strong to compete with gravity, but on cosmological scales because of the huge masses involved it becomes inevitably negligible with respect to the gravitational attraction.

One is thus led to postulate new vector fields, capable of modifying the large scale dynamics of the Universe while at the same time being undetectable at solar system scales. However, introducing a new vector field breaks the isotropy of space. Any vector defines a preferred direction, thereby immediately breaking this symmetry. In order to avoid this problem, which would invalidate the cosmological principle (see Sec.~\ref{chap:Cosmology}), some authors have considered a {\it triad} of vector fields invariant under SO(3) transformations, i.e.~three dimensional rotations \citep{Bento:1992wy,ArmendarizPicon:2004pm}. A dynamical systems analysis of such a model has been considered in \citet{Wei:2006tn} who also added an interaction with the matter sector. It was found that the fine tuning and cosmic coincidence problems are less severe within this framework.

Moreover, phantom behaviour can be attained without incurring a big rip scenario. \citet{Wei:2006gv} also studied vector fields interacting with scalar fields which can be motivated by Weyl geometries. A clear review and dynamical systems analysis of various vector field models of dark energy has been presented in \citet{Koivisto:2008xf}. Among other results, these authors showed that {\it space-like} vectors admit scaling solutions, while {\it time-like} vectors easily avoid anisotropies.

In~\citet{Koivisto:2014gia}, a dynamical system analysis of an anisotropic Bianchi type I cosmology with a massive disformally coupled vector field was considered. An interesting variety of fixed points was found, including anisotropic scaling solutions. Assuming an exponential potential, it was shown that anisotropic fixed points either do not describe acceleration or give rise to incompatible models with anisotropies being too large. However, for the isotropic case, the oscillations of the vector field become faster and hence some viable models can describe disformally interacting massive dark matter scenarios.

Using dynamical system techniques, \citet{Landim:2016dxh} studied a vector-like dark energy model in the presence of a barotropic fluid. An interaction term between the barotropic fluid and the dark energy was also considered. It was shown that some fixed points describe both matter and dark energy dominated epochs.

\subsection{Yang-Mills fields}

Other vector fields that appear in the standard model of particle physics are non-abelian Yang-Mills fields satisfying a more general gauge symmetry than the $U(1)$ symmetry of electromagnetism. These fields are important in the standard model since they describe any particle physics model with interactions mediated by gauge bosons. Hence, some studies have suggested that Yang-Mills fields might interact with other particles as cosmic components. Usually, a $SU(2)$ Yang-Mills field is employed in cosmology since it admits a homogeneous and isotropic energy-momentum tensor. Inflationary scenarios can be obtained by using this specific Yang-Mills field. One interesting result of these models is that the big rip can be naturally avoided.

Such fields were investigated as dark energy models \citep{Zhao:2006mk,Zhao:2005bu} and studied using dynamical systems techniques by \cite{Zhang:2006rw}. An interaction with both radiation and non-relativistic matter was considered. It was shown that a solution of the fine tuning problem can be achieved. \citet{Zhao:2008tk} studied a coupling of Yang-Mills fields with to the matter sector finding that phantom behaviour without a big rip scenario can be achieved.

\subsection{3-form fields}

Besides dark energy models built out of standard model particles, there are other approaches motivated by particle physics. In several models, the dynamics of the physical degrees of freedom are represented by {\it forms}. Without going into the mathematical details, forms are geometrical objects which can be seen as generalisations of scalars (0-forms) and vectors (1-forms), closely related to skew-symmetric tensors.

In a four dimensional spacetime there can be only forms up to order four. The $0$-forms correspond to scalars while $1$-forms correspond to vectors. $4$-forms correspond to the volume differential and cannot describe dynamical fields. $2$-forms are usually used to provide a more mathematical description of the Maxwell equation where one introduces the Faraday $2$-form. Perhaps unsurprisingly, cosmological applications based on $2$-forms show many similarities with vector field models, they share many properties and lead to similar results. This leaves us with $3$-forms which have been studied in cosmology. They lead to new phenomenology and interesting dynamics at large scales.

\citet{Koivisto:2009fb} employed $3$-forms to build models of inflation and dark energy, showing that stable accelerator attractors are present in the dynamics. The dynamical systems arising from three-form cosmology contain non-hyperbolic critical points and a complete analysis should make use of centre manifold theory to determine the stability, as it has been done by \citet{Boehmer:2011tp}. \citet{Ngampitipan:2011se} added a coupling between three-forms and dark matter, looking for solutions of the cosmic coincidence problem.
In addition, inflation in a five-dimensional model in Randall-Sundrum II braneworld scenario (cf.~Sec.~\ref{sec:string_and_brane}) has been studied using $3$-form fields \citep{Barros:2015evi}.
The authors used dynamical system techniques to compare the mentioned model with the standard four dimensional case.
They observed that the presence of the bulk influences the value of the spectral index and the ratio of tensor to scalar perturbations. \citet{Morais:2016bev} studied $3$-form dark energy models with different interactions terms with the aim to visualise which interaction could avoid  the evolution towards a little sibling of the Big Rip (known as LSBR). It was shown that dark matter interactions dependence will give rise to LSBR. In addition, by using the statefinder hierarchy and computing the growth rate of matter perturbations, the authors were able to observationally distinguish between linear and quadratic dark energy interactions.

\subsection{Unparticles}

Other dark energy models motivated by theoretical developments in particle physics consider the so-called {\it unparticle physics}. {\it Unparticles} are scale invariant low-energy degrees of freedom coming from effective field theories of high energy physics, see e.g.~\citet{Georgi:2007ek}. Unparticles have different properties to ordinary matter fields. For instance, they do not have a mass and their scaling dimensions do not necessarily need to be equal to an integer or a half integer. Some studies suggest that unparticles can be important to understand the physics of the early universe. Moreover, since they can only weakly interact with ordinary matter, they might be potential dark matter candidates. Their cosmological dynamics has been studied using dynamical systems techniques by \citet{Chen:2009ui} who found scaling solutions and showed that the fine tuning problem of initial conditions can be largely avoided.

\subsection{Chaplygin gas}
\label{sec:chaplygin_gas}

Not all dark energy models are motivated by particle physics. One of the best known models is the Chaplygin gas which is motivated by phenomenological considerations. The Chaplygin gas is a perfect fluid satisfying the unusual equation of state
\begin{align}
  p = -\frac{A}{\rho^\alpha} \,,
\end{align}
where $\alpha$ is a parameter and $A$ is a constant of suitable dimensions. The first Chaplygin gas models considered the case $\alpha=1$, but later generalisations with $\alpha\neq 1$ have been advanced.

In a cosmological context the Chaplygin gas was proposed by \citet{Kamenshchik:2001cp} as a unified model capable of accounting for both dark matter and dark energy. \citet{Wu:2007zzf,Li:2008uv,Avelino:2003ig} studied the dynamics of large scales of Chaplygin gas models interacting with dark matter, obtaining late time scaling, de Sitter and phantom attractor solutions. Subsequently, \citet{delCampo:2013vka} coupled the Chaplygin gas to a scalar field and performed a dynamical systems analysis using the approach we adopted in Sec.~\ref{sec:other_potentials}. Chaplygin gases have been considered in the context of other theories too. \citet{Ranjit:2014xga} investigated a modified Chaplygin gas in an Einstein-Aether theory (cf.~Sec.~\ref{sec:Horava_and_aether}). The dynamical system is complicated, and critical points could only be found numerically. A stable scaling solution was found, alleviating the cosmic coincidence problem, and a late time acceleration possible.

Furthermore, a study of Chaplygin gas with a coupling between dark energy and dark matter has been considered. \citet{Xi:2015qua} performed a dynamical systems analysis with a particular interaction in the dark sector.
This model did not fix the sign of the interaction hence allowing energy transfer in both directions. Following on from this, \citet{Khurshudyan:2015mpa} and again \citet{Khurshudyan:2015mva} studied additional Chaplygin gas models with different forms of interactions. Late time scaling attractors were found, and constraints on the possible interactions were presented.

\subsection{Viscous fluids}

Another interesting approach motivated by particle physics is to take into account that the universe is filled by a bulk viscous cosmological fluid. Such a fluid is characterised by having an effective pressure of the form
\begin{align}
  p = p_{\rm std} + p_{\rm vis} = w \rho - \xi(\rho) \nabla_{\mu} u^{\mu} \,,
\end{align}
where the first term is the standard pressure component and the second term is due to viscosity. Here $\nabla_{\mu}$ is the covariant derivative. To simplify the model, one usually assumes a linear equation of state for the standard pressure component. The term $\nabla_{\mu} u^{\mu}\equiv \Theta$ is the fluid's expansion scalar which for flat FLRW cosmology is reduced to $3H$. The viscosity coefficient $\xi(\rho)$ introduces dissipation and in general depends on the energy density. It is necessary that $\xi(p)>0$ to ensure that the second law of thermodynamic is satisfied. If the bulk viscosity is sufficiently large, it allows the fluid to be in the phantom region even if $w>-1$.

\citet{Colistete:2007xi} were the first to study viscous fluid models using dynamical system techniques. The background solution showed similarities with generalised Chaplygin gas models. It was shown that this model is free from instabilities or oscillations of small perturbations. It can successfully describe the expected accelerated expansion of the universe.
\citet{Szydlowski:2006ma} analysed cosmological models with a dissipative dust fluid  and showed that for a flat FLRW cosmology, the system can be treated as a conservative one with a potential function being the same as a Chaplygin gas type. Moreover, the authors showed that viscous models fit well with SNIa observations without evoking a cosmological constant. Further, using Bayesian information criteria, they concluded that viscous models with a fixed viscosity parametrisation are more consistent with observations than $\Lambda$CMD model.

\citet{Acquaviva:2014vga} studied the dynamics of a dark energy model using a viscous fluid with friction interactions. \citet{Avelino:2013wea} and \citet{Cruz:2014iva} assumed dark matter to be the viscous fluid component and added an interaction with scalar fields, which describe the dark energy. These models allow for late time accelerated expansion of the Universe. \citet{Odintsov:2017icc} also investigated (non)-interacting DE models with a general viscous fluid EoS given by $p = - \rho + f(\rho) + G(H)$, where $f$ and $G$ are general functions. They found that for some specific choices of these function and of the dark sector interaction, a unified cosmic history, starting from inflation and ending in a DE dominated phase, can be obtained. Moreover \cite{Sasidharan:2015ihq} considered a model with a bulk viscous coefficient being equal to $\xi=\xi_0+\xi_1H+\xi_2 (\dot{H}/H+H)$ where $H$ is the Hubble parameter. They found that only the case $\xi_1=\xi_2=0$ is viable for having a coherent description of the different phases of the universe. Finally \citet{Biswas:2016idx} performed an extended dynamical system investigation of a model of interacting DE in the framework of particle creation, which gives rise to an effective viscous dark interaction.

\subsection{Higgs field}

The Higgs field has also been considered for its cosmological applications. Because there is only one scalar field in the standard model of particle physics, some authors have considered using the Higgs for inflation and dark energy, which is appealing as it eliminates the need for introducing new fields into the universe. The Lagrangian of the Higgs boson at tree-level is given by
\begin{equation}
S=\int d^4x \sqrt{-g} \left[ \frac{1}{2} R + \frac{1}{2} \nabla_{\mu}{\cal H}^{\dagger} \nabla^{\mu}{\cal H}  -\frac{\lambda}{4} ({\cal H}^{\dagger} {\cal H}-v^2)^2 \right],
\end{equation}
where $ v\sim 10^2$ GeV is the vacuum expectation value (VEV) of the Higgs potential in the broken phase, ${\cal H} $ the complex Higgs doublet, and $ \nabla_{\mu} $ the covariant derivative of electroweak interaction.

\citet{Rinaldi:2014yta} studied the dynamical system of the cosmology  of a non-abelian Higgs field coupled to gravity. The ultra slow-roll regime is studied and critical points found. There are many non-hyperbolic points so the author resorted to numerical investigations.
\citet{Rinaldi:2015iza} considered also the Einstein Yang-Mills Higgs equations. An eleven dimensional dynamical system was found, which possesses infinitely many critical points, although many of them can be classified as physically uninteresting. Metastable phases of radiation and matter domination can occur before ending in a dark energy stable era.

\subsection{Holographic dark energy}
\label{sub:interacting_holographic_dark_energy}

Some popular models of interacting DE are based on the holographic principle.
Holographic DE has been introduced by \citet{Cohen:1998zx} on the basis that in quantum field theory a short-scale cut off can be related to a long-distance cut off implied by the limit set by the formation of a black hole (see \citet{Wang:2016och} for a recent review).
In other words, if $\rho_{\rm de}$ is the vacuum energy density, the total energy in a region of size $L$ cannot exceed the mass of a black hole with the same mass, namely $L^3 \rho_{\rm de} \leq L M_{\rm p}^2$, where $M_{\rm p}$ is the reduced Planck mass.
By requiring $\rho_{\rm de}$ to saturate this inequality one arrives at the following assumption for the energy density of DE:
\begin{equation}
\rho_{\rm de} \simeq \frac{M_{\rm p}^2}{L^2} \,.
\label{eq:HDE}
\end{equation}
One natural choice of the cut-off scale $L$ in cosmology is the size of the current universe, i.e.~the Hubble scale $L = 1/H$ \citep{Cohen:1998zx}.
Unfortunately, if there are no interactions in the dark sector, this choice leads to non-viable cosmological dynamics \citep{Hsu:2004ri}.
For this reason, \citet{Li:2004rb} proposed to use another cut-off scale given by the future event horizon
\begin{equation}
L = a \int_t^\infty \frac{dt}{a} \,,
\end{equation}
which indeed yields an accelerating universe as the future attractor, but with a present value of the DE EoS parameter of $w_{\rm de} \simeq -0.9$ which is at tension with current observations.

In order to overcome these issues, holographic DE has been investigated allowing an interaction with the matter sector.
\citet{Wang:2005jx} found that a coupling in the dark sector can in fact lower the value of the present dark energy EoS parameter for holographic DE with a future event horizon cut-off, and even the crossing of the phantom barrier, slightly favoured by observations, can be achieved.
Moreover \citet{Pavon:2005yx} showed that any arbitrary interaction in the dark sector leads to an accelerating scaling solution for holographic DE with $L = 1/H$, making this choice for the cut-off scale compatible with experiments as well.

Since these results where obtained, many authors have started analysing the dynamics of interacting holographic DE models with different couplings to the matter sector. For example \citet{Setare:2007we} considered $Q \propto (\rho_{\rm m}+\rho_{\rm de}) H $ and \citet{Karwan:2008ig} generalised this coupling to $Q \propto H \left(\alpha \rho_{\rm de} + \beta \rho_{\rm m}\right)$. Later \citet{Banerjee:2015kva} delivered a dynamical analysis for the coupling $Q \propto \Gamma \rho_{\rm de}$, assuming both $\Gamma$ being a constant and a generic function of $H$. Other couplings have been studied by \citet{Mahata:2015nga} who considered $Q \propto \rho_{\rm m} \rho_{\rm de} / H$ and $Q \propto H \rho_{\rm de}$, and by \citet{Golchin:2016yci} who investigated several non-linear interactions.

There are some proposals using a different cut-off scale for $L$ in Eq.~\eqref{eq:HDE}. A popular alternative is to assume $L$ to be the Ricci scalar curvature \citep{Granda:2008dk}
\begin{equation}
L = \left(\dot{H} + 2H^2 \right)^{-1/2} \,,
\end{equation}
motivating this choice since it represents the size of the maximal perturbation leading to the formation of a black hole.
The cosmological dynamics with this cut-off scale for three different interacting terms has been studied by \citet{Mahata:2015nga}.
Another scale which has been used is the age of the universe \citep{Cai:2007us}
\begin{equation}
L = \int_0^{t_0} dt = \int_0^{a_0} \frac{da}{Ha} \,,
\end{equation}
which defines the holographic DE models known as \textit{agegraphic} DE.
A dynamical system analysis for these models with an interaction in the dark sector has been performed by \citet{Lemets:2010qz} and \citet{Xu:2015ata}, finding accelerated late time attractors and transient periods of acceleration.
On a different footing \citet{Wei:2009kp} proposed a modification to Eq.~\eqref{eq:HDE} given by entropic corrections to the area relation due to quantum gravity effects.
The cosmic dynamics of these models, again with an interaction to the matter sector, has been studied by \citet{Darabi:2016igc}.

A unified analysis of all the models introduced above, including also a coupling in the dark sector, has been performed by \citet{Zhang:2010icb} who obtained scaling solutions with a possible alleviation of the cosmic coincidence problem.

Finally there are also DE models using ideas similar to the holographic one. In particular the so called \textit{ghost DE} model which exploits the Veneziano ghost from QCD to define a DE energy density evolution characterised by $\rho_{\rm de} \propto H$ or $\rho_{\rm de} \propto H + H^2$. The cosmological dynamics of these models has been studied by \citet{Golchin:2016yci}.

\subsection{Cosmological models with modified matter contributions}

In this final section it is worth mentioning also some other phenomenological models, where the matter components sourcing the Friedmann equation are modified as
\begin{equation}
\label{b_eq}
H^2 = \frac{8 \pi}{3 m_4^2} \rho\, \mathbb{L}^2(\rho)\,,
\end{equation}
where $\mathbb{L}$ is a function of the total energy density $\rho$, which defines the effective model at hand. This class of models distinguish itself from the effective models reviewed in Sec.~\ref{sub:interacting_holographic_dark_energy} since instead of proposing a phenomenological evolution for dark energy, it considers a phenomenological modification of the way matter sources the Friedmann equation.

The general analysis of the properties of Eq.~\eqref{b_eq}, like the one performed by \citet{Copeland:2004qe} focusing on scaling solutions, allows one to reach some general conclusions on a number of different models, including for example the Chaplygin gas treated in Sec.~\ref{sec:chaplygin_gas}. In the case $\mathbb{L}(\rho)=\rho^{\frac{n-1}{2}}$, an in-depth analysis was performed also by \citet{Tsujikawa:2004dp}. \citet{Sen:2008yt} subsequently delivered the same analysis focusing instead on tachyonic scalar fields. {\it Cardassian cosmologies} in which  $\mathbb{L}(\rho)=\left(A + B \rho^{n-1} \right)^{1/2}$ were studied by \citet{Lazkoz:2005bf} where moreover further phenomenological modifications of the Friedmann equation were assumed and late time de Sitter attractors were obtained.

%% file: chapters/08_othermodels/othermodels.tex
\section{Dark energy models beyond general relativity}
\label{chap:modifiedgrav}

In the previous sections we have always assumed that the gravitational interaction is  described by general relativity up to cosmological scales. In this perspective, the accelerated expansion at late times is due to some field sourcing the right hand side of the Einstein field equations. This, however, is not the only possible approach to achieve a theoretical description of cosmic acceleration. Another option is to consider cosmic acceleration as a breakdown of general relativity at cosmological scales. In other words, instead of introducing some new matter fluid, one changes the left hand side of the Einstein field equations, i.e.~the pure gravitational sector, in order to obtain the needed late time accelerated expansion. Modifications of general relativity have a history almost as long as the one of general relativity itself. \citet{Weyl:1918} introduced the first extension of Einstein's theory of gravitation with the goal of unifying the gravitational interaction and electromagnetism. In the early 1920's, \citet{Kaluza:1921tu} and \citet{Klein:1926tv,Klein:1926fj} developed a five dimensional version of General Relativity that encompassed electromagnetism through the geometry of the fifth compact dimension and has proven to be the motivation for many subsequent papers of gravity in extra dimensions. Much later, in the 1960's, \citet{Brans:1961sx} proposed a new theory of gravitation in which, in contrast to general relativity,  Mach's principle could be fully integrated via the introduction of a non-minimally coupled scalar field. Following this idea, many other models were proposed in which not only  non-minimal coupling would appear, but also  in which different aspects of Einstein theory were extended/modified.  With the development of the semiclassical approaches to quantum gravity and, successively, of unification schemes like supergravity and M-theory, it was realised that in many cases low energy versions of these theories correspond  to modifications of general relativity, leading to an increased interest in such models, such that nowadays the exploration of modifications of GR occupies a significant part of the research in relativistic gravitation.

In this section we will show how dynamical systems techniques have been applied to the analysis  of a number of modifications of GR. Given that the number of approaches taken to such modifications is enormous, here we will limit our presentation to theories for which a phase space analysis has been formulated\footnote{It is important to stress that many of the theories we are presenting in this chapter are chosen because of their interest in terms of applications of dynamical system tools, rather than their physical validity. Most of these theories are still under theoretical investigation and have not been tested extensively against observations. Therefore, their presence in this review does not imply any statement on their validity, neither an endorsement by the authors.}. For more thorough reviews of modified gravity see \citet{Sotiriou:2008rp}, \citet{Clifton:2011jh} and \citet{Joyce:2014kja}. In addition, we will consider only the simplest versions of a given class of theory referring the interested reader to more specific literature when necessary.

Some caveats are necessary before starting our discussion. In the literature it is a common practice to employ conformal transformations to reformulate the action of modified gravity in the so-called {\it Einstein frame} where the pure gravitational sector is recovered to be that of standard Einstein-Hilbert Lagrangian with a coupling between the scalar field and the matter fields arising in the matter sector. In the following we present an analysis of the dynamics derived directly from the original equations in what is known as the {\it Jordan frame}. The reason behind this choice is twofold. On one hand, not all theories we will consider can be conformally mapped to the Einstein frame; on the other hand it is known that the conformal frame might hide some parts of the phase space, like in the case considered by \cite{Alho:2016gzi}. Readers interested in the treatment of modified gravity in the Einstein frame are referred, for example, to \citet{Avelino:2016lpj}.

Another important point to stress concerns the asymptotic analysis of the phase space. In contrast to the examples of perfect fluid GR based cosmologies (cf.~Sec.~\ref{sec:LCDM_dynam}), the phase space in modified gravity cannot always be described in a compact way. This has the important consequence that the phase space asymptotic (i.e.~the part of the phase space which corresponds to one or more dynamical system variables tending towards being infinite) might hide attractors of the phase space. In this respect therefore the analysis in a non-compact phase space cannot be considered complete if the asymptotic is not considered. An asymptotic analysis can however present technical problems on is own and, in spite of its importance is rarely considered. We will see  in Sec.~\ref{sec:f(R)} an example in which such analysis is relevant.

The reader will notice that, in contrast to the other sections of the review, this section is considerably less detailed. This is due to the fact that at present a great number of different modifications/extensions of General Relativity are under study and a complete presentation of all of them would go well beyond the scope of this review.  The analysis of the phase spaces of these theories is usually performed  using variables which are inspired by the ones used for the analysis of scalar field cosmologies (see Sec.~\ref{chap:scalarfields}). This strategy works reasonably well on the case of scalar-tensor gravity. However for more complex theories there seems to be  no universal recipe for the definition of dynamical system variables. As we will see it is rather the case that  a given set of variables can be optimised to uncover or describe a specific aspect of the cosmology of a given theory. Whenever possible, we will present sets of variables which are tailored specifically for cosmic acceleration including a discussion of their advantages and disadvantages. Even in this case, however, the search for the ``best'' set of  variables is still a matter of debate and we will indicate the cases where more than one set of options are available.

\subsection{Brans-Dicke theory}
\label{sec:BD_theory}

In this section we consider  a class of theories which introduce a scalar degree of freedom non-minimally coupled to the gravitational sector. In standard general relativity, matter fields are only coupled to gravity via the metric $g_{\mu\nu}$ and this ensures the validity of the strong equivalence principle. In the original idea of \citet{Brans:1961sx}, a non-minimal coupling was introduced to obtain a relativistic theory implementing Mach's principle. Not surprisingly it is known as {\it Brans-Dicke theory} and it is one of  the most studied modifications of Einstein gravity.

The action of Brans-Dicke theory is given by
\begin{equation}
	S_{\rm BD} = \int d^4x \sqrt{-g} \left[\frac{\phi}{2} R -\frac{\omega_{\rm BD}}{2\phi}\partial\phi^2 +\kappa^2 \mathcal{L}_{\rm m}\right] \,,
\end{equation}
where  $\mathcal{L}_{\rm m}$ represents the matter Lagrangian. The constant $\omega_{\rm BD}$ is called the {\it Brans-Dicke parameter}. In later generalisations of this theory  a self-interacting potential $V(\phi)$ for the scalar field  was introduced in the action. Here we will consider the extended action
\begin{equation}
	S_{\rm BD} = \int d^4x \sqrt{-g} \left[\frac{\phi}{2} R -\frac{\omega_{\rm BD}}{2\phi}\partial\phi^2 -V(\phi) +\kappa^2 \mathcal{L}_{\rm m}\right] \,,
	\label{def:BD_action}
\end{equation}
which for simplicity will also be referred to as Brans-Dicke theory.
In the presence of matter, Brans-Dicke theory reduces to general relativity in the limit $\omega_{\rm BD}\rightarrow\infty$, and from Solar System experiments the strong bound $\omega_{\rm BD}\gtrsim 10^4$ can be obtained (see e.g.~\citet{Bertotti:2003rm}). In the following we will also assume $\omega_{\rm BD}>0$, so that the Brans-Dicke field $\phi$ is non-phantom.

One interesting aspect of the action above is that the scalar field $\phi$  changes the effective Newton's gravitational constant. This implies that now the strength of the gravitational interaction depends on the value of the scalar field, which in turn can depend on the spacetime position.

The cosmological equations following from action (\ref{def:BD_action}), derived with a spatially flat FLRW metric, are \citep{Brans:1961sx}
\begin{align}
3 \phi H^2 +3H \dot\phi -\frac{\omega_{\rm BD}}{2}\frac{\dot\phi^2}{\phi}-V &=\kappa^2 \rho \,,\label{eq:Friedmann_BD}\\
2 \phi \dot H -H \dot\phi +\omega_{\rm BD}\frac{\dot\phi^2}{\phi}+\ddot\phi &= -\kappa^2 (1+w)\,\rho \,,\label{eq:acc_BD}\\
\ddot\phi +3H\dot\phi -\frac{2}{3+2 \omega_{\rm BD}} \left(2V-\phi V_{,\phi}\right) &= \frac{\kappa^2 (1-3w)}{3+2 \omega_{\rm BD}} \rho \,,\label{eq:KG_BD}
\end{align}
together with the standard conservation law for the matter fluid: $\dot\rho + 3 H \rho (1+w) = 0$.
Note that only three out of these four equations are independent, since one can always derives the fourth from the remaining three. In the above equations the ``dot" is the derivative with respect to cosmic time $t$, $V_{,\phi}$ denotes the derivatives of $V(\phi)$ with respect to $\phi$ and $w$ stands for the matter EoS parameter $p=w \rho$.

In order to recast Eqs.~(\ref{eq:Friedmann_BD})--(\ref{eq:KG_BD}) into a dynamical system we define the dimensionless variables \citep{Hrycyna:2013hla,Hrycyna:2013yia}
\begin{equation}
x=\frac{\dot\phi}{H\phi} \,,\qquad y=\frac{\sqrt{V}}{\sqrt{3\phi}H} \,, \qquad \lambda=-\phi \frac{V_{,\phi}}{V} \,,\qquad \tilde{\Omega}_{\rm m} = \frac{\kappa \rho}{3 \phi H^2} \,.	
\label{def:BD_EN_vars}
\end{equation}
Note that apart from factors of $\phi$ these are equivalent, up to some constants, to the standard EN variables (\ref{def:ENvars}).
The variable $y$ is real only for $V>0$. If $V$ is negative then the definition of $y$ must be slightly changed. In what follows we will assume that the potential does not change sign dynamically. As we will see, this is possible because $y=0$ is an invariant submanifold for the system. Furthermore, the variable $\tilde{\Omega}_{\rm m}$ is only defined positive if $\phi>0$. This assumption is not satisfied in general and it indicates the differences that are introduced by the non-minimal coupling present in the Brans-Dicke theory (cf.~the standard definition of $\Omega_{\rm m}$ given in Eq.~\eqref{eq:defxyOmega}). It should be noted however that assuming an ever attracting gravitational force implies $\phi>0$, and thus in these phenomenologically relevant cases this problem does not arise.

Using the variables (\ref{def:BD_EN_vars}), the cosmological equations (\ref{eq:Friedmann_BD})--(\ref{eq:KG_BD}) can be rewritten as the following dynamical system
\begin{align}
x' &= \frac{1}{2 (2\omega_{\text{BD}}+3)} \Big\{x^3 \omega _{\text{BD}} \left[1-(w-1) \omega _{\text{BD}}\right]+x^2 \left[(9 w-7) \omega _{\text{BD}}-6\right] \nonumber\\
&\qquad\qquad \qquad \quad -6 x \left[\omega _{\text{BD}}\left(w y^2-w+y^2+1\right)+3 w-\lambda  y^2\right] \nonumber\\
&\qquad \qquad \qquad \quad +6 \left[y^2 (2 \lambda +3 w+3)-3 w+1\right]\Big\} \,,\label{eq:x_BD}\\
y' &= -\frac{y}{2 (2 \omega _{\text{BD}}+3)} \Big\{(w-1) x^2 \omega _{\text{BD}}^2+3 \left(x+\lambda x-2 \lambda  y^2-4\right) \nonumber\\
&\qquad -\omega _{\text{BD}} \left[2x (3 w- \lambda +2)-6 (w+1) \left(y^2-1\right)+x^2\right]\Big\} \,,\label{eq:y_BD}\\
\lambda'&= \lambda  x \left[ 1- \lambda(\Gamma-1)\right] \,,\label{eq:lambda_BD}
\end{align}
where we have defined (cf.~Sec.~\ref{chap:scalarfields})
\begin{equation}
	\Gamma = \frac{V\,V_{,\phi \phi}}{V_{,\phi}^2} \,.\label{eq:Gamma}
\end{equation}
In the system  above  we have eliminated the equation for $\tilde{\Omega}_{\rm m}$ using the Friedmann constraint
\begin{equation}
  \tilde{\Omega}_{\rm m}= 1-\frac{ \omega _{\text{BD}}}{6}x^2+x-y^2= 1- \Omega_\phi  \,.
  \label{109}
\end{equation}
Note that Eq.~(\ref{109}) defines $\Omega_\phi$ and that its definition depends on the Brans-Dicke parameter $\omega_{\rm BD}$: if $\omega_{\rm BD}$ is positive the phase space is compact, while if $\omega_{\rm BD}$ is negative the phase space is always non-compact and requires further asymptotic analysis. As mentioned earlier we consider the first case here. The constraint~(\ref{109}) restricts the physical phase space to be inside an ellipse, rather than a circle encountered in previous sections.

Note also that this system presents two invariant submanifolds $y=0$ and $\lambda=0$ and therefore, strictly speaking, a global attractor can only have coordinates $y=0$ and $\lambda=0$, i.e.~the original Brans-Dicke model.  Eqs.~(\ref{eq:x_BD})--(\ref{eq:y_BD}) are invariant under a combination of the reflection $(x,y)\rightarrow (x,-y)$ and time reversal $t\rightarrow -t$. We can then only study positive values of $y$ since negative values would lead to the same qualitative dynamics. It is not too difficult to prove that this part of the phase space represents expanding cosmologies (from definitions \eqref{def:BD_EN_vars} one has $y>0$ for $H>0$).

The scalar field EoS ($w_\phi = p_\phi/\rho_\phi$)  is given by
\begin{equation}\label{wphigen}
	w_\phi =\frac{ x^2 \omega _{\text{BD}} \left(2 \omega _{\text{BD}}+3 w+2\right)-6 x \left(2
   \omega _{\text{BD}}+9 w\right)+18 y^2 \left(-2 \omega _{\text{BD}}+2 \lambda +3
   w\right)-54 w+18}{2 \left(2 \omega _{\text{BD}}+3\right) \left(x^2 \omega _{\text{BD}}-6x+6 y^2\right)}\,,
\end{equation}
while the effective EoS of the combined fluid ($w_{\rm eff} = (p + p_\phi)/(\rho + \rho_\phi)$) is
\begin{equation}
	w_{\rm eff} = w \left(1-\frac{1}{6} x^2 \omega _{\text{BD}}+x-y^2\right)+w_\phi \left(\frac{ \omega _{\text{BD}}}{6}x^2-x+y^2\right) \,.
\end{equation}
It is interesting to compare the above expressions with Eq.~\eqref{eq:EoS_phi_xy}. The latter is such that when the energy density of the scalar field is zero $(x=0,y=0)$, $w_\phi$ converges to a constant. This is not the case for \eqref{wphigen}, which in fact diverges when the energy density of the scalar field is zero. As we will see this can happen along an orbit much in the same way as the case of phantom dark energy.
The number of orbits that present this behaviour decreases as $\omega _{\text{BD}}$ grows. In fact, in the limit $\omega _{\text{BD}}\rightarrow \infty$ Eq.~\eqref{wphigen} reduces to an expression with the same properties as Eq.~\eqref{eq:EoS_phi_xy}.

For a given $V(\phi)$, Eqs.~(\ref{eq:x_BD})--(\ref{eq:Gamma}) allows us to obtain the fixed points and then the deceleration parameter $q$ determines whether or not the scale factor is accelerating, giving by
\begin{equation}
  \label{qsurface}
  -1-\frac{\dot{H}}{H^2}=q=\frac{6 \lambda y^2+  \omega x [x (\omega +1)-2]-6 \omega(y^2+1)+12}{4 \omega +6}\,.
\end{equation}
Setting $q=0$ we obtain a surface in the phase space which represents the transition between accelerated and decelerated cosmology. An interesting approach to constraining models of dark energy has been proposed in Chapter 2 of \citet{Amin:2008nsa}, in which a tower of first order differential equation relationships between dimensionless kinematic variables like $q(t)$ (deceleration) and the Jerk parameter $j(t)$ lead to a closed system that flows to various fixed points. For example, $H=q=j=1$ is a natural fixed point corresponding to exponential expansion.

Eqs.~(\ref{eq:x_BD})--(\ref{eq:lambda_BD}) do not form an autonomous system of equations unless $\Gamma$ can be written as a function of $\lambda$. Again the situation here is similar to the one we encountered in Sec.~\ref{sec:other_potentials} and an analysis for a general $\Gamma(\lambda)$ could be performed. In the following, however, we will focus on a simple model in which the potential corresponds to
\begin{equation}
  V(\phi) = V_0\, \phi^{-2n} \,,
  \label{def:power_law_pot_BD}
\end{equation}
which implies $\lambda=2n= const$\footnote{Note the difference between this case and the minimally coupled quintessence model of Sec.~\ref{sec:other_potentials}. In that case an exponential potential would return the simplest 2D dynamical system, whereas in this case it is a power-law potential.}. For the potential (\ref{def:power_law_pot_BD}) and considering only the case $w=0$, Eq.~(\ref{eq:lambda_BD}) vanishes identically so that we are left with a two dimensional phase space  described by the system
\begin{align}
x' &= \frac{1}{2 (2\omega_{\text{BD}}+3)} \Big\{x \left[x^2 \omega _{\text{BD}} \left(\omega _{\text{BD}}+1\right)-x \left(7
   \omega _{\text{BD}}+6\right)-6 \left(2n- \omega _{\text{BD}}\right)y^2+ \omega _{\text{BD}}\right]\nonumber\\
   \qquad&+6 [4 n+3] y^2+6\Big\} \,,\label{eq:x_BD_ex}\\
y' &= -\frac{y}{2 (2 \omega _{\text{BD}}+3)} \Big[6 y^2
   \left(2n-\omega _{\text{BD}}\right)+x^2 \omega
   _{\text{BD}} \left(\omega _{\text{BD}}+1\right)-4x(n+1) \omega _{\text{BD}} \nonumber\\
&\qquad + 3x (2n+1)+6\left(\omega _{\text{BD}}+2\right)\Big] \,.\label{eq:y_BD_ex}
\end{align}
Note that in this case only the invariant submanifold $y=0$ exists. Therefore the only possible global attractor must have $y=0$.

The critical points of the dynamical system (\ref{eq:x_BD_ex})--(\ref{eq:y_BD_ex}) are listed in Tab.~\ref{tab:cp_BD} together with their phenomenological quantities. There can be up to five critical points in the phase space. Two of them ($A_\pm$  and $B$) represent scalar field dominated solutions. The others represent an interplay between scalar field and matter. Points $A_\pm$  and $C$, represent states in which the kinetic part of the scalar field is dominant. Note that there is no potential dominated fixed point. All the points have phenomenological and stability properties dependent on the parameters $\omega_{\rm BD}$ and $n$.

\begin{figure}
\centering
\includegraphics[width=0.7\columnwidth]{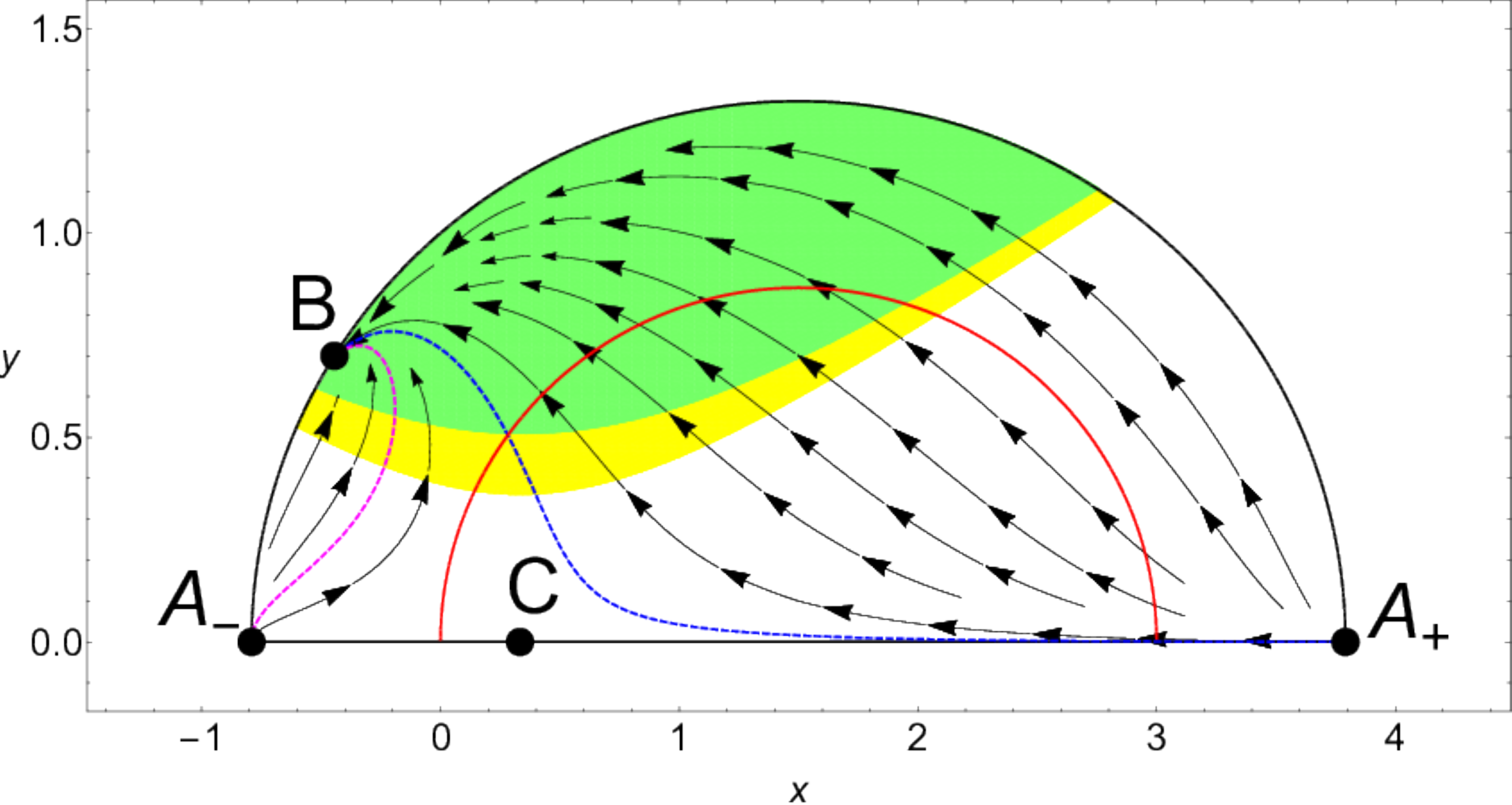}
\caption{Phase space for Brans-Dicke cosmology with a power-law potential and values $w=0$, $\lambda=-1$ and $\omega_{\rm BD}=2$. The red circle represents the divergence of the  EoS $w_\phi$ of the scalar field. The blue dashed line represents an example of orbit along which divergencies of $w_\phi$ occur, while the magenta dashed line represents example of orbit along which $w_\phi$ is regular. The yellow region denotes quintessence behaviour ($-1<w_{\rm eff}<-1/3$), while the green region denotes phantom behaviour ($w_{\rm eff}<-1$). Note that the value of $\omega_{\rm BD}$ is well outside of the current bounds and therefore this can only be considered a toy model given for illustrative purposes. (Note the scaling of the axes which distorts the ellipse into the shape of a circle.)
}
\label{fig:BD_phase_space}
\end{figure}

The solutions for the scale factor corresponding to the fixed points can be obtained from Eq.~(\ref{qsurface}) for the specific case of the potential (\ref{def:power_law_pot_BD}). The results are also summarised in Tab.~\ref{tab:cp_BD}.  An example of the structure of the phase space can be seen in Fig.~\ref{fig:BD_phase_space}.  In the following,  we limit ourselves to summarise the most important results.

\begin{figure}
\centering
\includegraphics[width=0.7\columnwidth]{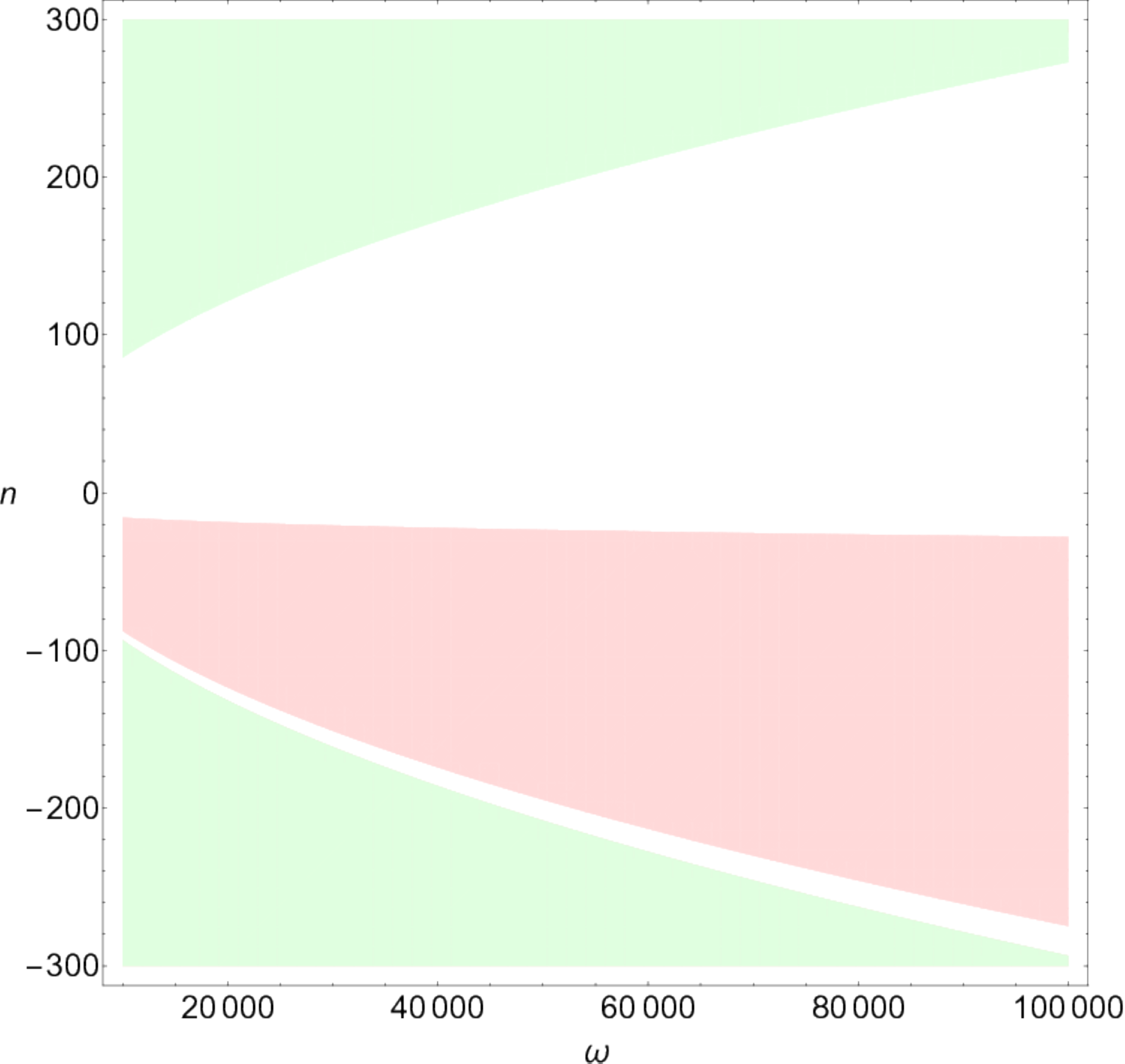}
\caption{A visual representation of the stability of Point~$D$ of the system \rf{def:power_law_pot_BD} in terms of the parameters $\omega=\omega_{\rm BD}$ and $n$. In red the parameter space in which the point is a repeller and in green the ones in which it is an attractor. }
\label{fig:DBStabPointD}
\end{figure}

The complete stability analysis of the fixed points is given in Tab.~\ref{tab:stab_BD} and Fig.~\ref{fig:DBStabPointD}.
We see that, although in principle all the fixed points, apart $A_\pm$,  could represent accelerated expansion attractors, in our hypothesis this can happen only for Point~$B$.
In fact, the other points which have the desired stability properties can never represent an accelerated expansion. Point~$B$ represents also a scaling solution. It is important to stress that Point~$B$ has $y>0$ and therefore cannot be considered a true global attractor due to the presence of the invariant submanifold $y=0$.  For the values of the parameters for which Point~$B$ is a repeller, Brans-Dicke theory can be used to model the graceful exit of Brans-Dicke inflation.

In conclusion  Brans-Dicke theory with a power-law potential can be used to study graceful exit scenarios,  and the onset of a dark era. However these two results are mutually exclusive: one can use the theory only to describe one of them. This is not an uncommon feature in these models, as well as in some other scalar-tensor theories of gravity which we will examine in the next section.

The case we have illustrated shows an example of the possible dynamical behaviour of a Brans-Dicke cosmology. It is clear, however, that  a complete study for different potentials or a general matter sector would return a richer and more complicated phenomenology.  For example, \citet{Kolitch:1994kr} considered the case of Brans-Dicke theory with a cosmological constant showing (perhaps not surprisingly) that a de Sitter solution  can exist if  $\omega_{\rm BD}>0$.

A detailed dynamical systems analysis for Brans-Dicke cosmology with a quadratic potential, given by (\ref{def:power_law_pot_BD}) with $n=-1$, has been delivered by \citet{Hrycyna:2013yia}. They showed that de Sitter solutions can be obtained in this special case, and extended the work to non-compact phase spaces ($\omega_{\rm BD}<0$). They also performed the analysis at infinity, and considered general values of the matter EoS parameter $w$ outside the physically allowed range $[0,1/3]$. The same authors studied de Sitter solutions for Brans-Dicke cosmology with a general potential, performing the stability analysis \citep{Hrycyna:2013hla} and comparing the results with observational data \citep{Hrycyna:2014cka}.  On the other hand, \citet{Garcia-Salcedo:2015naa} showed that the presence of a de Sitter attractor does not necessarily imply that the cosmology will be $\Lambda$CDM in a neighbourhood of this point.

The presence of a non-trivial matter field can change the features of the late time asymptotics. For example, \citet{Cid:2015pja} showed that in the case of the presence of a second, minimally coupled scalar field one can prove that a de Sitter solution is not the asymptotic state for the cosmological model.

More exotic cases have also been considered. For example, \citet{Farajollahi:2011zz} and \citet{Liu:2012kha} added an interaction between the scalar field and the matter sector depending on $H$ and proved the existence of late time accelerated solutions.

\begin{table}
\begin{center}
\resizebox{15cm}{!}{\begin{tabular}{|c|c|c|c|c|c|c|}
 \hline
 P & $\{x,y\}$ & $w_{\rm eff}$ & $w_{\phi }$ & $\tilde{\Omega}_{\rm m}$ & $a=a_0(t-t_0)^\alpha$ \\
 \hline & & & & & \\
 \multirow{2}*{$A_\pm$} & \multirow{2}*{$\left\{\displaystyle\frac{3\pm\sqrt{6 \omega +9}}{\omega },0\right\}$} & \multirow{2}*{$\displaystyle\frac{3 \omega+6 \pm2 \sqrt{6 \omega +9}}{ \omega }$} & \multirow{2}*{$\displaystyle\frac{3 \omega +6\pm2 \sqrt{6 \omega +9}}{\omega }$} & \multirow{2}*{0} & \multirow{2}*{$\alpha=\displaystyle\frac{\omega }{3 \omega \pm\sqrt{6 \omega +9}+3}$} \\ & & & & &\\ & & & & & \\
 \multirow{2}*{$B$} & $\left\{\displaystyle\frac{4 (n +1)}{1-2n +2 \omega },\right.$ & \multirow{2}*{$\displaystyle\frac{4 (n+1) (2 n+1)}{2 \omega -2 n+1}-3$} & \multirow{2}*{$\displaystyle\frac{4 (n+1) (2 n+1)}{2 \omega-2 n+1}-3$} & \multirow{2}*{0}& \multirow{2}*{$\alpha=\displaystyle\frac{1-2n+2 \omega }{4n^2+6 n+2}$} \\ &  $\left.\displaystyle\frac{\sqrt{(2 \omega +3) (-4n (n+2)+6 \omega +5)}}{\sqrt{3} \left| 2n-2 \omega -1\right|}\right\}$ & & & &\\& & & & &\\
 \multirow{2}*{$C$} & \multirow{2}*{$\left\{\displaystyle\frac{1}{\omega +1},0\right\}$} & \multirow{2}*{$-\displaystyle\frac{6 \left(\omega+1\right)}{5 \omega +6}$} & \multirow{2}*{$\displaystyle\frac{1}{ \omega +1}$} & \multirow{2}*{$\displaystyle\frac{(2 \omega +3) (3 \omega +4)}{6 (\omega +1)^2}$} & \multirow{2}*{$\alpha=\displaystyle\frac{2 (\omega +1)}{3\omega +4}$} \\ & & & & & \\ & & & & & \\
 \multirow{2}*{$D$} & \multirow{2}*{$\left\{\displaystyle\frac{3}{2 n},\frac{\sqrt{3 (\omega +1)-2 n}}{2 \sqrt{2} \left| n\right|}\right\}$} & \multirow{2}*{$\displaystyle\frac{12 n}{6 \omega -14
   n+3}$} & \multirow{2}*{$\displaystyle\frac{3}{2n }$} & \multirow{2}*{$\displaystyle\frac{2 n (4 n+7)-6 \omega
   -3}{8 n^2}$}& \multirow{2}*{$\alpha=\displaystyle\frac{4 n}{3 (2n+1)}$} \\ & & & & &\\
 \hline
\end{tabular}}
\end{center}
\caption{Critical points of the Brans-Dicke model with power-law potential \rf{def:power_law_pot_BD} and $w=0$ (dust). Here $\omega=\omega_{\rm BD}$.}
\label{tab:cp_BD}
\end{table}

\begin{table}
\begin{center}
\begin{tabular}{|c|c|c|c|c|}
 \hline
 P & Attractor & Repeller  \\
 \hline
{$A_+$} & {never} &  $\omega >0\land 2n+2+\sqrt{6 \omega +9}>0$  \\ &&\\
{$A_-$} & {never} &  $\omega >0\land 2n+2<\sqrt{6 \omega +9}$  \\ &&\\
\multirow{6}*{$B$} &$n\leq -\frac{5}{2}\land \omega >\frac{1}{6} \left(4 n^2+8 n-5\right)$ &  \multirow{4}*{$n<-\frac{5}{2}\land 0<\omega <\frac{1}{6} \left(4 n^2+8 n-5\right)$} \\
 &$-\frac{5}{2}<n<-1\land \omega >0$& \multirow{4}*{$n>\frac{1}{2}\land \frac{1}{2} (2 n-1)<\omega <\frac{1}{6} \left(4 n^2+8 n-5\right)$}\\
  & $ -\frac{1}{2}<n\leq \frac{1}{2}\land \omega >0$ &\\
 & $n>\frac{1}{2}\land \omega >\frac{1}{6}\left(4 n^2+8 n-5\right)$ &\\
  & $\omega >0\land -1<n<-\frac{1}{2}$ &\\
    & $\omega >0\land  n>\frac{1}{2} (2 \omega +1)$ &\\
 & &  \\
 \multirow{2}*{$C$} & $\omega >0\land n>\frac{3}{2} (\omega +1)$ & never  \\ & &  \\
  &&\\
 \hline
\end{tabular}

\end{center}
\caption{Stability of the critical points of the Brans-Dicke model with power-law potential \ref{def:power_law_pot_BD} and $w=0$. Here $\omega=\omega_{\rm BD}$. The stability of  Point~$D$ is represented graphically in Fig.~\ref{fig:DBStabPointD}.}
\label{tab:stab_BD}
\end{table}

\subsection{Scalar-tensor theories}
\label{sec:ST_theories}

A straightforward generalisation of Brans-Dicke theory is given by a class of gravitational theories known as {\it scalar-tensor theories}. The general action for these models is given by
\begin{equation}
	S_{\rm ST} = \int d^4x \sqrt{-g} \left[ \frac{F(\phi)}{2}R - \frac{\omega(\phi)}{2}\partial\phi^2 -V(\phi) +\kappa^2 \mathcal{L}_{\rm m} \right] \,,
	\label{def:ST_action}
\end{equation}
where now $F$, $\omega$ and $V$ are all arbitrary functions of $\phi$. A redefinition of $\phi$  could allow us to reduce the function $\omega(\phi)$  to a constant. In this way the scalar field kinetic term can be recast into its canonical form. In what follows, however, we will consider $\omega$ as a general function of $\phi$ in order to directly include all possible scalar-tensor models. An interesting property of this class of theories was pointed out by \citet{Damour:1993id,Santiago:1998ae,Mimoso:1998dn}. These papers show that in a wide variety of conditions scalar-tensor theories `relax' towards General Relativity in the sense that the scalar field tends to a constant asymptotically in time.

The cosmological equations derived from action (\ref{def:ST_action}) with a spatially flat FLRW metric are
\begin{align}
3 H^2 F +3 H \dot\phi  F_{,\phi} -\frac{1}{2} \omega \dot\phi^2 -V &=\kappa ^2 \rho \,,\label{eq:ST_Friedmann}\\
2 F \dot H+\dot\phi^2 F_{,\phi\phi} +(\ddot\phi-H \dot\phi) F_{,\phi} +\omega \dot\phi^2 &=-\kappa ^2 \rho (1+w) \,,\label{eq:ST_acc}\\
\ddot\phi+3 H \dot\phi+\frac{1}{\omega} \left[V_{,\phi}+\frac{\dot\phi^2}{2} \omega_{,\phi}-3  F_{,\phi} \left(2 H^2+\dot H\right)\right] &=0\,\label{eq:ST_KG}
\end{align}
together with the standard conservation equation for the matter fluid, which can also be derived from the equations above. Note that if one replaces $H^2$ and $\dot H$ in Eq.~(\ref{eq:ST_KG}) using Eqs.~(\ref{eq:ST_Friedmann}) and (\ref{eq:ST_acc}), then an explicit coupling between the scalar field and the matter sector appears. This situation is equivalent to Brans-Dicke theory (Sec.~\ref{sec:BD_theory}). The reader interested in reviews on scalar-tensor theories, including their applications to cosmology, can refer to the books by \citet{Fujii:2003pa} and \citet{Faraoni:2004pi}, as well as the review articles of  \citet{Clifton:2011jh} and \citet{Joyce:2014kja}.

In order to rewrite the cosmological equations (\ref{eq:ST_Friedmann})--(\ref{eq:ST_KG}) as a dynamical system, we introduce the following dimensionless variables
\begin{align}
x = \frac{\dot\phi}{H}\sqrt{\frac{\omega}{F}}\,,\quad y = \frac{1}{H}\sqrt{\frac{V}{3F}} \,, \quad \tilde{\Omega}_{\rm m} = \frac{\kappa^2 \rho}{3 F H^2} \,,
\label{110}
\end{align}
plus
\begin{align}
	\lambda_F &= -\frac{F_{,\phi}}{F}\sqrt{\frac{F}{\omega}}\,, & \lambda_V &= -\frac{V_{,\phi}}{V}\sqrt{\frac{F}{\omega}}\,, & \lambda_\omega &= -\frac{\omega_{,\phi}}{\omega}\sqrt{\frac{F}{\omega}} \,,\label{111}\\
	\Gamma_F &= \frac{F\,F_{,\phi \phi}}{F_{,\phi}^2} \,,& \Gamma_V &= \frac{V\,V_{,\phi \phi}}{V_{,\phi}^2} \,,& \Gamma_\omega &= \frac{\omega\,\omega_{,\phi \phi}}{\omega_{,\phi}^2} \,.\label{112}
\end{align}
Note that these variables are well defined only if $F/\omega>0$ and $V>0$. Therefore the variables (\ref{110})--(\ref{112}) can only be used if $F>0$ (always attractive gravitational force), $\omega>0$  (non-phantom scalar field) and $V>0$  (positive potential energy) for every value of $\phi$.

The variables (\ref{110})--(\ref{112}) are a straightforward generalisation of the variables used in Sec.~\ref{sec:BD_theory}. In fact if we take $F=\phi$ and $\omega=\omega_{\rm BD}/\phi$, corresponding to Brans-Dicke theory, the variables (\ref{110})--(\ref{112}) reduce, up to a constant factor, to the variables (\ref{def:BD_EN_vars}).

The cosmological equations (\ref{eq:ST_Friedmann})--(\ref{eq:ST_KG}) can now be rewritten as
\begin{align}
x' &= \nonumber \frac{1}{6 \lambda _F^2+4}\left\{x^3 \left[\left(2 \Gamma _F \lambda _F- \lambda _{\omega }\right) \lambda _F-w+1\right]+x^2 \lambda _F \left[3 \lambda _F \left(2 \Gamma _F \lambda
   _F+\lambda _F-\lambda _{\omega }\right)-9 w+7\right]\right.\\
   &\nonumber~~~-x \left[6 \left(1-w+3 w \lambda _F^2\right)+6 y^2 \left(\lambda _F \lambda _V+w+1\right)\right]+y^2 \left(12 \lambda _V-18 (w+1) \lambda _F\right)\\
   &~~~\left.+6 (3 w-1) \lambda _F \right\} \,,\label{eq:x_ST}\\
y' &=\nonumber \frac{y}{6 \lambda _F^2+4} \Big[6
   \left(2 \lambda _F^2+w+1\right)+x \left[\lambda _F \left(3 \lambda _F^2-6 w+4\right)-\left(3 \lambda _F^2+2\right) \lambda _V\right]\\
&~~~-x^2 \left[\lambda _F \left(\lambda _{\omega }-2 \Gamma _F \lambda _F\right)+w-1\right]-6 y^2 \left(\lambda _F \lambda _V+w+1\right)\Big] \,,\label{eq:y_ST}\\
\lambda_F' &= \frac{1}{2} x \lambda _F \left[\left(1-2 \Gamma _F\right) \lambda _F+\lambda _{\omega }\right] \,,\label{eq:LF_ST}\\
\lambda_V' &= -\frac{1}{2} x \lambda _V \left[\lambda _F-\lambda _{\omega }+2 \left(\Gamma _V-1\right) \lambda _V\right] \,,\label{eq:LV_ST}\\
\lambda_\omega' &= -\frac{1}{2} x \lambda _{\omega } \left[\left(2 \Gamma _{\omega }-3\right) \lambda _{\omega }+\lambda _F\right] \,,\label{eq:LO_ST}
\end{align}
with the Friedmann constraint
\begin{equation}
	\Omega_\phi = x \lambda _F+\frac{x^2}{6}+y^2=1-\tilde{\Omega}_{\rm m} \leq 1 \,.
	\label{113}
\end{equation}
Eqs.~(\ref{eq:x_ST})--(\ref{eq:LO_ST}) do not represent an autonomous system of equations unless the variables $\Gamma_F$, $\Gamma_V$ and $\Gamma_\omega$ can be written as functions of $\lambda_F$, $\lambda_V$ and $\lambda_\omega$. If this is the case then they constitute an autonomous 5D dynamical system. As in the case of the system (\ref{eq:x_BD})--(\ref{eq:lambda_BD}) these equations present a number of invariant submanifolds (e.g.~$y=0,\lambda_V=0, \lambda_F=0,\lambda_\omega=0$) and therefore no global attractor can exist which is not characterised by $y=0$, $\lambda_V=0$, $\lambda_F=0$, $\lambda_\omega=0$.

The scalar field EoS is given by
\begin{eqnarray}\label{wphiST}
	w_\phi &=& \frac{1}{\left(3 \lambda _F^2+2\right) \left(6 x \lambda _F+x^2+6 y^2\right)}\left\{2 \left(5 x^2-6 y^2\right)-6 \lambda _F \left[6 y^2 \lambda _V+x \left(x \lambda _{\omega }-6\right)\right]-18 x \lambda _F^3\right. \nonumber\\
	&&\left. 3 \lambda _F^2 x^2 \left[4 \Gamma _F+3 (w+1) \omega
   -1\right]+3 \lambda _F^2 \left[6 (3 w+2) y^2-18  (w+x+x w)+6\right]\right\} \,,
\end{eqnarray}
while the EoS of the universe is
\begin{equation}
	w_{\rm eff} = w \left(1-x \lambda _F-\frac{x^2}{6}-y^2\right)+w_\phi \left(1-x \lambda _F-\frac{x^2}{6}-y^2\right)\,.
\end{equation}
As in the Brans-Dicke case, Eq.~\eqref{wphiST} can diverge. This time, however, the denominator depends on $\lambda _F$ and, unless this variable is constant, there is no way to avoid that along the phase space orbits a singularity of $w_\phi$ will occur.

Eqs.~(\ref{eq:x_ST})--(\ref{eq:LO_ST}) can be the starting point for a general dynamical systems study of scalar-tensor cosmology. One can in fact assume $\Gamma_F$, $\Gamma_V$ and $\Gamma_\omega$ to be arbitrary functions of $\lambda_F$, $\lambda_V$ and $\lambda_\omega$ and then carry on a similar analysis to the one we performed in Sec.~\ref{sec:other_potentials}.
Such thorough analysis has never been performed in the literature and will be left for future investigations.
In what follows we will focus on a simple example with a suitable choice for the variables $\Gamma_F$, $\Gamma_V$ and $\Gamma_\omega$.

The dimension of the system can be reduced if one makes specific choices of the functions that characterise the theory. For example, in the case of a Brans-Dicke theory with a power-law potential, the variables $\lambda_F$, $\lambda_\omega$, $\Gamma_F$ and $\Gamma_\omega$ would be all constant and the system would be 3D (cf.~Sec.~\ref{sec:BD_theory}). Other interesting examples, include the case $F\propto\phi^m$, $V\propto\phi^n$ and $\omega\propto\phi^{m+2}$ (2D) or $F\propto\omega$ (3D).

Here we will consider briefly the case in which these functions are all exponentials i.e.
 \begin{equation}
 F= F_0 e^{m \phi},\quad \omega= \omega_0 e^{m \phi},\quad  V= V_0 e^
{n \phi},
 \end{equation}
where real $m$ and $n$ and $m\neq n$. This action corresponds also to the tree level action of string theory when the Kalb-Ramond fields are ignored  and an exponential potential is introduced (see Sec.~\ref{sec:string_and_brane} for a treatment of string inspired cosmological models).
Note that this particular case reduces the quantities $\lambda_F$, $\lambda_V$ and $\lambda_\omega$ to be all constants, and consequently the system \eqref{eq:x_ST}--\eqref{eq:LO_ST} to be 2D.

\begin{figure}
\centering
\includegraphics[width=0.7\columnwidth]{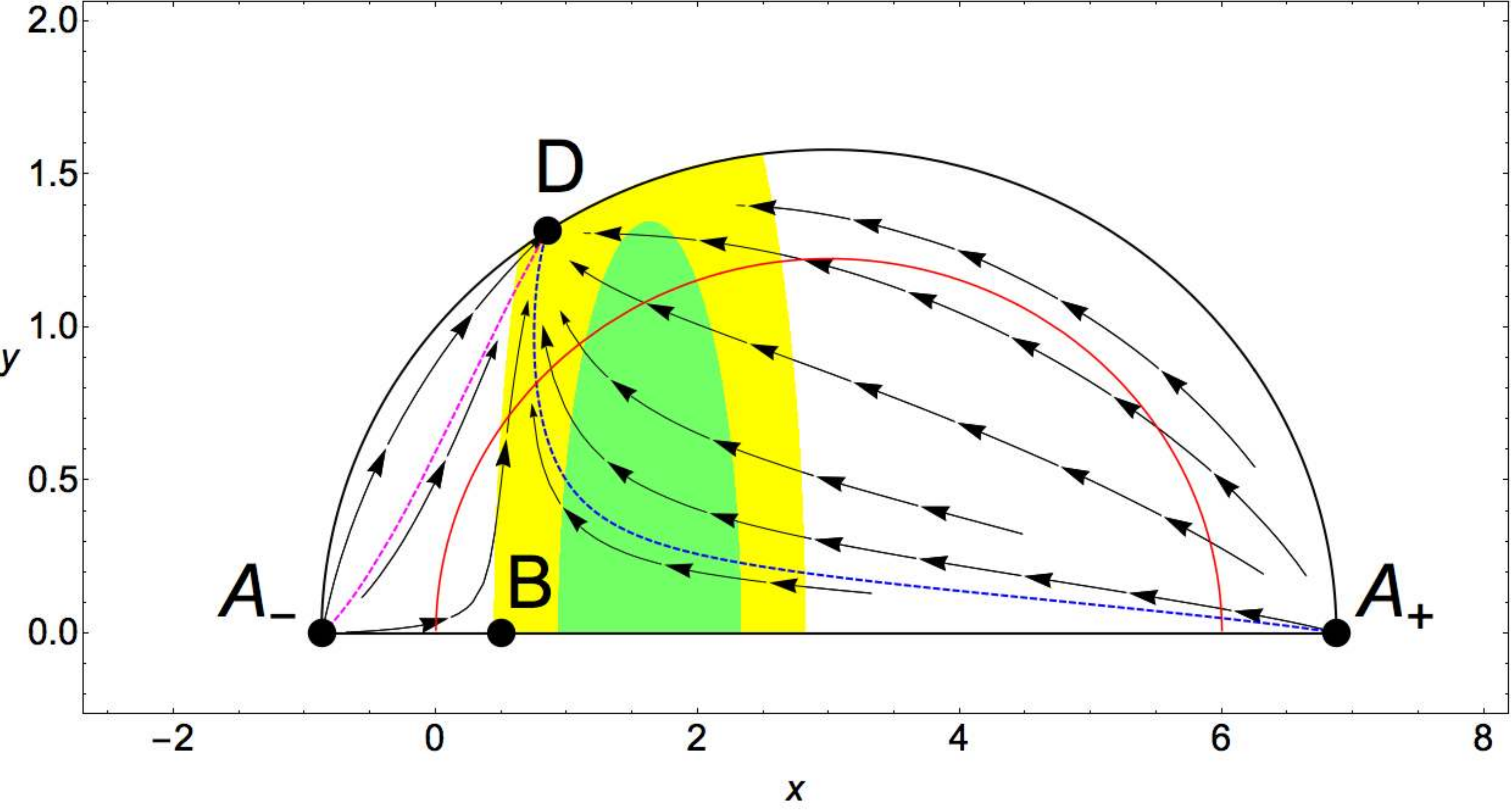}
\caption{Phase space for scalar-tensor cosmology described by Eq.~(\ref{eq:Sc-Tn_Dyn_Ex}) with  $F= F_0 e^{m \phi}$,\ $\omega= \omega_0 e^{m \phi}$, $V= V_0 e^{n \phi}$ and  $m=1$, $n=1/2$ and $w=0$. Point~$D$ is the future attractor with $-1<w_{\rm eff}<-1/3$ (yellow region), while the green region represents phantom acceleration.
The red circle represents the points of the phase space in which $w_\phi$ diverges. }
\label{fig:sctn_phase_space}
\end{figure}

\begin{figure}
\centering
\includegraphics[width=0.6\columnwidth]{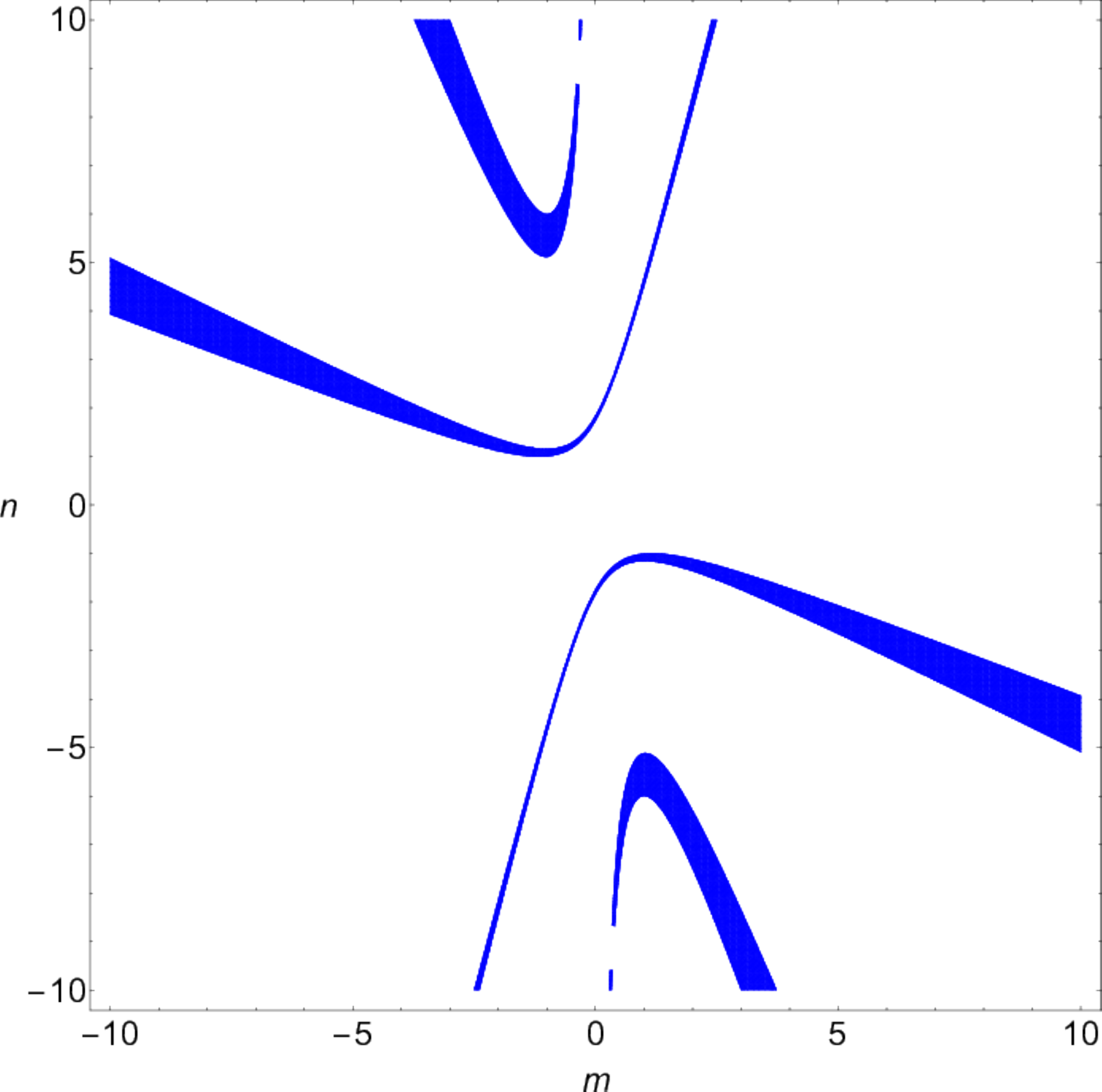}
\caption{Stability analysis of Point~$C$ in the phase space of the dynamical system \eqref{eq:Sc-Tn_Dyn_Ex}.  The region of the parameter space for which Point~$C$ is an attractor is in blue.
}
\label{fig:sctn_RegionC}
\end{figure}

\begin{figure}
\centering
\includegraphics[width=0.6\columnwidth]{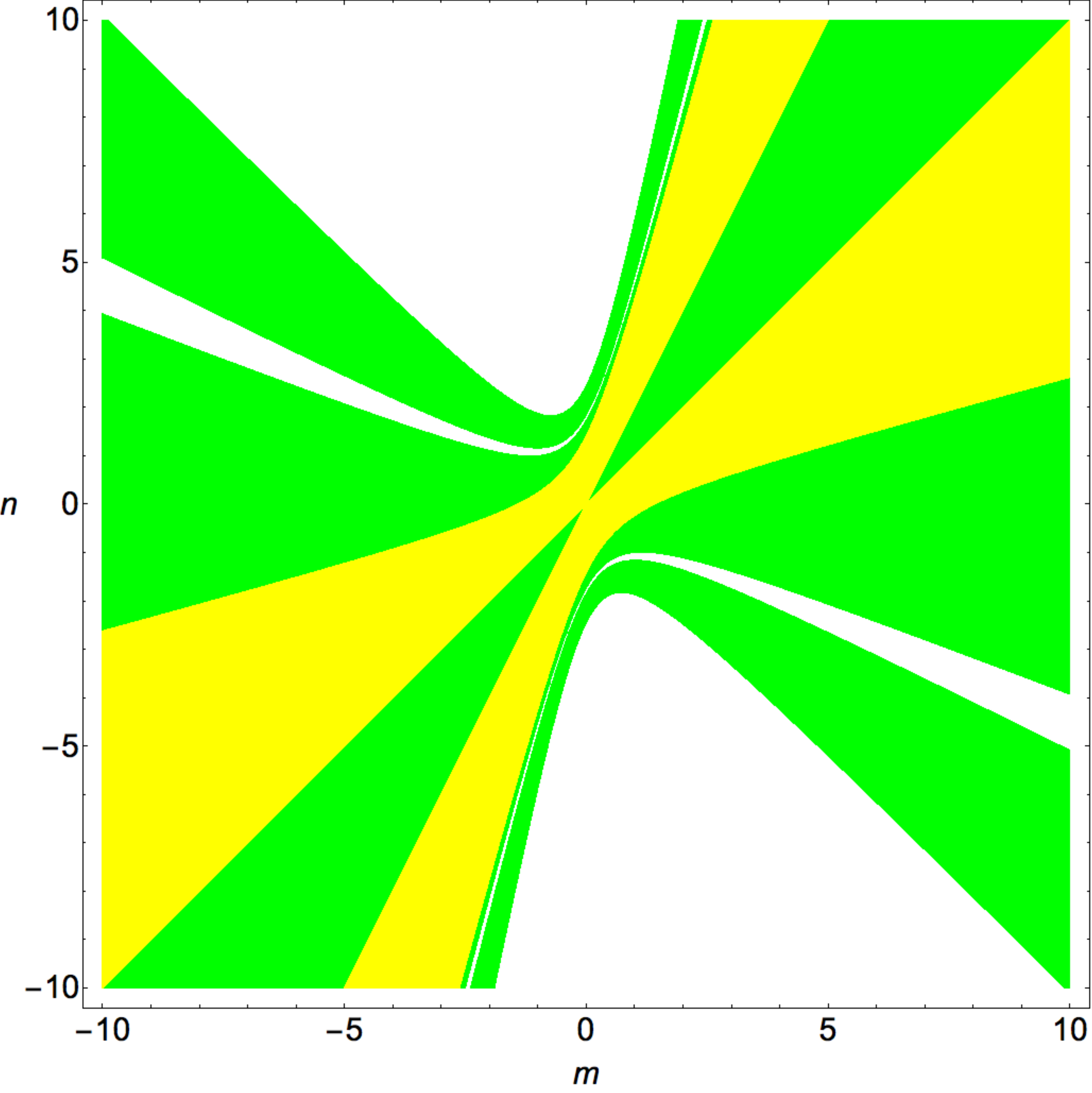}
\caption{Stability analysis of Point~$D$ in the phase space of the dynamical system \eqref{eq:Sc-Tn_Dyn_Ex}. The region of the parameter space for which Point~$D$ is an attractor is represented in green, while the yellow region the set of parameters for which point $D$ can be at the same time an attractor and represent accelerated expansion.}
\label{fig:sctn_RegionD}
\end{figure}

The dynamical systems equations read, considering only pressureless matter ($w=0$),
\begin{equation}
\begin{split}\label{eq:Sc-Tn_Dyn_Ex}
x' &= \frac{1}{6 m^2+4} \left\{18 m y^2+6 m-12 n y^2-6x \left[( m n+1) y^2+1\right]\right.\\
&~~~~\left.+\left(-6 m^2-7\right) mx^2+\left(m^2+1\right) x^3\right\},\\
y' &= \frac{y}{3 m^2+2}\left\{-3 m^3 x+m^2 \left(3 n x+x^2+12\right)-3 y^2 (m n+1)-4 m x+2 n x+x^2+6\right\}\,.
\end{split}
\end{equation}
As in the case of Brans-Dicke gravity, these equations present an invariant submanifold in $y=0$ and so, as we have anticipated, only points with $y=0$ can be true global attractors for the system.
Proceeding with the analysis, the system presents five fixed points listed in Tab.~\ref{tab:cp_SCTN}. Points~$C$ and $D$ only exist for specific values of the parameters $m$ and $n$. The fixed points all correspond to power law solutions which can be both decelerating and accelerating, although in the case of Points~$C$ and $D$  there are only specific intervals in which the points exist and represent accelerated expansion. The stability analysis, shown in Tab.~\ref{tab:stab_SCTN} and Fig.~\ref{fig:sctn_RegionC} and \ref{fig:sctn_RegionD},  indicates that only point $D$ can be at the same time an attractor and represent accelerated expansion. This happens in a region of the $(m,n)$ parameter space which is given in Fig.~\ref{fig:sctn_RegionD}. Therefore, a phase space attractor exists which can represent cosmic acceleration and it is a scaling solution. Again, since the coordinate $y$ of this point is not zero this attractor can never be the global one.  A plot of the phase space in the case $n=1/2$, $m=1$ is given in Fig.~\ref{fig:sctn_phase_space}.

Dynamical systems analyses of scalar-tensor cosmology have so far been considered in the literature only for specific models. \citet{Uzan:1999ch} examined scaling solutions and their stability in the model $\omega=1/2$ with both exponential and power-law potentials.
\citet{Gunzig:2000ce} showed that chaotic behaviour never arises for the non-minimal coupling $F(\phi)=1-\xi \phi^2$, with $\xi$ a constant, and subsequently \citet{Faraoni:2006sr} extended the same results to more general scalar-tensor models.

The model $F=1-\xi \phi^2$ has also been studied by \citet{Szydlowski:2008in}, who found new accelerated solutions for a general self-interacting scalar field potential.
\citet{Hrycyna:2010yv} further assumed $\omega=1/2$ and delivered an analysis for general potentials similar to the one we considered in Sec.~\ref{sec:other_potentials}. They found critical points corresponding to radiation, dark matter as well as dark energy dominated solutions and discussed the viability of this scalar-tensor theory as a dark energy model. Recently \citet{Szydlowski:2013sma} reconsidered the same model with a general potential and performed the analysis at infinity with a geometrical interpretation of the phase space of scalar-tensor theories.
The model $F=\xi\phi^2$ and $\omega=1/2$, which is equivalent to Brans-Dicke theory after a redefinition of the scalar field, has been studied by \citet{Maeda:2009js}, who focused their discussion on future cosmological attractors, and by \citet{CervantesCota:2010cb}, who analysed specific dark energy applications including interactions with the matter sector. The theory $F=\xi \phi^2$, $\omega=1/2$ and $V=V_0 \phi^n$ has been studied, in full detail including the behaviour at infinity, by \citet{Carloni:2007eu}.
A model with the same function $F$ but with $\omega=\varepsilon /2$  and a constant potential was similarly considered including the asymptotics by  \citet{Hrycyna:2015eta} using a different set of variables.

\citet{deSouza:2005ig} considered the linear coupling $F=\phi$ plus a linear or a quadratic self interacting potential, while \citet{Jarv:2010zc} assumed the same linear coupling but delivered a general work on potential dominated cosmological solutions. Finally \citet{Skugoreva:2014gka} looked at the model with power-law functions for both $F$ and $V$, while \citet{Agarwal:2007wn} performed a general analysis of scalar-tensor cosmological dynamics in both the Jordan and Einstein frame, finding transient and stable accelerated solutions.

More exotic scalar field theories include the non-minimal derivative coupling in which a scalar field is coupled (kinetically) to the Einstein tensor, to the Gauss-Bonnet term and to the matter sector. \citet{Mueller:2012kb} performed an analysis of  the phase space of this class of theories finding dynamical attractors for the cosmic evolution and cross-correlating with observations. More complicated non-minimally coupled theories have also been considered. \cite{Huang:2014awa}, assuming an exponential energy potential, studied the dynamical system of a scalar-tensor theory with an additional coupling between the Einstein tensor and the derivatives of the scalar field proportional to $G^{\mu\nu}\partial_{\mu}\phi\partial_{\nu}\phi$.
Within the same theory, \cite{Matsumoto:2015hua} studied the case of a Higgs-like potential showing that this kind of coupling can describe the acceleration of the universe. \cite{Granda:2016etr} studied a further generalisation introducing an additional coupling of the type $F_2(\phi)\mathcal{G}$ where $\mathcal{G}$ is the Gauss-Bonnet higher order term (see Sec.~\ref{otherhigher} for more details about this term). They also found a late-time accelerated attractor in this model.

In the context of Modified Gravity models  (MOG), \cite{Moffat:2005si} and \cite{Jamali:2016zww} studied the cosmology of a scalar-vector-tensor theory. In this theory, in addition to the metric tensor, there is also one additional vector field and two scalar fields. They studied the question of whether these additional fields could describe an accelerating expansion. Using dynamical system techniques, they showed that this theory describes  two radiation, two  matter and then two late time acceleration dominated eras. Moreover, the matter dominated epochs are very different to the standard $\Lambda$CDM model. These eras have a scalar factor behaving as $a(t)\sim t^{0.46}$ and then  $a(t)\sim t^{0.52}$ which are slower than the usual standard case $a(t)\sim t^{2/3}$. A generalised Proca theory was considered by \cite{DeFelice:2016yws}, who proved that stable de Sitter fixed points can exist.

Finally, considering cosmologies in which a non-minimally coupled condensate is present leads to equations similar (but not equivalent) to scalar-tensor equations.
\citet{Carloni:2014pha} analysed such systems  via phase space techniques, showing similarities and differences with the standard scalar-tensor case.

\begin{table}
\begin{center}
\resizebox{15cm}{!}{\begin{tabular}{|c|c|c|c|c|c|c|}
 \hline
 P & $\{x,y\}$  & $\tilde{\Omega}_{\rm m}$ & $a=a_0(t-t_0)^\alpha$ \\
 \hline
 & &&\\
$A_\pm$ & $\left\{ 3 m\pm\sqrt{9 m^2+6}, 0\right\}$ & 0& $\alpha=\displaystyle\frac{1}{3\pm m \left(\sqrt{9 m^2+6}-3 m\right)}$ \\
& & & \\
$B$ & $\left\{ \displaystyle\frac{m}{m^2+1}, 0\right\}$ & 0& $\alpha=\displaystyle\frac{2 \left(m^2+1\right)}{4 m^2+3}$ \\
 & & & \\
 \multirow{2}*{$C$} & \multirow{2}*{$\left\{-\displaystyle\frac{3}{n},\displaystyle\frac{\sqrt{m (3 m+n)+3}}{\sqrt{2} \left| n\right| }\right\}$} & \multirow{2}*{$\displaystyle\frac{2 n^2-3 m^2-7 m n-6}{2 n^2}$} & \multirow{2}*{$\alpha
   =\displaystyle\frac{2 n}{3 (n-m)}$} \\
   & & &  \\ & & & \\
 \multirow{2}*{$D$} & \multirow{2}*{$\left\{\displaystyle\frac{4 m-2 n}{m (m+n)+2}, \displaystyle\frac{\sqrt{3 m^2+2} \sqrt{(5 m-n) (m+n)+6}}{\sqrt{3} \left| m (m+n)+2\right|}\right\}$} & \multirow{2}*{0}& \multirow{2}*{$\alpha =\displaystyle\frac{m (m+n)+2}{2 m^2-3 m n+n^2}$} \\
 & & & \\
 \hline
\end{tabular}}\vspace{4mm}
\begin{tabular}{|c|c|c|c|c|c|c|}
 \hline
 P & $w_{\rm eff}$ & $w_{\phi }$  \\
 \hline & &\\
$A_\pm$ & $6 m^2+5\mp2 \sqrt{9 m^2+6} m$ & $6 m^2+5\mp2 \sqrt{9 m^2+6} m $ \\ & & \\
$B$ & $-\displaystyle\frac{m^2 \left(3 m^2+2\right)}{3
   \left(m^2+1\right)^2}$ & $1-\displaystyle\frac{1}{6 m^2+5}$ \\ & &\\
 \multirow{2}*{$C$} & $\displaystyle\frac{2 n^2-3 m^2-7 m n-6}{2 n^2}$& $\displaystyle\frac{3 m^2+4
   m n+6}{n^2}$
    \\ & &  \\
 \multirow{2}*{$D$} & $\displaystyle\frac{3 m^2-7 m n+2 n^2-2}{m (m+n)+2} $& $\displaystyle\frac{3 m^2-7 m n+2 n^2-2}{m (m+n)+2}$   \\ & & \\
 \hline
\end{tabular}
\end{center}
\caption{Critical points of the scalar-tensor model described by Eq.~(\ref{eq:Sc-Tn_Dyn_Ex}) with exponential potentials $F, V, \omega$ and  $w=0$.}
\label{tab:cp_SCTN}
\end{table}
\vspace{4mm}
\begin{table}
\begin{center}
\begin{tabular}{|c|c|c|c|c|}
 \hline
 P & Attractor & Repeller  \\
 \hline &&\\
{$A_+$} & {never} &  $ n<\sqrt{9 m^2+6}+2 m$  \\ &&\\
{$A_-$} & {never} &  $\sqrt{9 m^2+6}+n>2 m$  \\ &&\\
$B$ &$m (3 m+n)+3<0$ &  never \\
 & &  \\
 \hline
\end{tabular}
\end{center}
\caption{Stability of the critical points of the scalar-tensor model described by Eq.~(\ref{eq:Sc-Tn_Dyn_Ex}) with exponential potentials $F, V, \omega$ and $w=0$. Figs.~\ref{fig:sctn_RegionC} and \ref{fig:sctn_RegionD} give the stability analysis for Points~$C$ and $D$.
}
\label{tab:stab_SCTN}
\end{table}

\subsection{$f(R)$ gravity}
\label{sec:f(R)}
The general action of these theories is given by
\begin{equation}
	S_{f(R)} = \int d^4x \sqrt{-g} \left[f(R)+2\kappa^2 \mathcal{L}_{\rm m}\right] \,,
	\label{def:f(R)_action}
\end{equation}
where $f(R)$ is an arbitrary function of the Ricci scalar $R$. The reader broadly interested in the theory and applications of $f(R)$ gravity can find useful the well known reviews by \citet{Capozziello:2007ec}, \citet{Sotiriou:2008rp}, \citet{DeFelice:2010aj} and \citet{Nojiri:2010wj}.

The cosmological equations of $f(R)$ gravity, obtained by varying action (\ref{def:f(R)_action}) with respect to the metric tensor and then assuming homogeneity and isotropy of the spacetime (FLRW metric) can be written as
\begin{align}\label{eq:f(R)_Friedmann}
\begin{split}
& H^{2}+\frac{k}{a^2} =
\frac{1}{3F}\left\{\frac{1}{2}\left[FR-f\right]-3H\dot{F}+2\kappa^2\rho \right\}\,,\\
&2\dot{H}+H^{2}+\frac{k}{a^2} =
-\frac{1}{F}\left\{\frac{1}{2}\left[FR-f\right]+\ddot{F}-3H\dot{F}+\,2\kappa^2p\right\}\,,
\end{split}
\end{align}
where
\begin{equation}
	F=f_{,R}=\frac{\partial f}{\partial R} \,,
\end{equation}
\begin{equation}
R\,=\,6\left(\dot{H}+2H^{2}+\frac{k}{a^2}\right)\,,\label{R}
\end{equation}
and we recall that $H\equiv\dot{a}/a$ is the Hubble parameter and the ``dot" is the derivative with respect to cosmic time $t$. The two equations \eqref{eq:f(R)_Friedmann} are not independent: the second can be obtained by differentiating the first with respect to the cosmic time $t$ once the Bianchi identities for
$T^{\rm m}_{\mu\nu}$ are considered. In FLRW these identities take the form
\begin{equation}\label{BianchiMatt}
\dot{\rho}+3H(\rho+p)=0\;,
\end{equation}
which is the same as the GR energy conservation equation for the cosmic fluid. Note that in the above system of equations the Ricci scalar $R$ is considered as an effective additional scalar field. This has the advantage of hiding the higher derivative terms in the derivatives of the Ricci scalar. As we will see this approach has some limitations also in terms of the application of dynamical system analyses.

A common way to look at the equations above is to express them as GR plus effective fluids.
Motivated by the form of the two equations in (\ref{eq:f(R)_Friedmann}), we see that this can be done by defining
\begin{align}
\rho_{\rm R} &= \frac{R}{2} - \frac{f}{R} -3H\frac{\dot F}{F} \,,\label{rhoR} \\
p_{\rm R} &= \frac{\ddot F}{F} -2H\frac{\dot F}{F} -\frac{R}{2} +\frac{f}{F} \,, \label{pR}
\end{align}
from which the dark energy EoS can easily be obtained as
\begin{equation}
	w_{\rm R}= \frac{p_{\rm R}}{\rho_{\rm R}} \,.
\end{equation}
The fact that Eqs.~(\ref{rhoR}) and (\ref{pR}) are dependent on spacetime curvature terms, indicates that in this case dark energy can now be explained by the effects of spacetime curvature rather than by an external matter source. This feature sets apart this class of theories (and its generalisations) from the models we have encountered so far. The  EoS can now be generally defined as
\begin{equation}
	w_{\rm eff} = \Omega_{\rm R} w_{\rm R} + \tilde{\Omega}_{\rm m} w = -1-\frac{2\dot H}{3H^2} \,.
\end{equation}
The very first attempt to look at the phase space of $f(R)$ gravity can be traced back to \citet{Starobinsky:1980te}. \citet{Capozziello:1993xn} proposed a phase space analysis of several $f(R)$ models.
\citet{Miritzis:2003eu} considered the subclass given by $f(R)=R+\alpha R^2$ ({\it Starobinsky model}) and performed the stability analysis for accelerating solutions also in the presence of positive spatial curvature ($k=1$).
\citet{Carloni:2004kp} used expansion normalised variables for the first time in the case of the power-law  toy models $f(R)\propto R^n$.
This idea was successively generalised by \citet{Amendola:2006we} who proposed a general method applicable to a number of $f(R)$ theories in the case of flat Universes (see Sec.~\ref{Amendola_f(R)} below). Using this setting, the function $f(R)=R+\alpha R^{-n}$ was studied by \citet{Li:2007xn} in full detail. Subsequently a number of models not necessarily accessible with the method just described above have been considered. For example,  \citet{Goheer:2007wx,Goheer:2008tn}  considered variables suited to compactify the phase space. \citet{Sawicki:2007tf} showed that viable $f(R)$ cosmological models should satisfy $f_{,RR}>0$ and then made an example with the theory $f(R)=R+\alpha/R$. Models with (inverse) power-law corrections to the Einstein-Hilbert Lagrangian were also considered by \citet{Clifton:2008bn} who delivered an asymptotic analysis at both early and late times. An exponential potential model with $f(R)=\exp(-R/\Lambda)$ has instead been the subject of work by \citet{Abdelwahab:2007jp}, who found de Sitter solutions acting as both past and future attractors and thus capable of unifying inflation with dark energy.
More complicated models, including $f(R)=R\ln R$, have been studied by \citet{Guo:2013swa}, who provided a full dynamical systems analysis with the behaviour at infinity and found late time de Sitter solutions. A DBI $f(R)$ gravity theory with $f(R)=\sqrt{1- \alpha R}$ has been advanced by \citet{GarciaSalcedo:2009fb} who showed that rich phenomenological dynamics arise on cosmological scales. $f(R)$ Gamma functions which are generalisations of exponential models of $f(R)$ gravity have been also considered by \cite{Boko:2016mwr}. Some authors have also examined specific features of the dynamics of general $f(R)$ theories. \citet{Amendola:2007nt} focused their study on phantom late time attractors and crossing of the phantom barrier for viable $f(R)$ models. An alternative, geometrical analysis for the phase space of $f(R)$ theories has been explored by \citet{deSouza:2007fq}.

We briefly mention here also the attempts of an analysis of $f(R)$ theories in the context of anisotropic spacetimes. The model $f(R)\propto R^n$ has been studied by \citet{Leach:2006br} and \citet{Goheer:2007wu,Goheer:2007wx,Goheer:2008tn}, while \citet{Leon:2010pu} and again \citet{Leon:2013bra} analysed also general $f(R)$ anisotropic cosmologies, finding late time accelerated and bouncing solutions.

In order to understand more in depth the details of the dynamical system formulation of $f(R)$-gravity  we will summarise three different approaches that have appeared in the literature in chronological order. The first approach is clearly inspired by the variables we have defined in cosmologies with scalar fields, the second method uses a different set of variables to allow the treatment of any form of the function $f$, whereas the third is the only known construction for the phase space of $f(R)$ cosmology which is globally valid (although only constructed, so far, for a specific form of $f$). The reader will notice that the results of the three approaches are not always consistent with each other: the number of fixed points varies and stability properties may change. The reason behind these differences is still matter of debate and we will not delve into details here. However, let us mention that the lack of global variables certainly has an important role in these issues. It appears clear that choosing a suitable set of dynamical system variables is in fact more complicated than in the case of scalar-tensor gravity, and that there might not be an `all purpose' covering for the phase space of these theories.

\subsubsection{The approach of \citet{Amendola:2006we} }
\label{Amendola_f(R)}

We start summarising the idea of \citet{Amendola:2006we} which can be used as a template for most of the approaches above (and the ones in the following sections). In the case of spatially flat cosmologies sourced by a pressureless fluid, we define the following dimensionless variables

\begin{equation}
	x = \frac{\dot F}{HF} \,, \qquad y = \frac{R}{6H^2}\,, \qquad z= \frac{f}{6FH^2}   \,, \qquad\tilde{\Omega}_{\rm m}= \frac{2\kappa^2 \rho}{3FH^2}\,.
	\label{def:f(R)_variables}
\end{equation}
 The Friedmann equations (\ref{eq:f(R)_Friedmann}) now yields the constraint:
\begin{equation}
	\Omega_{\rm de} =   y - z - x= 1-\tilde{\Omega}_{\rm m}\,.
\end{equation}
Note that in contrast to the scalar-tensor case, $x$, $y$ and $z$ can also be negative, though their sum cannot exceed one. In addition,  the constraint above does not give a compact phase space. This fact implies that to achieve a complete treatment of this kind of phase space we will have to employ asymptotic analysis.

The cosmological equations can now be rewritten as
\begin{equation}\label{DynSysAm}
\begin{split}
x' &= 1-x^2-x y+y-3 z \,,\\
y' &= -\frac{xy}{m} -2y\left(y-2\right) \,,\\
z' &= \frac{x y}{m} +z\left(2y-x+4\right) \,,
\end{split}
\end{equation}
where we have defined
\begin{equation}
	m = \frac{\partial\log F}{\partial \log R} = \frac{R\,F_{,R}}{F} \,.
	\label{def:f(R)_m}
\end{equation}
Since $m$ is a function of $R$ only,  the problem of obtaining $m$ in terms of the dynamical system variables is reduced to the problem of writing $R$ in terms of those variables. This can be achieved from the definitions (\ref{def:f(R)_variables}) noting that:
\begin{equation}\label{Pre-r}
\frac{y}{z}=\frac{RF}{f}\,.\,
\end{equation}
Solving the above equation for $R$ allows one to write it in terms of $y$ and $z$ and close the system \rf{DynSysAm}. It is clear that the properties of \eqref{Pre-r} determine the possibility of closing (and therefore analysing) the system \eqref{DynSysAm}. This result implies that the system above can only be used for specific forms of the function $f$. Eq. \eqref{Pre-r} determines also some of the properties of this system i.e.~the differential structure. For example, some forms of $R=R(y,z)$ can introduce zeros of the quantity $m$ which correspond to  singularities (shock waves) of the system \eqref{DynSysAm}.

It should be remarked that in the original paper it was proposed to catalogue the fixed points in terms of the values of the quantity $m$. However such an approach was shown to lead in some cases to incorrect conclusions in terms of the fixed points and their stability \citep{Carloni:2007br}. We will not attempt such a general analysis here, referring to a single specific example.

Let us quickly analyse the case $f(R)=R+ \alpha R^n$ via Eq.~\eqref{DynSysAm} as it will be useful to have a concrete example to compare the properties of this method with the ones presented below. For brevity we consider the case in which the cosmic fluid has zero pressure: $p= w \rho = 0$. In this case, the characteristic function $m(r)$ reads\,:
\begin{equation}\label{m-R+Rngen}
m\,=\,\frac{n(z-y)}{y}\,,
\end{equation}
and the dynamical system equations become
\begin{equation}\label{dynsysAmRRn}
\begin{split}
x' &= 1-x^2-x y+y-3 z \,,\\
y' &= \frac{x y^2}{n(z-y)}-2y\left(y-2\right) \,,\\
z' &= \frac{x y^2}{n(z-y)} +z\left(2y-x+4\right) \,.
\end{split}
\end{equation}
The system is divergent on the hypersurface $y=z$ and  it admits only the invariant submanifolds  $y=0$ and $z=0$. The presence of these structures implies that a finite global attractor can only exist if it has $y=0$ and $z=0$, however such a point would also be singular for the system. We conclude that in this setting it is impossible to determine whether a global attractor can exist.

The fixed points with their stability in the specific case of $w=0$ are given in Tabs.~\ref{tab:cp_RRn} and \ref{tab:stab_RRn}. As in the previous section one can have an idea of the solution associated to these fixed points by calculating the values of the deceleration parameter which reads
\begin{equation}\label{q_f(R)_Am}
-1-\frac{\dot{H}}{H^2}=q=1+y\,,
\end{equation}
so that the variable $y$ is directly connected to $q$. The relation \eqref{q_f(R)_Am} can be used to calculate the solution at the fixed point, in the same way as in Sec.~\ref{sec:choosing-variables} and they are given in Tab.~\ref{tab:cp_RRn}.  On top of that, the system seems to present two additional special points which lay on the singular manifold with coordinates $z=y$. One can verify numerically that orbits flow from and around these point as if  they were actual fixed points. Their existence is however an artefact of the coordinates and have no physical meaning.

With this information one can conclude that this type of cosmology leads to either a de Sitter attractor or a power law one depending on the value of $n$. The only attractor that can be truly global is Point~$D$, but this happens only for $n=3/4$. In the other cases the asymptotic properties of the cosmology depend on the initial conditions.
\begin{table}
\begin{center}
\resizebox{15cm}{!}{\begin{tabular}{|c|c|c|c|c|c|}
 \hline
 P & $\{x,y,z\}$  & $\tilde{\Omega}_{\rm m}$ & Scale Factor & Energy Density \\
 \hline & &&&\\
$A$ & $\left\{0, -2, -1\right\}$ & 0& $a=a_0 \exp\left(\lambda t\right)$ &$\rho=\rho_0 \exp\left(-3\lambda t\right)$\\
& & & & \\
$B$ & $\{ 4, 0, 5\}$ & 0 & $a= a_0 \sqrt{t-t_0}$ &$\rho=\rho_0  \left(t-t_0\right){}^{-3/2}$\\ & & & & \\
$C$ &$\left\{ \displaystyle\frac{2 (n-2)}{2 n-1}, \displaystyle\frac{(5-4 n) n}{2 n^2-3 n+1}, \displaystyle\frac{5-4 n}{2 n^2-3 n+1}\right\}$ & 0 & $a= a_0 (t-t_0)^{\frac{2 n^2-3 n+1}{2-n}}$& $\rho=\rho_0\left(t-t_0\right){}^{\frac{6 n^2-9 n+3}{n-2}} $ \\ & & & &  \\
 \multirow{2}*{$D$} &  $\left\{ \displaystyle\frac{3}{n}-3,\displaystyle\frac{3}{2 n}-2,\displaystyle\frac{3}{2 n^2}-\displaystyle\frac{2}{n}\right\} $& $\displaystyle\frac{(13-8 n) n-3}{2 n^2}$ & $a= a_0\left(t-t_0\right){}^{2 n/3}$ & $\rho=\rho_0 \left(t-t_0\right){}^{-2 n}$\\ & & & & \\
 \hline &&&&\\
$E$ & $\{ -1, 0, 0\}$ & 0 & NA& NA\\ & & & & \\
$F$ & $\{ 1, 0, 0\}$ & 2 & NA& NA\\ & & & & \\
\hline
\end{tabular}}
\end{center}
\caption{Critical points of the $f(R)=R+\alpha R^n$ with $w=0$ for the system (\ref{dynsysAmRRn}). Here $\lambda = 2^{\frac{1}{n-1}} \left[\alpha  3^{n-1} 4^n (n-2)\right]^{\frac{1}{2-2 n}}$.}
\label{tab:cp_RRn}
\end{table}

\begin{table}
\begin{center}
\begin{tabular}{|c|c|c|c|c|}
 \hline
 P & Attractor & Repeller  \\
 \hline &&\\
$A$ & $\left\{\begin{array}{cc}
                  \mbox{Node} &  \frac{32}{25}\leq n < 2\\
                  \mbox{Focus} &  0 < n < \frac{32}{25}\\
                \end{array}\right.$ &   never \\ &&\\
{$B$} & never & never  \\ &&\\
$C$ &$n<\frac{1}{16}\left(13-\sqrt{73}\right)\lor n>2$ & $1<n<\frac{5}{4}$\\
 & &  \\
$D$&  $\left\{\begin{array}{cc}
                  \mbox{Node} & \frac{1}{16}\left(13-\sqrt{73}\right)<n<-1.3275
                  \\
                  \mbox{Node} &
                  0.7135 < n < \frac{3}{4}\\
                    \mbox{Focus} & 0< n < 1\\
                \end{array}\right.$  & never\\
&& \\
 \hline &&\\
 {$E$} & never & always  \\ &&\\
 {$F$} & never & never  \\ &&\\
  \hline
\end{tabular}
\end{center}
\caption{Stability of the critical points of the  $f(R)=R+\alpha R^n$ with $w=0$ for the system (\ref{dynsysAmRRn}). The (approximated) real numbers appearing in the inequalities above come from the numerical  analysis of the sign of the eigenvalues of the fixed point.}
\label{tab:stab_RRn}
\end{table}

One should note that the variables \eqref{def:f(R)_variables} contain $H$ in the denominator. This implies that their choice does not allow us to study static $H=0$ cosmologies as they are part of the asymptotics. This is, in fact, a common problem related to the use of EN variables. In Sec.~\ref{sec:standard_cosmology_beyond_spatial_flatness}, we have seen an example in which, in the case of GR with a positive spatial curvature, one can deform the time variable $\bar{\eta}=1/2\ln\left(H^2+k/a^2\right)$ and it is then able to consider $H=0$ in the phase space. This means that the $H=0$ divergence can be (partially) addressed considering  a similar ``shifting'' of the time variable definition. \citet{Abdelwahab:2011dk} proposed a  solution of this type to this problem for spatially flat cosmologies, functions $f(R)$ with certain characteristics and $R>0$. The method used variables similar to those of  \rf{def:f(R)_variables} but the authors defined a new time variable given by
\begin{equation}
\bar{\eta}^2= 9\left(H+\frac{1}{2}\frac{\dot{F}}{F}\right)^2+\frac{3}{2}\frac{f}{F} \,,
\end{equation}
which returns a compact phase space and allows one to look at $H=0$ states. Unfortunately this method has the disadvantage that $R=0$ can be proven not to be an invariant submanifold in every case and therefore the assumption $R>0$ cannot be considered valid in all cases.

\subsubsection{The approach of \cite{Carloni:2015jla}}
\label{Carloni_f(R)}

A different strategy, proposed in \cite{Carloni:2015jla}, is based on the following set of variables\footnote{In some cases it has proven more useful to define the variable $\bar{\mathbb{A}}=\mathbb{A}^{-1}$ instead of simply $\mathbb{A}$. In terms of the derivation of the dynamical equations this does not make much difference, but the choice is significant in terms of the physical meaning of the variable and the structure of the Jacobian of the definition of the dynamical system variables.}
\begin{align}  \label{DynVar2}
\begin{split}
\mathbb{R}=\frac{R}{6 H^2}\,,\quad \tilde{\Omega}_{\rm m} =\frac{2 \kappa^2\rho }{3H^2  F }\,,\\ \mathbb{J}=\frac{\mathfrak j}{4}\,,\quad \mathbb{Q}={\mathfrak q}\,,\quad \mathbb{A}=R_0H^{2}\,,
\end{split}
\end{align}
where
\begin{equation} \label{HubbleVarDS}
{\mathfrak q} =\frac{\dot{H}}{H^2}\,,\quad {\mathfrak j} =\frac{\ddot{H}}{ H^2}-\frac{\dot{H}^2}{H^3} \,,
\end{equation}
and $R_0$ is a dimensional constant used to make the function $f$ and any coupling constants appearing in it dimensionless.

The new variables allow us to write the general dynamical system as
\begin{align}\label{DynSys}
\begin{split}
&\mathbb{R}'=2 \mathbb{R} (\mathbb{R}+2)-\frac{4}{{\bf Y}} ({\bf X}-\mathbb{R}-\tilde{\Omega}_{\rm m}+1)\,,\\
&\tilde{\Omega}_{\rm m}'=\tilde{\Omega}_{\rm m}  (2-3w+{\bf X}-3 \mathbb{R}-\tilde{\Omega}_{\rm m} )\,,\\
&\mathbb{A}'=-2\mathbb{A} (2-\mathbb{R})\,,
\end{split}
\end{align}
together with the two constraints given by the Friedmann equation plus the definition of the Ricci scalar:
\begin{eqnarray}
&& 1=\tilde{\Omega}_{\rm m}+\mathbb{R}-{\mathbf X}-\frac{1}{4} {\mathbf Y} \left(4 \mathbb{J}+\mathbb{R}^2-4\right)\,,  \\
&&  \mathbb{R}=\mathbb{Q}+2\,,\label{Ricci Constr}
\end{eqnarray}
 where
\begin{align}
\begin{split}& {\bf X}\left(\mathbb{A},\mathbb{R}\right)= \frac{f\left(\mathbb{R}, \mathbb{A},\alpha,...\right)}{6 H^2  F\left(\mathbb{R}, \mathbb{A},\alpha,...\right)}\,,\qquad  {\bf Y}\left(\mathbb{A},\mathbb{R} \right)= \frac{24H^2  F'\left(\mathbb{R}, \mathbb{A},\alpha,...\right)}{  F\left(\mathbb{R}, \mathbb{A},...\right)}\,,
\end{split}
\end{align}
represent the part of the system which depends on the form of the Lagrangian. Unlike the case of the system \eqref{DynSysAm}, in this approach we do not need to invert the algebraic equations to obtain an autonomous system. This means that with this approach we can in principle analyse {\it all}  $f(R)$ models. The derivation of the solution for the fixed point  is also changed: in the variables \eqref{DynVar2} the second equation of \eqref{eq:f(R)_Friedmann} is an equation for the {\it third} derivative of the Hubble parameter as shown by \citet{Carloni:2015jla}
\begin{equation}\label{RAy3Ord}
\begin{split}
\frac{H'''}{H}=\mathfrak{s}^*= g( \mathbb{Q}_*, \mathbb{J}_*,\mathbb{R}_*, \mathbb{A}_*,\tilde{\Omega}_{\rm m})\,,
\end{split}
\end{equation}
where the asterisk indicates the value of a variable evaluated at a fixed point, and $g$ is given by
\begin{equation}
\begin{split}
  g( \mathbb{Q}, \mathbb{J},\mathbb{R}, \mathbb{A},\tilde{\Omega}_{\rm m})&=-4 (4 \mathbb{J}-5) \mathbb{R}-\frac{3 }{{\bf Y}} [{\bf Y}+4 (\mathbb{J}-1)
  {\bf  Z}]\mathbb{R}^2-\mathbb{R}^3-\frac{3 {\bf Z}}{2{\bf Y}}\mathbb{R}^4  \\
&~~+4 \left(3 \mathbb{J} -5\right)+\frac{4}{{\bf Y }} \left[\left(12 \mathbb{J}-6
   \mathbb{J}^2 -6 \right) {\bf Z}-2 {\bf X}-8\tilde{\Omega}_{\rm m}+2\right]\,,
\end{split}
\end{equation}
where
\begin{equation}
{\bf Z}\left(\mathbb{A},\mathbb{R}\right)= \frac{96 H^4  f'''\left(\mathbb{R}, \mathbb{A},\alpha,...\right)}{f'\left(\mathbb{R}, \mathbb{A},\alpha,...\right)}\,.
\end{equation}

\begin{figure}%
\centering
\includegraphics[width=0.7\columnwidth] {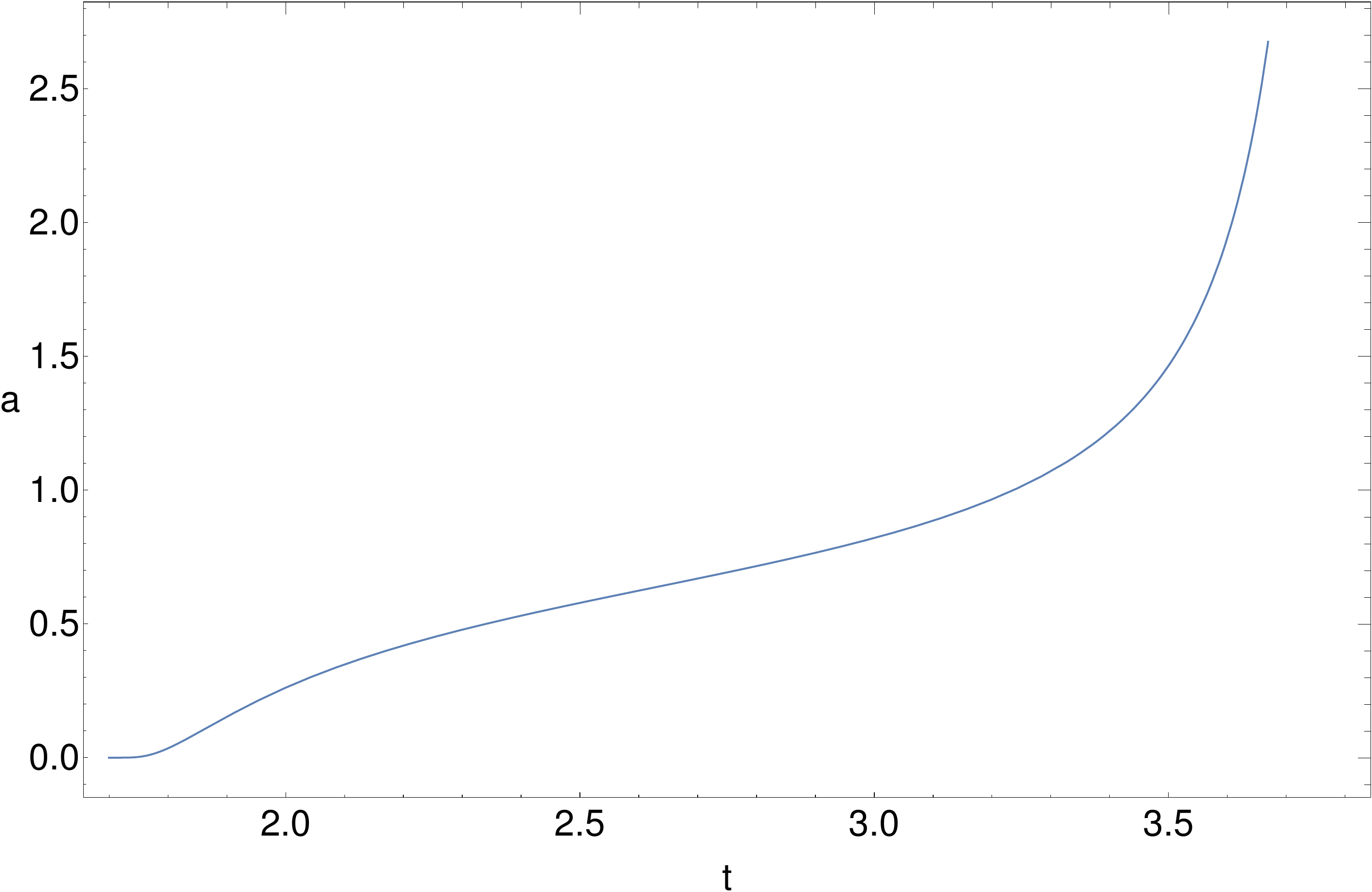}
\caption{Plot of a period of \rf{SolH} in the case $4 H_0 H_2-H_1^2>0$. Here $H_1=1$,  $H_2=3$, $H_3=2$  and $a_0=1$.} \label{PlotSolH}
\end{figure}

Eq.~\rf{RAy3Ord} has the following solutions
\begin{equation}\label{SolHGen}
H=\left\{
\begin{array}{ll}
H_0+H_1 \eta+ H_2 \eta^2 & \mathfrak{s}^*=0\\
H_0 e^{-p \eta}+ e^{\frac{p \eta}{2}}\left(H_1\cos \frac{p\eta\sqrt{3}}{2}+ H_2 \sin \frac{p\eta\sqrt{3}}{2}\right)&\mathfrak{s}^*\neq 0
\end{array}
\right. \,,
\end{equation}
where $p=-\sqrt[3]{\mathfrak{s}^*}$ and the $H_i$ are integration constants. Recalling that $\eta = \ln a$, the above equations translate to equations for the scale factor
\begin{equation}\label{Eqa}
\frac{\dot{a}}{a}=\left\{
\begin{array}{ll}
H_0+H_1 \ln a+ H_2 (\ln a)^2 & \mathfrak{s}^*=0\\
H_0 a^{-p}+ a^{p/2}\left[H_1\cos \left( \frac{p\sqrt{3}}{2}\ln a\right)+ H_2 \sin \left( \frac{p\sqrt{3}}{2}\ln a\right)\right]&\mathfrak{s}^*\neq 0
\end{array}
\right.\,.
\end{equation}
In the first case the equation can be solved exactly to give
\begin{equation}\label{SolH}
  a(t)=a_0 \exp \left\{\frac{\sqrt{4 H_2 H_0-H_1^2}}{2 H_2} \tan \left[\frac{1}{2} (t-t_0) \sqrt{4 H_2
   H_0-H_1^2}\right]-\frac{H_1}{2 H_2}\right\} \,,
\end{equation}
in the other a numerical solution is required. The behaviour of these last solutions, however, can be  a power law expansion at early times followed by  accelerated expansion and finally an approach to a static universe (see Fig.~\ref{PlotSolH}  for an example).  The solution \rf{SolH} is remarkable in many aspects: depending on the choices of the parameters, it includes an initial de Sitter phase a decelerated intermediate phase and an accelerated expansion phase. The issue with such a solution is that it also possesses a finite time singularity. In other words, the realisation of this interesting type of cosmic history is inextricably bound to the onset of a Big Rip type singularity in the future.

Let us now analyse again, this time with the above approach, the model $f(R)= R+\alpha R^n$  in the case of spatially flat cosmologies and a pressureless cosmic fluid. We first rewrite the action as $f(R_0R,\alpha)=R_0 R+\alpha R_0^{n}R^{n}$. Calculating  ${\bf X}$ and $\bf Y$,  we obtain the following dynamical equations
\begin{equation}\label{DynSysRR^n}
\begin{split}
 &\mathbb{R}'=\frac{6^{1-n} (\tilde{\Omega}_{\rm m} -1) }{\alpha n (n-1)}\mathbb{A}^{1-n} \mathbb{R}^{2-n}-\frac{\mathbb{R} }{n(n-1) }\left[\left(2 n^2-3 n+1\right) \mathbb{R}
-n \tilde{\Omega}_{\rm m} -4 n^2+5 n\right]\,,\\
 &\tilde{\Omega}_{\rm m} '=\tilde{\Omega}_{\rm m} \left[\tilde{\Omega}_{\rm m} -\frac{ 3 n-1}{n}\mathbb{R}-2+\frac{n-1}{n \left(1+6^{n-1}\alpha n \mathbb{A}^{n-1} \mathbb{R}^{n-1}\right)}\right]\,,\\
   &\mathbb{A}' =2 \mathbb{A} (\mathbb{R}-2)\,.
\end{split}
\end{equation}
This system admits three invariant submanifolds ($\tilde{\Omega}_{\rm m}=0, \A=0, \mathbb{R}=0$) and therefore also here we conclude that there is no global attractor.

The fixed points and their stability are presented in Tabs.~\ref{tab:cp_RRn1} and \ref{tab:stab_RRn1}. It is clear that many different attractors are possible depending on the value of $n$  chosen.

\begin{table}
\begin{center}
\resizebox{15cm}{!}{\begin{tabular}{|c|c|c|c|c|}
 \hline
 P & $\{\mathbb{R},\tilde{\Omega}_{\rm m},\mathbb{A}\}$  & $\{\mathbb{J},\mathbb{Q}\}$ & Scale Factor  \\
 \hline & &&\\
 \multirow{2}*{$O$} & \multirow{2}*{$\{ 0, 0, 0\}$ }& \multirow{2}*{$\left\{1, -2\right\}$} & \rf{Eqa} with  \\ & & & $\mathfrak{s}^*\neq 0, p=2$ \\& & &  \\
$A$ & $\{ 2, 0, 12 [\alpha(n-2)]^{\frac{1}{1-n}}\}$ & $\left\{0, 0\right\}$ & \rf{SolH} \\ & & &  \\
\multirow{3}*{$B$} &\multirow{3}*{$\left\{\displaystyle\frac{(5-4 n) n}{4 n^2-6 n+2}, 0, 0\right\}$} & \multirow{3}*{$\left\{\displaystyle\frac{
 n^2 (17 n-45)+33n-4-6^{-n}(5-4 n)^2 n}{4(1-2 n)^2 (n-1)^3}, \frac{n-2}{2 n^2-3 n+1}\right\}$} &  \rf{Eqa} with  \\& & & $\mathfrak{s}^*\neq 0,$ \\& & & $ p=\displaystyle\frac{2-n}{(n-1) (2 n-1)}$   \\
& & & \\ %
 \multirow{2}*{$C$} &   \multirow{2}*{$\left\{ -\displaystyle\frac{3-4 n}{2 n}, -\displaystyle\frac{8 n^2-13 n+3}{2 n^2}, 0\right\}$}&  \multirow{2}*{$\left\{ \displaystyle\frac{n (25 n-33)+9}{16 (n-1) n^3}-\displaystyle\frac{3^{-n} (3-4 n)^2}{16 (n-1) n^3}, -\displaystyle\frac{3}{2 n}\right\}$} & \rf{Eqa} with\\& & &  $\mathfrak{s}^*\neq 0, p=\displaystyle\frac{3}{2n} $ \\ & & &  \\
 \multirow{2}*{$D$} &  \multirow{2}*{$\left\{0, 2, 0\right\}$} &  \multirow{2}*{$\left\{1, -2\right\}$} & \rf{Eqa} with \\ & & & $\mathfrak{s}^*\neq 0, p=2$ \\& & &  \\
\hline
\end{tabular}}
\end{center}
\caption{Critical points of the $f(R)=R+\alpha R^n$ with $w=0$ for the system \rf{DynSysRR^n}.}
\label{tab:cp_RRn1}
\end{table}

\begin{table}
\begin{center}
\begin{tabular}{|c|c|c|c|c|}
 \hline
 P & Attractor & Repeller  \\
 \hline &&\\
 {$O$} & never & never  \\ &&\\
 {$A$} &$ \left\{\begin{array}{cc}
            \mbox{Node} & \frac{32}{25}<n<2 \\
               \mbox{Focus} & 0<n< \frac{32}{25}\\
               \end{array}\right. $& never  \\ &&\\
$B$ &$n<\frac{1}{16}\left(13-\sqrt{73}\right)\lor \frac{1}{16}\left(13+\sqrt{73}\right)<n<2$ & never\\
 & &  \\
$C$&$\left\{\begin{array}{cc}
             \frac{1}{16}\left(13-\sqrt{73}\right)<n\leq 0.3341
             \\
              0.7134< n < \frac{3}{4}\\
              1.3275< n <  \frac{1}{16}\left(13+\sqrt{73}\right)\\
               \end{array}\right. $& never\\
&& \\
$D$ & $3/4<n<1$ &   never \\ &&\\
  \hline
\end{tabular}
\end{center}
\caption{Stability of the critical points of the  $f(R)=R+\alpha R^n$ with $w=0$ for the system \rf{DynSysRR^n}.
The (approximated) real numbers appearing in the inequalities above come from the numerical  analysis of the sign of the eigenvalues of the fixed point.}
\label{tab:stab_RRn1}
\end{table}

\subsubsection{The approach of \citet{Alho:2016gzi}}

A promising new method, which, at the time of writing this review, has only been formulated  for the special case of $f(R)=R+\alpha R^{2}$ has been proposed by \citet{Alho:2016gzi}. The dynamical variables $\ts$ and $\xs$ in this case are given by the equations

\begin{subequations}
\begin{align}
\ts-\xs &= 2\sqrt{\frac{3}{\alpha}}H\,, \label{htx1}\\
\ts+\xs&=\sqrt{\frac{3}{\alpha}}\left(2\alpha \dot{R}+ 2\alpha HR + H\right)\label{htx2}\,,
\end{align}
\end{subequations}
with
\begin{subequations}\label{tsxseq}
\begin{equation}\label{txconstr}
\ts^2 = \xs^2 + R^2 \,,
\end{equation}
and the flow on this state space is determined by the following evolution equations:
\begin{align}\label{dimevol}
\dot{\ts} &= \frac{1}{4\sqrt{3\alpha}}\left[R - 2\alpha(\ts-\xs)^2\right]\,, \\
\dot{\xs} &= \frac{1}{4\sqrt{3\alpha}}\left[2\alpha(\ts-\xs)^2-3R\right]\,, \\
\dot{R} &= \frac{1}{4\sqrt{3\alpha}}\left[\ts + 3\xs - 2\alpha(\ts-\xs)R\right]\,.
\end{align}
\end{subequations}
where, as usual, a dot represents differentiation with respect to $t$. The constraint~\eqref{txconstr} makes it explicitly clear that the reduced vacuum state
space is a 2-dimensional double cone with a joint apex, which represents a Minkowski fixed point for the system. Note that the variables (\ref{htx1})--(\ref{htx2}) can be seen as a global homeomorphism \citep{Plastock:1974}. Therefore, using these variables, one can perform a global analysis of the phase space. This is impossible with the other approaches discussed.

Exploiting the structure of the constraint \rf{txconstr}, considering only  expanding cosmologies and eliminating the fixed point corresponding to the apex, one obtains
\begin{equation}
T = \frac{1}{1 + 2\alpha\ts}\,,\qquad \theta=\arccos \left(\frac{\xs}{\ts}\right)=-\arcsin\left(\frac{R}{\ts}\right)\,,
\end{equation}
so that the phase space can be analysed in an easier way. The dynamical equations become
\begin{subequations}\label{Jordan2DDynSys}
\begin{align}
\frac{d T}{d \bar{t}} &= T(1 - T)\left[T\sin{\theta} + (1-T)(1 - \cos{\theta})^2\right], \\
\frac{d \theta}{d \bar{t}} &= -T(3 + \cos{\theta}) - (1-T)(1 - \cos{\theta})\sin{\theta}\,,
\end{align}
\end{subequations}
where  $\bar{t}$ is defined as $\frac{dt}{d\bar{t}} = 2\sqrt{12\alpha}\,T$. Since the system is regular, one can perform both a local and global analysis. All the results are summarised in Tab.~\ref{tab:Alho_RRn2}.

\begin{table}[!hbt]
\begin{center}
\begin{tabular}{|c|c|c|c|}
 \hline
 P & $\{T,\theta\}$ & Behaviour & Stability \\
 \hline & & &\\
 $O$ & $\left\{0, 2 n \pi\right\}$ &  Asymptotic De Sitter & Non-hyperbolic saddle  \\
& & & \\
$A$ & $\left\{0, \pi +2 n \pi\right\}$ & $q=1$ & Repeller \\
& & & \\
\hline
\end{tabular}
\end{center}
\caption{Critical points and their stability of the $f(R)=R+\alpha R^n$  for the system \rf{Jordan2DDynSys}. Here $n$ is a generic integer.}
\label{tab:Alho_RRn2}
\end{table}

The behaviour of the scale factor can be deduced by the expression of the deceleration factor in terms of the dynamical systems variables
\begin{equation}\label{qJ}
q = 1 + 4\left(\frac{T}{1-T}\right)\frac{\sin\theta}{(1 - \cos\theta)^2}\,.
\end{equation}
Note that  for  Point~$O$,  $q$ appears to diverge. However, one can prove \citep{Alho:2016gzi} that $q$ can be expanded as
\begin{equation}
q = -1 + \frac{1}{12}\theta^2 + \frac{1}{120}\theta^4 + \dots \,,
\end{equation}
which indicates that the solution has $q=-1$ asymptotically.

\begin{figure}[!htb]
\centering
\includegraphics[width=0.7\columnwidth]{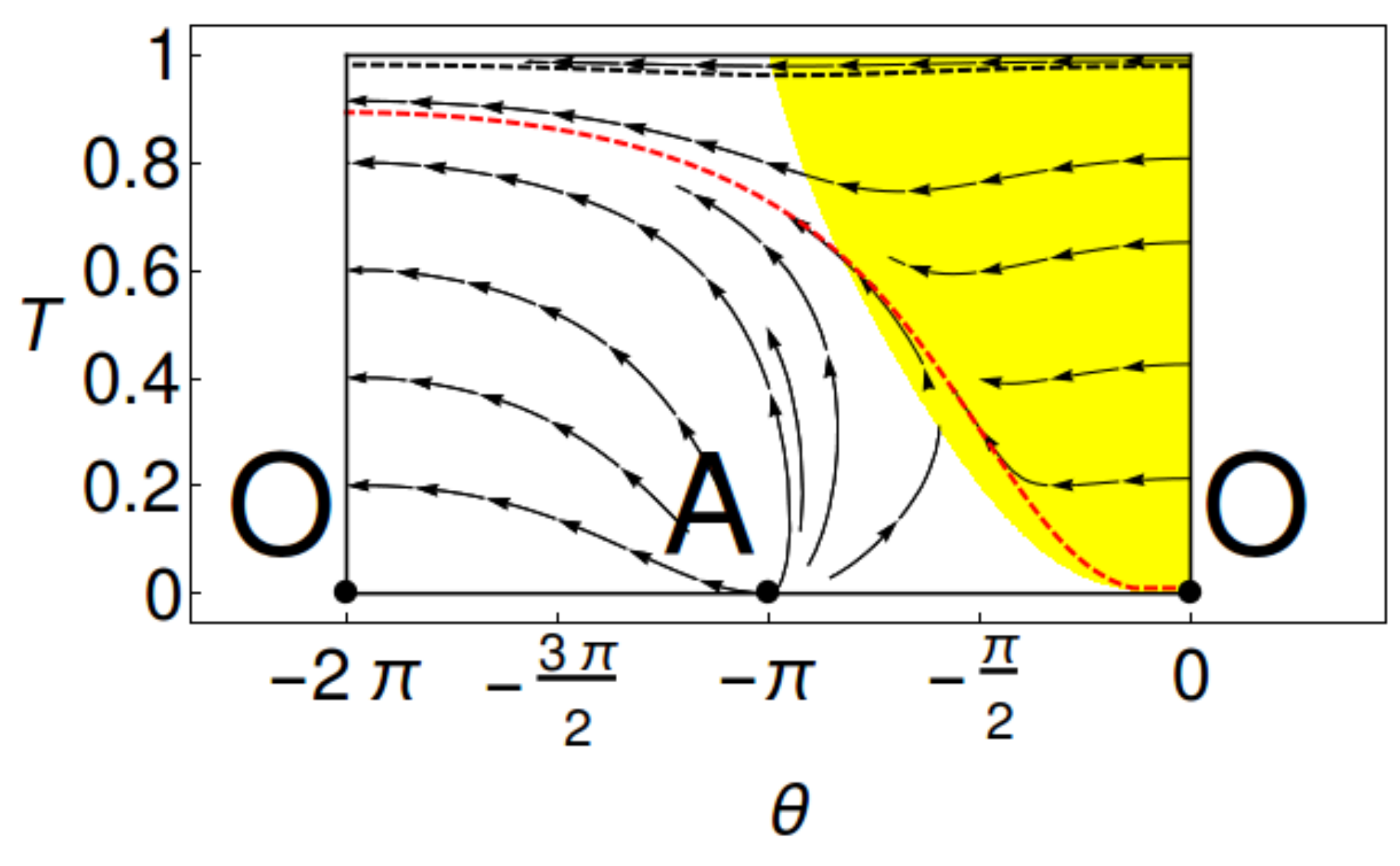}
\caption{A period of the phase space for $f(R)$ gravity using the approach \rf{Jordan2DDynSys}. The yellow region represents the $q<0$ part of the phase space. The dashed line represents the heteroclinic orbit starting in $A$ and arriving to the limit cycle in $T=1$ which represent the inflationary ``tracker''. The cosmology goes to a set of decelerated/accelerated phases whose number depends on the initial conditions before approaching  the final global Minkowski limit circle.}
\label{fig:f(R)_3}
\end{figure}

As already said, the system can also be analysed globally. A full analysis can be found in the work by \citet{Alho:2016gzi} and references therein. Here we limit ourselves to say that, since one can construct the monotonic function (see Sec.~\ref{globalPS} for some mathematical details)
\begin{equation}
\mathfrak{F} = \frac{(1-T)(3+\cos{\theta})}{T} > 0\,, \label{DefF}
\end{equation}
one can show that all orbits in the range $0<T<1$
originate from the subset $T=0$ and end at the subset $T=1$ and there are  no fixed
points or periodic orbits for these values of $T$.  On the  $T=0$ subset there is a single orbit that enters the physical state space from Point~$O$ while there is a 1-parameter set that originates from Point~$A$. These fixed points are the origins of \emph{all} solutions in the phase space. Finally $T=1$ represents a periodic orbit where $\theta$ is monotonically decreasing. This periodic orbit is a limit cycle that describes the future asymptotic
behaviour of \emph{all} solutions in $0<T<1$. A plot of the phase space is given in Fig.~\ref{fig:f(R)_3}.

\subsection{Higher-order theories and non-minimal coupling}
\label{otherhigher}

In the previous section we have considered the case of $f(R)$-gravity, one of the simplest and most studied example of theories of gravity in which higher order derivatives of the metric appear in the gravitational action. In recent years, however, several other Lagrangians containing other type of higher order terms have been proposed and analysed. For example, higher order actions can be built by contractions of the Riemann tensor or Ricci tensor by themselves or the metric tensor. The most popular are the square Ricci invariant $R_{\mu\nu}R^{\mu\nu}$ and the square Riemann invariant $R_{\mu\nu \alpha \beta}R^{\mu\nu \alpha \beta}$, but also scalars formed with the Weyl tensor are often considered.

A dynamical analysis of the late time cosmology of these models has been the subject of a work by \citet{Carroll:2004de}, who considered general Lagrangians as functions of $R^2$, $R_{\mu\nu}R^{\mu\nu}$ and $R_{\mu\nu \alpha \beta}R^{\mu\nu \alpha \beta}$. They found general late time accelerated attractors for theories with inverse power-law corrections of these higher order invariants. The same class of models was analysed in general also by \citet{Cognola:2007vq} and  \citet{Cognola:2008wy} with an approach inspired by minisuperspace formulation of these theories. \citet{Barrow:2006xb} instead performed a full analysis of models  in the context of Bianchi universes in a similar setting  to the one of Eqs.~\rf{DynSysRR^n}.
Another example has been given by \citet{Ishak:2008td} who studied the cosmological dynamics of a theory where the higher order corrections are given by the invariant $S_{\mu\nu}S^{\mu\nu}$, where $S_{\mu\nu}=R_{\mu\nu}-g_{\mu\nu}R/4$ is the traceless part of the Ricci tensor. They obtained accelerated attractors for a number of models.

A thoroughly studied, higher order curvature term is the {\it Gauss-Bonnet invariant} defined by
\begin{equation}
	\mathcal{G} = R^2 -4R_{\mu\nu}R^{\mu\nu} +R_{\mu\nu \alpha \beta}R^{\mu\nu \alpha \beta} \,.
\end{equation}
In four dimensions this quantity is a topological invariant of the spacetime, so that its linear contribution in the gravitational action can always be rewritten as a boundary term which is irrelevant for the equations of motion. Nevertheless  non-minimal couplings between  $\mathcal{G}$ and a scalar field or  nonlinear terms in $\mathcal{G}$ can modify the field equations and thus give rise to new dynamics. \citet{Koivisto:2009jn} studied the first setting computing the cosmological dynamics and finding late time stable de Sitter solutions, while \citet{Kim:2013xu} examined curved and anisotropic spacetimes. \citet{Tsujikawa:2006ph} studied instead scaling solutions for the same model with a generalised scalar field, showing that transition from the scaling regimes to dark energy domination is possible. \citet{Uddin:2009wp} analysed nonlinear actions in $\mathcal{G}$ considering the gravitational Lagrangian $\mathcal{L}_{\rm grav}=R+f(\mathcal{G})$. They worked in an equivalent scalar field representation of the theory and delivered an analysis on scaling solutions.  \citet{Zhou:2009cy} considered a similar model. Using a geometrical approach, they showed that for general $f(\mathcal{G})$ models there are two kinds of stable accelerated solutions, a de Sitter solution and a phantom-like solution. They co-exist with each other and which solution the universe evolves to depends on the initial conditions.

\cite{Carloni:2017ucm} proposed instead a general approach to treat Gauss-Bonnet cosmologies which is based on the same idea of the system \rf{DynSysRR^n}.
The general features of this latter approach are the following.
Consider the action
\begin{equation}
	S_{f({\cal G})} = \int d^4x \sqrt{-g} \left[\chi_0 R+ f\left({\cal G} \chi_0^2\right)+2\kappa^2 \mathcal{L}_{\rm m}\right] \,,
	\label{def:f(G)_action}
\end{equation}
where $\chi_0$ is a constant, and introduce the variables
\begin{equation}
\begin{split}
&\mathbb{G}=\frac{\G}{3 H^4}\,,\quad \tilde{\Omega}_{\rm m} =\frac{2\kappa^2\rho }{3H^2 \chi_0}\,,\\
& \mathbb{J}=\mathfrak j\,,\quad \mathbb{Q}=\mathfrak q\,,\quad \mathbb{A}=\chi_0H^2\,,
\end{split}
\label{eq:1000}
\end{equation}
where $\mathfrak q$ and  $\mathfrak j$ are defined in Eq.~(\ref{HubbleVarDS}).

The orbits are described by the autonomous system
\begin{equation}\label{DynSysG}
\begin{split}
&\DerN{\mathbb{G}}= \frac{\mathbb{G}}{2}  \left(8-\mathbb{G}\right)-\frac{8
   [\mathbf{X}-\mathbb{G} \mathbf{Y}-\tilde{\Omega}_{\rm m} +1]}{9 \mathbf{Z}}\,,\\
&\DerN{\tilde{\Omega}_{\rm m}} =\tilde{\Omega}_{\rm m}  \left[\frac{\mathbb{G}}{4}+(3 w+1)\right]\,,\\
&\DerN{\mathbb{A}}=\frac{\mathbb{A}}{4}  \left(\mathbb{G}-8\right)\,,
\end{split}
\end{equation}
together with the two constraints
\begin{eqnarray}
&& 0=9 \left[\mathbb{J}+\mathbb{Q} (3
   \mathbb{Q}+4)\right]+\mathbf{X}-\mathbb{G} \mathbf{Y}-\tilde{\Omega}_{\rm m} +1\,,  \\
&&  \mathbb{G}=8(1+\mathbb{\mathbb{K}})(1+\mathbb{Q})\,,\label{GB Constr}
\end{eqnarray}
where we have defined
\begin{align}
& {\bf X}= \frac{f}{3 \chi_0 H^2 }\,,\\
&  {\bf Y}= \frac{F H^2}{\chi_0}\,,\\
& {\bf Z}= \frac{3\, (2 H)^6  F'}{ \chi_0} \,,
\end{align}
and where, as usual $f'( \mathcal{G})=f_{,\mathcal{G}}=F$.
This approach not only allows us to explore the phase space of these cosmologies, but also to deduce general properties of the entire class of these theories.

More general cases of Gauss-Bonnet gravity were also considered. \citet{GarciaSalcedo:2009fb} studied a DBI modification of the type $\mathcal{L}_{\rm grav} = \sqrt{1- \alpha R+ \beta \mathcal{G}}$ which presents a rich phenomenology including matter and dark energy dominated solutions, scaling solutions, phantom and non-phantom late time attractors and even multiple future attractors. A general analysis on Gauss-Bonnet dark energy with the Lagrangian $\mathcal{L}_{\rm grav} = f(R,\mathcal{G})$ has been done by \citet{Alimohammadi:2009js} who employed the quantities $R$ and $H$ as dynamical systems variables and studied the stability of future accelerated attractors.

Another interesting fourth order model is the so-called {\it mimetic $f(R)$ gravity}. In this theory the metric is parametrised in terms of a scalar field $\phi$ and an additional metric $\hat{g}_{\mu\nu}$ via the relation
\begin{equation}
g_{\mu\nu}=-\hat{g}^{\rho\sigma}\partial_\rho \phi \partial_\sigma \phi \hat{g}_{\mu\nu}\,.
\end{equation}
Mimetic $f(R)$ gravity gives rise to new, but non-propagating, degrees of freedom, presents  interesting conformal properties and leads to a wider family of cosmological solutions. \citet{Leon:2014yua}  analysed mimetic $f(R)$ models with exponential, power law and arbitrary $f(R)$ functions using dynamical systems.
They found that the stable critical points are the ones corresponding to the standard $f(R)$ solutions, with the presence of non-stable additional critical points.
\cite{Odintsov:2015wwp} considered instead the dynamical approach to mimetic $f(R)$ gravity in the presence of a potential and a Lagrange multiplier constraint in both the Einstein and Jordan frame. They found among other things that the cosmology presents a number of stable and unstable de Sitter fixed points and that admit bouncing scenarios. Also \citet{Raza:2015kha} analysed mimetic gravity plus an interacting fluid, finding attractors for some specific forms of coupling.

A different and equally interesting possibility is to consider generic action functionals of the type $f(R, \mathcal{L}_{\rm m})$, where a general \textit{non-minimal coupling} between matter and curvature appears. \citet{Azevedo:2016ehy} considered a number of these models via dynamical system analysis. The particular case $f(R, \mathcal{L}_{\rm m})= f_1(R)+[1+\lambda f_2(R)]\mathcal{L}_{\rm m}$, suggested in~\citet{Bertolami:2007gv}, was also explored by~\citet{Ribeiro:2014sla} and \citet{An:2015mvw}, using different dimensionless variables. Note however that some particular models within the class of theories defined by the function $f(R, \mathcal{L}_{\rm m})$, might present instabilities at the cosmological perturbation level \citep{Tamanini:2013aca,Koivisto:2016jcu}. Another very popular theory where matter is non-minimally coupled to gravity, is the so-called $f(R,T)$ gravity where $T$ is the trace of the energy-momentum tensor. Dynamical systems investigations of these models has been provided by \cite{Shabani:2013mya} and \citet{Mirza:2014nfa}.

\subsection{Palatini $f(R)$ gravity and generalisations}
\label{palatini_f(R)+Gen}
\label{sec:Palatini_f(R)_gravity}

In General Relativity, and in all the modified theories considered above, the field equations are obtained from the action by a procedure of variation with respect to the metric.  This approach assumes that the connection is of the Levi-Civita type and therefore that the metric is the only relevant field of the theory. When dealing with more complicated types of spacetime (e.g.~when torsion or non-metricity are not assumed to vanish), the connection can be considered an independent field and therefore the variation of the action should be done with respect to the metric {\em and} the connection. This method of derivation of the equations is referred to as the  {\it Palatini variational principle} as opposed to the {\it metric approach}.
As one can expect, the two methods are equivalent for the Hilbert-Einstein Lagrangian, but this is not the case for modified gravity models such as $f(R)$ gravity \citep{Sotiriou:2008rp}.

In the Palatini picture, the $f(R)$ action (\ref{def:f(R)_action}) can be written as
\begin{equation}
  S_{\rm Palatini} = \int d^4x \sqrt{-g} \left[f(\mathcal{R})+2\kappa^2 \mathcal{L}_{\rm m}\right] \,,
  \label{120}
\end{equation}
where now $\mathcal{R}=g^{\mu\nu}\mathcal{R}_{\mu\nu}$ with $\mathcal{R}_{\mu\nu}$ the Ricci tensor generated by the independent Palatini connection $\tilde{\Gamma}^\lambda_{\mu\nu}$, while the matter Lagrangian $\mathcal{L}_{\rm m}$ is independent of the connection $\tilde{\Gamma}^\lambda_{\mu\nu}$. The variation of the Palatini action (\ref{120}) must be taken independently with respect to the metric $g_{\mu\nu}$ and the connection $\tilde{\Gamma}^\lambda_{\mu\nu}$.

The $f(R)$ equations of motion obtained with this procedure differ from the corresponding metric ones: most notably the Palatini equations are of order two, rather than four, in the derivative of the metric. This difference reflects also on the dynamics at cosmological scales giving rise to a  phenomenology completely different from the one of metric formulation.

A dynamical system formulation of Palatini $f(R)$ theories, based on the idea of Eqs.~\rf{DynSysAm},  was given by \citet{Fay:2007gg}. They considered a number of different  $f(R)$ models exploring the possibility of the presence of periods of matter domination followed by dark energy domination. Their results indicates that  such phases can be present also for $f(R)$ cosmologies which are non-viable in the metric formulation.

Another very well studied higher order model is the so called hybrid metric-Palatini models in which an $f({\cal R})$ term, constructed  \textit{\`{a} la Palatini}, is added to the Einstein-Hilbert Lagrangian \citep{Harko:2011nh}
\begin{equation}
  S_{\rm Hybrid} = \int d^4x \sqrt{-g} \left[R+f(\mathcal{R})+2\kappa^2 \mathcal{L}_{\rm m}\right] \,.
  \label{HmP}
\end{equation}
The analysis of this theory can be performed using a scalar field representation. In particular, setting
\begin{align}
&\phi= \frac{\partial f(\mathcal{R})}{\partial\mathcal{R}}\,,\\
&V(\phi)=-f(\mathcal{R}(\phi)) +\phi\,\mathcal{R}(\phi)\,,
\end{align}
one obtains the action
\begin{align}
  S=\int d^4x\sqrt{-g} \left[(1+\phi)\,R-\frac{3}{2 \phi
    }(\partial\phi)^2 -V(\phi)+2\kappa^2 \mathcal{L}_{\rm m}\right]\,.
  \label{ActMetPal}
\end{align}
whose cosmology can be studied, for example, with the variables we have presented in Sec.~\ref{sec:BD_theory} or a similar setting \citep{Capozziello:2015lza}.

A complete analysis of the phase space which does not require the introduction of scalar fields was performed by \citet{Carloni:2015bua} with a method based on the idea of Eqs.~\rf{DynSysAm}.
More specifically they found that, defining the auxiliary function $A=\sqrt{F(\mathcal{R})} \,a(t)$ where $F(\mathcal{R})=f'(\mathcal{R})$, one can write
\begin{eqnarray}
\mathcal{H} &=& \frac{\dot{a}}{a}+\frac{\dot{F}}{2 F} =\frac{\dot{A}}{A}\,, \label{A-R}\\
\mathcal{R} &=&
6F\left[\frac{1}{A} \frac{d^2A}{d \tau^2} +\left(\frac{1}{A}\frac{d A}{d \tau}\right)^2+\frac{k}{A^2} \right]\,, \label{A-R2}
\end{eqnarray}
where $\tau$ is defined by $d\tau=\sqrt{F(\mathcal{R})} dt$. Using these quantities defined in Eqs.~(\ref{A-R})-(\ref{A-R2}), the cosmological equations assume a particularly simple form:
\begin{align}
& \left(\frac{\dot{a}}{a}\right)^2 + F \mathcal{H}^2-\frac{1}{6}\left(F\mathcal{R}-f\right)-\frac{2\kappa^2\rho}{3}=0\,, \label{fr1} \\
&\frac{\ddot{a}}{a}-\mathcal{H}^2 F + \frac{f}{6}+\frac{\kappa^2}{3}\left(\rho+3p\right) =0 \,, \label{fr2} \\
& \dot{\rho}+3\frac{\dot{a}}{a}(1+w)\rho=0\,,\label{fr3}
\end{align}
 where $w$ is the barotropic factor of a perfect fluid, and for notational simplicity, we have dropped the dependence of $f$ and $F$ on $\mathcal{R}$. In this way, defining the variables
\begin{align}\label{DynVarP}
&X=\frac{\mathcal{H}}{H}\,,\qquad Y=\frac{\mathcal{R}}{6H^2}\,,\qquad Z=\frac{f}{6H^2}\,, \qquad\Omega_{\rm m}= \frac{2\kappa^2\rho}{3H^2}\,,
\end{align}
the dynamical system equations can be written as
 \begin{equation}
 \begin{split}\label{DynSysPalC}
& X'=\frac{1}{2} X \left\{\mathcal{F} \left[(3 w-1) X^2-(3 w+1) Y\right] \right.\\
&~~~~~~\left.+1+3 w+3 (w+1) Z-2X\right\}+Y\,,\\
& Y'=Y\left\{\mathcal{F} \left[(3 w-1)  X^2--(3 w+1) Y\right] \right.\\
&~~~~~~\left.+3 (w+1) (Z+1)+2 \mathcal{Q} (X-1)+2\right\}\,,\\
&Z'=Z \left\{\mathcal{F} \left[(3 w-1) X^2-(3 w+1) Y\right] \right.\\
&~~~~~~\left.+3(w+1) (Z+1)+2\right\}+2 \mathcal{F}\mathcal{Q}Y(X-1) \,,\\
&(X^2-Y)\mathcal{F}-\Omega_{\rm m} +Z+1=0\,.
\end{split}
\end{equation}
In the \eqref{DynSysPalC}, the quantities $\mathcal{F}= F(\mathcal{R})$ and $\mathcal{Q}(\mathcal{R})=F/(\mathcal{R} F')$ are only functions of $\mathcal{R}$,  and in order to close the system they have to be expressed in terms of the variables specified in Eq.~(\ref{DynVarP}). Given the form of $f$, this can be done by noticing that $Z/Y$ is function of $\mathcal{R}$ only, i.e.,
 \begin{equation}\label{KeyEq}
 \frac{Z}{Y}=\frac{f(\mathcal{R})}{\mathcal{R}}\,.
 \end{equation}
Inverting this relation for $Y\neq0$, one obtains $\mathcal{R}=\mathcal{R}(Z/Y)$ and the dynamical system can be closed. In this way a number of different models can be found using the same line of approach as in Eqs.~\rf{DynSysAm}.

 One can analyse the simple example $f(\mathcal{R})= \mathcal{R}^n$ with $n\neq2$ in the vacuum case ($\rho = 0$). The system \eqref{DynSysPalC} reduces to
\begin{align} \label{SysRn}
\begin{split}
& X'=-\frac{n X^4 -n X^3+(1-2 n) X^2 Y+X Y+(n-1) Y^2}{n
   \left(X^2-Y\right)+Y},\\
& Y'=\frac{2 Y \left[n X^3+ n X^2 (2 n-3)-(n-1)X-(n-1)^2\right)Y}{(n-1) \left(n \left[X^2-Y\right]+Y\right)}\,,\\
\end{split}
\end{align}
which admit a single non spurious fixed point $\left(X=1, Y=2, Z=\frac{2}{2-n}\right)$, representing de Sitter expansion, which is also an attractor. Other points appear if one considers the non-vacuum or the non spatially flat case.

\begin{figure}
\centering
\includegraphics[width=0.7\columnwidth]{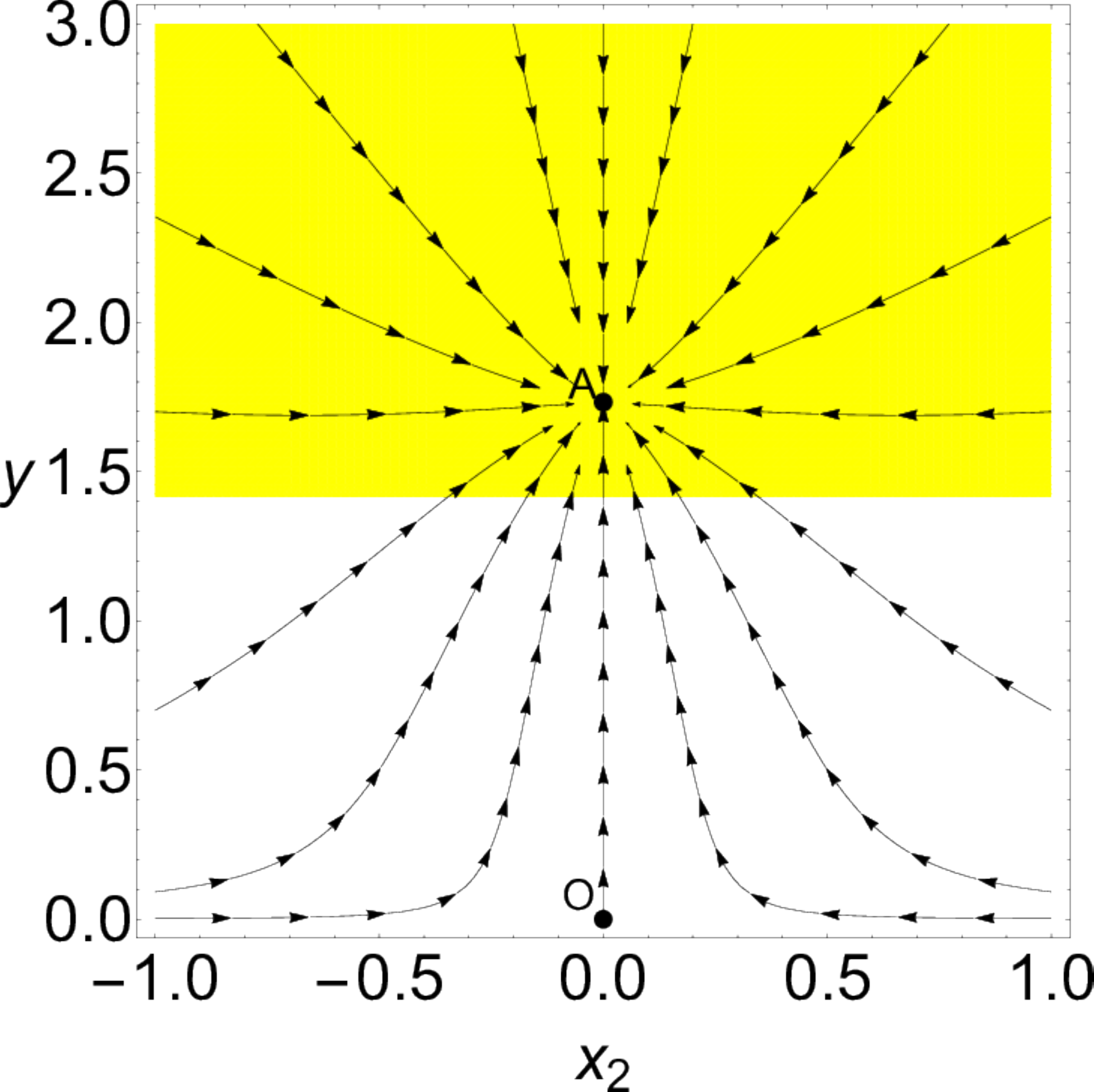}
\caption{A plot of the phase space for the spatially flat FLRW cosmology of generalised hybrid metric Palatini theory \eqref{121} in the form \eqref{ActGenMetPal} and for the potential $V(\phi,\psi)=V_0 \phi^\beta$ with $\beta=1/2$. The yellow area represents the accelerated expansion regime.}
\label{fig:GenHybridPS}
\end{figure}

The theory was further generalised to depend on a general function of both the metric and Palatini curvature scalars \citep{Tamanini:2013ltp}.  The action of this extended theory is given by
\begin{equation}
  S_{\rm hmP} = \int d^4x \sqrt{-g} \left[f(R,\mathcal{R})+2\kappa^2 \mathcal{L}_{\rm m}\right] \,,
  \label{121}
\end{equation}
where now $f$ is a general function of both the metric and Palatini Ricci scalars $R$ and $\mathcal{R}$. Similarly to the case of \eqref{ActMetPal}, the action (\ref{121}) can be proven to be equivalent to a non-minimally coupled bi-scalar field action \citep{Tamanini:2013ltp}. Such an action can be brought to the useful form
\begin{align}
  S=\int d^4x\sqrt{-g} \left[\phi\,R-\frac{1}{2
    }(\partial\psi)^2 -V(\phi,\psi)+2\kappa^2 \mathcal{L}_{\rm m}\right]\,.
  \label{ActGenMetPal}
\end{align}
where
\begin{gather}
\phi=\chi-\xi \,, \quad \psi=\sqrt{3} \log \xi\,, \quad V(\phi,\psi)=W(\chi(\phi,\psi),\xi(\psi)) \,, \nonumber \\
  \chi=\frac{\partial f(R,\mathcal{R})}{\partial R}\,, \quad \xi=-\frac{\partial f(R,\mathcal{R})}{\partial\mathcal{R}}\,, \\
   W(\chi,\xi)=-f(R(\chi),\mathcal{R}(\xi))+\chi\,R(\chi) -\xi\,\mathcal{R}(\xi) \,. \nonumber
\end{gather}
The cosmological equations associated to \eqref{ActGenMetPal} can be written, in the homogeneous, isotropic, and spatially flat case, as \citep{Rosa:2017jld}
\begin{align}
&6 \phi H^2  +6 H \dot{\phi} -V-\dot{\psi}^2=4\kappa^2 \rho\,,\\
&6 \phi \left(\dot{H}+H^2\right)-6 H \dot{\phi} -\phi  V_{\phi }+V+\dot{\psi}^2=-4\kappa^2 \rho\,,\\
&\ddot\psi+3H \dot\psi+\frac{\psi}{3}V_\psi=0\,,\\
&\ddot\varphi+3H\dot\varphi-\frac{1}{3}\dot\psi^2-\frac{1}{3}\left[2V-\varphi V_\varphi\right]=-\frac{\kappa^2}{3}(\rho-3p)\,.
\end{align}
In order to recast these equations into a dynamical system, one can define the variables
\begin{align}
x_1=\frac{\dot\phi}{H \phi }\,,\quad x_2=\frac{\dot\psi}{\sqrt{6\phi} H }\,,\quad y=\frac{\sqrt{V}}{\sqrt{6\phi} H }\,,\quad \tilde{\Omega}_{\rm m} =\frac{\rho }{3 H^2\phi } \,.
\end{align}
Setting\footnote{It is worth noticing that this choice of the potential corresponds to a very complicated (and non analytic) form of $f$. One should be always careful about this point in this case as in any other in which effective scalar field(s) are introduced. A simple choice of the effective potential $V$ might correspond to highly non-trivial $f$ or even to a function which is pathological.} $V(\phi,\psi)=V_0 \phi^\beta$ and considering only the vacuum case gives a particularly simple form of the dynamical equations
\begin{align}
&x_2'=-\frac{1}{2} x_2 \left[x_2^2+(2 \beta +1) y^2+1\right]\,,\\
&y'=\frac{1}{2} y \left[ (\beta -1)x_2^2-(\beta +1) y^2+5-\beta\right]\,,\\
&1+x_1-x_2^2-y=0 \,,
\end{align}
which we will analyse here as a simple example. The system presents two invariant submanifolds $x_2=0$ and $y=0$ and it is symmetric under the transformation $x_2\rightarrow -x_2$ and $y\rightarrow -y$. This implies that the $y<0$ phase space mirrors the $y>0$ and therefore in analogy with other scalar-tensor theories we can infer the property of the global phase space looking only at the $y>0$ part. In addition, global attractors can only exist if they have $(x_2=0,y=0)$, and a global attractor for orbits with a given sign of $y$ need to have $x_2=0$.
The phase space presents at most four fixed points, as shown in Tab.~\ref{tab:GmPVExp}.
Their existence, stability and associated solutions can change with the value of the parameters of the potential. It turns out, however, that for our specific choice of the potential only Point~$A$ can be an attractor associated to accelerated expansion. A plot of the phase space for a value of $\beta$ for which this scenario is realised is given in Fig.~\ref{fig:GenHybridPS}.

\begin{table}
\begin{center}
\begin{tabular}{|c|c|c|c|c|c|}
 \hline
 Point& Coordinates $\left\{x_1,x_2,y\right\}$& Scale Factor $a=a_0 (t-t_0)^\alpha$  \\ \hline
  & &\\
 $0$ & $\left\{ 0,0, 0\right\}$ & $\alpha =\displaystyle\frac{1}{2}$ \\ & &\\
 $A$ & $\left\{\displaystyle\frac{6}{\beta +1}-1, \displaystyle \frac{1}{2}, \displaystyle\sqrt{\displaystyle\frac{5-\beta }{\beta +1}}\right\}$ &  $\alpha =\displaystyle\frac{\beta +1}{\beta ^2-3
   \beta +2}$  \\  & &\\
 $B_\pm$ & $\left\{\displaystyle\frac{\beta -6}{\beta }, \pm\displaystyle\frac{\sqrt{(\beta -5) \beta -3}}{\beta }, \displaystyle\frac{\sqrt{3-\beta }}{\beta }\right\}$ &
   $\alpha =\displaystyle\frac{\beta }{3 (\beta -1)}$  \\  & &\\
   \hline
\end{tabular}\\ \vspace{1cm}
\begin{tabular}{|c|c|c|c|c|c|}
 \hline
 Point&  Attractor & Repeller \\ \hline  & &\\
 $0$ & $\beta >5$ & False \\  & &\\
 $A$  & $\beta <-1\lor \frac{1}{2} \left(5-\sqrt{37}\right)<\beta <5$ & $\beta >\frac{1}{2} \left(5+\sqrt{37}\right)$ \\  & &\\
 $B_\pm$   & $-0.5924<\beta<\frac{1}{2} \left(5-\sqrt{37}\right)$ & $5.465<\beta<\frac{1}{2} \left(5+\sqrt{37}\right)$ \\  & &\\
   \hline
\end{tabular}

\end{center}
\caption{Critical points and their stability for the generalised hybrid metric Palatini theory in the form \eqref{ActGenMetPal} and for the case $V(\phi,\psi)=V_0 \phi^\beta$. The (approximated) real numbers appearing in the inequalities above come from the numerical  analysis of the sign of the eigenvalues of the fixed point. }
\label{tab:GmPVExp}
\end{table}

\subsection{Teleparallel dark energy and $f(T)$ gravity}
\label{sec:f(T)_gravity}

Teleparallel gravity was first introduced by Einstein himself as a different mathematical setting for general relativity.  The reader interested in the idea can find more details in the book by \citet{Aldrovandi:2013wha}. Teleparallel gravity relies on the description of the gravitational field by means of {\it tetrad fields} $e_\mu^a$. These fields are defined by their relation to the metric tensor
\begin{equation}
	g_{\mu\nu} = e_\mu^a e_\nu^b \eta_{ab} \,,
\end{equation}
where $\eta_{ab}$ is the Minkowski metric of flat spacetime. In this section Latin indices as $a,b,...$ will run from 1 to 4 and refer to coordinates in the Minkowski space (tangent manifold). This relation makes it clear that all the degrees of freedom of the curvature of the spacetime can be integrated in the tetrad field and vice versa.  The tetrad field can be used to construct the torsion and the contorsion tensors
\begin{align}
T^\rho{}_{\mu\nu} &= e_a^\rho \left( \partial_\mu e_\nu^a -\partial_\nu e_\mu^a \right) \,,\\
K^{\mu\nu}{}_\rho &= - \frac{1}{2} \left( T^{\mu\nu}{}_\rho -T^{\nu\mu}{}_\rho -T_\rho{}^{\mu\nu} \right) \,.
\end{align}
The tensor $T^\rho{}_{\mu\nu}$ can also be shown to represent the antisymmetric part of the spacetime connection, which is known as the Weitzenb\"ock connection.
Note that we are following the standard approach in teleparallel gravity where the spin connection is set to zero (see e.g.~\citet{Maluf:2013gaa}), rather than the covariant approach where the spin connection plays a more active role \citep{Krssak:2015oua}.
Using  $T^\rho{}_{\mu\nu}$ one can define the {\it torsion scalar} $T$ (note the difference with the trace of the matter energy-momentum tensor $T_{\mu\nu}$) as
\begin{align}
T = S_\rho{}^{\mu\nu} T^\rho{}_{\mu\nu} \,,
\end{align}
with
\begin{equation}
S_{\rho}{}^{\mu\nu} = \frac{1}{2} \left( K^{\mu\nu}{}_\rho +\delta^\mu_\rho T^{\lambda\nu}{}_\lambda -\delta^\nu_\rho T^{\lambda\mu}{}_\lambda \right) \,.
\end{equation}

The definitions above suggest that one can characterise the same spacetime either in terms of curvature (and zero torsion) or in terms of torsion (and zero curvature). This last choice corresponds to the so-called  {\em teleparallel gravity} (see \citet{Aldrovandi:2013wha} for more details).  Teleparallel gravity and general relativity are therefore dual to each other since in the first one curvature disappears and torsion characterises the dynamics, while in the second one torsion vanishes and curvature describes the evolution of the gravitational field.

By analogy with General Relativity the action of teleparallel gravity is given by
\begin{equation}
	S_{\rm TG} = \int d^4x\, e \left(-T +2\kappa^2\mathcal{L}_{\rm m}\right) \,,
	\label{def:TEGR_action}
\end{equation}
where $e=\det e_\mu^a=\sqrt{-g}$ and $\mathcal{L}_{\rm m}$ is the usual matter Lagrangian. Note that in teleparallel gravity one tends to work with $\eta_{\mu\nu}=(+,-,-,-)$, changing the sign of the torsion scalar $T \mapsto -T$. Accordingly, a minus sign in the action needs to be introduced. For flat FLRW cosmology in Cartesian coordinates, the tetrad (co-frame one forms) can be written as $e_{\mu}^a={\rm diag}(1,a(t),a(t),a(t))$ and one has $T=6H^2$, or $T=-6H^2$ in the usual sign convention.

The equations of motion derived from the teleparallel action (\ref{def:TEGR_action}) are equivalent to the Einstein field equations and thus the two theories classically are physically indistinguishable. Mathematically this is due to the fact that $T$ can be rewritten, up to a total derivative, as the Ricci scalar $R$. The teleparallel action (\ref{def:TEGR_action}) can thus be recast into the Einstein-Hilbert action  plus a boundary term which does not influence the equations of motion. However, if $T$ couples to other (matter) fields inside the gravitational action or it does not appear linearly  (i.e.~higher order corrections are added), then new theories can be defined which differ from the usual higher order metric theories.

One of the most popular of such modifications we discuss is called {\it teleparallel dark energy} and it is based on the same idea of scalar-tensor theories (Sec.~\ref{sec:ST_theories}) of coupling the gravitational Lagrangian to a scalar field.  A non-minimal coupling between $T$ and a scalar field $\phi$ is thus the key ingredient of these models whose action is usually written as
\begin{equation}
	S_{\rm TDE} = \int d^4x\, e \left[-\frac{1}{2} F(\phi)T -\frac{1}{2}\partial\phi^2 -V(\phi) +2 \kappa^2\mathcal{L}_{\rm m}\right] \,,\label{def:TDE_action}
\end{equation}
where, as usual, $\partial\phi^2 \equiv g_{\alpha\beta} \partial^\alpha\phi\partial^\beta\phi$. The cosmological dynamics of teleparallel dark energy in the form (\ref{def:TDE_action}) has been studied in detail using a set of variables similar in structure to \rf{110}. \citet{Wei:2011yr} showed that in the case $F=1+\xi\phi^2$ scaling solutions do not appear in the phase space of teleparallel dark energy even if a coupling with the matter sector is introduced. He also found similarities between the dynamics of the teleparallel dark energy model (\ref{def:TDE_action}) and ELKO spinors dark energy models (see Sec.~\ref{sec:non-scalar_models}).
\citet{Xu:2012jf} analysed the same theory focusing  on late time accelerated attractors and proved that dynamical crossing of the phantom barrier can be obtained within teleparallel dark energy. Instead \citet{Skugoreva:2014ena} and again \citet{Skugoreva:2016bck} proposed a formalism to treat theories with a generic $F$ and made a direct comparison between the cosmological evolution in these theories and metric scalar-tensor theories with the same form of non-minimal coupling.

Generalisations of the action (\ref{def:TDE_action}) have been considered, and analysed with dynamical systems methods. \citet{Otalora:2013tba,Otalora:2013dsa} extended the theory both with an arbitrary non-minimal coupling of $\phi$ to $T$ and with a tachyonic kinetic term for the scalar field, obtaining in both cases scaling solutions.\citet{Bahamonde:2015hza} instead coupled the scalar field both with the torsion and the boundary term $B=(2/e)\partial_{\mu}(eT^{\mu})$ which connects it to the Ricci scalar via $R=-T+B$, see~\citep{Bahamonde:2015zma}. It was found that for an exponential potential, when the coupling is positive, the cosmology evolves to a dark energy phase without requiring any fine tuning of the parameters. \cite{Marciu:2017sji} studied dynamical properties of scaling solutions in the later theory for both inverse power-law and exponential potentials.

The teleparallel dark energy model (\ref{def:TDE_action}) is not the only gravitational theory capable of yielding an accelerated expansion within the teleparallel approach.
Higher order terms in $T$ are also considered as possible models for dark energy.
Such theories are known under the name of {\it $f(T)$ gravity} and are described by the action
\begin{equation}
	S_{f(T)} = \int d^4x\, e \left[ f(T) + 2 \kappa^2 \mathcal{L}_{\rm m} \right] \,,
\end{equation}
where $f$ is an arbitrary function of the torsion scalar $T$.

At cosmological scales a rich and interesting phenomenology can be obtained from $f(T)$ gravity.
The radiation-matter-dark energy sequence, with a late time stable accelerated attractor, arises in $f(T)$ models with a power-law correction of the type $f(T)=-T+ \alpha T^n$, as shown by \citet{Wu:2010xk}. Some tracker behaviours were found in the cases  $f(T)=-T+ \alpha T^{1/2}$  \citep{Jamil:2012yz}  and  $f(T)=-2a\sqrt{T}+b T+c$  \citep{Jamil:2012nma} in which also an interaction between different type of cosmic fluid was considered. \citet{Zhang:2011qp} considered also a logarithmic correction of the type $f(T)=-T + \beta \log T/T_0$, finding de Sitter late time attractors in both cases. \citet{Mirza:2017vrk} found that three conditions of a generic function $f(T)$ must be valid to describe matter and dark energy dominated eras. \citet{Feng:2014fsa} delivered a  dynamical systems analysis of general $f(T)$ theories employing the nullcline method to study the bifurcation phenomenon and then applied their results to the power-law and logarithmic models, studying them in full detail. \cite{Hohmann:2017jao} studied generic properties of the dynamical system of $f(T)$ gravity. Also \citet{Carloni:2015lsa}, using a method inspired by \rf{DynSysAm}, proposed a dynamical system approach applicable to a number of different methods when there is a  non-minimal coupling between the torsion scalar and the matter given by $f_1(T)-f_2(T)\mathcal{L}_{\rm m}$. \citet{Biswas:2015cva} analysed the dynamics of $f(T)$ gravity with interactions between the dark energy and dark matter and for different models. \citet{Kofinas:2014aka} considered the extended model with Lagrangian $f(T,T_\mathcal{G})$ where $T_\mathcal{G}$ is the teleparallel equivalent of the Gauss-Bonnet invariant (see Eq.~(48) in \cite{Kofinas:2014owa}).  They found scaling solutions and late time accelerated attractors of the quintessence, de Sitter and phantom kinds and also analysed the behaviour at the infinities of the phase space. \citet{Jamil:2012vb} instead analysed a generalised  version of \eqref{def:TDE_action} in which $T$ is substituted with $f(T)=-2a\sqrt{T}+b T+c$ and an exponential potential is considered. Higher order terms in the torsion action were considered instead by \citet{Otalora:2016dxe} who found that the corresponding cosmologies can result in a dark energy dominated and accelerated universe, where the dark energy EoS parameter lies in the quintessence regime, with the scale factor behaving asymptotically as a power law.

Finally, it is worth mentioning  the so-called {\it Poincar\'{e} gauge theory of gravity} where both torsion and curvature appear as dynamical quantities. \citet{Li:2009zzc} used dynamical system analysis to find  late time de Sitter attractors and quasi-periodic solutions. In the same spirit, the anisotropic cosmology of the theory with torsion, a nonlinear Lagrangian and a spin fluid source was analysed by  \citet{Carloni:2013hna}.


\subsection{String and brane cosmology}
\label{sec:string_and_brane}
Certainly one of the most important and most studied  modifications of general relativity are the ones which have the structure of the tree  level action of {\it string theory}. String theory is defined in a higher dimensional spacetime and compactifications of the extra dimensions usually give rise to new effective four dimensional quantities capable of playing the role of dynamical fields. String theory provides a highly rich cosmological phenomenology especially regarding the very early universe, but also related to the late time issues of dark matter and dark energy.  In fact, the subject is so wide that it has been named {\it string cosmology}. For details on the theory and applications of string-motivated cosmologies we refer the reader to \citet{Lidsey:1999mc,Quevedo:2002xw,Langlois:2002bb,McAllister:2007bg,Baumann:2014nda}.

Ignoring the contribution of  the Kalb-Ramond fields, the effective four dimensional action at tree level approximation   can be written as \citep{Billyard:1999wu}
\begin{equation}
	S = \int d^4x \sqrt{-g} e^{-\phi} \left[ R + \partial\phi^2 -6\partial\beta^2 -\frac{1}{2} e^{2 \phi} \partial\sigma^2 -2\Lambda \right] \,,
\end{equation}
where $\Lambda$ is a constant (central charge) and $\phi$, $\beta$ and $\sigma$ are scalar fields, called the {\it dilaton}, {\it modulus} and {\it axion} fields, respectively. These three scalar fields are naturally provided by string theory and can all be used as cosmological entities in order to address the problems of dark energy at late times and inflation at early times, subject to the condition that they must not vary too much at late times due to the fact that they couple to standard model particles and this would lead to temporal variations of the the gauge coupling constants, variations that have never been observed in nature. In the work of \citet{Billyard:1999wu}, for example, bouncing solutions and accelerated attractors have been found with applications to inflationary physics. \cite{Sonner:2006yn} showed using dynamical systems methods that in the context of dilaton-axion cosmology oscillations between periods of acceleration and deceleration in the Einstein frame is a fairly generic feature of these rather simple two-scalar models.  One of the most pressing problems in string cosmology is the stabilisation of the evolving moduli. Generally the non-perturbative potentials associated with these fields are very steep, with the result that the evolving moduli and dilaton fields overshoot their stabilising minima and head off to large field values, leading to decompactification and a breakdown of the model as pointed out by \citet{Brustein:1992nk}. A possible way of alleviating this problem was proposed by \citet{Barreiro:1998aj}, who showed, using a dynamical systems approach, that the presence of radiation in the early universe increased the contribution of the Hubble damping term to the evolving rolling dilaton (in the context of  gaugino condensate string models) slowing it down and allowing it to settle in the minimum of its potential. This was extended to the case of stabilising the cosmological evolution of the volume moduli in the KKLT \citep{Kachru:2003sx,Kachru:2003aw} class of Inflationary Universe models arising out of Type IIB string theory \citep{Brustein:2004jp,Barreiro:2005ua} and to models where the volume of the compact dimensions can be large \citep{Conlon:2008cj}.

Several authors have considered some of these effective low-energy fields to be constant, so that they do not contribute to the dynamics. \citet{Catena:2007jf} analysed the evolution of a universe filled with the axion and dilaton field, showing that scaling solutions as well as recurrent accelerated phases can be obtained.
Dilaton cosmology, where both the modulus and axion field are neglected, is largely studied not only for its easier dynamics, but also for its relation with scalar-tensor theories (see Sec.~\ref{sec:ST_theories}).
For this set up \citet{Fang:2006zq} found transient accelerated solutions, while \citet{Huang:2006cz} considered both standard and a phantom dilaton fields, showing that a late time attractor always exists if the dilaton potential has a non-vanishing minimum in the standard case and a non-vanishing maximum in the phantom case. They also provided examples with a Mexican hat potential.

String theory phenomenology can also motivate more general effective scalar fields and couplings to higher order curvature invariants. An extended (phantom) scalar field, equivalent to $k$-essence models, has been investigated by \citet{Piazza:2004df} who delivered a detailed analysis on scaling solutions. \citet{Sami:2005zc} considered higher order terms coupled to the dilaton and moduli fields and obtained de Sitter attractor solutions, unless a phantom matter fluid appears.

In order to guarantee a physical spectrum of particles string theory must live in more than three spatial dimensions. In this setting the dynamics of the system can be characterised also in terms of other objects called  {\it branes}. It is possible therefore to suppose that our four-dimensional universe is contained in one of these branes which is evolving in a higher dimensional spacetime (the bulk), such a setting corresponds to the so-called {\it brane-worlds} scenarios. There are several possible brane-world scenarios, but only two of them are largely applied to cosmology and for these two a standard dynamical systems approach can be constructed in full generality \citep{Leyva:2009zz}.

The first of these scenarios is known as {\it Randall-Sundrum type II brane-world model} (RS II) \citep{Randall:1999ee,Randall:1999vf}. On cosmological scales this scenario is characterised by the cosmological equations
\begin{align}
3H^2 &= \kappa^2 \rho_{\rm tot} \left(1 + \frac{\rho_{\rm tot}}{2 \lambda}\right) \,,\label{eq:brane_1}\\
2\dot H &= -\kappa^2\left(1+\frac{\rho_{\rm tot}}{\lambda}\right)\left(\rho_{\rm tot}+p_{\rm tot}\right) \,,\label{eq:brane_2}
\end{align}
where $\lambda$ is the brane tension, a constant associated with the properties of the brane in the higher dimensions, and $\rho_{\rm tot}$ and $p_{\rm tot}$ are respectively the total energy density and pressure of the matter fields living on the brane. The cosmological equations in the RS II model can be rewritten in the form of a dynamical system by a suitable choice of variables \citep{Gonzalez:2008wa,Leyva:2009zz}.
\citet{Campos:2001pa} assumed perfect fluid matter living on a brane and studied attractor solutions with a bifurcation analysis, while \citet{Goheer:2002bq} repeated the same work for a scalar field on the brane with an exponential potential (for earlier related work see \citet{Copeland:2000hn}). \citet{Campos:2001cn} also studied the cosmology of a Randall-Sundrum brane-world which possess a non-vanishing projection of the Weyl tensor onto the three-brane. Within this theory, \cite{Coley:2001va} showed that these models do not exhibit chaotic behaviour.
\cite{Leeper:2003dd} studied a brane-world of Randall-Sundrum type II where the interaction of particles at high energies implies  that the 5-dimensional gravitons escape into the bulk. In the latter model, the Weyl tensor behaves as radiation only at late times.

Since there is no evidence that matter can escape the brane, standard fluids are assumed to be bound on the brane. This is not true however for other fields. The case in which a scalar field can live in the bulk was analysed by \citet{Mizuno:2002wa}, where a scalar field with both an exponential and (inverse) power-law potential could generate accelerated expansion.
The same set up was used by \citet{Mizuno:2004xj} who presented a dynamical analysis in the regime $\rho_{\rm tot}\gg \lambda$ ($H\propto \rho_{\rm tot}/\lambda$)  assuming no particular form for the scalar field potential and analysing in detail scaling solutions.

The case where both a scalar field and matter live on a brane was considered by \citet{Huey:2001ae}  in which the scalar field was assumed to be the inflaton. The phase space analysis revealed the presence of accelerated expansion solutions which could be applied to early universe inflation. The case where both a (non inflaton) scalar field and matter live on a brane was considered with focus on scaling solutions by \citet{Savchenko:2002mi} and with different variables by \citet{Dutta:2015jaq}.
Instead \citet{Gonzalez:2008wa}, using a constant and exponential potential for the scalar field, showed that late time solutions are not influenced by the brane dynamics.
The same system has been considered by \citet{Escobar:2011cz} where  the stability of de Sitter solutions employing the centre manifold theory was examined  finding de Sitter stable solutions for a scalar field exponential potential.
The same authors subsequently studied anisotropic spacetimes and put dark radiation on the brane \citep{Escobar:2012cq,Escobar:2013js}, while \citet{Bohmer:2010re} applied the Jacobi stability analysis to this brane-world model.
A general analysis for a scalar field and matter on a brane in the RS II scenario has been given by \citet{Leyva:2009zz}, who studied general scalar field potentials with a similar method to the one we presented in Sec.~\ref{sec:other_potentials}. They also provided examples with $\cosh$-like and $\sinh$-like potentials.
Within this framework a coupling between matter fields and scalar field dark energy has been considered by \citet{Biswas:2015zka}.

Finally some authors even assumed more general scalar fields living on a RS II brane.
\citet{Chingangbam:2004xe} investigated a system with tachyons on the brane and applied the results to early universe inflation.
\citet{Bhadra:2012qj} employed instead a DBI scalar field (see Sec.~\ref{sec:tachyons}) showing that quintessence-like accelerated behaviour can be obtained at late times.

The second relevant brane-world scenario is called {\it Dvali-Gabadadze-Porrati} (DGP) model \citep{Dvali:2000hr}. This is characterised by a modification of the Friedmann equation given by
\begin{equation}
	3\left(H^2\pm \frac{H}{r_c}\right) = \kappa^2 \rho_{\rm tot} \,,
\end{equation}
where $r_c$ is a constant (cross-over scale).
Suitable variables can be defined in order to rewrite the cosmological equations of the DGP model as a dynamical system \citep{Quiros:2008hv,Leyva:2009zz}.
A scalar field is usually taken to live on the brane, as considered by \citet{Quiros:2008hv} who assumed a constant or exponential potential finding that phantom behaviour can be achieved at late times.
\citet{Leyva:2009zz} delivered instead a general analysis similar to the one of Sec.~\ref{sec:other_potentials} for a scalar field with a general potential in a DGP brane-world scenario.
A DBI scalar field on a DGP brane has been analysed by \citet{Bhadra:2012qj} who also obtained phantom behaviour.
The presence of late time accelerated scaling solutions for DGP models with a scalar field has been investigated by \citet{Dutta:2016dnt}. Although the DGP scenario has generated a great deal of interest amongst cosmologists and astronomers, searching for observational features of modified gravity, we should remind the reader of the theoretical hurdles it faces from the particle physics standpoint. \citet{Charmousis:2006pn} and \citet{Gregory:2007xy} have pointed out that the model suffers from a number of pathologies. By generalising the 5D geometry from Minkowski to Schwarzschild, they found that when the bulk mass is large enough, the brane hits a pressure singularity at finite radius, moreover on the self-accelerating branch, the five-dimensional energy is unbounded from below, implying that the self-accelerating backgrounds are unstable.

Other brane-world scenarios were only marginally considered for applications to cosmology from a dynamical systems point of view.
The so-called {\it D-brane model}, where a DBI scalar field arises from geometrical properties of the brane, has been analysed by \citet{Saridakis:2009uk} who found quintessence-like and phantom-like late time behaviour.
\citet{Koivisto:2013fta} then put matter on a D-brane and obtained accelerated scaling solutions capable of solving the cosmic coincidence problem and even discussed possible alleviations of the fine tuning problem of initial conditions.

Extensions of the brane-world scenarios using ideas from modified gravity have also been advanced.
For example \citet{Haghani:2012zq} employed $f(R)$ gravity generalisations showing that matter to dark energy transitions can be achieved in some specific examples. Finally, we mention a couple of examples where higher dimensional dynamics have been studied outside of the contest of string and brane cosmology: \citet{Chang:2004pbb} and \citet{Liu:2005uq} considered an exponential potential scalar field in general 5D spacetimes, analysing scaling solutions and quintessence attractors.

\subsection{Quantum gravity phenomenology}
\label{sec:quantum_grav}

{\it Quantum gravity} is envisaged as a theory able to connect general relativity and quantum mechanics, providing a consistent description of Nature at very high energies.
To date there have been many attempts to define such a theory, although it is fair to say that none of them are able to address all the theoretical problems related to the subject and there has been no definitive experimental observations preferring one model over another one. String theory is the approach to Quantum Gravity that has received most attention and its cosmological implications have been discussed in Sec.~\ref{sec:string_and_brane}. In this section we turn our attention to some of the other quantum gravity models focusing on their cosmological phenomenology. One of the most famous approaches to quantum gravity is {\it loop quantum gravity}  (LQG). LQG is a non-perturbative and background independent Hamiltonian approach to quantisation of general relativity \citep{Bojowald:2006da,Ashtekar:2011ni}.
By reducing the field content of the underlying model, a version has emerged known as {\it Loop Quantum Cosmology} (LQC), and the subsequent cosmology of this has been investigated by a number of authors. This simplification is based on symmetry reduction, which poses challenges on its own.

The dynamics of LQC is characterised by a modification of the cosmological equations given by
\begin{align}
3H^2 &= \kappa^2 \rho_{\rm tot} \left(1 - \frac{\rho_{\rm tot}}{\rho_c}\right) \,,\\
2\dot H &= -\kappa^2\left(1-\frac{2\rho_{\rm tot}}{\rho_c}\right)\left(\rho_{\rm tot}+p_{\rm tot}\right) \,,
\end{align}
where $\rho_c$ is a constant called the {\it critical loop quantum density}.
Note that these equations are basically equivalent (up to a crucial minus sign) to the cosmological equations (\ref{eq:brane_1})--(\ref{eq:brane_2}) arising in the Randall-Sundrum type II brane-world model.
In this sense the cosmic dynamics of the two theories have similarities which have been noticed by several authors (see for example \citet{Copeland:2004qe,Copeland:2005xs} and \citet{Singh:2006sg}).

Usually also in LQC a scalar field plus a standard fluid are employed to describe the late time phenomenology at cosmic distances. The possibility of an interaction between more than one type of fluid was considered in \citet{Jamil:2011iu}. Early time inflation without matter has been considered by \citet{Copeland:2007qt}. \citet{Chen:2008ca} studied a LQC quintessence scenario with a coupling to the matter sector and found possible accelerated scaling attractors capable of solving the cosmic coincidence problem, although the allowed region in the parameter space for this to happen is narrower than the corresponding general relativity result. \citet{Liu:2013dcw} analysed a scalar field in LQC with both positive and negative potentials. They showed that a cosmological bounce can be triggered by quantum effects or by negative values of the potential. A general dynamical systems analysis for arbitrary scalar field potentials similar to the one we presented in Sec.~\ref{sec:other_potentials} has instead been delivered by \citet{Xiao:2011nh}.

The phantom paradigm has largely been used in LQC. \citet{Samart:2007xz} considered a phantom scalar field with exponential potential and found that the big rip is always avoided by a bounce and then an oscillatory regime. Some authors \citep{Wu:2008db,Li:2010ju,Xiao:2010cy} have instead assumed phantom dark energy in the form of a perfect fluid interacting with matter, finding accelerated scaling solutions too, but showing also that they are more difficult to obtain than in standard general relativity.

The quintom and hessence scenarios (see Sec.~\ref{sec:quintom}) in LQC have been investigated by \citet{Wei:2007rp} who showed that no dynamical crossing of the phantom barrier can be obtained and that late time stable accelerated solutions are less frequent than in general relativity. Finally a DBI scalar field (see Sec.~\ref{sec:tachyons}) has been assumed in an LQC framework by \citet{Bhadra:2012qj} who showed that quintessence-like behaviour can be achieved at late times.

Though string theory phenomenology and LQC have received most attention as quantum gravity approaches in the literature, there are many other quantum gravity theories which can be applied to cosmology, although only a few of them have been analysed with dynamical systems methods.
{\it Renormalisation group flow techniques}, employed in consistent quantum field theories, have been applied to cosmological dynamics by some authors.
\citet{Ahn:2011qt} showed that possible dark energy dominated solutions can be used to accelerate the late time expansion of the universe within this framework.
\citet{Bonanno:2011yx} assumed an instability-induced renormalisation triggered by the low-energy quantum fluctuations in a universe with a positive cosmological constant, which effectively yields time-dependent gravitational and cosmological constants with a possible consistent description of matter to dark energy transition.
Also \citet{Hindmarsh:2011hx} considered renormalisation group flow techniques applied to cosmology finding accelerated solutions depending on the quantum correction parameters.
Subsequently \citet{Yang:2012tm} obtained a possible avoidance of the big rip in the same set up with a phantom field.

Finally \citet{EscamillaRivera:2010py} studied the phenomenology of {\it supersymmetry} and found that only transient accelerated phases are possible, while \citet{Obregon:2010nt} showed that effects of {\it noncommutative geometry} can drive the late time cosmic accelerated expansion.

\subsection{Other modified theories of gravity}
\label{sec:other_modified_theories}

The difficulties in finding a theoretical understanding of dark phenomenology has resulted in the proposal and the analysis of  a number of other  modifications of General Relativity.  We give here a brief description of the ones that  have been analysed with dynamical systems techniques.

\subsubsection{Massive gravity}
An interesting gravitational theory which has recently received much attention is {\it massive gravity}. In this theory the {\it graviton}, which in General Relativity is exactly massless,  is assumed to have a small mass.  This idea offers a natural way to reduce the gravitational force at large distances and therefore  change the cosmological dynamics.  However, the introduction of a mass for the graviton has been problematic until recently when a suitable ghost free modification  has been proposed (see \citet{deRham:2014zqa} for a recent review). At present there are only a few massive gravity cosmologies that have been analysed in detail via dynamical systems. An early analysis, based on a theory with tachyonic massive gravitons, has been provided by \citet{Sami:2002qg} who found late time accelerated attractor solutions. More recently \citet{Koorambas:2013nta} worked with an alternative theory where massless gravitons couple to massive gravitons, showing that late time phantom behaviour can arise. Also \citet{Heisenberg:2016spl} performed a phase space analysis  of these models, but without EN variables.

A ghost-free version of massive gravity has been studied by \citet{Gumrukcuoglu:2012aa} who considered anisotropic spacetimes, but proved that there is always a possible late time isotropic attractor. A mass varying theory of massive gravity, where the mass of the graviton depends on a scalar field, has been examined with dynamical systems techniques by a few authors. \citet{Gannouji:2013rwa} identified the scalar field with the dilaton field predicted by string theory finding that a matter to dark energy transition, with a late time de Sitter solution, is possible in such a model. They also derived bouncing solutions. \citet{Wu:2013ii} and \citet{Leon:2013qh} showed that for a general scalar field coupled to the gravitational mass term, quintessence, phantom and de Sitter behaviours are all late time cosmological evolutions which can be obtained. \citet{Bamba:2013aca} instead assumed the scalar field to be a Brans-Dicke field (Sec.~\ref{sec:BD_theory}), which in turn couples to the matter sector in the Einstein frame. They found de Sitter future attractors and a transient phantom accelerated period. Extensions of the standard massive gravity theories have also been considered as dark energy applications. For example \citet{Heisenberg:2014kea} analysed the phase space of a proxy theory of massive gravity obtained by  covariantisation of the decoupling limit of a more general theory proposed by \citet{PhysRevLett.106.231101}. They showed that in this context no accelerated solutions can be obtained and that the future attractor is always Minkowski spacetime. \citet{Wu:2014hva} mixed the $f(R)$ and massive gravity approaches delivering a similar analysis to the one we presented in Sec.~\ref{sec:f(R)}. He also provided some examples with $f(R)\propto R^n$ and $f(R)\propto \log R$, showing that a rich cosmological phenomenology arises in this theory. \cite{Anselmi:2017hwr} analysed the extended quasidilaton massive gravity model around a FRW cosmological background, and presented a stability analysis of asymptotic fixed points. Intriguingly they found that the traditional fixed point cannot be approached dynamically, except from a perfectly fine-tuned initial condition involving both the quasidilaton and the Hubble parameter, whereas the fixed-point solution, where the time derivative of the zeroth Stuckelberg field vanishes, encounters no such difficulty, and the fixed point is an attractor in some finite region of initial conditions.

\subsubsection{Ho\v{r}ava-Lifshitz gravity and Einstein-aether theory}
\label{sec:Horava_and_aether}

The gravitational theory known as {\it Ho\v{r}ava-Lifshitz gravity} proposes to modify general relativity in order to obtain a renormalisable theory, at the cost of losing Lorentz invariance \citep{Horava:2009uw}.
The cosmological applications of such theory have been reviewed by \citet{Mukohyama:2010xz}, and have been analysed with dynamical systems techniques by \citet{Carloni:2009jc}, \citet{Leon:2009rc}. They derived late time bouncing-oscillatory behaviour and showed the presence of future accelerated attractors. In that context the dynamical system analysis was useful to select classes of theories able to give rise to cosmic acceleration. 

Ho\v{r}ava-Lifshitz gravity can also be connected to another interesting modified theory of gravity: the so-called {\it Einstein-aether theory}, first proposed by \citet{Jacobson:2000xp}.
In such a theory a dynamical aether is restored in a relativistic environment and, although the Lorentz invariance is lost, applications to cosmology show that accelerated expansion can be obtained.
Dynamical systems investigations in Einstein-aether theory usually focus on a scalar field coupled to the expanding aether \citep{Arianto:2008aa,Zen:2008cn,Donnelly:2010cr,Sandin:2012gq,Wei:2013sya,Coley:2015qqa,Latta:2016jix}.
The scalar field is taken with either an exponential or a power-law potential and in both cases late time accelerated solutions can be obtained. Scaling solutions can be found coupling the scalar field to the matter fluid and in some models also late time phantom behaviour can be realised. A broader review on the properties of these theories in cosmological and non cosmological spacetimes, which also uses dynamical systems tools, can be found in \citet{Coley:2015qqa}.

Finally another modified gravity approach, known as \textit{time asymmetric extension of general relativity} \citep{Cortes:2015ola}, proposes changes to the Einstein-Hilbert action which break the time asymmetric invariance of general relativity, similarly to how Ho\v{r}ava-Lifshitz gravity break the Lorentz invariance. The cosmological dynamics of this theory has been investigated by \citet{Leon:2015via} with dynamical system techniques, showing that late time acceleration can be attained within this scenario.

\subsubsection{Nonlocal gravity}
\label{nonlocal}

Inspired by loop quantum corrections, different nonlocal theories of gravity have been proposed in order to understand the current cosmic acceleration of the universe.  One interesting nonlocal theory is defined by the following action (\citet{Deser:2007jk})
\begin{eqnarray}
S_{\rm NL}&=&\int d^{4}x\, \sqrt{-g} \Big[ R\left(1 + f(\square^{-1}R) \right)+2\kappa^2\mathcal{L}_{\rm m}\Big]\,, \label{1bNL}
\end{eqnarray}
where $f$ is a function which depends on the inverse of the d`Alembertian acting on the Ricci scalar. This operator acting on any function $F$ can be defined using the Greens function as follows
\begin{eqnarray}
(\square^{-1}F)(x)=\int d^4x'\,\sqrt{-g(x')} F(x')G(x,x')\,.\label{GNL}
\end{eqnarray}
These kind of theories could introduce constants of roughly the order of the Planck mass and some studies suggest that they could be good candidates to solve the black hole paradox and the cosmological fine-tuning problem. Moreover, they are ghost-free, stable and they correctly describe the observed acceleration of the universe at late times \citep{Deser:2013uya}. One problem of this theory is that when one is varying the nonlocal operator which appears in the action, one obtains an advanced Green's function which leads to the problem of acausality \citep{Zhang:2016ykx}. An interesting aspect of some nonlocal theories are that some of them are super-renormalisable \citep{Biswas:2010zk}.

Due to the difficulty of treating the nonlocal terms, it is possible to rewrite the above action by introducing two auxiliary fields $\phi$ and $\xi$ as follows \citep{Odintsov:2015wwp}
\begin{eqnarray}
S&=&\int d^{4}x\, \sqrt{-g}\Big[ \, R\left(1 + f(\phi) \right)-\partial_{\rho}\xi\partial^{\rho}\phi-\xi R+2\kappa^2\mathcal{L}_{\rm m}\Big]\,, \label{action2NL}
\end{eqnarray}
where $\square\phi=R$ (or $\phi=\square^{-1}R$) and $\square \xi=R\,df/d\phi$. In this way, the action is local and the acausality issue does not arise. However, the introduction of these scalar fields generally introduces a ghost problem and for only some special functions of $f(\phi)$, does one have a ghost-free theory \citep{Nojiri:2010pw,Zhang:2011uv}.

Using the scalar field approach, it is easier to study the cosmological properties of this theory. \cite{Koivisto:2008xfa} used dynamical systems techniques to analyse power-law and exponential forms of the nonlocal term in FLRW cosmology. He found that for Planck scale constants, the theory exhibits a late time acceleration of the universe without changing early cosmology. \cite{Jhingan:2008ym} found stable dark energy solutions at late times for an exponential nonlocal function. 

Another interesting particular case of this theory is the \cite{Maggiore:2014sia} non-local theory where $f(\square^{-1}R)=m^2\square^{-2}R$ with $m$ being a mass parameter. This theory  passes solar system and lab scale observations for $m=\mathcal{O}(H_0)$ and differs from GR on cosmological scales. Moreover, fixing the parameter $m$, this theory naturally produces the current acceleration of the universe with an equation of state of $w_{\rm de}\sim -1.14$. \cite{Nersisyan:2016hjh} studied the dynamical system of this theory finding that it requires $m < 1.2 H_0$.

\subsubsection{Effective field theories}

We finally mention a unified approach which aims at producing a model independent framework that encompasses all single-field dark energy and modified gravity models, characterising the evolution of the background cosmology and of perturbations
with a finite number of time functions introduced at the level of the action.
The idea is to consider every possible Lagrangian term allowed by the symmetries of the system, in our case the isotropy and homogeneity of the FLRW universe.
The models emerging from this approach have been collectively named {\it effective field theories of dark energy} \citep{Gubitosi:2012hu,Bloomfield:2012ff}, and at the background level the action can be expressed as
\begin{equation}
	S = \int d^4x \sqrt{-g} \left[ \varOmega(t) R + \Lambda(t) - c(t) \delta g^{00} \right] \,,
\end{equation}
where $\delta g^{00}$ is the perturbation to the upper time-time component of the metric and $\varOmega(t)$, $\Lambda(t)$, $c(t)$ are arbitrary function of time.

These theories have been investigated employing dynamical systems methods by \citet{Frusciante:2013zop} who defined a hierarchical dynamical system by Taylor expanding the functions $\varOmega(t)$, $\Lambda(t)$, $c(t)$ and performing the stability analysis at each successive order. They showed that interesting late time (scaling) cosmological solutions can be obtained, and they also recognised a recursive structure in the features of critical points at each order.

%% file: chapters/09_conclusions/conclusions.tex
\section{Concluding remarks}
\label{chap:conclusions}

Dynamical systems approaches provide a very powerful mathematical technique with applications covering a wide range of fields from biology, through to epidemiology, climate and economic forecasting and of course physics. It allows us to take a complicated set of higher order non-linear differential equations, and through judicious choosing of new variables, to write the same system as a set of non-linear first order ordinary differential equations. The new variables can then be analysed, fixed points of this new system obtained, where the variables take constant values, and the corresponding linear stability of the fixed point then easily be determined. It offers a geometrical way of analysing a complicated network of equations and of determining analytically the key properties of the system in a way that may well be unobtainable from the initial higher order differential equations. In this review we have brought the power of dynamical systems to the field of cosmology, in particular on cosmological models that aim to explain the current acceleration of the universe, whether that be from the standpoint of modifying the matter part of the scenario or by modifying the gravitational part.

Our approach has been to include as much of the full mathematical armoury as required to fully develop the analysis and determine the stability of the solutions. In doing so we have gone beyond what the majority of physicists and cosmologists are used to when analysing dynamical systems. The inclusion of this material in Sec.~\ref{sec:dynamicalsystems} has allowed us to formally go beyond linear stability theory, extending it for example to include the method due to Lyapunov and more generally centre manifold theory. We were therefore able to analyse cosmological models which have critical fixed points whose eigenvalues have zero real parts. Following this formal section with an introductory cosmology Sec.~\ref{chap:Cosmology}, our intention has been to enable physicists and mathematicians from a number of backgrounds to get to grips with the basic essentials required in order to follow some of the more technical sections that follow. In each section we have tried to go into sufficient mathematical detail to allow the interested reader to rederive the key equations, at least for the most straightforward of the examples presented. We have not attempted to provide a complete bibliography of every paper published in the field of dark energy and modified gravity. That is impossible, there are currently over 5500 Inspires publications with the words `dark energy' or `modified gravity' in the title. Instead, we have concentrated on reviewing work that has been based on a dynamical systems approach to the subject, still some 500 papers.

In Sec.~\ref{chap:scalarfields} we described the most common form of scalar field driven dark energy, namely the Quintessence potential. It allowed us to also introduce the general form of the dynamical variables we end up solving for, and motivate the form they have. What is remarkable is that even though the precise form of the variables changes, depending on the scenario being investigated in each section, the methodology behind choosing `good' variables remains similar. Namely they are the terms involved in the Friedmann constraint equation. As an example for standard Quintessence with a single scalar field we chose the variables $x$ and $y$ defined in Eq.~(\ref{def:ENvars}) which corresponds to the constraint equation~(\ref{eq:Friedmann_constr_xy}), whereas in Sec.~\ref{chap:noncanonicalscalarfields} where we consider non-canonical scalar fields, the corresponding variables for the case of say Tachyon models are given by Eq.~(\ref{def:nonENvars}) with the Friedmann constraint being given by Eq.~(\ref{eq:tachyon_Friedmann_const}).
In Sec.~\ref{chap:IDE} we considered the case of interacting dark energy models and introduced the variables~(\ref{eq:vars_IDE_fluids}) with the corresponding Friedmann constraint being Eq.~(\ref{Friedmann-mixed}).
Although this is by no means the only way to establish the set of dynamical variables that we should use, it does provide a good starting point and is physically well motivated.
Obviously, there will be cases where better variables can be identified, however, the Friedmann equation should always be the first reference point.
This approach can be summarised with the following sentence: look at the form of the Friedmann equation and use it to determine suitable variables to use in the dynamical system!

Having chosen the variables we solved the system of first order equations numerically to obtain the full evolution and analytically to obtain the class of fixed points. We then went on to describe the behaviour of the fixed points and their stability. The use of the phase plane diagrams proves extremely useful as a means of quickly establishing the nature and stability of the attractor solutions we obtained. The power of the method can be seen as we apply it to a number of situations in the various sections. In fact, every set of cosmological equations which are based on ordinary differential equations can be recast into the form of a dynamical system. Of note is the fact that the same underlying method is applicable to so many approaches to dark energy: not just to standard canonically normalised scalar fields (Sec.~\ref{chap:scalarfields}), but to a wide class of non-canonical fields (Sec.~\ref{chap:noncanonicalscalarfields}), multi-interacting dark energy-dark matter fields (Sec.~\ref{chap:IDE}), models going beyond scalar fields, introducing fermionic and gauge fields (Sec.~\ref{sec:non-scalar_models}) and  crucially for us, situations where we have modified curvature terms present in the underlying Lagrangian (Sec.~\ref{chap:modifiedgrav}). For example in the latter case, a wide class of scalar-tensor theories of gravity defined by the action in Eq.~(\ref{def:ST_action}) are described by the dynamical variables~(\ref{110})-(\ref{112}) satisfying the Friedmann constraint~(\ref{113}). It is testament to the power of Dynamical Systems techniques that such different models can be described and analysed using the same framework.

One of the key benefits of our approach to dark energy models is simply our ability to easily discard models based on the absence of stable or transient late-time accelerating solutions. We can reduce the number of viable dark energy models further by only considering those which also allow for a matter dominated epoch at early times. This greatly reduces the number of candidate models which should be studied further. When coupled with observational constraints it opens up a powerful new way of constraining models.

As we finish this review, the new field of multi-messenger astronomy is beginning to take off and could well revolutionise the way we constrain models \citep{TheLIGOScientific:2017qsa,GBM:2017lvd}. Indeed it is already happening as the detection of just one pair of merging neutron-stars has led to a wide class of modified gravity models being ruled out including a class of the Horndeski models we have discussed in Sec.~\ref{chap:noncanonicalscalarfields}. It bodes well for the future of constraining wide classes of models which differ from General Relativity on large scales and it could well be that dynamical systems approaches will provide a way in which to describe those constraints.